\documentclass[12pt]{iopart}
\usepackage{epsfig}
\usepackage{graphicx}
\usepackage{iopams}
\usepackage{bm}
\usepackage{amssymb}

\def\be{\begin{eqnarray}}
\def\ee{\end{eqnarray}}
\def\bm{\boldmath}
\def\bb{\bbox}

\def\bbox#1{\mbox{\boldmath $#1$}}
\def\O18p{$^{18}$O+p\ }
\def\O20p{$^{20}$O+p\ }
\def\O22p{$^{22}$O+p\ }
\def\O24p{$^{24}$O+p\ }
\def\beqy{\begin{eqnarray}}
\def\eeqy{\end{eqnarray}}
\def\bmlet{\begin{mathletters}}
\def\emlet{\end{mathletters}}

\newcommand{\lp}{\left}
\newcommand{\rp}{\right}

\newcommand{\nn}{\nonumber}
\newcommand{\iunit}{{\rm i}}
\newcommand{\Pop}{\hat{P}}
\newcommand{\rvec}{\boldsymbol{\mathbf r}}
\newcommand{\kvec}{\boldsymbol{\mathbf k}}

\begin{document}

\title[Exotic modes of excitation in atomic nuclei far from stability]
{Exotic modes of excitation in atomic nuclei far from stability} 

\author{Nils Paar}
\address{Institut f\" ur Kernphysik, Technische Universit\" at Darmstadt, Schlossgartenstrasse 9,
D-64289 Darmstadt, Germany}
\author{Dario Vretenar}
\address{Physics Department, Faculty of Science, University of Zagreb, 
Bijeni\v cka 32, 10000 Zagreb, Croatia}
\author{Elias Khan}
\address{Institut de Physique Nucl\'eaire, IN$_{2}$P$_{3}$-CNRS/Universit\'e
Paris-Sud, 91406 Orsay, France}
\author{Gianluca Col\`o}
\address{Dipartimento di Fisica dell'Universit\`a degli Studi and INFN, Sezione di Milano,
via Celoria 16, 20133 Milano, Italy}

\begin{abstract}
We review recent studies of the evolution of collective excitations in 
atomic nuclei far from the valley of $\beta$-stability. 
Collective degrees of freedom govern essential aspects 
of nuclear structure, and for several 
decades the study of collective modes such as rotations and vibrations 
has played a vital role in our understanding of complex properties of 
nuclei. The multipole response of unstable nuclei and the
possible occurrence of new exotic modes of excitation in 
weakly-bound nuclear systems, present a rapidly growing
field of research, but only few experimental studies of 
these phenomena have been reported so far. Valuable data 
on the evolution of the low-energy dipole response in unstable 
neutron-rich nuclei have been gathered in recent experiments, but the
available information is not sufficient to determine 
the nature of observed excitations. Even in stable 
nuclei various modes of giant collective oscillations 
had been predicted by theory years before they were observed, 
and for that reason it is very important to perform detailed 
theoretical studies of the evolution of collective modes of 
excitation in nuclei far from stability. We therefore discuss  
the modern theoretical tools that have been developed in recent 
years for the description of collective excitations in weakly-bound 
nuclei. The review focuses on the applications of these 
models to studies of the evolution of low-energy dipole modes from 
stable nuclei to systems near the particle emission threshold, 
to analyses of various 
isoscalar modes, those for which data are already available, as well 
as those that could be observed in future experiments, to 
a description of charge-exchange modes and their 
evolution in neutron-rich nuclei, and to studies of the 
role of exotic low-energy modes in astrophysical processes. 
\end{abstract}



\maketitle

\pagestyle{myheadings}
\tableofcontents
\pagestyle{headings}
\newpage

\section{Introduction}

Studies of nuclear vibrational modes over several past decades have 
provided valuable information on the structure of the nucleus and the
forces of cohesion that are responsible for the nuclear binding. 
Collective degrees of freedom characterize many aspects of nuclear 
structure. The response of a nucleus to external forces often exhibits 
a degree of simplicity associated with collective modes: rotations 
and vibrations. Even the simplest excitations, giant vibrations or giant 
resonances characterized by a coherent oscillation of all the 
nucleons, probe not only global nuclear properties such as the size, 
the shape, the distributions of protons and neutrons, 
the compressibility of nuclear matter, but also the details of 
the in-medium modification of the nucleon-nucleon interaction, and 
the interplay between different degrees of freedom in dissipative 
processes that determine the damping mechanism. The excitation 
energies and decay properties of giant resonances have been 
measured mostly for nuclei along the valley of $\beta$-stability, 
and the extension of these studies to regions of unstable nuclei 
is still in its infancy. 

The multipole response of nuclei far from the $\beta$-stability line and the
possible occurrence of exotic modes of 
excitation presents a rapidly growing
field of research. Characteristic ground-state properties 
(weak binding of the outermost nucleons,
coupling between bound states and the particle continuum, 
nuclei with very diffuse neutron densities, formation of neutron skin and
halo structures) will also have a pronounced effect on the multipole 
response of unstable nuclei. 
For instance, the dipole (E1) response of neutron-rich
nuclei is characterized by the fragmentation of the strength distribution
and its spreading into the low-energy region, and by the mixing of isoscalar
and isovector components. While in light nuclei the onset of dipole strength
in the low-energy region is caused by non-resonant independent
single-particle excitations of the loosely bound neutrons, several
theoretical analyses have predicted the existence of the pygmy dipole
resonance (PDR) in medium-mass and heavy nuclei, i.e., the resonant
oscillation of the weakly-bound neutron skin against the isospin saturated
proton-neutron core. The interpretation of the dynamics of the observed
low-energy E1 strength in nuclei with a pronounced neutron excess is
currently very much under discussion. 

Of course, not only pygmy states, but also other possible exotic modes are
being studied: isoscalar dipole, toroidal, giant pairing vibrations, 
low-energy monopole and quadrupole, and spin-isospin excitations in 
unstable nuclei. The isoscalar giant dipole resonance (ISGDR) 
corresponds to a second order high-energy compression mode and 
therefore provides information on the nuclear matter compression modulus, 
but the existence of a low-energy component has recently been experimentally 
confirmed. Several theoretical studies have predicted that the low-energy 
isoscalar dipole vibration is not sensitive to the nuclear compressibility 
and that, in fact, it could correspond to the toroidal dipole resonance. 
The toroidal dipole mode is a transverse zero-sound wave and its 
experimental observation would invalidate the hydrodynamical picture 
of the nuclear medium, since there is no restoring force for such modes 
in an ideal fluid. Pairing vibrations are induced in the nucleus 
by the addition or removal of a pair of neutrons, and can be associated with
the fluctuation of the pairing field.
High-energy collective pairing modes -- Giant Pairing Vibrations (GPV), 
have been predicted and analyzed theoretically, 
but have never been observed in studies of reactions induced by beams 
of stable isotopes. Entirely new types of collective excitations might 
arise in nuclei near the particle emission threshold: 
di-neutron vibrations close to the 
neutron-drip line, and proton pygmy resonances in proton-rich nuclei.  
Several new theoretical approaches have recently been developed, which 
provide a fully microscopic description of low-energy collective excitations 
in weakly bound nuclei. This review presents an
opportunity to compare the results and predictions of various 
models, and to discuss the development of modern theoretical 
tools based on the interacting shell-model, the 
time-dependent non-relativistic and relativistic self-consistent 
mean-field framework, and the extensions of the latter models 
beyond the mean-field approximation. 

Theoretical predictions of exotic modes have also prompted the design 
of experiments with radioactive beams, and a number
of studies of low-energy multipole response in unstable nuclei
have been reported in recent years. Low-lying E1 strength has been observed
in neutron-rich oxygen isotopes, exhausting about 10\% of the classical
dipole sum rule below 15 MeV excitation energy. In heavier systems data have
recently been reported on the concentration of electric dipole strength
below the neutron separation energy in $N=82$ semi-magic nuclei. The
experimental information which is presently available, however, 
is not sufficient to determine the dominant structure of 
the observed states. The Sn isotopes
present another very interesting example of the evolution of the low-lying
dipole strength with neutron number. Very recently the dipole strength
distribution above the one-neutron separation energy has been measured in
the unstable $^{130}$Sn and the doubly-magic $^{132}$Sn. 
In addition to the giant dipole resonance (GDR), evidence has been 
reported for a PDR structure at excitation energy around 
10 MeV both in $^{130}$Sn and $^{132}$Sn, exhausting a few percent 
of the E1 energy-weighted sum rule. Obviously this is a rapidly 
expanding field and many new experiments are being planned and designed 
at existing or future radioactive-beam facilities, which will
allow the study of the evolution of collective modes in 
nuclei far from stability, and the discovery of new exotic modes 
of excitation. 

Besides being intrinsically interesting as new structure phenomena, exotic
modes of excitation might play an important role in nuclear astrophysics. 
For example, the occurrence of the PDR could have a pronounced effect
on neutron capture rates in the r-process nucleosynthesis, and consequently
on the calculated elemental abundance distribution. Even though its strength
is small compared to the total dipole strength, the PDR significantly
enhances the radiative neutron capture cross section on neutron-rich nuclei,
as shown in recent large-scale QRPA calculations of the E1 strength for the
whole nuclear chart. The latest theoretical and computational advances in 
nuclear structure modeling have had a strong impact on nuclear astrophysics. 
More and more often calculations of stellar nucleosynthesis, nuclear aspects 
of supernova collapse and explosion, and neutrino-induced reactions, are based
on microscopic global predictions for the nuclear ingredients, rather than
on phenomenological approaches. The ability to model the Gamow-Teller 
response, for instance, is  essential for
reliable predictions of $\beta$-decay rates in neutron-rich nuclei along the
$r$-process path. The calculation of GT strength, however, can also be used
to constrain the spin-isospin channel of energy density functionals. When
approaching the neutron (proton) drip lines, an increasing fraction of the
GT$_-$ (GT$_+$) strength is found within the $\beta$-decay window, and a
consistent study of this phenomenon has yet to be carried out in the
framework of microscopic self-consistent models. 

The low-energy E1 strength could also play a role in
the photodisintegration of Ultra-High Energy Cosmic Rays (UHECR). Under the
assumption that UHECR are extra-Galactic nuclei accelerated to energies up
to 10$^{21}$ eV, their interaction with the 2.7~K Cosmic Microwave
Background (CMB) leads to photoabsorption reactions, followed by nucleon
emission. Recent calculations have shown that the photo-disintegration path
proceeds through regions of unstable nuclei, and that the nucleon emission
rate is very sensitive to the low-energy dipole strength.

The ultimate exotic modes far from stability could be collective excitations
in nuclei far beyond the drip-line. Such systems can be expected to exist in
the inner crust of neutron stars, where nuclear clusters are immersed in a
dilute gas of neutrons and electrons. Model calculations have predicted the
existence of super-giant resonances (SGR) at very low energies, typically
around 3 MeV, and exhausting more than 70\% of the EWSR. The SGR can have 
a pronounced effect on the specific heat of the crust, and could therefore 
affect the cooling time of the neutron star.

The most accurate description of nuclear vibrations is provided by the 
time-dependent mean-field theory, and thus we begin this review with an 
outline of the theoretical tools which are based on the self-consistent 
theory of small-amplitude vibrations (Sec.~\ref{Sec2}), and its extension 
beyond the mean-field approximation (Sec.~\ref{Sec3}). An extensive review 
of recent studies of the evolution of low-energy dipole vibrations, 
and the possible occurrence of pygmy modes in 
nuclei far from stability, is presented in Sec.~\ref{Sec4}. Various 
isoscalar modes, those already observed in experiments, as well 
as those that so far have only been predicted in theoretical studies, 
are reviewed in Sec.~\ref{Sec5}. A discussion of charge-exchange modes and 
their evolution in neutron-rich nuclei is included in Sec.~\ref{secsi}. 
The possible role of exotic low-energy modes in astrophysical processes 
is described in Sec.~\ref{Sec7} and, finally, Sec.~\ref{Sec8} contains   
the concluding remarks and ends with an outlook for future studies.

\section{Self-Consistent Theory of Small Amplitude Vibrations}
\label{Sec2}
Modern nuclear structure theory has evolved from macroscopic and 
microscopic studies of phenomena in stable nuclei towards regions 
of exotic, short-lived nuclei far from the valley of stability, 
and nuclear astrophysics applications. The principal challenge is
to build a consistent microscopic theoretical framework that will
provide a unified description of bulk properties, nuclear excitations
and reactions. 

The {\em ab-initio} approach, which starts from accurate two-nucleon and 
three-nucleon interactions, adjusted to nucleon-nucleon
scattering data and spectroscopic 
data on few-nucleon systems, respectively, provides the basis for a
quantitative description of ground-state properties, excited states and
transitions in relatively light nuclei with $A\leq16$. Improved shell-model 
techniques, which employ accurately adjusted effective interactions 
and sophisticated truncation schemes, are used in large-scale calculations 
of structure phenomena in medium-mass nuclei, including properties which
are relevant for astrophysical applications. The structure of 
heavy nuclei with a large number of active nucleons, however,
is best described in the framework of self-consistent mean-field models. 
A vast body of data, not only in medium-heavy and heavy stable nuclei, 
but also in regions of exotic nuclei far from the line of $\beta$-stability, 
has been successfully analyzed with mean-field 
models based on the Skyrme and Gogny non-relativistic interactions, 
and on relativistic meson-exchange effective Lagrangians. 
The self-consistent mean-field approach to nuclear structure represents 
an approximate implementation of Kohn-Sham density functional theory, 
which enables a microscopic description of the nuclear many-body problem 
in terms of a universal energy density functional. When compared with 
{\em ab-initio} and shell-model approaches, important advantages 
of the mean-field framework include the use of global effective nuclear 
interactions, the ability to describe arbitrarily heavy systems including 
superheavy nuclei, and the resulting intuitive picture of intrinsic 
nuclear shapes. 

The unique structure properties which characterize highly 
unstable nuclei as, for instance, the weak binding of the 
outermost nucleons and the coupling between bound states and 
the particle continuum, the modification of the effective nuclear 
potential and the formation of nuclei with very diffuse 
neutron densities, the occurrence of neutron skin and halo structures, 
will also affect the multipole response of these systems, and 
new modes of excitation could arise in nuclei at the limits of stability.   
Therefore a quantitative description of properties of ground and excited 
states in weakly-bound nuclei, and especially studies of exotic modes far 
from stability, necessitate using the time-dependent self-consistent 
mean-field framework. In this section we present an outline of the 
theoretical tools that have been employed in most studies included in 
this article. We start with the non-relativistic Hartree-Fock-Bogoliubov 
theory, extend this approach to the relativistic mean-field framework, and
derive the (continuum) non-relativistic and relativistic quasiparticle 
random phase approximations in the small-amplitude limit of the 
self-consistent time-dependent mean-field theory. For a more detailed 
introduction we refer the reader to several excellent 
monographs~\cite{ring80,BR.86,wale04,BB.94,borti98,harak01}, and 
recent review articles~\cite{BHR.03,Vre.05}.   

\subsection{The Hartree-Fock-Bogoliubov Method with Effective Nuclear
Forces}
\label{firstHFB}

In addition to the self-consistent mean-field single-nucleon potential,
the inclusion of pairing correlations is essential for a quantitative
description of structure phenomena in open-shell spherical  
and deformed nuclei. In weakly-bound systems far from stability, in 
particular, the Fermi surface for one type of nucleons is found close
to the particle continuum. The single-nucleon separation energies become 
comparable to the pairing gaps, and this results in the lowest 
particle-hole ($ph$) and particle-particle ($pp$) modes being embedded 
in the continuum. A unified and self-consistent treatment of both the 
mean-field and pairing correlations becomes necessary, and the
coupling between bound and continuum states has to be taken into
account explicitly.  
 
The Hartree-Fock-Bogoliubov (HFB) theory~\cite{ring80,BR.86} 
provides a unified description of $ph$- and $pp$-correlations in 
nuclei and, when the self-consistent HFB equations are formulated in
coordinate space, allows for a treatment of continuum effects in 
the presence of pairing. In the HFB framework two average potentials 
are taken into account: the self-consistent Hartree-Fock 
field $\hat{\Gamma}$ which encloses all the $ph$ correlations, 
and the pairing field $\hat{\Delta}$ which sums up the effects of the
$pp$ interaction. The ground state of a given nucleus is described
by a generalized Slater determinant $|\Phi\rangle$ of single-quasiparticle
self-consistent solutions of the HFB equations, and represents
the vacuum with respect to independent quasiparticles. The
quasiparticle operators are defined by the unitary Bogoliubov
transformation of the single-nucleon creation and annihilation
operators:
\begin{equation}
\alpha_{k}^{+}=\sum\limits_{l}U_{lk}^{{}}c_{l}^{+}+V_{lk}^{{}}c_{l}^{{}}\;,
\label{HFB_basis}
\end{equation}
where $U_{lk}$, $V_{lk}$ are single-quasiparticle wave functions that 
satisfy the HFB equation. The index $l$ 
denotes an arbitrary basis, for instance
the harmonic oscillator states. In the coordinate space
representation $l \equiv (\mathbf{r,}\sigma,\tau)$, with the spin-index
$\sigma$ and the isospin index $\tau$. The HFB wave
functions determine the hermitian single-particle density matrix%
\begin{equation}
\hat{\rho}_{ll^{\prime}}=\langle\Phi|c_{l^{\prime}}^{+}c_{l^{{}}}^{{}}%
|\Phi\rangle=(V^{\ast}V^{T})_{ll^{\prime}},%
\label{rho0}%
\end{equation}
and the antisymmetric pairing tensor
\begin{equation}
\hat{\kappa}_{ll^{\prime}}=\langle\Phi|c_{l^{\prime}}^{{}}c_{l^{{}}}^{{}}%
|\Phi\rangle=(V^{\ast}U^{T})_{ll^{\prime}}. \label{kappa0}%
\end{equation}
These two densities can be combined into the generalized density matrix
\begin{equation}
\mathcal{R}=\left(
\begin{array}
[c]{cc}%
\rho & \kappa\\
-\kappa^{\ast} & 1-\rho^{\ast}%
\end{array}
\right)  \;.
\end{equation}
For a nuclear Hamiltonian of the form
\begin{equation}
\hat{H}=\sum\limits_{l}\varepsilon_{l}^{{}}c_{l^{{}}}^{+}c_{l}^{}
+\frac{1}{4}\sum\limits_{ll^{\prime}mm^{\prime}}\tilde{v}_{lm^{\prime
}l^{\prime}m}^{{}}c_{l^{{}}}^{+}c_{m^{\prime}}^{+}c_{m^{{}}}^{{}}c_{l^{\prime
}}^{{}} \;,
\label{hamiltonian}%
\end{equation}
where $\tilde{v}$ denotes a general nucleon-nucleon interaction, 
the expectation value $\langle\Phi|\hat{H}|\Phi\rangle$ can be
expressed as a function of the hermitian density matrix $\hat{\rho}$,
and the antisymmetric pairing tensor $\hat{\kappa}$. The
minimization of this energy functional with respect to $\hat{\rho}$ and
$\hat{\kappa}$ leads to the Hartree-Fock-Bogoliubov equations
\begin{equation}
\left(
\begin{array}
[c]{cc}%
\hat{h}-\lambda & \hat{\Delta}\\
-\hat{\Delta}^{\ast} & -\hat{h}^{\ast}+\lambda
\end{array}
\right)  \left(
\begin{array}
[c]{l}%
U_{k}\\
V_{k}%
\end{array}
\right)  =E_{k}\left(
\begin{array}
[c]{l}%
U_{k}\\
V_{k}%
\end{array}
\right)  \;. \label{HFBnonrel}%
\end{equation}
The single-nucleon Hamiltonian reads
$\hat{h}=\hat{\varepsilon}+\hat{\Gamma}$, and the two
self-consistent potentials $\hat{\Gamma}$ and $\hat{\Delta}$ are
defined by
\begin{equation}
\hat{\Gamma}_{ll^{\prime}}=\sum\limits_{mm^{\prime}}\tilde{v}_{lm^{\prime
}l^{\prime}m}\hat{\rho}_{mm^{\prime}}\;, \label{gammapot}%
\end{equation}
and
\begin{equation}
\hat{\Delta}_{ll^{\prime}}=\sum\limits_{m<m^{\prime}}\tilde{v}_{ll^{\prime
}mm^{\prime}}\hat{\kappa}_{mm^{\prime}}  \; .
\label{deltapot}
\end{equation}
The chemical potential $\lambda$ is determined by the particle
number subsidiary condition, in such a way that the expectation value
of the particle number operator in the ground state equals the given
number of nucleons. The column vectors denote the quasiparticle wave
functions, and $E_{k}$ are the corresponding quasiparticle energies.

In the framework of Kohn-Sham density functional theory 
(DFT)~\cite{KS.65,Koh.99,DG.90}, of 
which the self-consistent HFB represents a particular implementation,
the nuclear many-body problem is defined in terms of a universal energy 
density functional. Self-consistent mean-field models approximate the 
exact energy functional, which includes all higher-order correlations, 
with powers and gradients of ground-state nucleon densities. Although 
it models the effective interaction between nucleons, a general 
density functional is not necessarily related to any given 
microscopic nucleon-nucleon potential, i.e. it is rather 
the density functional that defines the effective 
nuclear interaction. This means that in the DFT formulation of the 
HFB framework one does not start with a Hamiltonian defined by a
two-body interaction as in Eq. (\ref{hamiltonian}), but rather
from the energy functional $E[\mathcal{R}]=E[\hat{\rho},\hat{\kappa}]$
that depends on the densities $\hat{\rho}$ and $\hat{\kappa}$. The
generalized Hamiltonian $\mathcal{H}$ is then obtained as a functional
derivative of the energy with respect to the generalized density: 
\begin{equation}
\mathcal{H}~=~\frac{\delta E}{\delta\mathcal{R}}~=~\left(
\begin{array}
[c]{cc}%
\hat{h} & \hat{\Delta}\\
-\hat{\Delta}^{\ast} & -\hat{h}^{\ast}%
\end{array}
\right)\; ,
\label{first_der_HFB}
\end{equation}
where the single particle Hamiltonian $\hat{h}$ results from the 
variation of the energy functional with respect to the hermitian 
density matrix $\hat{\rho}$
\begin{equation}
\hat{h}=\frac{\delta E}{\delta\hat{\rho}},\;
\end{equation}
and the pairing field is obtained from the variation of the energy
functional with respect to the pairing tensor
\begin{equation}
\hat{\Delta}~=~\frac{\delta E}{\delta\hat{\kappa}}.
\end{equation}
In a compact form the stationary HFB equation is given in terms
of the generalized density:
\begin{equation}
\left [ \mathcal{H}, \mathcal{R} \right ]=0\; .
\label{statHFB}
\end{equation}

In principle, a universal energy density functional can be built as 
an expansion in terms of local densities and currents, including all 
terms allowed by the underlying symmetries, and without direct reference 
to any specific nucleon-nucleon interaction. By employing global 
effective nuclear interactions, with a small set of parameters
adjusted to reproduce empirical properties of symmetric and asymmetric 
nuclear matter, and bulk properties of few stable spherical nuclei, the current
generation of self-consistent mean-field models has achieved a high level of 
accuracy in the description of ground states and properties of excited states
in arbitrarily heavy nuclei, including rare isotopes with a large 
neutron to proton asymmetry. In the non-relativistic framework, 
in particular, two classes of effective nucleon-nucleon interactions have 
become standard in self-consistent Hartree-Fock and Hartree-Fock-Bogoliubov 
calculations. The first is the finite-range Gogny force~\cite{DG.80,berg91}:
\begin{eqnarray}
\label{Gogny}
\hat{v}_{\rm Gogny} (\rvec_{12}) 
& = & \sum_{j=1}^2 e^{-(\rvec_{12}/\mu_j)^2}
      \big(   W_j
            + B_j \hat{P}_{\sigma}
            - H_j \hat{P}_{\tau}
            - M_j \hat{P}_{\sigma} \hat{P}_{\tau}
      \big)
      \nn \\
&   & + t_3 \big( 1+x_0 \hat{P}_{\sigma} \big)
       \delta (\rvec_{12}) \;
       \rho^\alpha \left( \frac{\rvec_1 + \rvec_2}{2} \right)
      \nn \\
&   & + \iunit W_{ls} ( \hat{\sigma}_1 + \hat{\sigma}_2 ) \cdot
        \hat{\kvec}^\dagger \times
        \delta (\rvec_{12}) \; \hat{\kvec}
\end{eqnarray}
where \mbox{$\hat{P}_\sigma = \frac{1}{2}(1 + \hat{\sigma}_1 \cdot
\hat{\sigma}_2$)} is the spin-exchange operator,
\mbox{$\hat{P}_\tau = \frac{1}{2}(1 + \hat{\tau}_1 \cdot \hat{\tau}_2$)}
the isospin exchange operator, \mbox{$\rvec_{12}=\rvec_1 - \rvec_2$}, 
and \mbox{$\hat{\kvec} = - \frac{\iunit}{2} (\nabla_1 - \nabla_2)$}. 
The interaction includes the sum of two Gaussians with space, spin and 
isospin exchange, the term which includes an explicit density dependence, 
and the spin-orbit term. $W_j$, $B_j$, $H_j$, $M_j$, $\mu_j$, $t_3$, $x_0$, 
$\alpha$ and $W_{ls}$ are adjustable parameters of the interaction. 

The second class of effective interactions is based on the zero-range, 
momentum-dependent Skyrme force:
\begin{eqnarray}
\label{Skyrme}
\hat{v}_{\rm Sk} (\rvec_{12}^{\mbox{}})
& = & t_0 \, ( 1 + x_0 \Pop_\sigma ) \, \delta (\rvec_{12}^{\mbox{}}) 
      \nn \\
&   & + \frac{1}{2} \, t_1 \, ( 1 \!+\! x_1 \Pop_\sigma ) 
        \Big(   \hat{\kvec}{}^\dagger{}^{2} \; 
                \delta (\rvec_{12}^{\mbox{}})
                \!+\! \delta (\rvec_{12}^{\mbox{}})  
                \; \hat{\kvec}^2 \Big)
      \nn \\
&   & + t_2 \ ( 1 \!+\! x_2 \Pop_\sigma ) \,
        \hat{\kvec}{}^\dagger \!\cdot\!
        \delta (\rvec_{12}^{\mbox{}}) \; \hat{\kvec}
      \nn \\
&   & + \frac{1}{6} \, t_3 \, ( 1 \!+\! x_3 \Pop_\sigma ) \, 
        \delta (\rvec_{12}^{\mbox{}}) \; 
        \rho^\alpha \left( \frac{\rvec_1 \!+\! \rvec_2}{2} \right) 
       \nn \\
&   & + \iunit W_0 \,
        ( \hat{\sigma}_1 \!+\! \hat{\sigma}_2 ) \!\cdot\! 
        \hat{\kvec}{}^\dagger \times 
        \delta (\rvec_{12}^{\mbox{}}) 
        \hat{\kvec} \; .
\end{eqnarray}
Standard Skyrme interactions include ten adjustable parameters which determine 
the central term, the velocity dependent terms, the density dependent term, 
and the spin-orbit term~\cite{BHR.03,vaubrink,beiner,chab98}. 
The Skyrme energy density functional can be derived from
the Hartree--Fock expectation value of the zero-range momentum 
dependent two-body force Eq.~(\ref{Skyrme}), or it can be parameterized 
directly without reference to an effective two-body force \cite{BHR.03}. 
In the latter case the universal functional contains systematically all 
possible bilinear terms in the local densities and currents 
(plus derivative terms) which are invariant with respect to parity, 
time-reversal, rotational, translational and isospin transformations.

The pairing field in Eq.~(\ref{deltapot}) is determined by the effective 
interaction in the pairing channel. In applications of the HFB model
with the Gogny force, the same effective interaction is 
used both in the $ph$ and $pp$ channels, with the exception of the 
density-dependent zero-range term which, with the choice of the parameter
$x_0=1$, does not contribute to the pairing channel. There is no physical
reason, however, to use identical terms of the energy density 
functional in the the mean-field and 
pairing channels. Skyrme-type forces, for instance, generally exhibit 
unrealistic pairing properties, and thus an additional effective pairing 
interaction has to be specified in Skyrme-HFB calculations of open-shell 
nuclei. A standard choice for the pairing interaction is a zero-range 
local force, often including an explicit density-dependence: 
\begin{equation}\label{v_pair_dd}
v_{\rm pair} (\rvec_{12})
= \frac{V_0}{2} \, (1-\hat{P}_\sigma) \,
  \left[ 1-\eta\left(\frac{\rho(\rvec_1)}{\rho_c}\right)^\beta \right] \,
  \delta(\rvec_1-\rvec_2)
.
\end{equation}
Depending on the value of the parameters $\eta$, $\rho_c$ and $\beta$, 
pairing is more active in
the volume of the nucleus, or on its surface. The strength $V_{0}$ 
is adjusted to reproduce the odd-even staggering of 
binding energies in selected isotopic chains. Usually this results in 
slightly different values of the pairing strengths for protons and 
neutrons, thus breaking the isospin invariance of the pairing energy 
functional.

For nuclear systems with time-reversal invariance, the HFB method can be
considerably simplified by employing the Bardeen-Cooper-Schrieffer (BCS) 
approximation. In the BCS approximation the pairing potential $\hat d$, 
defined by the relation:
\begin{equation}
\hat{\Delta}=\left(\begin{array}{cc}
0&\hat{d} \\ -\hat{d}~^T &0
\end{array}\right)\; ,
\end{equation}
where $\hat{\Delta}$ is the pairing field of Eq.~(\ref{deltapot}), is diagonal 
in the basis of the eigenstates of the mean-field Hamiltonian $\hat{h}$
\begin{equation}
d_{n\bar{m}}
= \delta_{nm} d_{n \bar{n}}
, \quad
\hat{h}\varphi_n
= \varepsilon_n\varphi_n \; ,
\end{equation}
and $\bar{m}$ ($\bar{n}$) denotes the time-conjugate partner 
of the single-particle 
state $m$ ($n$). The resulting two-component HFB wave functions read 
$U_n=u_n\varphi_n$ and $V_n=v_n\varphi_n$, and 
the occupation amplitudes ($u_n$,$v_n$) are determined by the gap equation
\begin{equation}
(\varepsilon_n-\lambda)(u_n^2-v_n^2)+ 2 d_{n\bar{n}} u_n v_n
= 0 ,
\end{equation}
and the normalization condition:
\begin{equation}\label{eq:no}
   u_{n}^2 + v_{n}^2 = 1 \;.
 \end{equation}
 
In the case of well-bound nuclei close to the stability line, pairing 
correlations are often treated in the BCS approximation, with the 
strength of the pairing force adjusted to the experimental odd-even mass
differences. This approach, however, presents only a poor approximation 
for weakly-bound nuclei far from the valley of $\beta$-stability. 
In particular, for nuclei at the limits of particle stability (drip-line 
nuclei), the Fermi level lies close to the continuum, and the
coupling between bound and continuum states has to be taken into
account explicitly. The BCS model does not provide a correct
description of the scattering of nucleonic pairs from bound states
to the positive energy continuum, and as a result levels high in the 
continuum become partially occupied. Including the system in a box of 
finite size, i.e. solving the HFB equations in coordinate space, 
leads to unreliable predictions for quantities that crucially 
depend on the size of the box, e.g. nuclear radii. The reason is that in 
the BCS approximation the pairing field does not vanish 
asymptotically. Thus for weakly-bound nuclei the full HFB theory, 
including the continuum, has to be employed~\cite{DFT.84,DNW.96}. 
 
\begin{figure}[htb]
\centerline{\includegraphics[width=9cm,height=8.cm]{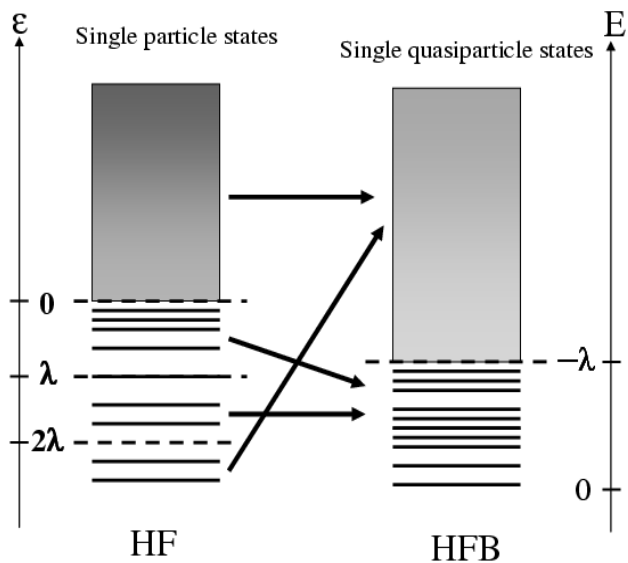}}
\caption{Illustrative representation of single-particle (left) and 
single-quasiparticle (right) spectra. $\lambda$ is the chemical potential. 
The arrows depict the correspondence between the single-particle states and 
the single-quasiparticle states in the discrete and continuous regions of 
the spectra.
\label{fig:conti}}
\end{figure}

The asymptotic behavior of the HFB wave function is determined
by the physical condition that at large distances from a nucleus 
the mean field $\Gamma(\rvec)$ and the pairing field 
$\Delta(\rvec)$ vanish. For a bound system (negative chemical potential 
$\lambda < 0$), two distinct regions characterize
the quasiparticle spectrum. This is illustrated in Figure \ref{fig:conti}.
Between 0 and $-\lambda$ the quasiparticle spectrum is discrete
and both the upper and lower components of the radial HFB wave function
decay exponentially at $r \rightarrow \infty$.
In the continuum region above $-\lambda$ one finds
two types of HFB quasiparticle resonant states. First, those  
which correspond to the single-particle resonances of the mean field. 
The low-lying single-particle resonances are particularly important in 
the treatment of pairing correlations of weakly-bound nuclei, because the
pairing interaction scatters pairs of nucleons into positive-energy states 
above the particle threshold. A second type of resonant states is specific 
to the HFB method, and originates from bound single-particle states 
which, in the absence of pairing correlations, are found at energies
$\epsilon < 2\lambda$. With the inclusion of the pairing
field these bound states couple with the continuum single-particle 
states and, therefore, acquire finite widths. 

The HFB equations are usually solved by imposing box boundary conditions, i.e.  
the HFB wave functions are assumed to vanish beyond some distance, usually 
chosen to be a few times the nuclear radius. The energy continuum is thus
replaced by a spectrum of discrete energies, whose density increases with 
the size of the box. Continuum HFB calculations can be performed either in the 
complex energy plane by employing Green's function techniques~\cite{bely87}, 
or on the real energy axis~\cite{DFT.84,DNW.96}. It is possible to treat the continuum exactly 
in HFB calculations on the real energy axis, by imposing the correct boundary 
conditions for the HFB wave functions $U_k$ and $V_k$~\cite{grass01}. 
Far from the nuclear potential, the $U_k$ and $V_k$ which belong to the single
quasiparticle continuum ($E_k$ $>$ $-\lambda$), take the asymptotic form of a
scattering state and a function that exponentially decreases at infinity,
respectively. These asymptotic forms must, of course, be matched with the 
corresponding radial functions in the inner region of the potential. By 
comparing the exact treatment of the continuum with different levels 
of approximation (discretization in a box, HF-BCS including the
resonant part of the continuum), 
it has been shown that the different ways in which the coupling to the 
continuum is treated in HFB strongly affect the resulting pairing 
correlations in nuclei close to the particle drip-lines.   


\subsection{The Relativistic Hartree-Bogoliubov model (RHB)}\label{subsec_RHB}


Self-consistent mean-field models based on the 
relativistic mean-field theory (RMF)~\cite{SW.86,SW.97},
have been very successfully employed in 
analyses of a variety of nuclear structure phenomena, 
not only in nuclei along the valley of $\beta$-stability, 
but also in exotic nuclei with extreme isospin 
values and close to the particle drip-lines. RMF-based models 
have reached a level of sophistication and 
accuracy comparable to the non-relativistic 
Hartree-Fock-Bogoliubov framework based on Skyrme and 
Gogny effective interactions~\cite{Vre.05}. 

Most applications have used the finite-range 
meson-exchange representation of the RMF theory, 
in which the nucleus is described as a system of Dirac 
nucleons coupled to effective mesons and the electromagnetic field. 
A medium dependence of the effective interaction can be introduced 
either by including non-linear meson self-interaction terms
in the Lagrangian, or by assuming an explicit density dependence 
for the meson-nucleon couplings~\cite{Vre.05}.
An alternative RMF representation is formulated in terms 
of point-coupling (PC) (contact) nucleon-nucleon 
interactions~\cite{BMM.02}, without the inclusion of meson fields.
The medium dependence of the interaction can be taken into 
account by terms of higher-order in the nucleon fields, or it 
can be encoded into the effective couplings, i.e. in the strength 
parameters of the interaction in the isoscalar and isovector channels. 
On the phenomenological level, when applied in studies of properties 
of finite nuclei, the meson-exchange and nucleon point-coupling  
representations produce results of comparable 
quality. The applications reviewed in this article are 
based on the finite-range meson-exchange picture of 
effective nuclear interactions, and in the following we outline 
its basic features. 

The isoscalar scalar
$\sigma$-meson, the isoscalar vector $\omega$-meson, and the
isovector vector $\rho$-meson form the minimal set of meson
fields which, together with the electromagnetic field, are necessary
for a quantitative description of bulk nuclear properties. The model is
defined by the Lagrangian density
\begin{equation}
\mathcal{L}=\mathcal{L}_{N}+\mathcal{L}_{m}+\mathcal{L}_{int}.
\label{Lagrangian}
\end{equation}%
$\mathcal{L}_{N}$ is the Lagrangian of the free nucleon
\begin{equation}
\mathcal{L}_{N}=\bar{\psi}\left( i\gamma ^{\mu }\partial _{\mu
}-m\right) \psi \; ,
\end{equation}
$m$ is the bare nucleon mass and $\psi$ denotes the Dirac spinor.
$\mathcal{L}_{m}$ is the Lagrangian of the free 
$\sigma$, $\omega$ and $\rho$ meson fields, and
the electromagnetic field
\begin{eqnarray}
\mathcal{L}_{m} &=&\frac{1}{2}\partial_{\mu }\sigma
\partial^{\mu }\sigma - \frac{1}{2}m_{\sigma }^{2}\sigma
^{2}-\frac{1}{4}\Omega_{\mu \nu
}\Omega ^{\mu \nu }+\frac{1}{2}m_{\omega }^{2}\omega _{\mu }\omega ^{\mu }
\nonumber \\
&&-\frac{1}{4}\vec{R}_{\mu \nu }\vec{R}^{\mu \nu
}+\frac{1}{2}m_{\rho }^{2} \vec{\rho}_{\mu }\vec{\rho}^{\mu
}-\frac{1}{4}F_{\mu \nu }F^{\mu \nu } ,
\end{eqnarray}
with the corresponding masses $m_{\sigma }$, $m_{\omega }$, $m_{\rho }$.
The field tensors
$\Omega_{\mu\nu}$, $\vec{R}_{\mu\nu}$, $F_{\mu\nu}$ read
\begin{equation}
\begin{array}{rcl}
\Omega _{\mu \nu } & = & \partial _{\mu }\omega _{\nu }-\partial _{\nu
}\omega _{\mu } \\
\vec{R}_{\mu \nu } & = & \partial _{\mu }\vec{\rho}_{\nu }-\partial _{\nu }
\vec{\rho}_{\mu } \\
F_{\mu \nu } & = & \partial _{\mu }{A}_{\nu }-\partial _{\nu }{A}_{\mu }\; ,
\end{array}%
\end{equation}%
where arrows denote isovector quantities. 
The minimal set of interaction terms is contained in 
\begin{equation}
\mathcal{L}_{int}=-\bar{\psi}\Gamma _{\sigma }\sigma \psi %
-\bar{\psi}\Gamma _{\omega }^{\mu}\omega_{\mu} \psi -\bar{\psi}
\vec{\Gamma}_{\rho }^{\mu}%
\vec{\rho}_{\mu}\psi -\bar{\psi}\Gamma _{e}^{\mu}A_{\mu}\psi \; ,
\end{equation}%
where the vertices read%
\begin{equation}
\Gamma_{\sigma }=g_{\sigma },\;\;\;\Gamma_{\omega }^{\mu
}=g_{\omega }\gamma ^{\mu },\;\;\;\vec{\Gamma}_{\rho }^{\mu
}=g_{\rho }\vec{\tau}\gamma ^{\mu },\;\;\;\;\Gamma_{e}^{\mu}%
=e\frac{1-\tau _{z}}{2}\gamma ^{\mu },
\end{equation}%
with the coupling parameters $g_{\sigma }$, $g_{\omega }$,
$g_{\rho}$ and $e$. The phenomenological $\sigma$-meson approximates
a large attractive scalar field that results from complicated
microscopic processes, such as uncorrelated and correlated two-pion
exchange. The $\omega$-meson describes the short-range repulsion
between the nucleons, and the $\rho$-meson carries the isospin
quantum number. The latter is required by the large empirical
asymmetry potential in finite nuclear systems. Because of parity
conservation there is no direct contribution from the pion field. 
The self-consistent RMF approach represents a particular 
realization of the relativistic Kohn-Sham density functional 
theory~\cite{Koh.99,DG.90}, in which one attempts to 
effectively include in the nuclear
energy density functional effects which go beyond
the Hartree approximation (Fock terms, short-range correlations, 
vacuum-polarization effects). The many-body 
correlations of the energy density functional can be represented by a 
medium dependence of the corresponding effective nuclear interaction. 
An effective density dependence can be included, for instance, through 
meson self-interaction terms. Over the years a number of non-linear 
meson-exchange interactions have been adjusted to the nuclear
matter equation of state and bulk properties of a set of spherical
closed-shell nuclei, and applied in the description of nuclear 
properties along the $\beta $-stability line. One of the most 
successful phenomenological interactions of this type is the 
non-linear parameter set NL3~\cite{LKR.97}, which has been employed 
in many studies of ground-state properties and collective 
excitations both in stable nuclei and in exotic systems far from
the line of $\beta$-stability.

Another class of medium-dependent effective interactions is 
characterized by an explicit baryon-density dependence of the 
meson-nucleon vertices. Such an approach retains the basic
structure of the relativistic mean-field framework, 
but can be more directly related to the
underlying microscopic description of nuclear interactions.
The functional form of the meson-nucleon vertices 
can be deduced from in-medium Dirac-Brueckner interactions, obtained
from realistic free-space NN interactions, or a 
phenomenological approach can be adopted, with the density dependence  
for the $\sigma$, $\omega$ and $\rho$ meson-nucleon
couplings adjusted to properties of nuclear matter and 
a set of spherical nuclei. The former represents an {\it ab-initio} 
description of nuclear matter and finite nuclei, and the corresponding 
density-dependent relativistic mean-field model has also been applied to
asymmetric nuclear matter and exotic nuclei~\cite{HKL.01}. In the 
latter approach very accurate phenomenological density-dependent 
relativistic effective interactions have recently been 
adjusted~\cite{TW.99,NVFR.02,LNVR.05}, and employed in analyses of 
both bulk nuclear properties and collective excitations. A number of 
recent studies have shown that, in comparison with non-linear meson 
self-interactions, relativistic models with an explicit density dependence
of the meson-nucleon couplings provide an improved description of asymmetric
nuclear matter, neutron matter and nuclei far from stability.
 
An extensive review of the relativistic extension of the HFB theory --
the relativistic Hartree-Bogoliubov (RHB) model -- including numerous 
applications, can be found in Ref.~\cite{Vre.05}. In this framework 
the ground state of an open-shell nucleus is determined by the 
single-quasiparticle solutions of the RHB equations (cf. also 
Eq. (\ref{HFBnonrel})):\\
\begin{equation}
\left(
\begin{array}
[c]{cc}%
\hat{h}_{D}-m-\lambda & \hat{\Delta}\\
-\hat{\Delta}^{\ast} & -\hat{h}^{\ast}_{D}+m+\lambda
\end{array}
\right)  \left(
\begin{array}
[c]{c}%
U_{k}(\mathbf{r})\\
V_{k}(\mathbf{r})
\end{array}
\right)  =E_{k}\left(
\begin{array}
[c]{c}%
U_{k}(\mathbf{r})\\
V_{k}(\mathbf{r})
\end{array}
\right)  \;. \label{eqhb} \\
\end{equation}
The self-consistent mean field $\hat{h}_{D}$ is the Dirac
Hamiltonian determined by the Lagrangian density of Eq.~(\ref{Lagrangian}),
$\hat{\Delta}$ denotes the pairing field, and $(U_{k} ,V_{k} )$ are 
quasiparticle spinors. Pairing effects in nuclei are restricted to 
a narrow window of a few MeV around the Fermi level, and their scale
is well separated from the scale of binding energies which
are in the range of several hundred to thousand MeV. 
There is no empirical evidence for any relativistic effect in the 
nuclear pairing field and, therefore, pairing can be treated as a 
non-relativistic phenomenon. In most applications of the 
RHB model for spherical and deformed nuclei the 
Gogny force (cf. Eq.~(\ref{Gogny})) has been employed in the
$pp$ channel. 

The RHB equations are solved self-consistently, with mean-field potentials 
in the particle-hole channel determined from solutions
of the stationary Klein-Gordon equations
\begin{eqnarray}
\left[ -\Delta +m_{\sigma }^{2}\right] \,\sigma (\mathbf{r}) &=&-g_{\sigma
}(\rho _{v})\,\rho _{s}(\mathbf{r}) \label{messig} \\
\left[ -\Delta +m_{\omega }^{2}\right] \,\omega ^{0}(\mathbf{r})
&=&g_{\omega }(\rho _{v})\,\rho _{v}(\mathbf{r})  \label{mesome} \\
\left[ -\Delta +m_{\rho }^{2}\right] \,\rho ^{0}(\mathbf{r}) &=&g_{\rho
}(\rho _{v})\,\rho _{3}(\mathbf{r})  \label{mesrho} \\
-\Delta \,A^{0}(\mathbf{r}) &=&e\,\rho _{p}(\mathbf{r})  \label{photon} 
\end{eqnarray}
for the $\sigma $-meson, the $\omega $-meson, the $\vec{\rho}$-meson and the
photon field, respectively. Because of charge conservation, only the 3-rd
component of the isovector $\rho $-meson contributes. For the ground-state
solution of an even-even nucleus there are no contributions 
from currents (time-reversal
invariance) and the spatial components ${\mbox{\boldmath $\omega $}}$, ${
\mbox{\boldmath $\rho$}}_{3}$, ${\mbox{\boldmath $A$}}$ of the vector fields
vanish. The source terms in equations (\ref{messig})
to (\ref{photon}) are sums of bilinear products of baryon amplitudes 
\begin{eqnarray}
\rho _{s}(\mathbf{r}) &=&\sum\limits_{k>0}V_{k}^{\dagger }(\mathbf{r})\gamma
^{0}V_{k}(\mathbf{r})  \label{densities} \\
\rho _{v}(\mathbf{r}) &=&\sum\limits_{k>0}V_{k}^{\dagger }(\mathbf{r})V_{k}(%
\mathbf{r}) \\
\rho _{3}(\mathbf{r}) &=&\sum\limits_{k>0}V_{k}^{\dagger }(\mathbf{r})\tau
_{3}V_{k}(\mathbf{r})_{{}} \\
\rho _{\mathrm{em}}(\mathbf{r}) &=&\sum\limits_{k>0}V_{k}^{\dagger }(\mathbf{%
r}){\frac{{1-\tau _{3}}}{2}}V_{k}(\mathbf{r})\;, 
\end{eqnarray}
where the summation is performed only over occupied orbitals of positive 
energy. This is the  \textit{no-sea} approximation, in which the Dirac sea
of negative-energy states does not contribute to the densities and currents 
in an explicit way.
The self-consistent solution of the Dirac-Hartree-Bogoliubov
integro-differential equations and Klein-Gordon equations for the meson
fields determines the ground state of a nucleus.

\subsection{Continuum QRPA}
\label{sec2_cqrpa}

Small amplitude collective excitations of arbitrarily heavy nuclei 
can be accurately described by the random phase approximation 
(RPA) or, in the case of open-shell nuclei, by the quasiparticle random 
phase approximation (QRPA)~\cite{ring80,BR.86,BB.94,borti98,harak01}.
As it has been already emphasized in the introduction to this section, 
a quantitative description of ground-states and excitations in 
weakly-bound nuclei characterized by the closeness of the Fermi surface 
to the particle continuum, must take into account the effects of 
the coupling between bound states and the particle continuum.
Here we derive the QRPA response based on the continuum 
HFB description of the nuclear ground state in the coordinate space 
representation, which is naturally suitable for the 
treatment of the coupling to continuum states~\cite{kh02,ma01}. 
The QRPA represents the small amplitude limit of the general 
time-dependent (TD) HFB theory, and we therefore start from the 
HFB equation which describes the 
response of the generalized density matrix to an external periodic 
perturbation~\cite{ring80}:
\begin{equation}\label{eq:tdhfb}
i\hbar\frac{\partial{\cal R}}{\partial t}=[{\cal H}(t) +
{\cal F}(t),{\cal R}(t)] \; .
\end{equation}
${\cal R}$ and ${\cal H}$ are the time-dependent generalized density and
HFB Hamiltonian, respectively, and ${\cal F}$ is the external 
periodic perturbation given by
\begin{equation}\label{eq:pert}
{\cal F} = F e^{-i\omega t} + {\rm h.c.} \; 
\end{equation}
In the presence of pairing correlations, the fluctuating field $F$
in Eq. (\ref{eq:pert}) is a generalized one-body operator which 
includes both particle-hole and two-particle transfer
operators 
\begin{equation}\label{eq:extpart}
F=\sum_{ij} F^{11}_{ij} c_{i}^{\dagger}c_{j}+\sum_{ij}
(F^{12}_{ij} c_{i}^{\dagger}c_{j}^{\dagger}+ F^{21}_{ij} c_{i}c_{j})\;,
\end{equation}
and $c_{i}^{\dagger}$, $c_{i}$ are the single-particle 
creation and annihilation operators, respectively.
We assume that the external field induces small oscillations around the
stationary solution of the HFB equation (\ref{statHFB}), which we
denote here as ${\cal R}^0$ (${\cal H}^0$ being the 
corresponding Hamiltonian). Accordingly, 
\begin{equation}\label{eq:pertr}
{\cal R}(t) = {\cal R}^0+ {\cal R}' e^{-i\omega t} + {\rm h.c.}, 
\end{equation}
\begin{equation}\label{eq:perth}
{\cal H}(t) = {\cal H}^0+ {\cal H}' e^{-i\omega t} + {\rm h.c.}, 
\end{equation}
and the TDHFB equation (\ref{eq:tdhfb}) becomes
\begin{equation}\label{eq:lin}
        \hbar\omega{\cal R}'=[{\cal H}',{\cal R}^0] + [{\cal H}^0,{\cal
	R}']+[F,{\cal R}^0] \;.
\end{equation}
The variation of the generalized density has the form
\begin{equation}\label{eq:denspart}
	{\cal R}'=\left(
	\begin{array}{cc}
	\rho' & \kappa' \\
	\bar{\kappa}' & -\rho' \\
	\end{array}
	\right) \; ,
\end{equation}
where $\rho'_{ij} = \langle 0\vert c^{\dagger}_jc_i\vert ' \rangle$
is the variation of the particle density,
$\kappa'_{ij} = \langle 0\vert c_jc_i \vert ' \rangle$ 
and $\bar{\kappa}'_{ij} =
\langle 0\vert c^{\dagger}_jc^{\dagger}_i \vert ' \rangle$ are the corresponding
fluctuations of the pairing tensor, and
$\vert ' \rangle$ denotes the change in the ground-state wavefunction
$\vert 0 \rangle$ caused by the external field. Instead of the variation of 
just one quantity ($\rho'$) as in RPA, in QRPA we must
specify the variations of three
independent quantities, namely $\rho'$, $\kappa'$ and $\bar{\kappa}'$.
%
If we use the notation 
\begin{equation}\label{eq:rhodef}
{\bb{\rho'}} \equiv \left(
	\begin{array}{c}
	\rho' \\
        \kappa' \\
	\bar{\kappa}' \\
        \end{array}
	\right) \;,
	\end{equation}
for the density variations, the corresponding variation 
of the HFB Hamiltonian reads
\begin{equation}\label{eq:hvar}
\bb{{\cal H}}'=	\bf{V}\bb{\rho}' \;,
\end{equation}
where $\bf{V}$ is the matrix of the residual interaction, 
expressed in terms of the second derivatives of the HFB energy 
functional $E$, defined in Sec. \ref{firstHFB} 
\begin{equation}\label{eq:vres}
{\bf{V}}^{\alpha\beta}({\bf r}\sigma\tau,{\bf r}'{\sigma}'{\tau}')=
\frac{\partial^2 E}{\partial{\bf{\rho}}_\beta({\bf r}'{\sigma}'{\tau}')
\partial{\bf{\rho}}_{\bar{\alpha}}({\bf r}\sigma\tau)},~~~\alpha,\beta = 1,2,3.
\end{equation}
Here the notation $\bar{\alpha}$ means that whenever $\alpha$ 
is 2 or 3, then $\bar{\alpha}$ is 3 or 2. In this three dimensional space, 
the first dimension represents the particle-hole ($ph$) subspace, 
the second one the particle-particle ($pp$), 
and the third one the hole-hole ($hh$) subspace. 

The QRPA Green's function $\bf{G}$ relates the perturbing external 
field to the density variations 
\begin{equation}\label{eq:g}
{\bb{\rho'}}=\bf{G}\bf{F}~\boldmath \everymath{ } \; ,
\end{equation}
where $\bf{F}$ is the three dimensional column vector 
\begin{equation}\label{eq:f}
\bf{{F}}=	\left(\begin{array}{c}
	 {F}^{11} \\
	 {F}^{12}\\
	 {F}^{21}   \\
	\end{array}
	\right) \; .
\end{equation}
$\bf{G}$ is the solution of the Bethe-Salpeter equation
\begin{equation}\label{eq:bs}
\bf{G}=\left(1-\bf{G}_0\bf{V}\right)^{-1}\bf{G}_0=
\bf{G}_0+\bf{G}_0\bf{V}\bf{G} \; ,
\end{equation}
and the unperturbed Green's function $\bf{G}_0$ reads
\begin{equation}
\label{eq:g0}
\hspace{-2cm}
{\bf{G}_0}^{\alpha\beta}({\bf r}\sigma\tau,{\bf r}'{\sigma}'{\tau}';\omega)=
\sum_{ij} \hspace{-0.5cm}\int ~ \frac{{\cal U}^{\alpha 1}_{ij}({\bf r}\sigma\tau)
\bar{{\cal U}}^{*\beta 1}_{ij}({\bf r}'\sigma'\tau')}{\hbar\omega-(E_i+E_j)+i\eta}
-\frac{{\cal U}^{\alpha 2}_{ij}({\bf r}\sigma\tau)
\bar{{\cal U}}^{*\beta 2}_{ij}({\bf r}'\sigma'\tau')}{\hbar\omega+(E_i+E_j)+i\eta}, 
\end{equation}
where $E_i$ are the quasiparticle energies and ${\cal U}_{ij}$ 
are 3 by 2 matrices
expressed in terms of the two components $U$ and $V$ of the HFB wave functions
\begin{equation}\label{eq:ubig}
{\cal U}_{ij}({\bf r}\sigma)=\left(
 \begin{array}{cc}
 U_i({\bf r}\sigma)V_j({\bf r}\sigma)~~ & 
 U^*_j({\bf r}\sigma)V^*_i({\bf r}\sigma)
\\
 U_i({\bf r}\sigma)U_j({\bf r}\bar{\sigma})~~ & 
 V^*_i({\bf r}\sigma)V^*_j({\bf r}\bar{\sigma})
\\ 
-V_i({\bf r}\sigma)V_j({\bf r}\bar{\sigma})~~ &
-U^*_i({\bf r}\sigma)U^*_j({\bf r}\bar{\sigma}) 
\\
 \end{array}
\right) \; ,
\end{equation}
with the notation 
$f({\bf r}\bar{\sigma})$=$-2\sigma f({\bf r}-\sigma)$.
The $\sum$ \hspace{-0.45cm}$\int$ ~ symbol in Eq. (\ref{eq:g0})
indicates that the summation is taken both over discrete and 
continuum quasiparticle states, i.e., the unperturbed Green's function $\bf{G}_0$ 
takes into account the resonant states. This can be done either in the complex
plane by integrating on a contour~\cite{ma01}, or on the real axis by
integrating the resonant states with the corresponding widths~\cite{kh02}. 

$\bf{G}_0$ in Eq. (\ref{eq:g0}) is constructed from
the solutions of the HFB equations, i.e. 
the quasiparticle energies and the corresponding wave functions $U$ and $V$.
For a given residual interaction in Eq. (\ref{eq:vres}), $\bf{G}$ 
is then calculated from the Bethe-Salpeter equation (\ref{eq:bs}). 
$\bf{G}$ can be used for calculating the strength functions associated 
with various external perturbations. For instance,
in the case of transitions from the ground-state to excited states within
the same nucleus, only the ($ph$,$ph$) component of $\bf{G}$ is acting. 
If the interaction does not depend on spin variables, the strength 
function is then given by
\begin{equation}\label{eq:stren}
S(\omega)=-\frac{1}{\pi}{\rm Im} \int  F^{11*}({\bf r}){\bf{
G}}^{11}({\bf r},{\bf r}';\omega) F^{11}({\bf r}')
d{\bf r}~d{\bf r}'
\end{equation}
where ${\bf{G}}^{11}$ is the ($ph$,$ph$) component of the QRPA Green's function.
Examples of such calculations can be found in Refs.~\cite{kh02,ma01}.

It should be noted that generally the notation includes the
neutron-proton formalism, i.e. each supermatrix ($\bf{G}_0$, $\bf{G}$,
$\bf{V}$) consists of nine blocks which correspond to the $ph$, $pp$, $hh$ 
channels, and each of these blocks is further divided into four sub-blocks 
corresponding to the neutron-neutron, neutron-proton, 
proton-neutron and proton-proton channels, respectively.

An important point that must be emphasized is the issue of
self-consistency. Namely, since the characteristic ground-state properties 
of weakly-bound nuclei strongly influence the multipole response of these
systems, it is particularly important that for the QRPA residual 
interactions, both in $ph$ and $pp$ channels, the same interactions are 
used which also determine the ground-state solutions of the HFB equations.   

It is, however, difficult to achieve full self-consistency in continuum RPA or
QRPA calculations with Skyrme forces. While the zero-range terms of 
the interaction do not present any particular problem, the velocity-dependent 
terms introduce serious technical difficulties that are often avoided 
by approximating the residual interaction in the ($ph,ph$) subspace by its 
Landau-Migdal limit, in which the momenta of the interacting particle and 
hole are equal to the Fermi momentum, and the transferred momentum is zero.  
Taking the Landau-Migdal form for the $ph$ interaction
simplifies the numerical task considerably, however at the cost of losing 
the full self-consistency. This deficiency of the QRPA approach is cured
by renormalizing the residual interaction. It should be noted, however,
that an exact treatment of continuum effects within the general framework of
the fully self-consistent QRPA is presently not available.

\subsection{Discrete QRPA}

In addition to the linear response formalism based on Green's functions,
the QRPA can also be formulated in a discrete basis.
We therefore start 
from Eq.~(\ref{eq:lin}), and choose a discrete basis
of two-quasiparticle configurations
$|kl\rangle\equiv\alpha^\dagger_k\alpha^\dagger_l|0\rangle$, where
$\alpha^\dagger$ is a quasiparticle creation operator and $|0\rangle$ 
is the quasiparticle vacuum. The energy of the configuration $|kl\rangle$ 
is $E_k+E_l$. The quasiparticle space consists of discrete states, even 
when it describes configurations in the continuum. This discretization 
is achieved either by enclosing the nuclear system in a finite
box, or by expanding the HFB wave functions in a
discrete basis, e.g., in terms of eigenfunctions of a harmonic oscillator 
potential. In a QRPA description of transitions to low-lying excited states 
in open-shell weakly-bound nuclei, in particular, the two-quasiparticle 
configuration space must include states with both nucleons in discrete 
bound levels, states with one nucleon in a bound level and one nucleon 
in the continuum, and also states with both nucleons in the continuum. 

The QRPA equation is obtained by taking the matrix elements of 
Eq.~(\ref{eq:lin}) between $\langle kl|$ and $|0\rangle$, and 
then also between $\langle 0|$ and $|kl\rangle$. Inserting 
a completeness relation in terms of a sum over all configurations 
labeled by $k'l'$, the QRPA matrix equation finally reads
\begin{equation}\label{eq:matrixqrpa}
\left( \begin{array}{cc} A_{kl,k'l'} & B_{kl,k'l'} \\
-B_{kl,k'l'}^* & -A_{kl,k'l'}^* \end{array} \right) 
\left( \begin{array}{c} X_{k'l'} \\ Y_{k'l'} 
\end{array} \right) =
E \left( \begin{array}{c} X_{kl} \\ Y_{kl} 
\end{array} \right) \;.
\end{equation}
$A$ and $B$ contain matrix elements of the HFB or HF-BCS Hamiltonian. 
In the simplest case in which pairing is treated in the BCS approximation,
the explicit expressions for the QRPA matrices read 
\begin{eqnarray}\label{eq:ABmatrices}
A_{kl,k'l'} & = & (E_k+E_l)\delta_{kk'}\delta_{ll'} + \nonumber \\
            &   & + V_{klk'l'} (u_{k}u_{l}u_{k'}u_{l'}
                  +v_{k}v_{l}v_{k'}v_{l'}) + \nonumber \\
            &   & + W_{klk'l'} (u_{k}v_{l}u_{k'}v_{l'}
                  +v_{k}u_{l}v_{k'}u_{l'}), \nonumber \\
B_{kl,k'l'} & = & - V_{klk'l'} (u_{k}u_{l}v_{k'}v_{l'}
                  +v_{k}v_{l}u_{k'}u_{l'}) +  \nonumber \\
            &   & + W_{klk'l'} (u_{k}v_{l}u_{k'}v_{l'}
                  +v_{k}u_{l}v_{k'}u_{l'}),
\end{eqnarray}
where $V$ and $W$ denote the matrix elements of the interactions in the $pp$
and $ph$ channels, respectively, and $u$ and $v$ are the 
corresponding BCS occupation factors of single-nucleon states. 

The solution of the eigenvalue problem of Eq.~(\ref{eq:matrixqrpa}) 
determines the energies $E_n$ of the excited vibrational states $|n\rangle$, 
and the corresponding wave functions expressed in terms of 
the forward and backward amplitudes 
$X^{(n)}_{kl}$ and $Y^{(n)}_{kl}$ 
\begin{equation}\label{eq:phonon}
|n\rangle = \sum_{kl} \left( X_{kl}^{(n)} \alpha^\dagger_k\alpha^\dagger_l  
+ Y_{kl}^{(n)} \alpha_k\alpha_l \right) |0\rangle.
\end{equation}

If instead of the simpler HF-BCS model, the full HFB framework is 
used as a basis on which the QRPA is formulated, the matrix equation 
(\ref{eq:matrixqrpa}) remains 
valid, but the matrices $A$ and $B$ display a more complicated structure. 
In place of the simple $u$ and $v$ occupation factors which depend on a 
single index, one finds the Bogoliubov matrices 
$U$ and $V$.  One way to circumvent the technical difficulties of working 
in the quasiparticle basis is to formulate the QRPA in the canonical 
single-nucleon basis  which diagonalizes the density matrix. 
The corresponding matrix equations require only 
the evaluation of matrix elements of the residual $ph$ and pairing 
$pp$ interactions in this basis, multiplied by certain combinations 
of occupation factors. The fully self-consistent QRPA, 
formulated in the HFB canonical single-particle basis, was 
introduced in Ref.~\cite{Eng.99}. Self-consistency requires the 
QRPA residual interaction to be derived from the same force or 
energy functional that determines the HFB solution. This is 
crucial for the decoupling of spurious states, associated with 
symmetry breaking by the mean field solution, from the 
spectrum of physical excitations. The details of the implementation of 
the fully self-consistent HFB+QRPA framework in the 
canonical basis, and accurate tests using Skyrme energy density 
functionals and density-dependent pairing functionals, have 
recently been reported in Ref.~\cite{Ter.05}. 

The principal source
of arbitrariness in the matrix representation of QRPA is the truncation 
of the basis. In realistic applications it is, therefore, necessary to 
verify the stability of the results with respect to variations of 
parameters that determine the discretization and basis truncation 
(the size of the box in coordinate space or the harmonic oscillator parameter, 
and the upper limit for unperturbed energies of two-quasiparticle 
configurations). 

Vibrational states can be excited by acting on the nucleus with an 
external operator $\hat{F}({\bf r})$. The corresponding strength function 
is defined as
\begin{equation}\label{eq:strengthf1}
S(\omega) = \sum_n |\langle n | \hat{F} | 0\rangle|^2 \delta(\hbar\omega-E_n).
\end{equation}
If the explicit wave function of the state $|n\rangle$ (Eq.~(\ref{eq:phonon}))
is inserted in this expression, the strength of each peak can be 
evaluated in terms of the single-particle
matrix elements of the operator $\hat{F}({\bf r})$, the coefficients of
the BCS or Bogoliubov transformations, and the forward $X$ and backward $Y$ 
amplitudes.

%
\subsection{Relativistic QRPA}
%

Relativistic RPA calculations have been performed since the 
early 1980s, but it is only more recently that non-linear
meson self-interaction terms or density-dependent meson-nucleon 
couplings have been included in the RRPA 
framework~\cite{Ma.97,VWR.00,NVR.02}. As in the case of 
ground-state properties, the inclusion of a medium dependence in 
the residual interaction is necessary for a quantitative 
description of collective excited states.   
Another essential feature of the RRPA is the fully consistent
treatment of the Dirac sea of negative energy states.
Within the no-sea approximation, in addition to
the usual particle-hole pairs, the RRPA configuration
space must also include pair-configurations built from 
positive-energy states occupied in the ground-state solution, 
and empty negative-energy states in the Dirac sea~\cite{Rin.01}. 
Collective excitations in open-shell nuclei 
can be analyzed with the relativistic quasiparticle 
random-phase approximation (RQRPA), which in Ref.~\cite{PRNV.03} 
has been formulated in the canonical single-nucleon basis of the 
relativistic Hartree-Bogoliubov (RHB) model. An alternative 
derivation of the RQRPA in the response function formalism, but 
with pairing correlations treated only in the BCS approximations, 
has recently been formulated in Ref.~\cite{CaoMa.05}.

The RQRPA represents the small amplitude limit of the time-dependent 
relativistic Hartree-Bogoliubov (RHB) framework. The RQRPA matrix 
equations in the quasiparticle basis are, however, rather complicated
and require the evaluation of the matrix elements of the Dirac 
Hamiltonian in the basis of the Hartree-Bogoliubov spinors 
$U_{k}(\mathbf{r})$ and $V_{k}(\mathbf{r})$. A considerably simpler 
representation is provided by the canonical single-nucleon basis.
Namely, any RHB wave function can be expressed either in the 
quasiparticle basis as a product of independent quasiparticle states, 
or in the canonical basis as a highly correlated BCS-state. 
The canonical 
basis is specified by the requirement that it diagonalizes the single-nucleon
density matrix. The transformation to the canonical basis
determines the energies and occupation probabilities of
single-nucleon states that correspond to the self-consistent
solution for the ground state of a nucleus. Since it diagonalizes
the density matrix, the canonical basis is always localized. It describes
both the bound states and the positive-energy single-particle
continuum. 

Taking into account the rotational invariance of the nuclear system,
the matrix equations of the RQRPA read~\cite{PRNV.03}:
\begin{equation}
\vspace{0.3cm}
\label{rrpaeq} \left(
\begin{array}{cc}
A^J & B^J \\
B^{^\ast J} & A^{^\ast J}
\end{array}
\right) \left(
\begin{array}{c}
X^{\nu,JM} \\
Y^{\nu,JM}
\end{array}
\right) =\omega_{\nu}\left(
\begin{array}{cc}
1 & 0 \\
0 & -1
\end{array}
\right) \left(
\begin{array}{c}
X^{\nu,JM} \\
Y^{\nu,JM}
\end{array}
\right)\; .
\vspace{0.3cm}
\end{equation}
For each RQRPA energy $\omega_{\nu}$, $X^{\nu}$ and $Y^{\nu}$ denote
the corresponding forward and backward two-quasiparticle
amplitudes, respectively. The coupled RQRPA matrices in the
canonical basis read
\begin{eqnarray}
A^{J}_{kk'll'} & = &
H^{11(J)}_{kl}\delta_{k'l'}-H^{11(J)}_{k'l}\delta_{kl'}
-H^{11(J)}_{kl'}\delta_{k'l}+H^{11(J)}_{k'l'}\delta_{kl}
\label{amat}\nonumber \\
& &
+\frac{1}{2}(\xi^{+}_{kk'}\xi^{+}_{ll'}+\xi^{-}_{kk'}\xi^{-}_{ll'})V^{J}_{kk'll'}
\nonumber \\
& & +\zeta_{kk'll'} \tilde{V}^{J}_{kl'k'l}
\\
B^{J}_{kk'll'} & = &
\frac{1}{2}(\xi^{+}_{kk'}\xi^{+}_{ll'}-\xi^{-}_{kk'}\xi^{-}_{ll'})V^{J}_{kk'll'}
\label{bmat}
\nonumber\\
& & +\zeta_{kk'll'}(-1)^{j_{l}-j_{l'}+J}\tilde{V}^{J}_{klk'l'}\; .
\end{eqnarray}
$H^{11}$ denotes the one-quasiparticle terms
\begin{eqnarray}
H^{11}_{kl} & = & (u_{k}u_{l}-v_{k}v_{l})h_{kl}-
(u_{k}v_{l}+v_{k}u_{l})\Delta_{kl}\; ,
\end{eqnarray}
i.e. the canonical RHB basis does not diagonalize either the Dirac
single-nucleon mean-field Hamiltonian $\hat h_D$, or the pairing
field $\hat\Delta$. The occupation amplitudes $v_{k}$ of the
canonical states are eigenvalues of the density matrix. $\tilde{V}$
and $V$ are the particle-hole and particle-particle residual
interactions, respectively. Their matrix elements are multiplied by
the pairing factors $\xi^{\pm}$ and $\zeta$, defined by the
occupation amplitudes of the canonical states. The relativistic
particle-hole interaction $\tilde{V}$ is defined by the same
effective Lagrangian density as the mean-field Dirac single-nucleon
Hamiltonian $\hat h_D$. $\tilde{V}$ includes the exchange of the
isoscalar scalar $\sigma$-meson, the isoscalar vector
$\omega$-meson, the isovector vector $\rho$-meson, and the
electromagnetic interaction. The two-body matrix elements include
also contributions from the spatial components of the vector fields.
The pairing factors read
\begin{eqnarray}
\zeta _{\kappa ^{{}}\kappa ^{\prime }\lambda ^{{}}\lambda ^{\prime
}}=\left\{
\begin{array}{ll}
\eta _{\kappa ^{{}}\kappa ^{\prime }}^{+}\eta _{\lambda ^{{}}\lambda
^{\prime }}^{+} & \textrm{for $\sigma$, $\omega ^{0}$
,$\rho^{0}$, $A^{0}$; if J is even} \\
& \textrm{for ${\bb\omega}$, ${\bb\rho}$, ${\bb A}$;
if J is odd} \\
\eta _{\kappa ^{{}}\kappa ^{\prime }}^{-}\eta _{\lambda ^{{}}\lambda
^{\prime }}^{-} & \textrm{for $\sigma$, $\omega^{0}$,
$\rho^{0}$, $A^{0}$; if J is odd} \\
& \textrm{for ${\bb\omega}$, ${\bb\rho}$, ${\bb A}$;
if J is even} 
\end{array}
\right.
\end{eqnarray}
with the $\eta$-coefficients defined by
\begin{eqnarray}
\eta_{kk'}^{\pm} & = & u_{k}v_{k'}\pm v_{k}u_{k'}\; ,
\end{eqnarray}
and
\begin{eqnarray}
\xi_{kk'}^{\pm} & = & u_{k}u_{k'}\mp v_{k}v_{k'}\; .
\end{eqnarray}
$\sigma$, $\omega ^{0}$,
$\rho^{0}$, and $A^{0}$ denote the time-like components, and
${\bb \omega}$, ${\bb\rho}$, ${\bb A}$ the spatial components
of the meson and photon fields, respectively.

The RQRPA configuration space must also include the Dirac sea of negative
energy states, i.e. pair-configurations formed from the fully or
partially occupied states of positive energy and the empty
negative-energy states from the Dirac sea. The inclusion of
configurations built from occupied positive-energy states and empty
negative-energy states is essential for current conservation and the
decoupling of spurious states, as well as for a
quantitative comparison with the experimental excitation energies of
giant resonances.

The RQRPA model is fully self-consistent: the same interactions, in
the particle-hole and particle-particle channels, are used both in
the RHB equations that determine the canonical quasiparticle basis,
and in the RQRPA equations.
The parameters of the effective interactions are
completely determined by the RHB calculations of ground-state
properties, and no additional adjustment is needed in the RQRPA
calculations. This is an essential feature of the RHB+RQRPA approach
and it ensures that RQRPA amplitudes do not contain spurious
components associated with the mixing of the nucleon number in the
RHB ground state, or with the
center-of-mass translational motion.


\subsection{Multipole Transition Strength and Transition Densities}
\label{trans}

We conclude this section with a brief overview of the operators 
associated with nuclear collective excitations, and provide a 
summary of basic definitions and useful relations.

In the simplest case the number of protons and neutrons does not change
when an external perturbation acts on the nucleus. The one-body operator
$F({\bf r})$ associated with the external field can then induce,  
even in a superfluid system, only $ph$ excitations. Later on we will 
also consider more general cases, including pairing vibrations and 
charge-exchange excitations.

The case which is best defined is the electromagnetic excitation of 
a nucleus with real photons. A detailed
discussion of this process can be found in many textbooks, for instance in 
Refs.~\cite{BB.94,borti98,harak01}. The multipole decomposition of the
photon field plane wave leads to terms which contain Bessel functions 
$j_J(kr)$, where $k$ is the photon momentum. In the cases of 
interest here the photon energy is of the order of MeV, and
$r$ can be the size of the nuclear radius, so that $kr\approx$ 
10$^{-2}$ and the Bessel function is approximated by ${(kr)^J\over
(2J+1)!!}$. In this approximation the dominant
part of the matrix element which corresponds to the 
electric multipole $J$, is proportional to
\begin{equation}\label{eq:elemul}
\int\ \rho({\bf r}) r^J Y_{JM}(\hat r)\ d{\bf r}
\end{equation}
where $\rho$ is the charge density. The transition amplitude 
is therefore the matrix element of the operator
\begin{equation}\label{eq:eleop}
\hat{F}_{JM}({\bf r})=\sum_{i=1}^Z e r_i^J Y_{JM}(\hat r_i)
\end{equation}
between the initial and the final state (the sum is over
protons). For states with good angular momentum, the response 
to the electric multipole transition operator is described 
by the reduced transition probability
\begin{eqnarray}
B(EJ, J_i \to J_f) & = & \frac{1}{2J_i+1}\big\vert \langle
f || \hat{F}_J || i \rangle \big\vert^2 \;.
\label{trgen}
\end{eqnarray}
This relation is, of course, also valid for magnetic multipole 
transition operators.

In many cases studies of excited states with
electromagnetic probes are not possible, either because forbidden by 
selection rules, like for example the isoscalar monopole 
resonance which has zero angular momentum and therefore 
cannot be excited by photons, or because
strongly hindered by nuclear excitations. Transitions
induced by the strong interaction are typically studied by means of 
hadron inelastic scattering. In analogy with the 
electromagnetic case, the strong field is described by a plane
wave and, in the limit of small momentum transfer, the nuclear 
multipole transition operator reads
\begin{equation}\label{eq:fisos}
\hat{F}_{JM}({\bf r})=\sum_{i=1}^A r_i^J Y_{JM}(\hat r_i)
\end{equation}
when neutrons and protons are excited in phase (isoscalar
excitation), or 
\begin{equation}\label{eq:fisov}
\hat{F}_{JM}({\bf r})=\sum_{i=1}^A r_i^J Y_{JM}(\hat r_i) \tau_z(i),
\end{equation}
when neutrons and protons are excited with opposite phases (isovector
excitations), and where $\tau_z(i)$ is the third component of the 
isospin operator. 

There is no compelling theoretical justification for the choice of 
the ``effective'' nuclear operators of Eqs. (\ref{eq:fisos}) and 
(\ref{eq:fisov}). In those cases when the transferred momentum is 
not small, excitations of ``overtones'' induced by the component of 
the operator proportional to $r_i^{J+2}$ are also observed. Nevertheless, 
it has been shown that properties of nuclear giant resonances can be 
consistently described with the excitation operators 
Eqs. (\ref{eq:fisos}) and (\ref{eq:fisov})~\cite{borti98,harak01}. 
Concerning the isospin degree of freedom, some nuclear reactions are 
selective and either isoscalar or isovector states are excited. 
This is the case, for instance, in $(\alpha,\alpha')$ inelastic 
scattering which predominantly excites isoscalar states. In other 
cases, like (p,p$^\prime$) inelastic  scattering, 
the reaction itself is not isospin selective but, 
if the target is a light nucleus with good isospin quantum number, 
then isoscalar and isovector excitations can be separated. 
In those nuclei where isospin is no longer a good quantum 
number, e.g. light neutron-rich nuclei far from stability, 
nuclear excitations will correspond to a mixture of isoscalar and 
isovector modes.

When excitations include the spin degree of freedom, the corresponding 
isoscalar and isovector multipole operators read 
\begin{equation}\label{eq:fisos_spin}
\hat{F}_{JM}({\bf r})=\sum_{i=1}^A r_i^L \left[ Y_{L}(\hat r_i) 
\otimes \vec\sigma(i) \right]_{JM} \;,
\end{equation}
and 
\begin{equation}\label{eq:fisov_spin}
\hat{F}_{JM}({\bf r})=\sum_{i=1}^A r_i^L \left[ Y_{L}(\hat r_i) 
\otimes \vec\sigma(i) \right]_{JM} \tau_z(i)\; ,
\end{equation}
respectively, with $J=L,L\pm 1$. 

The strength function $S(\omega)$ associated with the 
transition operator $\hat{F}$ is defined by Eq.(\ref{eq:strengthf1}).
%
%
It is often useful to consider moments of the strength function:
\begin{equation}\label{eq:moments}
m_k(\hat{F}) = \int\ (\hbar\omega)^k S(\omega) d(\hbar\omega) =
\sum_n E_n^k |\langle n | \hat{F} | 0\rangle|^2 \delta(\hbar\omega-E_n).
\end{equation}
The first moment, or the energy-weighted sum rule (EWSR), is very
important because its value equals the ground-state
expectation value of the double commutator $[\hat{F}, [\hat{H},\hat{F}] ]$, 
and therefore it can be evaluated without actually calculating the 
strength function. This is the well known Thouless theorem, 
and its non-trivial extension
to the HFB-QRPA case has been proven in Ref.~\cite{kh02}. 
The ratio $m_1/m_0$ is the quantity often compared with the 
experimental excitation energy of the corresponding resonance 
although, of course, this is only correct if there are no
multiple peaks within the energy interval over which the 
integration in Eq.~(\ref{eq:moments}) is performed. One often
finds the notation $E_0$ ($E_{-1}$) for $m_1/m_0$ 
($\sqrt{m_1/m_{-1}}$). 

When $\hat{F}$ represents the isovector electric dipole operator, and the
residual interaction does not include velocity-dependent 
or exchange terms, a very simple result is obtained for the EWSR:
\begin{equation}
m_1 = {2\pi^2e^2\hbar\over mc}{NZ\over A} \simeq 60 {NZ\over A}\ 
{\rm MeV}\cdot{\rm mb}.
\label{TRK}
\end{equation}
This is the well known Thomas-Reiche-Kuhn (TRK) sum rule, which
only includes the numbers of neutrons and protons, and is 
therefore completely model independent. For Skyrme and 
Gogny forces, which exhibit velocity-dependent and exchange terms, 
respectively, the value of $m_1$ in the above expression is multiplied 
by $(1+\kappa)$, where $\kappa$ is typically $\approx$ 
0.2-0.3. In the general case of a spin-independent 
isoscalar operator of multipole $L$, one can derive the 
following result:
\begin{equation}
m_1 = {\hbar^2\over 2m}{L(L+1)^2\over 4\pi} A \langle r^{(2L-2)} 
\rangle,
\end{equation}
where $\langle r^{(2L-2)} \rangle$ denotes the ground-state 
expectation value. 

Essential information on the dynamics of a nuclear 
collective mode is contained in the transition density.
For the state $|\nu \rangle$ this quantity is defined as 
the matrix element of the density operator:
\begin{eqnarray}
\delta\rho_{J}^{\nu}(\mathbf{r})  
=  \langle\nu | \sum_i \delta(\mathbf{r}-\mathbf{r}_i) | 0 \rangle\; ,
\label{trdens1}
\end{eqnarray}
where $| 0 \rangle$ denotes the ground state. 
The proton (neutron) transition density includes summation 
over protons (neutrons) in Eq.~(\ref{trdens1}), and 
the isoscalar $(T=0)$ and isovector $(T=1)$ transition densities
are defined by:
\begin{eqnarray}
\delta\rho_{J}^{T,\nu} 
= \delta\rho_{J}^{n,\nu} +(-1)^T \delta\rho_{J}^{p,\nu}.
\end{eqnarray} 
Assuming spherical symmetry, the transition density reads 
\begin{eqnarray}
\delta\rho^{n(p),\nu}_J(\mathbf{r}) & = & 
\delta\rho_{J}^{n(p),\nu}(r) Y^*_{JM}(\hat{r}),
\end{eqnarray}
where the radial factor
can be expressed in terms of the forward ($X$) and
backward ($Y$) QRPA amplitudes, and the single-nucleon radial wave functions 
multiplied by the corresponding occupation factors:
\begin{equation}\label{eq:td1}
\delta\rho_{J}^{n(p),\nu}({\bf r}) = \sum_{kl\in n(p)} \left( X_{kl}^{\nu}+Y_{kl}^{\nu} \right) 
\left( u_kv_l + v_ku_l \right) \langle k \vert\vert Y_J \vert\vert l \rangle 
R_k(r) R_l(r).
\end{equation}

\section{Beyond the Mean-Field Approximation}
\label{Sec3}

\subsection{Extensions of the (Q)RPA}\label{subsec_pvc}

The mean field theories which have been discussed in the 
previous Section, cannot provide a complete
description of the physical phenomena associated with 
collective nuclear excitations. The stationary models based
on effective interactions, e.g. the HF or HFB, certainly include 
a large amount of correlations in their phenomenological 
parameters, and predict nuclear binding energies with 
remarkable accuracy. However, these models fail to
reproduce the empirical features of single-particle 
levels around the Fermi surface~\cite{BBB,Mah.85}, 
which is essential for a complete description of 
collective excitations. 

Time-dependent theories, and their small amplitude limit such
as the RPA and QRPA, provide an accurate description of the 
two principal integral characteristics of giant resonances: 
the total strength $m_0$ and the energy-weighted sum rule $m_1$.
Accordingly, the calculated centroid energies $m_1/m_0$ will not 
be modified by including additional correlations. 
On the other hand, the second moment of the strength distribution, 
which in the case of a single peak is associated with its full width at 
half-maximum (FWHM), or other details of the response 
function, cannot be described at the level of a mean-field theory. 
(Q)RPA calculations do not reproduce the experimental
values of the total width of vibrational states.

At the (Q)RPA level, nuclear collective motion is represented
as a coherent superposition of $1p-1h$ (or two-quasiparticle)
states. The energy and angular momentum of
these vibrations can be released to other degrees of
freedom, because vibrational states are embedded in a
dense background of excited states. When the energy of
a vibrational state lies above the particle emission threshold, 
the state can decay by neutron or proton emission. This damping 
mechanism is associated with the escape width $\Gamma^\uparrow$,
which can be taken into account within the framework of 
continuum-(Q)RPA (see  Sec.~\ref{sec2_cqrpa}).
The spreading width $\Gamma^\downarrow$ 
arises because the energy and angular momentum of coherent
vibrations can be transferred to more complicated nuclear states,
of $2p-2h$ (and eventually $3p-3h$ $\ldots$ n$p$-n$h$) character.
In order to describe $\Gamma^\downarrow$ a theoretical framework 
must include the coupling to these complex configurations. 

A vast amount of data on widths of giant resonances in stable 
nuclei has been accumulated. Escape widths are large in light nuclei,
but become less important in medium-heavy and heavy systems,
where typical values are of the order of 1 MeV or less. 
On the other hand, spreading widths of several MeV 
are typical for resonances in heavier nuclei. For an
extensive review on this subject the reader is referred to
Ref.~\cite{BBB}. Much less is experimentally known 
about transition strengths of giant resonances in unstable 
nuclei, however, it has been pointed out that in neutron-rich 
nuclei far from stability spreading widths could be enhanced 
with respect to stable nuclei \cite{ghielmetti}. 

In the following we will briefly review several models which
go beyond the mean-field approximation in the description of 
collective excitation phenomena. 
Simple $1p-1h$, or two-quasiparticle states are 
coupled by the residual interaction to more complicated configurations. 
A straightforward approach to this problem would be the 
diagonalization of the effective Hamiltonian in the 
configuration space which includes at least the $2p-2h$, 
or four-quasiparticle states. This avenue, however, is
usually not feasible because the number of $2p-2h$ configurations
can become very large ($\approx$ 10$^2$ or 10$^3$ per MeV).
The second RPA (SRPA) is based on non-interacting $2p-2h$ 
configurations and, in addition to interaction terms acting in the 
$1p-1h$ space, only interactions between $2p-2h$ and $1p-1h$ 
configurations are explicitly taken into 
account. The SRPA retains basic properties of the RPA, e.g. 
the conservation of the EWSR \cite{Yan.87}. 
Practical implementations of this framework 
have been restricted to relatively light nuclei, and only
with radical truncations of the configuration space (for a
detailed discussion, cf.~\cite{Dro.90} and references therein). 
In addition, the current SRPA models are not self-consistent, 
e.g. in Ref.~\cite{Dro.90} the Woods-Saxon potential is used to 
determine the single-particle energies and wave functions, whereas 
the RPA residual interaction and terms that mix complex 
configurations are derived from a G-matrix.
A more complete SRPA description of collective excitations
necessitates a fully self-consistent implementation of the
effective interaction, both in the ground-state and in the 
construction of the SRPA matrix, and also the inclusion of 
contributions from complex configurations that have been
omitted in the current versions of the SRPA.

Another theoretical framework which takes into account the 
spreading width is based on the concept of particle-vibration 
coupling. Since the nucleus is a highly correlated
system, a nucleon which propagates in the nuclear medium can 
excite the whole nucleus. The low-lying nuclear
excitations correspond predominantly to surface vibrations,  
and theories which are based on particle-vibration coupling 
take into account the fact that the nucleons are confined inside 
the nucleus because of the surface, but at the same time they 
strongly couple to the dynamical fluctuations of the surface. 
\begin{figure}
\centering
\includegraphics[scale=0.7]{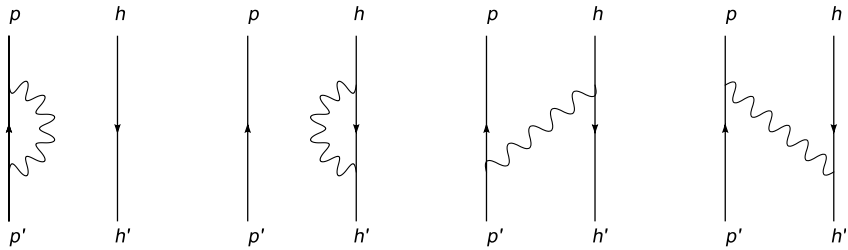}
\caption{
Diagrams which correspond to the coupling of the $p-h$ components 
of a giant resonance with phonon states. 
}
\label{diagrams1}
\end{figure}

The particle-vibration coupling concept lies at the basis of
Nuclear Field Theory (NFT)~\cite{BM.II}. At the lowest order
of NFT the nucleons and the collective vibrations are taken as
independent degrees of freedom, and nuclear dynamics is determined
from the hierarchy of their couplings. These couplings are described
by a (non-relativistic) field theory. Diagrams which correct for 
the violation of the Pauli principle and appear in lowest order, 
because vibrations are microscopically built from $p-h$ pairs, 
are also included. In the NFT framework the natural extension of the 
RPA is a model in which the $1p-1h$ configurations are 
coupled with states composed of $1p-1h$ pair plus a phonon 
(instead of $2p-2h$ states). This coupling is expressed by
the sum of the four diagrams depicted in Fig.~\ref{diagrams1}.

\begin{figure}
\centering
\includegraphics[scale=0.8]{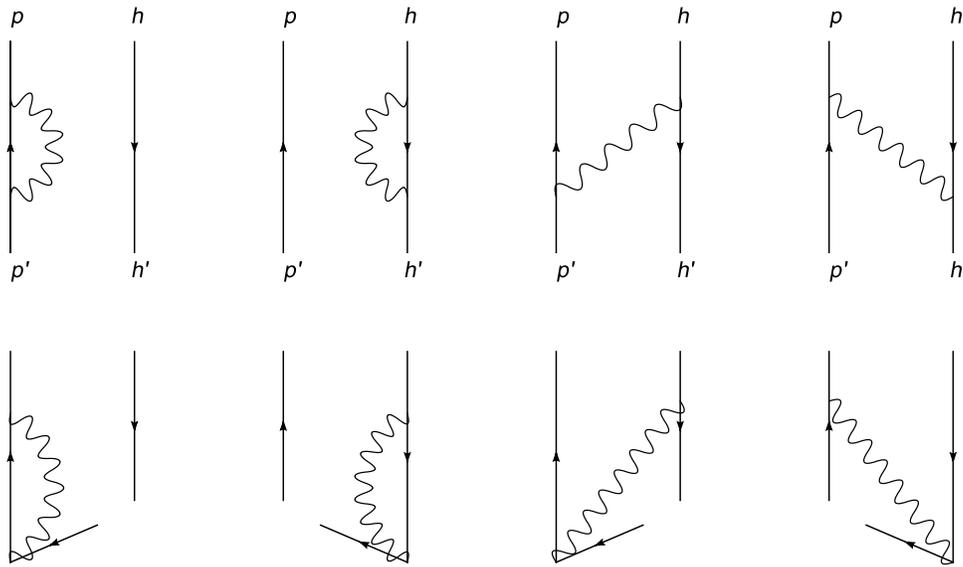}
\caption{Diagrams included in the model of Ref. \cite{kam.93}. The upper
row contains the same diagrams as Fig. \ref{diagrams1}, 
whereas the lower row displays the ``ground-state
correlations'' diagrams. 
}
\label{diagrams_k}
\end{figure}

The more recent Extended Theory of Finite Fermion Systems 
(ETFFS) \cite{kam.93, Kam.04} takes into account $1p-1h$, 
complex $1p-1h \otimes$phonon configurations, the single-particle 
continuum and ground-state correlations. The starting point is 
the exact Bethe-Salpeter equation for the $p-h$ Green's function, 
which is approximated in such a way that it basically corresponds
to the RPA solution, plus the contribution of the coupling of
$1p-1h$ configurations with phonon configurations. 
These couplings are depicted by the diagrams
in Fig.~\ref{diagrams_k}. The ETFFS approach 
includes diagrams associated with the 
presence of $1p-1h$ plus phonon components in the ground-state 
(bottom row of Fig.~\ref{diagrams_k}), i.e.  ``ground-state correlations''.

\begin{figure}
\centering
\includegraphics[scale=0.6]{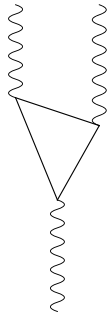}
\caption{Basic diagram which gives rise to the spreading width of 
one-phonon states in the QPM.
}
\label{diagrams2}
\end{figure}

Another theoretical framework based on the idea that the
interaction between nucleons and phonons determines
nuclear dynamics, is the Quasiparticle-Phonon Model  
(QPM) \cite{QPM,Ber.99}. Within the QPM the excited states
of even-even nuclei are treated as phonons built
from two-quasiparticle pairs, i.e. they correspond to
solutions of QRPA equations. The ground state is considered
as a phonon vacuum and the QRPA yields several collective phonons
of each multipolarity, as well as many non-collective
(or almost pure two-quasiparticle) solutions; for simplicity 
all these solutions are referred to as phonons. 
The main advantage of the QPM approach is that it accounts for
the coupling between simple (one-phonon) and complex (two- or
three-phonon) configurations, even in cases which are
numerically very demanding. The wave function of the excited 
states represents a combination of one-, two- and three-phonon
configurations,  with the one-phonon configurations corresponding to the 
set of all  QRPA solutions (of given multipolarity $J^\pi$). 
The two- and three-phonon configurations are composed of the 
phonons of different multipolarities coupled to the given $J^{\pi}$.
The model Hamiltonian is diagonalized in this basis, and the
result are the eigenvalues and the microscopic structure
of each excited state (i.e., the amplitudes of one-,
two- and three-phonon configurations in the wave function).
The NFT and QPM produce similar results. In the QPM the basic
process which gives rise to the spreading width of giant
resonances, is the coupling of one- and two-phonon configurations.
The leading diagram is depicted in Fig.~\ref{diagrams2}. It has
been shown that the largest contributions come
from configurations in which one of the two phonons 
in the intermediate state is non-collective. In this case
the equivalence with the NFT diagrams in Fig.~\ref{diagrams_k} 
can be demonstrated \cite{BBB}. 
The model Hamiltonian contains terms which correspond to 
the mean-field for protons and neutrons (with some simplified form 
of pairing in the case of open-shell nuclei), plus a residual effective
nucleon-nucleon interaction. In most applications the mean-field is 
a phenomenological Woods-Saxon potential $U$, and the residual 
nucleon-nucleon interaction has a separable form. A separable 
interaction is characterized either by 
a form factor $r^{L}$, or by the Bohr-Mottelson form factor
${\partial U\over \partial r}$, and is
used both in the (Q)RPA and as the particle-vibration coupling 
interaction: 
\begin{equation}
\hspace{-1.75cm}
H_{\rm PVC} \equiv \sum_{k,k';n,LM}
{\beta_{nL}\over \sqrt{2L+1}} 
\langle k \vert R_0 {\partial U\over \partial r} Y^*_{LM}
\vert k' \rangle \left[ \Gamma^\dagger_{nLM} +
(-)^{L+M}\Gamma_{nL-M} \right]
c^\dagger_k c_{k'},
\label{eq:deltaU}
\end{equation}
where $k$ and $k'$ ($n$, $\lambda$ and $\mu$) label single-nucleon states
(phonons). $R_0$ is the nuclear radius, and $\beta_{nL}$ is proportional
to the square root of the reduced transition probability Eq.~(\ref{trgen}),
i.e. it is a measure of the collectivity of the phonon state.

The time-dependent density-matrix (TDDM) model \cite{GT.90}  
is an extension of the time-dependent Hartree-Fock theory beyond 
the mean-field level. The model describes the time evolution of 
one-body and two-body density matrices and, therefore, includes the 
effects of both a mean-field potential and two-body correlations. 
TDDM has originally been formulated to describe large-amplitude 
collective motion, but it has also been applied in studies of small-amplitude 
oscillations, in particular low-energy excitations in unstable oxygen 
isotopes \cite{Toh.01}.  

Concluding, we emphasize that the basic advantage of 
mean-field models lies in the fact that they can be formulated in
a fully self-consistent way, and easily applied to nuclei
all over the periodic table. They reproduce  
global properties of nuclear collective excitations, 
but fail to describe specific 
phenomena in which particular orbitals around the Fermi
surface play a special role, or when the inclusion of 
damping mechanism becomes
essential. The models that we have described above extend 
the physical picture beyond the mean-field approximation. 
They can be considered microscopic in the sense that they are
based on rigorous formal schemes but, on the other hand, in most 
applications the particle-vibration coupling vertex is 
phenomenological. 


\subsection{Illustrative calculations}\label{nft}

%
There are, of course, many applications of the NFT theory~\cite{Bor.77}.
We would like to mention, in particular, a recent study of the effective 
pairing interaction which arises from the exchange of vibrations 
between nucleons close to the Fermi surface~\cite{indint}. 
Here we review another NFT-based approach to 
nuclear collective excitations, which uses Skyrme effective
interactions~\cite{col.early}. Starting from the self-consistent 
Skyrme HF-RPA, this model includes the couplings 
associated with the diagrams shown in
Fig.~\ref{diagrams1}. Both the phonons in the intermediate 
states, and the particle-vibration couplings
are consistently calculated using the same Skyrme force.
%

\begin{figure}
\centering
\includegraphics[scale=0.4,angle=0]{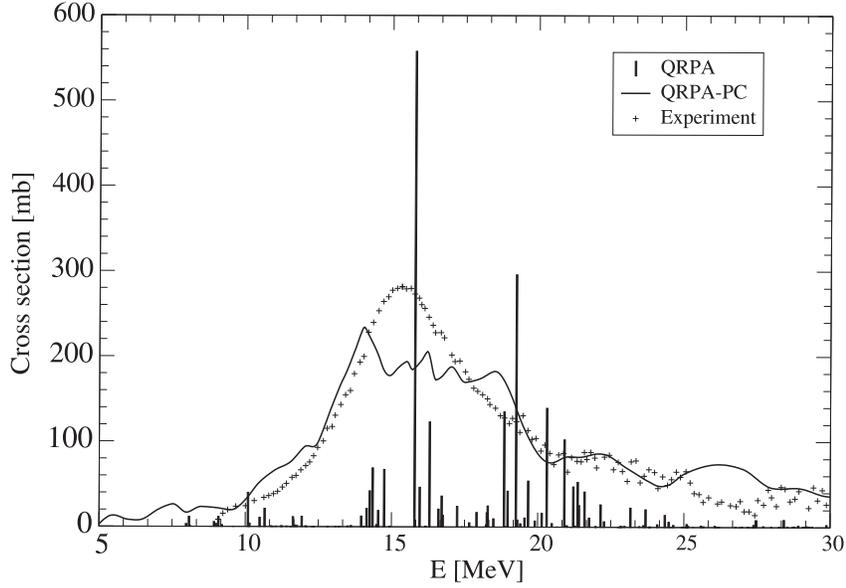}
\caption{Photoabsorbtion cross section for
$^{120}$Sn, calculated with the QRPA (vertical bars) and 
QRPA-PC (solid curve). The theoretical results are shown 
in comparison with experimental values.
}
\label{sn_2p2h}
\end{figure}

In the original formulation of the model, which does not include 
pairing correlations, the coupling with the one-particle 
continuum is considered. In this way both the escape and the 
spreading width of the vibrations are taken into account. In some
cases~\cite{Col.94} it has been possible to describe the
particle decay of giant resonances (i.e.  partial escape
widths to definite hole channels in the residual $A-1$ system). 
More recently this model has been extended to include pairing, 
but without considering the coupling to the continuum.
As an illustrative example, in Fig.~\ref{sn_2p2h} we display the
calculated IVGDR strength distribution in $^{120}$Sn, obtained with
the Skyrme force SIII. Both the QRPA discrete peaks (vertical bars),
and the result obtained with the inclusion of phonon coupling -- (Q)RPA-PC, 
are shown. We notice that the dipole strength is significantly redistributed
by the phonon coupling, but the position of the centroid energy $E_0$ 
does not change from the QRPA value. The (Q)RPA-PC result is in very good 
agreement with the experimental photoabsorbtion cross
section for $^{120}$Sn, which is also included in the figure. 
The most important QRPA-PC additional contributions originate 
from the coupling to the low-energy density vibrations with 
$J^\pi$ equal to 2$^+$, 3$^-$, 4$^+$ and 5$^-$. When these
phonons are collective, corrections associated with the violation
of the Pauli principle are less important. In addition, 
the coupling to giant resonances can also be taken into account, 
but this gives minor contributions. The reason can be
understood from the diagrams shown in Fig.~\ref{diagrams1}: 
the coupling to low-energy phonons is 
associated with smaller energy denominators.

%
In the QPM~\cite{Ber.99} the model Hamiltonian
is composed of terms corresponding to the mean-fields for
protons and neutrons, a monopole pairing, and a residual 
nucleon-nucleon interaction.
The mean field is a phenomenological
Woods-Saxon potential $U$. The residual nucleon-nucleon
interaction is separable, and can be characterized either by
a form factor $r^{\lambda}$, or by a Bohr-Mottelson form factor.
The strength parameters of the residual interaction are
adjusted in each particular nucleus to reproduce the energy
position and the $B(E\lambda)$ value of the low-lying $2^+_1$
and $3^-_1$ levels. The spurious $1^-$ state is excluded
from the excitation spectra by adjusting the isoscalar
strength of the residual interaction for $\lambda^{\pi}=1^-$,
so that this state has zero energy. 


\begin{figure}
\centering
\includegraphics[scale=0.5]{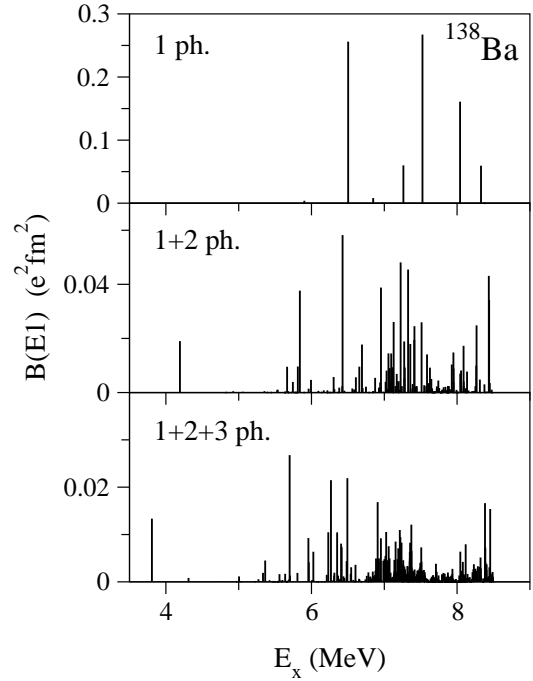}
\caption{
Fragmentation of the low-lying electric dipole strength
in $^{138}$Ba.
Calculations are performed
in the one-phonon approximation (top panel), and taking into 
account the coupling to two-phonon  
configurations (middle panel), or to two- and three-phonon
configurations
(bottom panel).}
\label{qpmpygmy}
\end{figure}

Here we present an example of the fragmentation of
one-phonon configurations in QPM calculations~\cite{pono}.
In Fig.~\ref{qpmpygmy} the low-energy portion of the  
electric dipole transition strength of $^{138}$Ba is plotted 
for three QPM calculations. In the upper panel
the $B(E1)$ strength distribution corresponds to the one-phonon
approximation, and only five QRPA states carry visible $E1$ strength. 
When the coupling to two-phonon
configurations is included, the $E1$ strength becomes
strongly fragmented, and is distributed over a hundred of 
states (middle panel). The fragmentation becomes
even stronger when three-phonon configurations are included in
the wave function (bottom panel). The $B(E1)$'s of the 
strongest $1^-$ states in the bottom part of Fig.~\ref{qpmpygmy},
are in quantitative agreement with the experimental values \cite{Zil.02}. 
In the calculation presented in the bottom panel all two- and three-phonon
configurations with an energy below 8.5 MeV are included.
These configurations include phonons with multipolarity from 
$1^{\pm}$ to $9^{\pm}$, and their total number
is about 1200. At higher energies the density of complex
configurations increases rapidly, and for a feasible calculation
truncation become necessary.

\section{Low-Energy Electric Dipole Strength}
\label{Sec4}

\subsection{Low-Energy Response of Light Nuclei}
\label{dipole_light}

Evidence of unusually strong dipole response at low-energy in light  
nuclei was first reported in fragmentation experiments of halo nuclei
on heavy targets with a large number of protons. In the reaction 
$^{208}$Pb($^{11}$Li,$^9$Li) at 800 MeV/nucleon~\cite{Kob89} a 
large Coulomb excitation cross section of $0.9~b$ 
was extracted (see also~\cite{Han87}). This cross section 
is associated with a large peak in the $B(E1)$ distribution that
appears around the two-neutron separation energy, which is of the 
order of 300 keV. Such a small value of the separation energy
could be correlated with the enhancement of the low-lying 
dipole strength. Namely, the wave functions of the two weakly bound 
neutrons which form the halo structure are extended far beyond 
the $^9$Li core. As a result, the excitation of the halo 
neutrons to the continuum is intensified, and the low-lying 
dipole response is decoupled from the IV GDR. Initial 
calculations could reproduce this phenomenon only qualitatively. 
In Ref.~\cite{Ber90} 
for instance, a simple RPA calculation was 
performed on top of a Woods-Saxon mean-field 
potential, whose parameters were adjusted in such a way that the 
$p_{1/2}$ neutron state was located at $-0.2$ MeV. Such an 
approximation is not satisfactory, of course, because recent experiments
have shown that, in addition to the $(p_{1/2})^2$ orbital, 
the two halo neutrons outside the $^9$Li core distribute their 
spectroscopic amplitudes between the $(s_{1/2})^2$ and
$(d_{5/2})^2$ configurations~\cite{spec_Li}.

\begin{figure}[t]
\begin{center}
\includegraphics[scale=0.5]{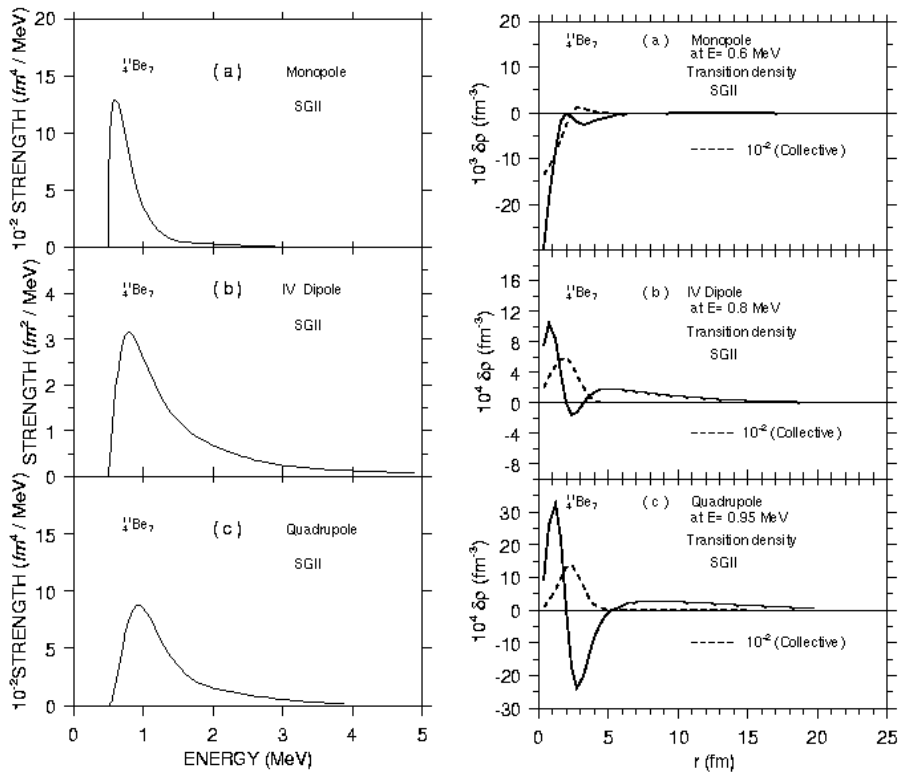}
\vspace*{0.2cm}
\caption{Left panel: low-energy isoscalar monopole, isovector 
dipole and isoscalar quadrupole strength distributions in $^{11}$Be,
calculated with the self-consistent continuum RPA~\protect\cite{Ham99}. 
For the peak states the radial transition
densities are shown in the panel on the right, and  
compared with collective transition densities.}
\label{11be}
\end{center}
\end{figure}

The enhancement of the low-lying multipole strength is expected to be 
a general phenomenon in nuclear systems characterized by small 
values of particle (e.g. neutrons) separation energies. For a 
simple spherical potential well, the $1p-1h$ transition strength 
can be evaluated analytically, and it has been shown that weakly-bound 
particles produce a non-resonant concentration of strength 
just at the threshold energy~\cite{Cat96}. The origin of this
strength is in the optimal matching of the continuum wave functions 
with the tail of the wave functions of the outermost weakly-bound 
orbitals (orbitals are said to be weakly bound when
the corresponding value of $k \equiv \sqrt{2m\vert \varepsilon 
\vert}/\hbar$, associated to the single-particle energy $\varepsilon$,  
is much smaller than the inverse of the nuclear radius). 
Such a simple model could explain the experimentally observed
``threshold effect'' in the response of one-neutron halo nuclei,
like $^{11}$Be~\cite{Nak94}. The occurrence of threshold strength 
is also predicted by more realistic continuum 
RPA calculations. For instance, RPA transition densities 
and currents in $^{11}$Be have been studied in Ref.~\cite{Ham99}.
In Fig.~\ref{11be} we plot the strength distributions and 
transition densities for the lowest isoscalar monopole, 
isovector dipole and isoscalar quadrupole modes.
The transition densities differ from the prediction
of the macroscopic model, not only because of the long tail
associated with the excitations of the weakly-bound neutrons, 
but also due to the node of the 2s$_{1/2}$
radial function. Macroscopic models, by definition,
do not include single-particle shell structures. 
For $^{11}$Li, $^{11}$Be, and other light 
nuclei that we mention below, the term 
"threshold effect" probably applies much better to the phenomenon 
under study than "soft mode". The latter expression is
generic, and has also been used for low-lying states 
in stable nuclei, which absorb only a tiny fraction of the 
total strength for a given multipole. Even though the occurrence
of pronounced threshold strength can be expected also for 
other multipoles when the separation energies become small, 
in Ref.~\cite{Ber92} it has been pointed out 
that the contributions of $L\neq$ 1 to the cross 
section are negligible in experiments with Coulomb excitations 
on high-Z targets. 

An interesting question is whether the response of a two-neutron 
halo system like $^{11}$Li, is very different from that of a one-neutron
halo, e.g. $^{11}$Be. Two-neutron correlations are,  
of course, essential for the binding of $^{11}$Li, considering that
$^{10}$Li is not a bound system. 
In Ref.~\cite{Ber91} the ground state 
and the dipole response of $^{11}$Li were studied 
by solving a Hamiltonian that includes a Woods-Saxon potential
for the two neutrons outside the structureless $^{9}$Li core, 
plus a density-dependent zero-range pairing force. 
With a careful treatment of the continuum~\cite{Ber_II}, 
it was shown that in this model the dipole response
of the correlated system differs from the free response by 
15\%-20\%. In the description of the dipole response the 
recoil of the $^9$Li was not taken into account. This term
has been shown to play a role, however, in the quantitative
description of the ground state of $^{11}$Li~\cite{Ber_III}.
More generally, one expects that not only the recoil, 
but also the polarization of the $^9$Li core is important for 
a quantitative description of the dynamical response~\cite{barranco_two}. 

Electromagnetic dissociation measurements have been performed for 
a series of neutron halo nuclei: $^6$He~\cite{6he_3}, 
$^8$He~\cite{8he_4}, $^{11}$Li~\cite{11li_5,Shi.95,Naka.06},
$^{12}$Be~\cite{12be_7}, $^{14}$Be~\cite{14be_8}, 
$^{19}$C~\cite{19c_9}, and for the proton halo nucleus 
$^{8}$B~\cite{8b_10}. Appreciable E1 strength is generally found 
already at low excitation energy, far below the domain of the GDR.
For $^{11}$Li, in particular, the $E1$ strength observed below 4 MeV 
excitation energy corresponds to $\approx 8\%$ of the TRK 
sum rule, and can be decomposed into at least two broad structures 
with peak energies at $\approx 1.0$ MeV and at $\approx 2.4$ MeV. 
In a $(p,p^\prime)$ experiment, the corresponding 
peaks have been observed at 1.3 and 2.9 MeV~\cite{Li_pp}. 
A long-standing issue in $^{11}$Li is the description of the 
low-energy E1 excitations in terms of its halo structure and, 
particularly, the question of whether the two halo neutrons are 
subject to strong correlations and eventually form a dineutron cluster. 
While several early experiments \cite{11li_5,Shi.95} did not find evidence
for strong correlations between the halo neutrons, a new and significantly 
improved measurement of the low-lying B(E1) distribution \cite{Naka.06} 
has revealed a strong low-energy E1 excitation peaked at 
$E_x = 0.6$ MeV with B(E1) = 4.5(6) Weisskopf units, which is 
the largest low-lying E1 strength ever observed in nuclei, and 
the B(E1) distribution could only be reproduced by a three-body 
model with a pronounced two-neutron correlation.

The structure and excitations of these light systems are best 
described in a shell-model approach, and calculations have been 
reported~\cite{sm_light} which include model spaces of 
2$\hbar\omega$, or even 3$\hbar\omega$ configurations. 
Many more transition amplitudes are included than in 
ordinary mean-field, e.g. RPA calculations, and the coherence 
between these amplitudes is crucial in enhancing the dipole
strength at low energy. In order to obtain
realistic results for the $B(E1)$ values, extended 
Woods-Saxon single-particle wave functions have to be used in 
calculations, adjusted in such a way that separation energies 
reproduce the experimental values. This procedure introduces 
spurious components that have to be carefully removed from the 
wave functions.  

\begin{figure}[]
\begin{center}
\includegraphics[scale=0.6]{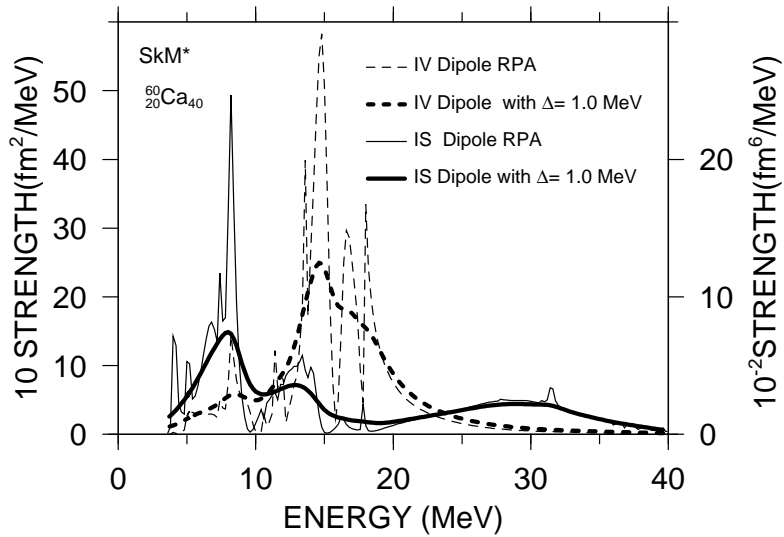}
\vspace*{0.2cm}
\caption{Continuum RPA IS and IV dipole strength functions in the 
nucleus $^{60}$Ca. The scale of the IS dipole strength is shown on 
the right-hand side, whereas on the left we plot the scale
of the IV dipole strength. The thick curves are obtained
by averaging the calculated RPA strength with Lorentzian functions
of 1 MeV width.}
\label{ca60}
\end{center}
\end{figure}

Let us also mention that the continuum RPA model with Skyrme 
interactions has been used in studies of the isovector and 
isoscalar dipole response in $^{34}$Ca, $^{28}$O, $^{60}$Ca and 
$^{22}$C~\cite{HSZ.98}. These nuclei could be considered 
as benchmark systems for the dipole response at the 
nucleon drip-line. It has been shown that the low-lying 
neutron excitations are characterized by the mixing of 
isovector and isoscalar modes, and that the threshold 
strength is predominantly of isoscalar nature. 
For the nucleus $^{60}$Ca the dipole strength functions
calculated with the SkM* force are shown in Fig.~\ref{ca60}. 
The isoscalar strength  found at high 
excitation energies above the IV GDR can be associated 
with the dipole compressional mode. 
Such extremely neutron-rich systems are,
of course, not yet accessible in experiments, but the 
calculation nicely illustrates the features of the dipole 
response at the limit of neutron binding. 

\subsection{Low-Energy Dipole Excitations in Oxygen Isotopes}\label{low_oxy_sec}

In the oxygen isotopic chain the drip-line nucleus $^{24}$O is 
located only eight neutrons away from the stable isotope $^{16}$O 
and, therefore, the evolution of the low-energy dipole strength can 
in principle be traced up to the neutron drip-line. The possible 
occurrence of low-lying collective dipole strength in neutron-rich 
oxygen isotopes has attracted considerable experimental and theoretical 
interest in recent years. 

While in the stable nucleus $^{16}$O earlier 
experiments~\cite{Ful.85,Ves.74,Ber2.75} did not find any appreciable 
low-lying dipole strength, more recent electromagnetic excitation 
experiments (at beam energies around 600 MeV/nucleon) 
have confirmed the expected occurrence of low-energy dipole strength in the
isotopes $^{18-22}$O~\cite{Aum.99,Lei.01}.
Low-energy transition strength has been observed in all 
neutron-rich oxygen isotopes that were studied, 
exhausting up to 12\% of the TRK sum rule (\ref{TRK}) at excitation energies
below 15 MeV, i.e., below the region of giant resonances.
The low-energy $E1$ structure in $^{18}$O and $^{20}$O was also 
studied by intermediate-energy Coulomb excitation at 100 MeV/nucleon,
and new low-energy dipole states were found 
in $^{20}$O~\cite{Try.01,Try.02,Try.03}. The dynamics of these 
low-lying dipole excitations, however, has not been resolved 
in experiments. In particular, it is not clear
whether some of these states correspond to a collective 
soft mode, or they all simply result from incoherent 
single-particle excitations. 

Several modern theoretical approaches have recently been employed 
in the description of the evolution of low-lying dipole 
strength in oxygen isotopes. In one of the first studies,  
large-scale shell-model calculations were performed for 
neutron-rich oxygen isotopes with up to $3\hbar\omega$ excitations
in the 0p-1s0d-1p0f model space. Pronounced low-lying dipole 
transition strength below 15 MeV was predicted for
$^{17}$O, $^{18}$O, $^{20}$O and $^{22}$O, 
exhausting $\approx$ 10\% of the TRK sum-rule~\cite{Sag.99,Sag.01b}. 
The calculated low-lying strength in $^{17}$O and $^{18}$O was 
found to be consistent with the experimental photoreaction cross sections.
The continuum QRPA on top of the coordinate-space
HFB~\cite{ma01,Mat.02}, with a Woods-Saxon single-particle
potential, a Skyrme force as the residual interaction in
the $ph$ channel, and a density-dependent $\delta$-force in the 
pairing channel, has also been employed to analyse the low-energy 
modes in oxygen isotopes. Pairing is not very strong in these
isotopes, in particular the pairing gap is considerably below the 
empirical 12/$\sqrt{\rm A}$ estimate in $^{22,24}$O. However, since 
the residual pairing interaction in QRPA generates
dynamical correlation effects on the response function through pair density
fluctuations, and therefore provides a contribution to the
low-lying multipole strength, it is important to include 
a consistent treatment of pairing correlations 
within the HFB+QRPA framework. Moreover, the energy weighted sum rules are 
fulfilled only if the pairing interaction is consistently included both 
in the solution for the stationary ground state, and in the dynamical linear 
response~\cite{ma01,PRNV.03}. In Ref.~\cite{PRNV.03} the self-consistent 
RHB+RQRPA has been applied in the study of multipole excitations of
neutron-rich oxygen isotopes and, in particular, in the analysis of the
evolution of the low-lying isovector dipole strength. 

The overall picture emerging from all these calculations is that the onset 
of dipole strength in the low-energy region is caused by nonresonant 
independent single-particle excitations of the last bound neutrons. 
This is similar to the case of light nuclei discussed 
in the previous subsection. The difference, however, is that 
for the oxygen isotopes the neutron separation energies are 
larger, i.e. 3.61 MeV for $^{24}$O, 
and thus the low-lying strength is much less pronounced than
for the threshold effect in light systems. 

\begin{figure}[]
\vspace*{1.5cm}
\begin{center}
\includegraphics[scale=0.5]{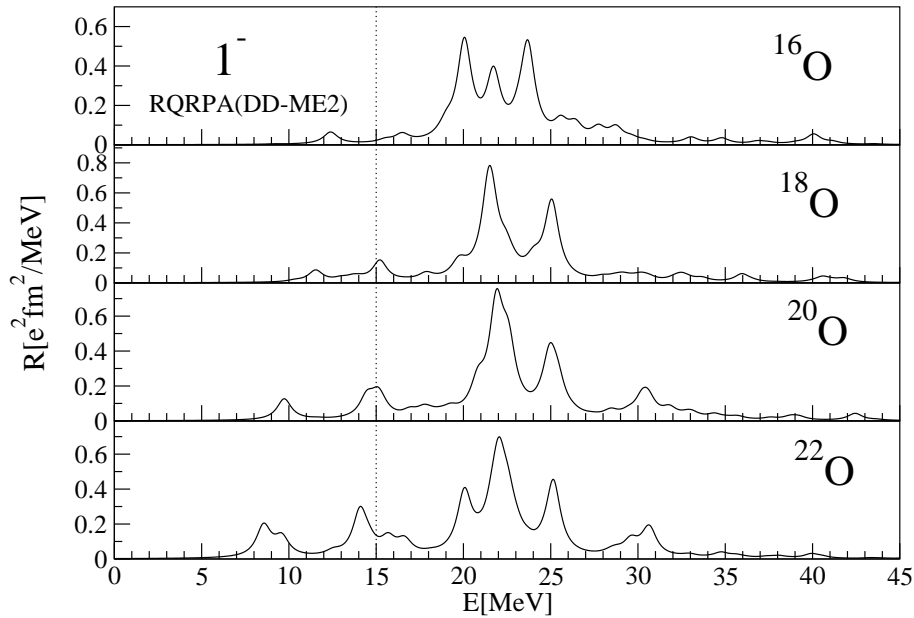}
\vspace*{0.2cm}
\caption{Evolution of the isovector dipole strength distribution in oxygen
isotopes, calculated within the RHB+RQRPA model using the 
DD-ME2 effective interaction.}
\label{oxevol}
\end{center}
\end{figure}

In order to illustrate the evolution of low-lying dipole strength 
along the chain of neutron-rich oxygen isotopes, we show the results 
of the self-consistent RHB+RQRPA calculation~\cite{PRNV.03}, 
based on the density-dependent 
effective interaction DD-ME2~\cite{LNVR.05} plus the Gogny D1S 
force in the pairing channel. The strength distributions associated 
with the dipole operator Eq. (\ref{eq:fisov}) are displayed 
in Fig.~\ref{oxevol}. With the increase of the 
number of neutrons one finds a pronounced fragmentation of the dipole
strength, and low-lying strength appears below 15 MeV.
The centroid energy of the low-lying $E1$ states is lowered with the 
increase of neutron excess, whereas the total strength is enhanced.

\begin{figure}[]
\vspace*{1cm}
\begin{center}
\includegraphics[scale=0.4,angle=-90]{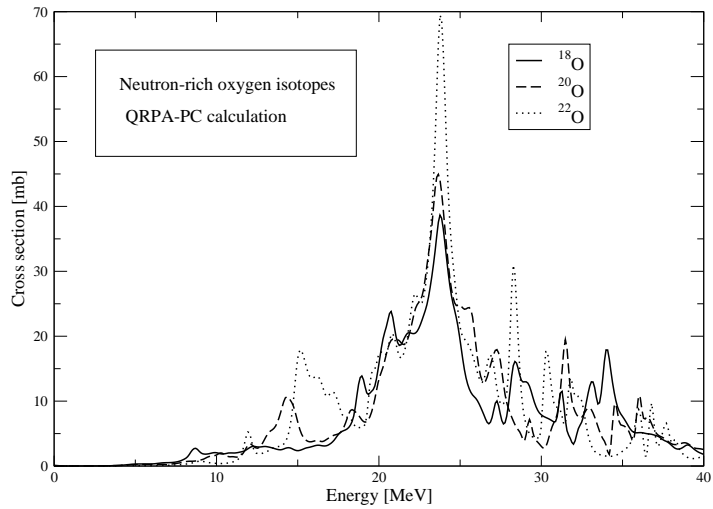}
\vspace*{0.2cm}
\caption{Total photoabsorbtion cross section for the isotopes
$^{18,20,22}$O, calculated using the full QRPA plus phonon
coupling~\protect\cite{CB.01}.}
\label{oxyphon}
\end{center}
\end{figure}

The electric dipole strength distributions in $^{18}$O, $^{20}$O, 
and $^{22}$O have also been analyzed in calculations which go beyond 
the mean-field level by including the coupling of single-quasiparticle 
states to vibrational modes~\cite{CB.01}. By employing the QRPA-PC 
model with up to four-quasiparticle configurations (two uncorrelated 
quasiparticles plus a collective phonon), it has been shown that
the calculated total photoabsorption cross section below 15 MeV 
is in very good agreement with experiment. While the simple QRPA 
analyses predict values which are systematically 
below the data, the coupling with phonons increases the cross 
section in the low-energy region. Because of the repulsion 
between the simple two-quasiparticle states and 
the complex configurations that include a phonon, the former 
are shifted to lower energy and this increases the total 
QRPA strength in the low-energy region. The QRPA-PC 
photoabsorption cross sections are shown in Fig.~\ref{oxyphon}.
We note that the calculation predicts the spreading widths, 
both for the low-energy dipole strength and for the 
giant dipole resonance.   

The QRPA-PC analysis of Ref.~\cite{CB.01} is self-consistent, in 
the sense that the only input is the Skyrme force which determines  
the ground state, and no further adjustment of parameters 
is necessary in the calculation of the response function. A more 
phenomenological model which also emphasizes 
the role of phonon coupling is the quasiparticle representation 
of the phonon damping model~\cite{Dang.01}. In Ref.~\cite{Toh.01}
the time-dependent density-matrix (TDDM) model, which is an extension 
of the time-dependent Hartree-Fock theory beyond the mean-field 
level, was used to calculate the isovector 
dipole strength functions of the even-A isotopes $^{18-24}$O. 
By adjusting the strength of the residual
interaction, the observed isotopic dependence of 
low-lying dipole strength was reproduced. 

In Table~\ref{tableo} the predictions of several
models for the low-lying $E1$ strength in $^{18-24}$O are 
summarized and compared with data~\cite{Lei.01,Woo.79}. The sum 
of the energy-weighted $E1$ transition strength below 15 MeV is 
given in units of the classical TRK sum rule. Even though all  
models agree on the overall effect of the neutron excess
on the E1 transition strength, significant differences can be noted 
in isotopes close to the drip-line. In particular, the inclusion
of particle-vibration coupling brings the results in closer 
agreement with experiment. 
\begin{table}[h]
\centering
\begin{tabular}{lcccc} \hline
A         &  18  &  20  &  22  & 24 \\ \hline\hline
Shell model~\cite{Sag.01b}     &   0.06      &    0.11 &  0.10 & 0.09  \\ \hline
continuum QRPA~\cite{Mat.02}    &   0.07      &   0.11 &  0.16 & 0.21  \\ \hline
QRPA-PC~\cite{CB.01}    &   0.07      &    0.09 &
0.07 &    \\ \hline
RHB + RQRPA(DD-ME2)             &   0.04      &    0.06 &  0.15 & 0.18  \\ \hline
Exp.~\cite{Lei.01}    &   0.08  &  0.12  &  0.07  &    \\ \hline
Exp.~\cite{Woo.79}       &   0.11  &        &        &    \\ \hline
\end{tabular}
\caption{Sum of the energy-weighted dipole strength for $^{18-24}$O up to
15 MeV excitation energy, in units of the TRK sum rule.}
\label{tableo}
\end{table}

\begin{figure}[]
\vspace*{1cm}
\centering
\includegraphics[scale=0.50]{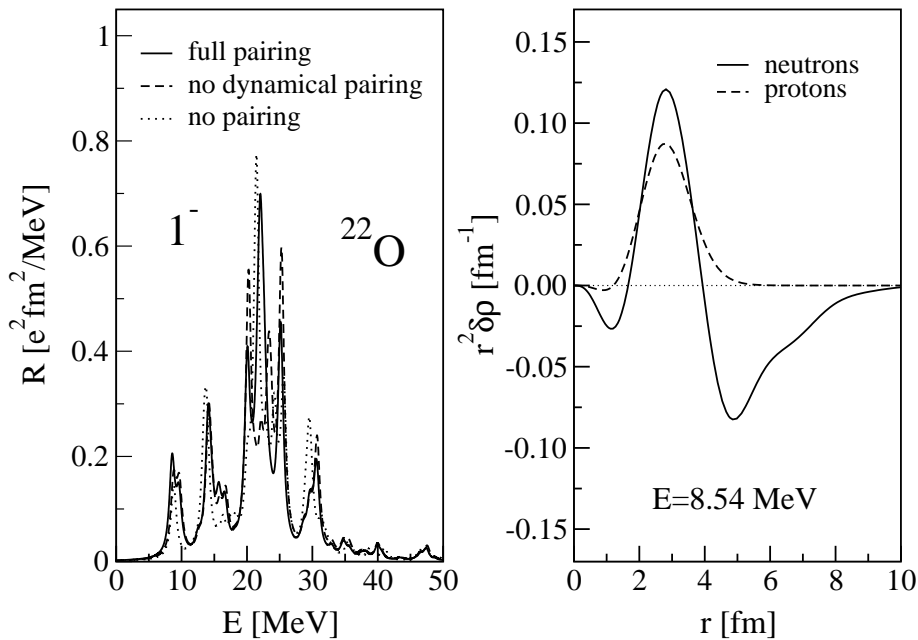}
\caption{The strength function of the IV dipole operator in
$^{22}$O (left). The fully self-consistent RHB+RQRPA response
(solid line) is compared with the RMF+RRPA calculation without
pairing (dotted line), and with the RHB+RRPA calculation which
includes pairing correlations only in the ground state (dashed
line). The proton and neutron transition densities for the peak at
$E=8.54$ MeV are shown in the right panel.}
\label{figO22dip}
\end{figure}

The role of dynamical pairing correlations is illustrated 
in the example of $^{22}$O. 
The RHB+RQRPA isovector dipole transition strength functions
are plotted in the left panel of Fig.~\ref{figO22dip}
for three different calculations: a) the RMF+RRPA calculation
without pairing, b) pairing correlations included in the RHB calculation
of the ground state, but not in the RQRPA residual interaction (no dynamical
pairing), and c) the fully self-consistent RHB+RQRPA calculation.
The residual pairing interaction in the 
RQRPA generates pronounced dynamical correlation effects on the 
responses through pair density fluctuations. Moreover, the energy-weighted 
sum rules are only satisfied if the pairing interaction is consistently 
included both in the static RHB and in the dynamical linear response.
Pairing is, of course, particularly important for the low-lying
strength. The inclusion of pairing correlations in the full
RHB+RQRPA calculation enhances the low-energy dipole strength near the
threshold~\cite{ma01,PRNV.03}.

For the main peak in the low-energy region ($\approx$ 8.54 MeV),
in the right panel of Fig.~\ref{figO22dip} we display the proton and neutron
transition densities. In contrast to the well known radial dependence of the
IVGDR transition densities (proton and neutron densities oscillate with
opposite phases, the amplitude of the isovector transition density is much
larger than that of the isoscalar component), the proton and neutron
transition densities for the main low-energy peak are in phase in the
nuclear interior, there is no contribution from the protons in the surface
region, the isoscalar transition density dominates over the isovector one in
the interior, and the strong neutron transition density displays a long tail
in the radial coordinate. However, a detailed analysis of R(Q)RPA amplitudes 
associated with the low-lying states in oxygen
isotopes indicated that they originate mainly from the single-nucleon
transitions from the loosely bound neutron orbits~\cite{PRNV.03,Vrepyg2.01}.
Similar results have been obtained with Skyrme (Q)RPA calculations. 
For instance, in the study of $^{28}$O performed 
with the self-consistent Hartree-Fock-Skyrme plus RPA, it was shown 
that the strength around the threshold originates essentially 
from uncorrelated excitations of neutrons with small binding 
energies~\cite{HSZ.98}. Even though in open-shell nuclei the number of 
partially occupied configurations increases because of the smearing of 
the Fermi surface, both non-relativistic and relativistic 
QRPA calculations do not predict the occurrence
of pronounced collectivity for the low-lying dipole states 
in neutron-rich oxygen isotopes.

\subsection{Pygmy Dipole Resonances in Heavier Neutron-Rich Nuclei}
\label{nepygmy}
Medium-heavy and heavy neutron-rich isotopes are characterized by the
appearance of a neutron skin, i.e. a layer of excess neutrons on 
the nuclear surface~\cite{Fuk.93,HaZh.95,Miz.00,Amo.04}. 
When approaching the neutron drip-line, in particular, the large 
proton -- neutron asymmetry leads to a pronounced 
difference between the corresponding Fermi energies, and
neutron orbitals just above the Fermi surface can become
unbound. The radial wave functions of very weakly-bound or unbound 
neutron states are extended far beyond the nuclear surface 
and this results in the formation of diffuse surface neutron density distributions: 
skin and halo structures. Experimental evidence for the formation of 
neutron skin is available from antiproton absorption~\cite{Trz.01}, 
heavy-ion reaction cross sections~\cite{Oza.01}, and from  
studies of isovector dipole and spin-dipole resonances~\cite{Kra.01}. 
Estimates about the size of the neutron skin can be deduced 
from experimental radii of charge distributions~\cite{Who.81,MCC.86} 
and mirror displacement energies~\cite{Duf.02}.

The question whether the excess neutrons in the skin can be excited 
to perform collective oscillations against the
rest of the nucleus, or they only contribute to the non-collective
threshold strength, has attracted considerable interest in recent years. 
In the former case one expects that, because the outer neutron orbitals 
are weakly bound, the resulting dipole mode will be rather soft,
i.e. its excitation energy will be far below the giant resonance region.
From the theoretical point of view, such a mode also provides a unique test
of the isospin-dependent components of effective nuclear interactions, 
which are particularly pronounced in nuclei with a large proton -- neutron 
asymmetry. Besides being intrinsically interesting as an exotic mode of 
excitation, the occurrence of low-lying dipole strength plays an important 
role in predictions of neutron capture rates
in the r-process nucleosynthesis, and consequently in the calculated elemental
 abundance distribution. Namely, although its  
transition strength is small compared to the total dipole strength, 
the low-lying collective dipole state located close to the neutron 
threshold can significantly enhance the radiative neutron capture 
cross section on neutron-rich nuclei, as shown in recent 
large-scale QRPA calculations~\cite{Gor.02,Gor.04}. This issue will be 
discussed in more details in Sec.~\ref{ssec_isopygmy}.

The possible occurrence of a soft dipole mode, or Pygmy Dipole Resonance (PDR) 
in neutron-rich nuclei, has been analyzd using a variety of theoretical 
approaches. Early studies of the PDR were based on
rather simple hydrodynamical models which involve classical oscillations
of the nucleon fluids. These include the three-fluid (protons, 
neutrons in the same orbitals as protons, and excess neutrons)
hydrodynamical model~\cite{Moh.71}, the Steinwedel-Jensen~\cite{Suz.90} 
and Goldhaber-Teller models~\cite{Isa.92}. 
The low-lying mode has been qualitatively described as a collective oscillation 
of the neutron enriched surface layer against the core nucleons. It was also 
suggested, however, that the PDR could arise in nuclear systems with 
only moderate neutron excess, for instance in Ca isotopes \cite{Cha.94}.

More recently, microscopic calculations based on Skyrme effective
interactions have been employed in studies of the isovector
dipole response in neutron-rich nuclei:
the Hartree-Fock+RPA~\cite{Cat.97,Vit.98,Rei.99}, the
continuum RPA approaches~\cite{HSZ.98,Ham.96,HSZ.96,Kam.98,SE.01,Ham.02}, 
and the self-consistent HFB+QRPA framework
formulated in the canonical basis \cite{Ter.05,TE.06}. 
By employing the continuum RPA, rather large escape widths for direct 
neutron decay from low-energy dipole states were estimated, implying
a pronounced coupling to the continuum~\cite{Lei.04}.
The results of these studies can be summarized as follows: 
(a) the dipole strength distributions in neutron-rich nuclei 
are more fragmented than in stable nuclei; 
(b) the centroids are calculated at significantly lower energies; 
(c) the ratio of neutron to proton particle-hole amplitudes of low-lying 
dipole states is much higher than in stable nuclei and, accordingly, 
the isoscalar (IS) transition densities do not vanish 
and isoscalar probes can excite these states. 

\begin{figure}[]
\centering
\vspace*{1cm}
\includegraphics[scale=0.4]{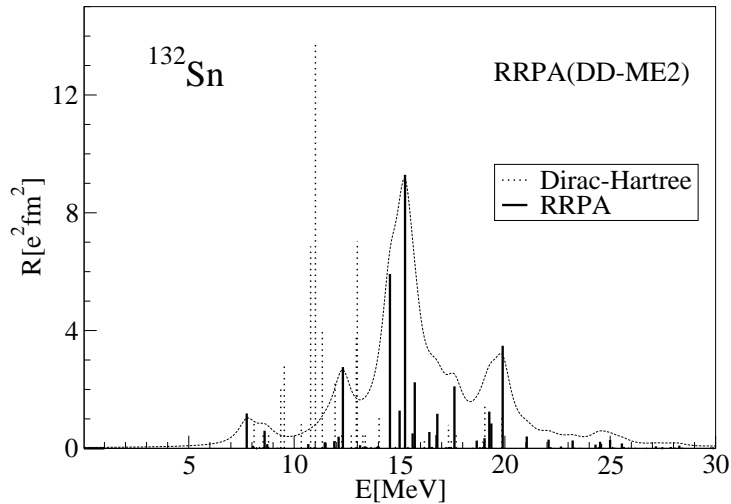}
\vspace*{0.5cm}
\caption{The discrete RRPA dipole strength distribution in $^{132}$Sn, 
in comparison with the unperturbed Dirac-Hartree response.  
}
\label{figSn132-discrete}
\end{figure}

The mixing of isoscalar and isovector states in the
low-lying dipole response has been analyzed in 
several studies~\cite{Vrepyg2.01,Cat.97,Das.97,TE.06}.
More information about the isospin structure of the PDR can
be obtained from a comparison between the RPA dipole strength distribution 
and the unperturbed response. For $^{132}$Sn this is illustrated
in Fig.~\ref{figSn132-discrete}, where we display the discrete dipole 
spectra for the unperturbed Dirac-Hartree response and the 
relativistic RPA response. When the residual
interaction is turned on, most of the unperturbed strength
is pushed towards higher energies, as one expects for
isovector states. The pygmy states, however, are shifted below the 
Dirac-Hartree response. Since the residual interaction is attractive 
in the isoscalar channel, it appears that the structure of the PDR 
is predominantly isoscalar. Experimentally the isospin structure 
of the low-lying E1 states could be, at least in principle, probed by a 
complementary study of $(\alpha,\alpha')$ and $(p,p^\prime)$ 
scattering~\cite{Zil.05}. Assuming a simplified picture, only isoscalar 
modes should be excited when the scattered $\alpha$ particle is detected 
under extreme forward angles.

RPA calculations with zero-range Skyrme forces have shown that, 
when the densities of the core nucleons and the excess  
neutrons are well separated, more like in halo nuclei, 
oscillations of these densities give rise to pronounced 
low-energy dipole strength. On the other hand, if 
the two densities overlap, which is the case in 
neutron-skin nuclei, then the coupling between
the low-lying excitations and the GDR depletes the 
strength of the former~\cite{SE.01}. This result has 
also been confirmed in studies which have used the 
self-consistent Hartree-Fock+RPA with the finite-range
Gogny interaction~\cite{Per.04,Per.05}. 

In open-shell neutron-rich nuclei, where pairing correlations play an 
important role also for low-lying excitations, a fully self-consistent QRPA 
approach is essential. Two such frameworks have been developed recently: 
the relativistic QRPA formulated in the canonical basis of the Relativistic 
Hartree-Bogoliubov model~\cite{PRNV.03}, and the 
HFB+QRPA based on Skyrme 
energy-density functionals \cite{Ter.05,YKG.02,YG.04}. These models
consistently employ for the QRPA residual interactions,
both in the $ph$ and $pp$ channels, the same effective interactions 
which determine the nuclear ground state. In this way a direct relation
is established between the unique ground-state properties of exotic 
nuclei and low-lying collective excitations. In addition, the 
fully self-consistent formulation of QRPA is particularly important
for excitations in $1^-$ channel, because it ensures the separation 
of spurious center-of-mass motion without introducing additional
adjustable parameters. The relativistic (Q)RPA has been
employed in several studies of low-lying dipole strength in 
neutron-rich nuclei. These include the analysis of transition 
densities and velocity fields associated with the PDR in 
$^{208}$Pb~\cite{VPRL.01,PPhD.03}, the evolution 
of the PDR in exotic isotopes far from the valley
of $\beta$-stability~\cite{Vrepyg2.01}, the study 
of the effects of pairing correlations on the low-lying E1 strength 
in exotic nuclei, the isotopic dependence of PDR excitation
energies and transition strength distributions~\cite{PRNV.03}, 
and the relationship between the PDR excitation energies and 
one-neutron separation threshold~\cite{PNVR.05}.
An alternative approach to the RQRPA, based on the response function 
formalism and the BCS approximation for the description of 
pairing correlations, has recently been used in
studies of low-lying E1 modes in $^{26,28}$Ne~\cite{CaoMa.05} and
Ni isotopes~\cite{CaoMa.04}.

The fully self-consistent non-relativistic HFB+QRPA, based on 
Skyrme density functionals and density-independent delta pairing, 
has recently been employed in an extensive analysis of strength functions 
and transition densities in the $J^{\pi} = 0^+$, $1^-$, and $2^+$ channels 
for the even Ca, Ni and Sn isotopes from the proton to the neutron 
drip-lines \cite{TE.06}. It has been shown that the low-energy strength 
increases with neutron number in all multipoles. However, in all 
channels the correlation between strength and collectivity is found to 
be much weaker than in stable nuclei.

The spreading width of the low-lying dipole transition strength in 
neutron-rich nuclei has been evaluated with the phonon damping 
model~\cite{DSA.00}, the consistent
Skyrme Hartree-Fock + QRPA with phonon coupling~\cite{CB.01},
the quasiparticle phonon model (QPM)~\cite{Rye.02,TLS.04,TLS2.04},
and the Extended Theory of Finite Fermion Systems 
(ETFFS)~\cite{Kam.04,Har.04}. These models do not agree 
on the effects of the coupling to complex configurations 
on low-lying E1 strength, e.g. for $^{48}$Ca the
ETFFS reduces the pygmy strength by 31\% with respect to the QRPA 
result~\cite{Har.04}, whereas for neutron-rich Sn isotopes the QPM 
predicts a low-lying dipole strength enhanced by a 
factor 1.5-2 with respect to the QRPA value~\cite{TLS2.04}. 
The differences result from different approaches to the coupling 
with complex phonon configurations.
A consistent description of the fine structure of low-lying 
dipole strength would, of course, necessitate a consistent
implementation of a nuclear effective interaction in the calculation 
of ground state properties, in the (Q)RPA residual interaction, 
and in the interaction terms which describe the coupling
to complex configurations.  

On the experimental side, extensive studies of low-lying electric 
dipole excitations have been performed in recent years.
Low-lying E1 states were observed in neutron-capture $\gamma$-ray 
spectra~\cite{Bar.72,Iga.86}, and in resonant
scattering of real photons~\cite{Kne.96}. 
The latter method, although mainly restricted to nuclei 
with moderate proton -- neutron asymmetry, provides detailed information 
about the fine structure of dipole transition spectra below the neutron
threshold. In particular, pronounced low-energy E1 strength has
been observed in $^{44,48}$Ca~\cite{Har.04,Har.00}, $^{56}$Fe and
$^{58}$Ni~\cite{Bau.00}, $^{88}$Sr~\cite{Kau.04}, $^{112}$Sn~\cite{Oze.06},
$^{116,124}$Sn~\cite{Gov.98},
N=82 isotones~\cite{Zil.02,Her.97,Her.99}, and
$^{204,206,207,208}$Pb~\cite{Rye.02,Cha.80,End.00,End.03}.
Recent advances in studies of low-lying dipole
modes by photon scattering have been reviewed 
in Ref.~\cite{KPZ.06}.    
Radioactive nuclear beams provide new opportunities for 
studies of low-lying dipole excitations in heavier nuclei
with large proton -- neutron asymmetry~\cite{Aum.05}. In a 
recent experiment of the Coulomb dissociation of secondary 
Sn beams produced by in-flight fission of a primary $^{238}$U beam,
the dipole strength distribution above the one-neutron 
separation energy was measured in the unstable $^{130}$Sn and 
the doubly-magic $^{132}$Sn~\cite{Adr.05}. In addition to the 
giant dipole resonance (GDR), evidence was reported for a PDR 
structure at excitation energy around 10 MeV both in $^{130}$Sn
and $^{132}$Sn, exhausting a few percent of the E1 
energy-weighted sum rule.

\begin{figure}[]                                                                  
\centering                                                                  
\vspace*{1cm}                                                                
\includegraphics[scale=0.4]{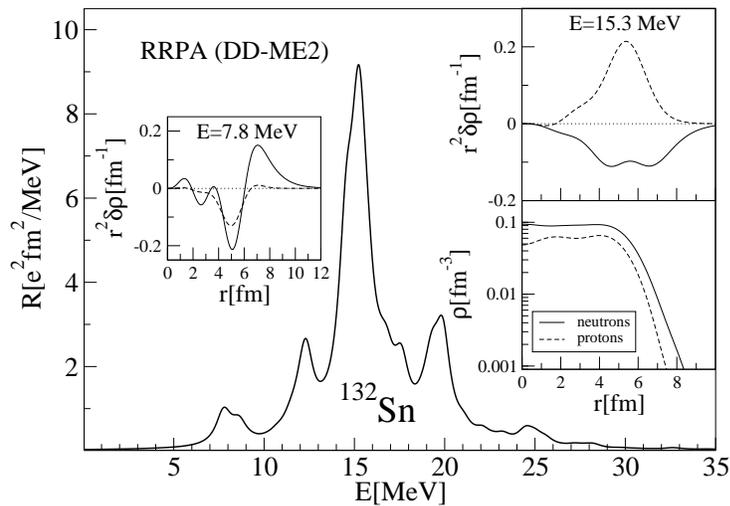}                        
\vspace*{0.5cm}                                                              
\caption{The RRPA dipole strength distribution in $^{132}$Sn,   
calculated with the DD-ME2 effective interaction. In the insertions 
we plot the ground-state proton and neutron density profiles, and       
the proton and neutron transition densities for the peaks at            
7.8 MeV and 15.3 MeV excitation energy.     
}                                                                               
\label{figSn132}                                                               
\end{figure}                                                             
Similar to the results obtained with nonrelativistic models, the 
relativistic QRPA dipole response of neutron-rich nuclei is characterized by 
the fragmentation of the strength distribution and its spreading into the
low-energy region. Fully consistent R(Q)RPA calculations have shown that 
with the increase of the number of neutrons along an isotopic chain, a 
relatively strong E1 peak appears below 10 MeV. The dynamics of this 
peak is very different from that of the isovector giant dipole 
resonance (IV GDR)~\cite{PRNV.03,Vrepyg2.01}.
This is illustrated in Fig.~\ref{figSn132} with the example of $^{132}$Sn, 
where we plot the RRPA strength distribution which corresponds to the 
isovector dipole operator, and is calculated with the
DD-ME2 effective interaction~\cite{LNVR.05}. 
In the inserted panels we display the neutron and 
proton ground-state density distributions, and the
neutron and proton transition densities for the low-lying state at 7.8 MeV,
and for the IV GDR at 15.3 MeV. For the main peak at 15.3 MeV the 
transition densities display a radial dependence which is characteristic 
for the isovector dipole mode (IV GDR): the proton and neutron densities 
oscillate with opposite phases. The dynamics of the state at 7.8 MeV is 
completely different: the proton and neutron transition densities are 
in phase in the bulk of the nucleus, whereas only neutron excitations 
contribute to the transition density in the surface region. 
Thus the low-lying pygmy state does not belong to statistical 
E1 excitations sitting on the tail of the IV GDR, but rather represents 
a new mode -- the PDR: the neutron skin oscillates against the core. 
The neutron skin, i.e. the difference between the neutron and proton
density distributions in the ground state (shown in the right panel in
Fig.~\ref{figSn132}) basically determines the properties of 
the PDR~\cite{TLS.04}. Therefore, for a quantitative description of PDR 
dynamics it is essential to use effective interactions that 
reproduce available data on the neutron skin. This is the case, for 
instance, of the relativistic density-dependent interactions 
DD-ME1~\cite{NVFR.02} and DD-ME2~\cite{LNVR.05}, which have been 
specifically designed to reproduce the differences between the 
$rms$-radii of neutron and proton density distributions.

In light nuclei the low-energy dipole strength predominantly
originates from non-resonant independent single particle
excitations of the loosely bound neutrons. However, the structure of 
the low-lying strength changes with mass. As has been shown in the 
RRPA analysis of Ref.~\cite{Vrepyg2.01},
in heavier nuclei some of the low-lying dipole states display a
more distributed structure of the RRPA
amplitudes. Among several peaks characterized by single particle transitions,
a single collective dipole state is identified below 10 MeV and its RRPA
amplitude presents a coherent superposition of many
neutron particle-hole configurations. For instance, in the case of 
of $^{132}$Sn (see the dipole strength 
distribution in Fig.~\ref{figSn132}) the following neutron $ph$ transitions 
principally contribute to the RRPA amplitude of the state at 7.8 MeV:
$3s_{1/2} \rightarrow 3p_{3/2}$ (51\%), $2d_{3/2} \rightarrow 3p_{3/2}$ (19\%),
$2d_{3/2} \rightarrow 3p_{1/2}$ (11\%), $3s_{1/2} \rightarrow 3p_{1/2}$ (7\%),
$1h_{11/2} \rightarrow 1i_{13/2}$ (4\%), $1g_{7/2} \rightarrow 1h_{9/2}$ (0.9\%), 
$2d_{5/2} \rightarrow 3p_{3/2}$ (0.4\%),
$2d_{5/2} \rightarrow 2f_{7/2}$ (0.3\%), 
$2d_{3/2} \rightarrow 4p_{1/2}$ (0.2\%), $1g_{7/2} \rightarrow 2f_{5/2}$ (0.1\%), etc.
On the other hand, the total contribution from all proton transitions to 
the state at 7.8 MeV is small: $\approx$ 3\%, thus the ratio of neutron 
to proton contribution is much higher than the value N/Z, typical for the
IV GDR state. Such a rich structure of the RRPA amplitude is in contrast 
to the situation found in light neutron-rich nuclei, where the low-lying 
dipole peaks below 10 MeV are usually dominated by just one or two 
neutron $ph$ transitions. The level of collectivity can be further 
enhanced in open-shell nuclei, where because of pairing correlations 
many additional neutron states become partially occupied and, therefore, 
many more $2qp$ transitions contribute to the RRPA amplitude. 

A similar analysis of neutron particle-hole components of strong 
low-energy $1^-$ excited states in $^{132}$Sn has also been 
performed in Ref.~\cite{TE.06}, for the self-consistent HFB+RPA 
calculation with the Skyrme SkM$^*$ interaction. The distribution 
of the largest neutron $ph$ components and the degree of collectivity 
for the most pronounced low-energy states is comparable to 
the results of the relativistic RPA, but the two models differ in the 
integrated energy-weighted strength in the low-energy region. 
While relativistic RPA calculations typically predict 
$\approx 5$\% of the classical TRK sum rule in the energy 
region below 10 MeV, only about 1\% is obtained in the calculation 
with the Skyrme SkM$^*$ interaction. In Ref.~\cite{TE.06} it has 
been suggested that this difference is related to the larger neutron 
skin typically calculated with relativistic mean-field models, which 
may be responsible for the more pronounced pygmy resonances.
However, this does not seem to be the case for the DD-ME2 
relativistic interaction (see Fig.~\ref{figstrSn}), which does not 
overestimate the empirical values for the neutron skin in Sn isotopes 
and, at the same time, reproduces the experimental results for 
the integrated energy-weighted dipole strength in the low-energy 
region. 

The RPA and QRPA analyses of the dynamics of low-lying 
E1 strength distributions have mostly been performed on the mean-field level, i.e.
without taking into account the spreading effects which arise from the
coupling of single-nucleon states to the collective low-lying 
excitations (phonons). The principal effect of the particle-vibration 
coupling is an increase of the nucleon effective mass at the Fermi 
surface, and this is reflected in an increase of the density of 
single-nucleon states close to the Fermi energy. It has been argued 
that the inclusion of particle-vibration coupling in (Q)RPA calculations, 
i.e. extending the (Q)RPA model space to include selected 
two-quasiparticle $\otimes$ phonon states, 
would not only improve the agreement 
between the calculated and empirical widths of the GDR structures, 
but it could also have a pronounced effect on the low-lying E1 strength. 
For instance, the coupling to low-lying phonons could fragment the PDR 
structure over a wide region of excitation energies. As a result of 
this fragmentation only an enhancement of the E1 strength would be 
observed in the low-energy region, rather than a prominent PDR peak. 
The importance of particle-vibration coupling effects for the multipole 
response of neutron-rich nuclei has particularly been emphasized 
in studies that have used the QRPA plus phonon coupling model based 
on the Hartree-Fock (Q)RPA with Skyrme effective 
forces~\cite{CB.01,Sar.04}. In Ref.~\cite{Sar.04} the QRPA plus 
phonon coupling model was applied in the analysis of dipole 
excitations in $^{208}$Pb, $^{120}$Sn and $^{132}$Sn. In 
contrast to the results obtained in the relativistic (Q)RPA 
framework, the QRPA plus phonon coupling model predicts low-lying 
E1 strength of non-collective nature in all three nuclei. 

\begin{figure}[]                                                                  
\centering                                                                  
\vspace*{1cm}                                                                
\includegraphics[scale=0.5,angle=270]{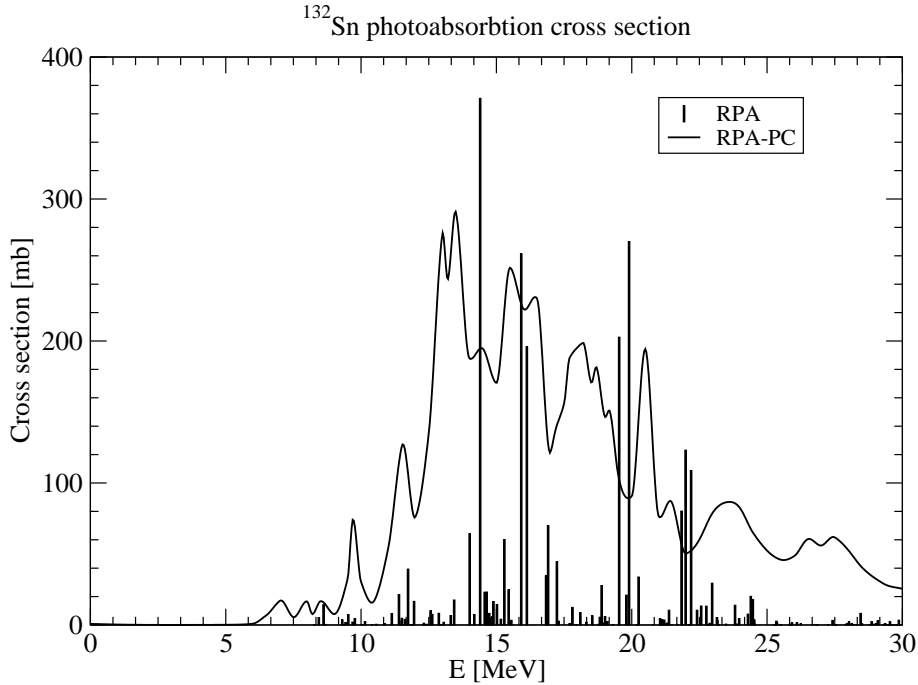}                        
\vspace*{0.5cm}                                                              
\caption{Photoabsorption cross section for $^{132}$Sn, 
calculated with the RPA and RPA-PC models. The effective 
interaction is Skyrme SIII.  
}                                                                               
\label{fig:132sn}                                                               
\end{figure}

\begin{figure}[]                                                                  
\centering                                                                  
\vspace*{1cm}                                                                
\includegraphics[scale=0.55,angle=270]{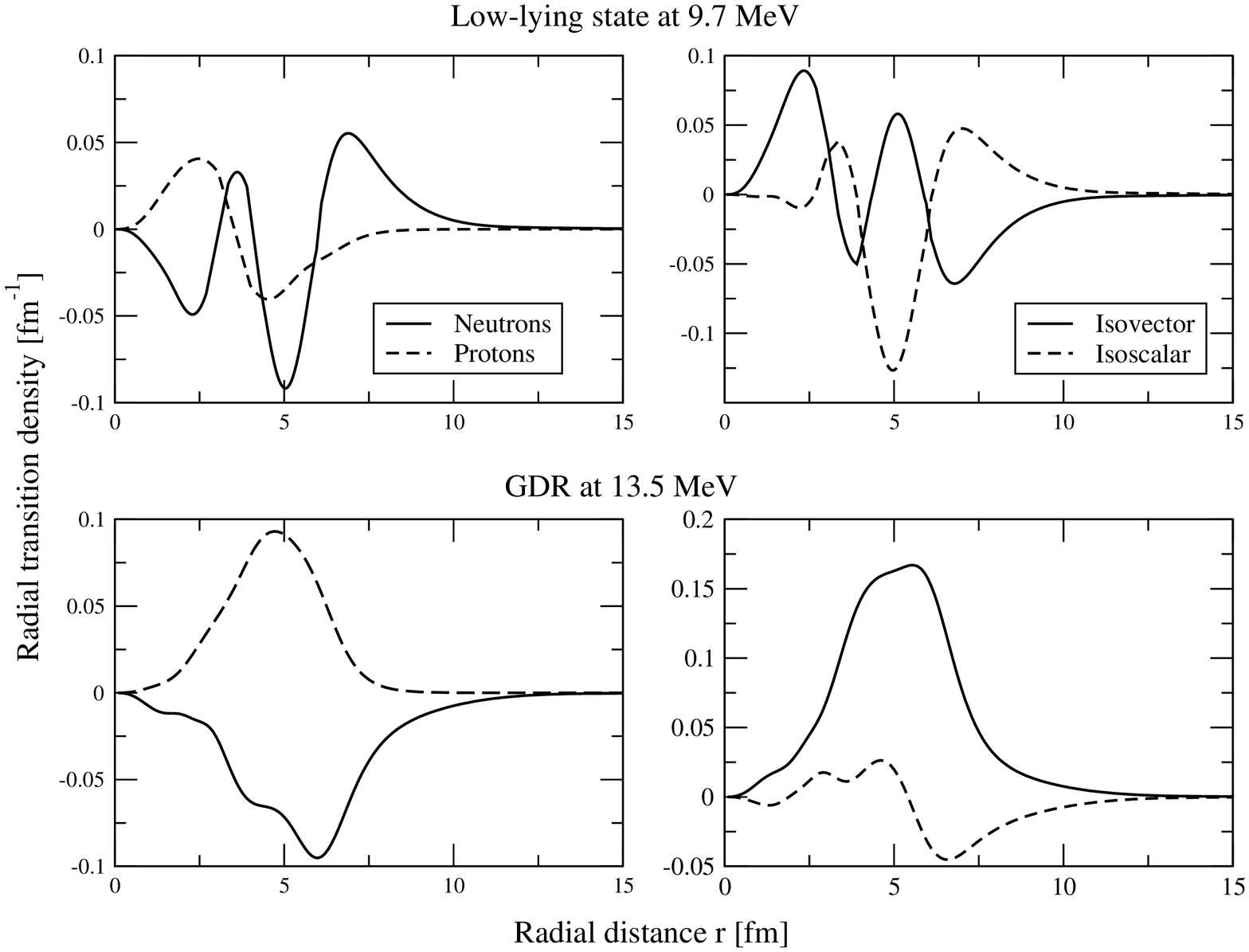}                        
\vspace*{0.5cm}                                                              
\caption{RPA-PC transition densities for $^{132}$Sn. In the upper 
panels the proton and neutron (right panel), and isoscalar and 
isovector (left panel) transitions densities for the state at 
9.7 MeV are shown. For comparison in the lower panels we plot 
the corresponding transition densities for the IV GDR 
at 13.5 MeV.}                                                                               
\label{fig:td_132_rpapc}                                                               
\end{figure}

In Fig.~\ref{fig:132sn} we display the photoabsorbtion cross 
section for $^{132}$Sn, calculated with the fully 
consistent RPA and RPA-PC models using the Skyrme force SIII. 
The corresponding RPA-PC transition densities for the 
GDR state at 13.5 MeV and for the most pronounced low-energy peak 
at 9.7 MeV are shown in Fig.~\ref{fig:td_132_rpapc}. 
Even though the transition densities, both for the 
GDR and for the low-lying peak at 9.7 MeV, are similar to those 
calculated with the relativistic RPA (see Fig.~\ref{figSn132})
the analysis of the structure of RPA (RPA-PC)  
amplitudes shows  that none of the peaks below 10 MeV contain 
contributions of more than two or three different neutron 
particle-hole ($ph$) configurations. 
Predominantly these peaks correspond to just a single-neutron 
transition, and each of them exhausts less than 0.5\% of the 
energy-weighted sum rule. 


Low-lying E1 excitations in neutron-rich 
Sn isotopes have also been studied in the Quasiparticle Phonon 
Model~\cite{TLS.04}, in a model space that included up to 
three-phonon configurations built from a basis of QRPA states, 
and with separable multipole-multipole residual interactions. 
The single-nucleon spectra were calculated for a 
Woods-Saxon potential with adjustable parameters. 
Empirical couplings were
used for the QPM residual interactions. In the QPM spectra for 
$^{120-132}$Sn the low-energy dipole strength was found 
concentrated in a narrow energy interval such that the PDR 
could be identified. 
It was shown that, despite significant multi-phonon contributions 
to the mean-energy and transition strength, the PDR states 
basically retain their one-phonon character.
         
Because of its relatively large neutron excess, the stable nucleus 
$^{208}$Pb has also been investigated for a possible occurrence of 
pygmy dipole resonant states. Experimental evidence has been reported 
in elastic photon~\cite{SAC.82} and photoneutron scattering~\cite{BCA.82}, 
and in electron scattering~\cite{Kuh.81}: pronounced E1 strength has been
observed in the energy region between 9 and 11 MeV, several MeV below 
the IV GDR in $^{208}$Pb. On the theoretical side, one of the first 
microscopic analysis was performed in the Hartree-Fock
plus RPA model based on the Skyrme interaction SGII~\cite{ACS.96}. 
Two pronounced peaks were calculated at 8.7 MeV and 9.5 MeV, which
appeared as likely candidates for the PDR.
In a recent self-consistent relativistic RPA study based on the NL3
effective interaction, a pronounced low-energy dipole 
peak was calculated at 7.29 MeV~\cite{VPRL.01}. The structure of 
the RRPA amplitude, the corresponding transition densities and 
velocity fields indicate that this state can be interpreted as a
collective PDR mode. The RRPA prediction for the PDR state 
has been confirmed in a subsequent $(\gamma,\gamma')$ experiment, which 
disclosed a resonance-like structure centered at 7.37 MeV, approximately 
at the neutron emission threshold~\cite{Rye.02}. 

\begin{figure}[]
\centering
\vspace*{1cm}
\includegraphics[scale=0.4]{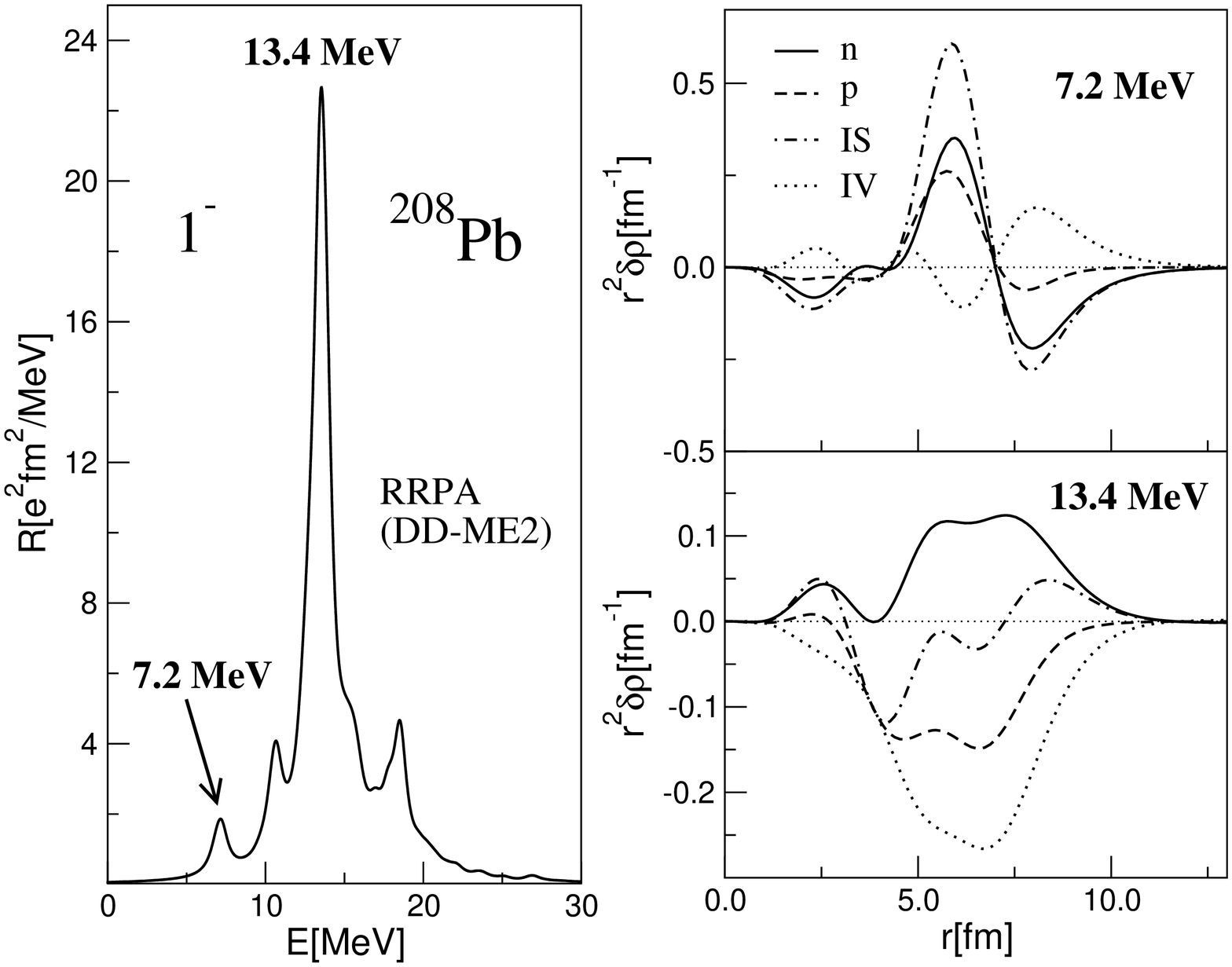}
\vspace*{0.5cm}
\caption{The RRPA dipole strength distribution in $^{208}$Pb,
calculated with the DD-ME2 effective interaction.
The neutron, proton, isoscalar(IS), and isovector(IV) transition
densities, for the pygmy state at 7.2 MeV and the IV GDR state at 13.4 MeV  
are plotted in the panels on the right.} 
\label{figPb208}
\end{figure}
In Fig.~\ref{figPb208} we display the isovector dipole strength
distribution in $^{208}$Pb, evaluated with the self-consistent RRPA model 
employing the DD-ME2 effective interaction~\cite{LNVR.05}.
The calculated energy of the main peak at 13.4 MeV is in excellent
agreement with the experimental value of the excitation energy of IV GDR:
$13.3\pm 0.1$ MeV~\cite{Rit.93}. The pronounced low-energy peak
at 7.2 MeV is close to the experimental centroid of the low-lying
dipole strength at 7.37 MeV~\cite{Rye.02}. 
In the panels on the right in Fig.~\ref{figPb208} we plot the RRPA 
transition densities for the low-energy state
at 7.2 MeV, and the IV GDR state at 13.4 MeV, respectively.
Obviously, the dynamics of the low-lying mode is very different from that 
of the isovector giant resonance: the proton and neutron transition densities
are in phase in the nuclear interior and 
there is large contribution from the neutrons in the surface region.
For the IV GDR state the total isovector transition density is much
stronger than the isoscalar component. 
On the other hand, for the state at 7.2 MeV
the isoscalar transition density dominates
over the isovector one in the interior, and
the large neutron component in the surface
region contributes to the formation of the node in the isoscalar
transition density~\cite{Vrepyg2.01,Cat.97,Das.97}.

\begin{figure}[hb]
\centering
\vspace*{1.2cm}
\includegraphics[scale=0.5]{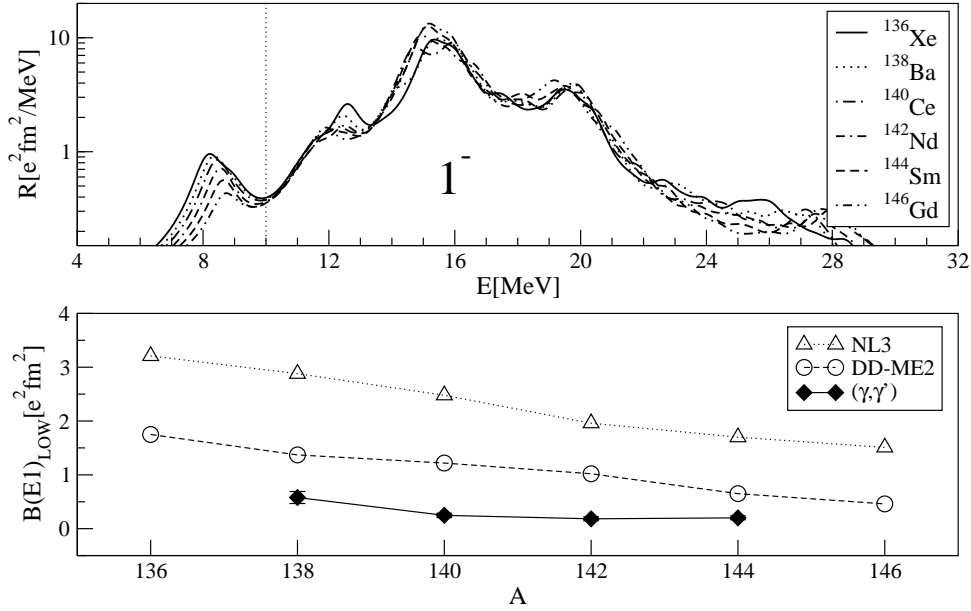}
\vspace*{0.5cm}
\caption{The isovector dipole strength distributions 
in N=82 isotones calculated in the RHB+RQRPA model, with
the DD-ME2 effective interaction (upper
panel). The integrated B(E1) transition strength below
10 MeV, for the NL3 and DD-ME2 interactions, is compared with
the data from $(\gamma,\gamma')$ 
scattering~\protect\cite{Zil.02,Zil.05}(lower panel).}
\label{figN82}
\end{figure}

Low-lying E1 excitations have also been observed below 10 MeV 
in $(\gamma,\gamma')$ scattering on the N=82 
isotones: $^{138}$Ba, $^{140}$Ce, $^{142}$Nd, 
and $^{144}$Sm~\cite{Zil.02,Zil.05}.
The subsequent analysis of the RHB+RQRPA transition densities
for the calculated low-lying states in these nuclei has shown 
that a collective PDR mode indeed develops. However, the 
calculated PDR are located $\approx$ 1-2 MeV above the experimental
centroids~\cite{PRNV.03}.
The evolution of the isovector dipole strength distribution in 
N=82 isotones, evaluated in the fully self-consistent RHB+RQRPA with 
the DD-ME2 effective interaction, is illustrated in Fig.~\ref{figN82}.  
The dotted vertical line separates the low-energy region below 10 MeV 
from the region of giant resonances. In contrast
to the IV GDR, which weakly decreases in excitation
energy with the increase of the proton number, i.e. with mass number,  
the centroids of the low-lying structure increase in energy, whereas 
the total low-energy strength decreases when the proton -- neutron 
asymmetry is reduced. One notices that the
low-lying states are far more sensitive to the variations of the 
proton number, than the IV GDR structure. These observations
are consistent with the interpretation of the low-energy peaks
in terms of the PDR, because
the reduction of the asymmetry between the neutron and proton
density distributions in the ground state should generally
result in higher PDR excitation energies and 
in the suppression of its strength~\cite{PNVR.05}. 
The RHB+RQRPA B(E1) strength in the low-energy region 
below 10 MeV decreases with mass number along the N=82
isotone chain (lower panel in Fig.~\ref{figN82}), but the 
calculated values are systematically above the data~\cite{Zil.02,Zil.05}. 
On the other hand, the quasiparticle phonon model (QPM) 
predicts a constant summed E1 strength in all the measured 
N=82 nuclei~\cite{End.05}. The reason is that, in contrast to
the fully self-consistent RHB+RQRPA approach, the QPM 
calculations employ the same single-particle
spectrum for all the N=82 isotones~\cite{End.04},
and therefore cannot describe the details in nuclear structure
which result
from the variation of the proton number. 

The theoretical analysis of low-lying excitations and the 
comparison with data, can be used as a sensitive
tool to constrain the isovector channel of effective
nuclear interactions. In the lower panel of Fig.~\ref{figN82}
we compare the theoretical B(E1) strength in the low-energy region 
below 10 MeV with data. The theoretical values have been calculated 
in the consistent RHB+RQRPA model with the very popular non-linear 
meson-exchange effective interaction 
NL3~\cite{LKR.97}, and with the new meson-exchange interaction 
DD-ME2~\cite{LNVR.05}, which explicitly includes a medium dependence of 
the meson-nucleon couplings. Obviously
the NL3 interaction, which is known to overestimate the size of the neutron
skin not only in exotic neutron-rich nuclei but also in $^{208}$Pb,
predicts too much low-lying  B(E1) strength. On the other hand, 
an interaction like DD-ME2 which has been adjusted to the empirical
differences between the radii of neutron and proton density distributions,
significantly improves the agreement of the calculated low-energy 
dipole strength with data. The remaining
difference might be caused by the coupling with more complex
phonon configurations~\cite{TLS2.04}, not taken into account in 
the RHB+RQRPA models, or in the missing  
E1 strength in $(\gamma,\gamma')$ scattering which may be
quite considerable 
when dealing with end-point energies close to the neutron separation
threshold~\cite{Gov.98}.


\subsection{\label{ssec_isopygmy} 
Isotopic Dependence of Pygmy Dipole Resonances}

The evolution of low-lying dipole transition strength along an isotopic 
chain provides useful information about the underlying
dynamics of soft modes in exotic nuclei. In particular, 
an important question is the location of PDR with respect to 
the neutron separation threshold~\cite{PNVR.05}. This  
is important not only for the possible detection of PDR in 
experiments, but also for modeling the
r-process nucleosynthesis~\cite{Gor.04}. 
Most of the recent photon scattering experiments provide data 
on the dipole strength only below the neutron separation energy, i.e. 
only a portion of the overall low-lying E1 strength is 
observed~\cite{Zil.02,Har.04,Har.00,Bau.00,Kau.04,Gov.98,
Her.97,Her.99,Cha.80,End.00,End.03}. Of course for
a more complete understanding of the structure 
of the low-lying dipole strength and its relation to
the PDR mode, data on transition strength above the neutron 
threshold are necessary~\cite{PNVR.05}.

\begin{figure}[ht]
\centering
\vspace*{1cm}
\includegraphics[scale=0.4]{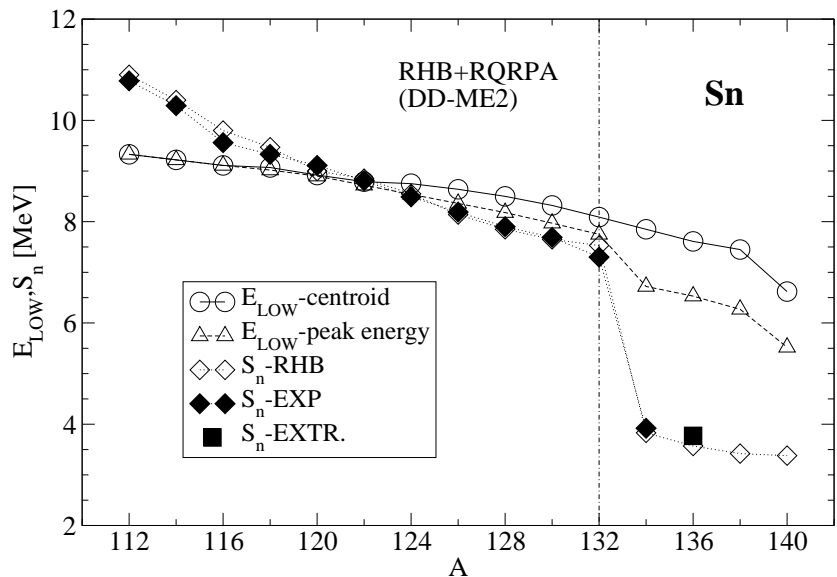}
\vspace*{0.5cm}
\caption{The calculated PDR peak and centroid energies, and the
one-neutron separation energies for the sequence of Sn isotopes, as
functions of the mass number. The DD-ME2 effective interaction
has been used in the RHB+RQRPA calculations. The RHB results for the
neutron separation energies are compared with the
experimental and extrapolated values~\protect\cite{AW.95}.
}
\label{figSnenergy}
\end{figure}
This is illustrated in Fig.~\ref{figSnenergy},
where we display the RHB+RQRPA results for the
peak and centroid energies of the PDR in a series of Sn isotopes.
The RQRPA predicts a monotonic decrease of the PDR with mass number, 
and only a small kink in the peak excitation energies is 
calculated at the $N=82$ shell closure. In the same plot we have 
also included the calculated one-neutron separation energies, in 
comparison with the data and the extrapolated value~\cite{AW.95}.
The self-consistent RHB calculation, with the DD-ME2 mean-field 
effective interaction in the $ph$ channel and the D1S Gogny force 
in the pairing channel, reproduces in detail the one-neutron
separation energies in Sn nuclei. We notice
that the separation energies decrease faster than the
calculated PDR excitation energies. At the doubly closed-shell 
nucleus $^{132}$Sn a sharp reduction of the one-neutron 
separation energy is observed and reproduced by the RHB calculation,
whereas the shell closure produces only a much weaker effect on the 
PDR peak energies. The increased fragmentation of the low-lying strength
in heavier Sn isotopes results in larger differences between the
PDR peak and centroid energies. The important result here is that 
for $A < 122$ the PDR excitation energies are below the 
corresponding one-neutron separation energies, whereas 
for $A\geq 122$ the pygmy resonance is located above the neutron 
emission threshold. This means, of course, that in the latter case 
the observation of the PDR in $(\gamma,\gamma^\prime)$
experiments will be strongly hindered~\cite{PNVR.05,Gov.98}.

\begin{figure}[ht]
\centering
\vspace*{1cm}
\includegraphics[scale=0.45]{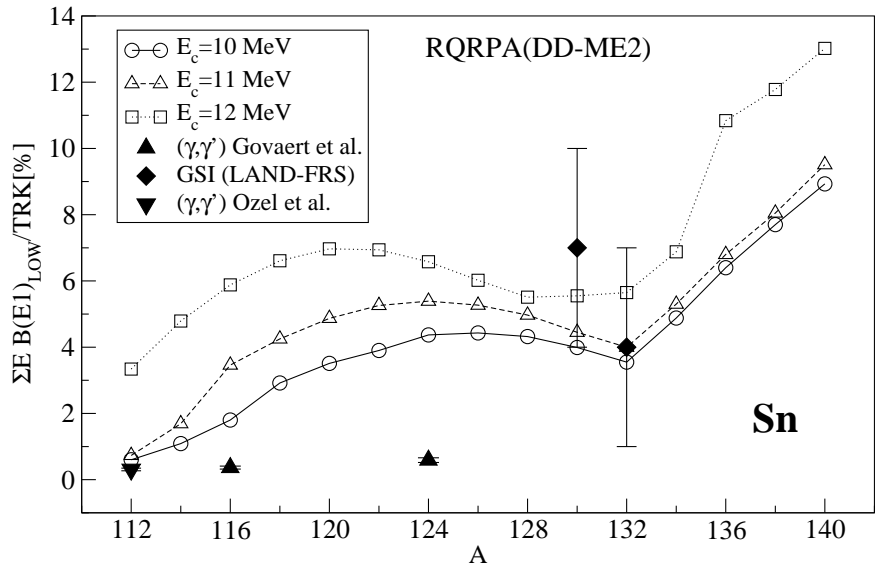}
\vspace*{0.5cm}
\caption{The RHB+RQRPA energy-weighted dipole strength,
integrated up to the energy cut-off $E_c$=10, 11, and 12 MeV, 
respectively, and plotted in percents of 
the TRK sum rule. The experimental results are from
Refs.~\protect\cite{Oze.06,Gov.98,Adr.05}.
}
\label{figstrSn}
\end{figure}
The presently missing data on dipole strength above the neutron threshold 
could be obtained in the near future by using tagged photons at 
S-DALINAC~\cite{Zil.05}. In addition, photon scattering with high intensity 
beams at energies below and above the neutron separation threshold
are planned at the superconducting electron accelerator 
ELBE~\cite{Sch.05,Wag.05}. The first studies at ELBE include 
photon-scattering on $^{92,98,100}$Mo~\cite{Rus.05}. 
It is interesting to note that the data show an enhancement of the 
dipole transition strength around 9 MeV: in $^{92}$Mo the pygmy strength 
is located below the neutron separation energy, whereas in $^{100}$Mo 
it shifts above the neutron threshold. 

In Fig.~\ref{figstrSn} we display the isotopic dependence 
($^{112}$Sn-$^{140}$Sn) of the energy weighted dipole strength in 
the low-energy region, integrated  up to the cut-off energy
$E_c$=10, 11, and 12 MeV, respectively, and plotted in 
percents of the classical TRK sum rule. Model calculations 
are performed in the RHB+RQRPA with the DD-ME2 plus Gogny D1S 
interactions, and the results are compared with the
available data from photon scattering~\cite{Oze.06,Gov.98}, and
Coulomb dissociation of secondary Sn beams from in-flight fission~\cite{Adr.05}.
The calculated low-lying E1 strength is in excellent agreement with 
the recent experimental data for $^{112}$Sn~\cite{Oze.06} and
$^{130,132}$Sn~\cite{Adr.05}, whereas it 
overestimates the $(\gamma,\gamma')$ data for $^{116,124}$Sn~\cite{Gov.98}. 
When considering the evolution of low-lying dipole strength along an
isotopic chain, in a first approximation one could expect that
the relative strength of the PDR increases monotonically with the number 
of neutrons, at least within a major shell. In the case of Sn isotopes 
the RHB+RQRPA calculations predict, however, that the PDR peak is 
most pronounced around $^{124}$Sn (depending on the cut-off, 
see Fig.~\ref{figstrSn})~\cite{PRNV.03}. A combination of
shell effects and reduced pairing correlations, leads to a reduction of
the strength of the PDR in heavier Sn nuclei below $N=82$. The local minimum 
in the low-lying E1 strength is calculated 
for $^{132}$Sn, whereas in the neighboring isotopes the transition strength
increases because of enhanced collectivity, i.e. the increase in the
number of two-quasiparticle pairs contributing to the RQRPA configuration space. 
We also notice  the pronounced difference in the pygmy strength between 
nuclei close to the valley of $\beta$-stability and exotic
nuclei: while below the N=82 shell closure the integrated transition strength 
is at most $\approx$ 4\% of the TRK sum rule value (for $E_c$=10 MeV), 
beyond $^{132}$Sn the PDR strength exhibits a strong enhancement. 

\begin{figure}[ht]
\centering
\vspace*{1.5cm}
\includegraphics[scale=0.45]{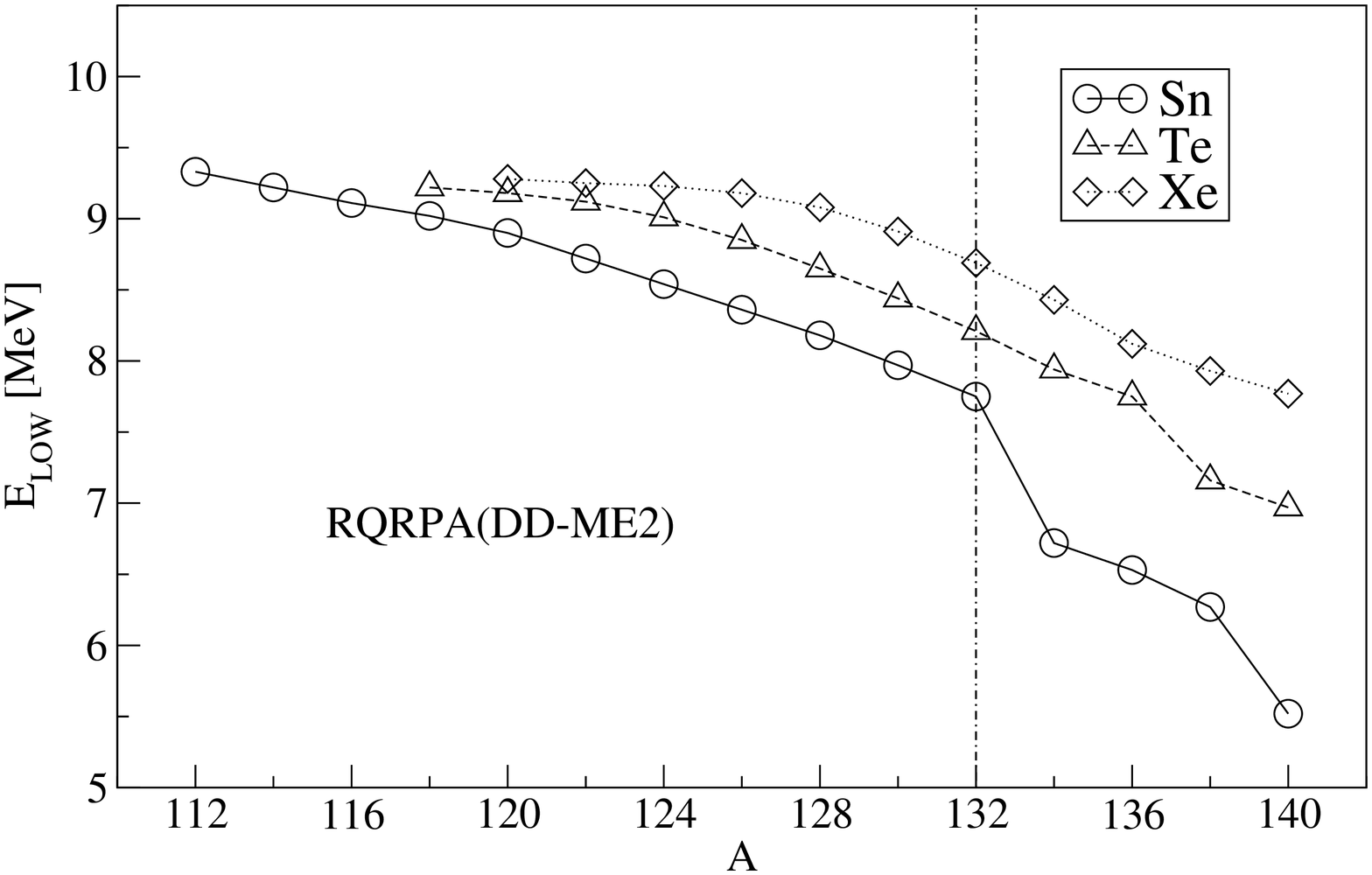}
\vspace*{0.5cm}
\caption{The RHB+RQRPA calculated PDR peak energies 
for Sn, Te, and Xe isotopes.}
\label{figenSnTeXe}
\end{figure}
\begin{figure}[]
\centering
\vspace*{1cm}
\includegraphics[scale=0.45]{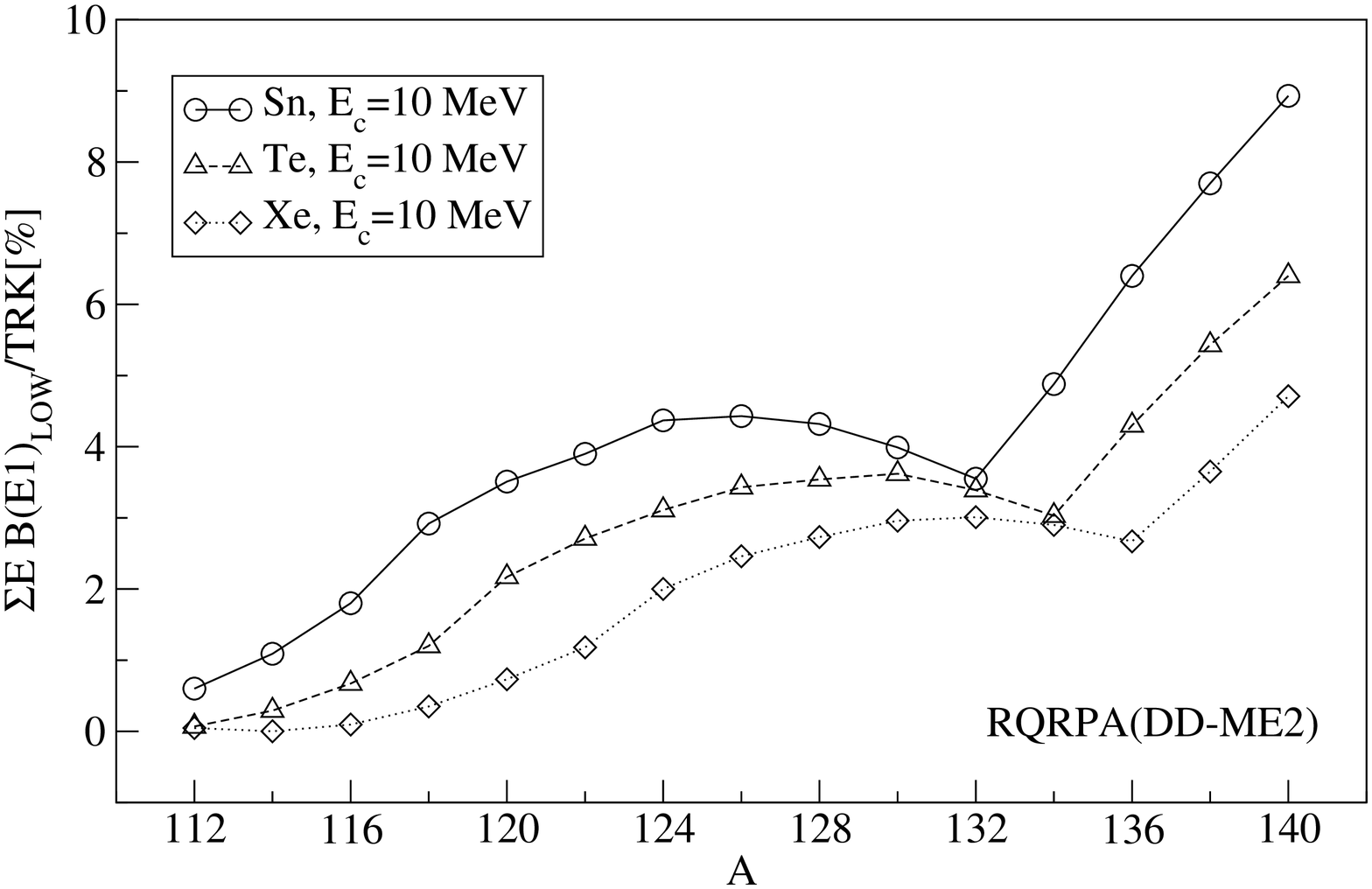}
\vspace*{0.5cm}
\caption{Same as Fig.~\ref{figstrSn}, but for the Sn, Te, and Xe isotopes.
The cut-off energy for the low-lying E1 strength is set at $E_c$=10 MeV.
}
\label{figstrSnTeXe}
\end{figure}

It is, of course, interesting to explore other isotopic chains of
spherical nuclei where one expects the occurrence of the PDR in the E1 
excitation spectrum. In Figs.~\ref{figenSnTeXe} and~\ref{figstrSnTeXe}
we plot the calculated PDR peak energies and the integrated
low-lying E1 strength for Sn, Te, and Xe isotopes. 
The cut-off energy for the low-lying strength is arbitrarily 
set at $E_c$=10 MeV. The calculated low-lying dipole strength appears 
rather sensitive to small variations in the number of nucleons. The 
PDR excitation energies are lowest in the isotopic chain with the smallest 
number of protons, i.e. in Sn nuclei. This behavior reflects the nature of the 
PDR: a larger neutron excess should result in lower PDR excitation energy. 
In the region beyond A=132, the slope of the PDR peak energies becomes 
steeper than for stable nuclei, because the neutrons in outer orbitals are 
more loosely bound and thus the restoring force in the oscillation of the skin 
against the core becomes weaker. For Te and Xe isotopes we plot the 
PDR peak energies starting from $^{118}$Te and $^{120}$Xe, respectively. 
In the lighter systems the PDR could not be uniquely identified.
The energy-weighted dipole transition strength in the region below 10 MeV 
(Fig.~\ref{figstrSnTeXe}) is strongest for the Sn chain, and somewhat weaker 
for Te and Xe. This is, of course, to be expected because the
PDR strength must be proportional to the neutron excess.
In all the three chains the local minima in the integrated transition 
strength are obtained at N=82, and the PDR strength rapidly
increases beyond the neutron shell closure. 
\begin{figure}[]
\centering
\vspace*{1cm}
\includegraphics[scale=0.45]{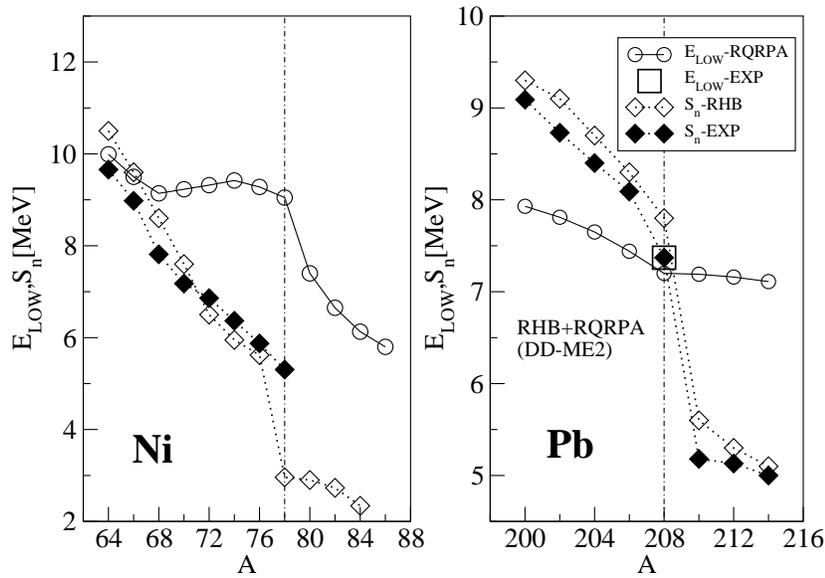}
\vspace*{0.5cm}
\caption{Same as in Fig.~\protect\ref{figSnenergy}, but for the Ni and
Pb isotopic chains. The large  square denotes the experimental position
of the PDR in $^{208}$Pb~\protect\cite{Rye.02}.
}
\label{figNiPben}
\end{figure}

Finally, in Fig.~\ref{figNiPben} the RHB+RQRPA results 
for the PDR in Ni and Pb isotopes are shown. 
The RHB neutron separation energies, calculated with the DD-ME2 plus 
Gogny D1S interactions, are compared with the experimental values~\cite{AW.95}.
In the sequence of Ni nuclei the crossing between
the theoretical curve of one-neutron separation energies and the PDR
excitation energies is calculated already at $A=66$. In heavier
Ni isotopes the excitation energy of the PDR is predicted high above the 
neutron emission threshold. One should notice that for the lighter Ni isotopes
the agreement between the calculated and experimental neutron
separation energies is not as good as for the Sn nuclei and, therefore,
the actual point of crossing between the PDR and the one-neutron separation
energy could occur for $A<66$. The Ni nuclei are not very rigid and,
for a more quantitative description, one would have to go beyond the
simple mean-field plus QRPA calculation and include correlation effects.
For the Pb isotopes the crossing point is calculated at $A=208$, in excellent 
agreement with the data on low-lying E1 excitations in $^{208}$Pb~\cite{Rye.02}.

Motivated by the experimental results on the 
PDR in $^{130}$Sn and $^{132}$Sn \cite{Adr.05}, the relativistic RPA 
has recently been applied in the study of the isotopic dependence of the 
PDR in tin \cite{Pie.06}, focused on the following questions: (a) is there 
a correlation between the development of a neutron skin and the 
emergence of low-energy dipole strength? and (b) can the data be used 
to discriminate among effective interactions that predict different 
values for the neutron skin in heavy nuclei? The results of the RPA analysis 
are not conclusive because, although a strong linear correlation between 
the neutron skin and the fraction of the energy-weighted sum rule at low 
energy was observed, an anti-correlation actually developed beyond 
 $^{120}$Sn, and it was attributed to the filling of the neutron 
 $1h_{11/2}$ orbital. It should be pointed out, however, that the 
 analysis was performed on the RPA level, without considering the 
 effect of pairing correlations. Comparing different effective interactions, it was 
 found that the centroid energy of the PDR is not sensitive to the 
 density dependence of the symmetry energy. The fraction of the 
 energy-weighted sum rule exhausted by the PDR, on the other hand, 
 increases sharply with increasing neutron skin. Although the experimental 
 error bars are large, the data seem to disfavor effective interactions with 
 stiff symmetry energy, i.e. those which predict excessively large neutron skins.

\subsection{The Proton Electric Pygmy Dipole Resonance}

Because the proton drip-line is 
much closer to the line of $\beta$-stability than the 
neutron drip-line, bound nuclei with an excess of protons 
over neutrons can be only found in the region of light 
$Z\leq 20$ and medium mass $20 < Z \leq 50$ elements. 
For $Z > 50$, nuclei in the region of the proton drip-line 
are neutron-deficient rather than proton-rich.    
In contrast to the evolution of the neutron skin 
in neutron-rich systems, because of the presence of the Coulomb 
barrier, nuclei close to the proton 
drip-line generally do not exhibit a pronounced proton 
skin, except for very light elements. Since in light nuclei 
the multipole response is generally less collective, all 
these effects seem to preclude the formation of the 
pygmy dipole states in nuclei close to the proton drip-line.
Nevertheless, a recent analysis based on the RHB+RQRPA approach 
has shown that proton 
pygmy dipole states can develop in light and medium mass 
proton-rich nuclei~\cite{PVR.05}.

\begin{figure}
\vspace*{1cm}
\centering
\includegraphics[scale=0.4]{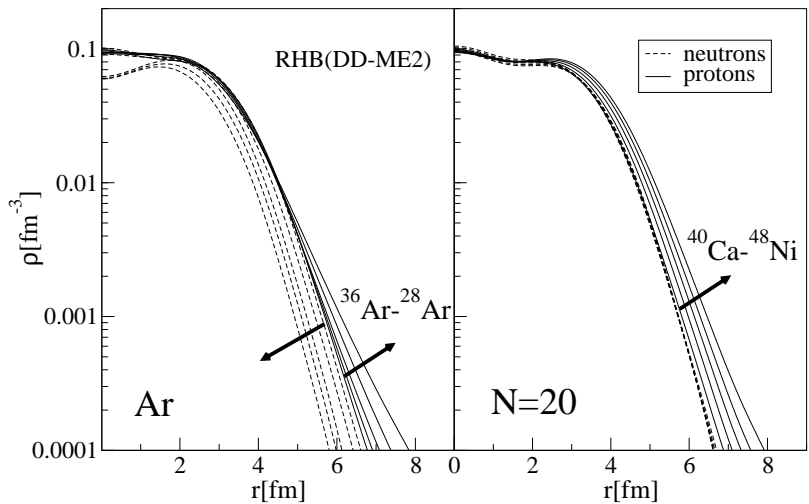}
\vspace*{0.5cm}
\caption{Neutron and proton ground-state density profiles for Ar
isotopes (left panel), and N=20 isotones (right panel),
evaluated in the RHB model with the
DD-ME2 effective interaction. The arrows denote the increasing (decreasing) 
proton (neutron) density distributions along the isotopic, and isotonic chains.}
\label{figprskin}
\end{figure}

In Fig.~\ref{figprskin} we plot the
ground state density profiles for the Ar isotopes and for the 
N=20 isotones, respectively, evaluated in the RHB model using the 
DD-ME2 effective interaction, and the Gogny D1S force in the pairing 
channel. In the both examples we observe the formation of the proton 
skin on the surface of nuclei which have a higher ratio of protons over 
neutrons. The proton skin is, of course, not so pronounced as neutron 
skin in neutron-rich nuclei, because of  the
Coulomb barrier which tends to localize the protons in the nuclear
interior. Evidence for a possible formation of the proton skin in 
neutron-deficient or proton-rich nuclei has been reported in 
recent experimental studies~\cite{Oza.02}, and is supported by 
model predictions~\cite{Naz.96,LVR.01}. 

\begin{figure}
\vspace*{1cm}
\centering
\includegraphics[scale=0.4]{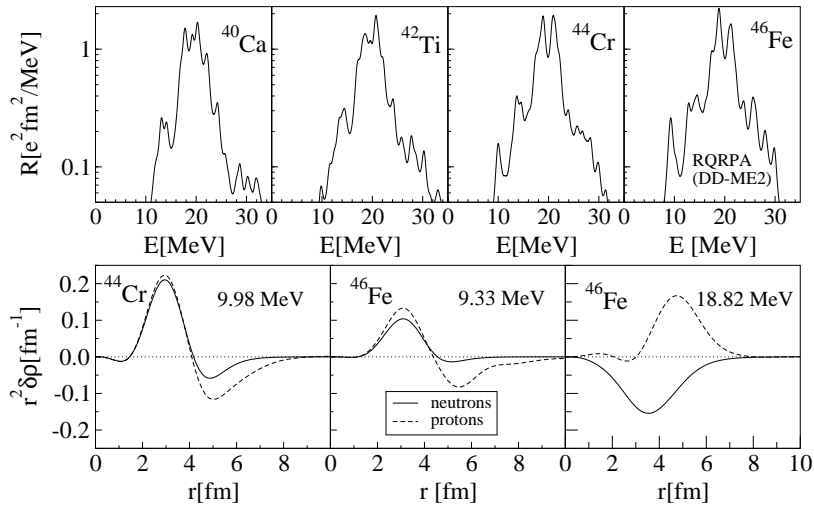}
\vspace*{0.5cm}
\caption{The RHB+RQRPA isovector dipole strength distributions
in the N=20 isotones (upper panel). The proton and neutron transition 
densities for the low-lying states in $^{44}$Cr and $^{46}$Fe,
and for the IV GDR state in $^{46}$Fe are shown in the lower panel.}
\label{figN20}
\end{figure}

Only few studies of dipole excitations in proton-rich nuclei have 
been reported so far. The isovector dipole response 
of the proton drip-line nucleus $^{34}$Ca has been analyzed with the 
continuum  RPA based on Skyrme interactions, and a  
multiple peak structure has been predicted between the low-energy 
isoscalar dipole response and the IV GDR~\cite{HSZ.98}.
In the large-scale shell-model calculations for $^{13}$O~\cite{SSII.01},
pronounced E1 strength has been found in the low-energy
region below 3 MeV, and related to the coherence in 
the transition amplitudes between the loosely-bound valence
nucleons, and also between the core and the valence
nucleons. The relativistic RPA calculation of the dipole response 
in Ar isotopes \cite{GiaiMa.99}, has shown a concentration of strength 
in the proton-rich nuclei $^{30}$Ar and $^{32}$Ar, which has been 
attributed to excitations from weakly-bound single-particle states into
the continuum. A recent study based on the self-consistent
RHB+RQRPA framework has shown that a new collective mode -- the
Proton Pygmy Dipole Resonance(PPDR) could arise in medium-heavy
nuclei close to the proton drip-line~\cite{PVR.05}.
In Fig.~\ref{figN20} the RQRPA dipole strength 
distributions in the N=20 isotones $^{40}$Ca, $^{42}$Ti, 
$^{44}$Cr, and $^{46}$Fe are shown, calculated with 
the fully consistent RHB+RQRPA with the DD-ME2 plus Gogny D1S 
effective interactions. The strength distributions are dominated by the 
IV GDR at $\approx 20$ MeV excitation energy.
With the increase of the number of protons,  
low-lying dipole strength appears in the region below the GDR and, 
for $^{44}$Cr and $^{46}$Fe, a pronounced 
low-energy peak is found at $\approx 10$ MeV excitation energy. 
In the lower panel of Fig.~\ref{figN20} we plot the proton and neutron 
transition densities for the peaks at 9.98 MeV in $^{44}$Cr and 
9.33 MeV in $^{46}$Fe, and compare them with the transition densities
of the GDR state at 18.82 MeV in $^{46}$Fe. Obviously the dynamics 
of the two low-energy peaks is very different from that
of the isovector GDR: the proton and neutron transition densities
are in phase in the nuclear interior and 
there is almost no contribution from the neutrons
in the surface region. As in the case of the PDR in neutron-rich nuclei, 
obviously the low-lying state does not belong to
statistical E1 excitations sitting on the tail of the GDR, but could 
indeed represent a fundamental mode of excitation: the proton electric pygmy 
dipole resonance (PPDR).

\begin{figure}
\vspace*{1cm}
\centering
\includegraphics[scale=0.45]{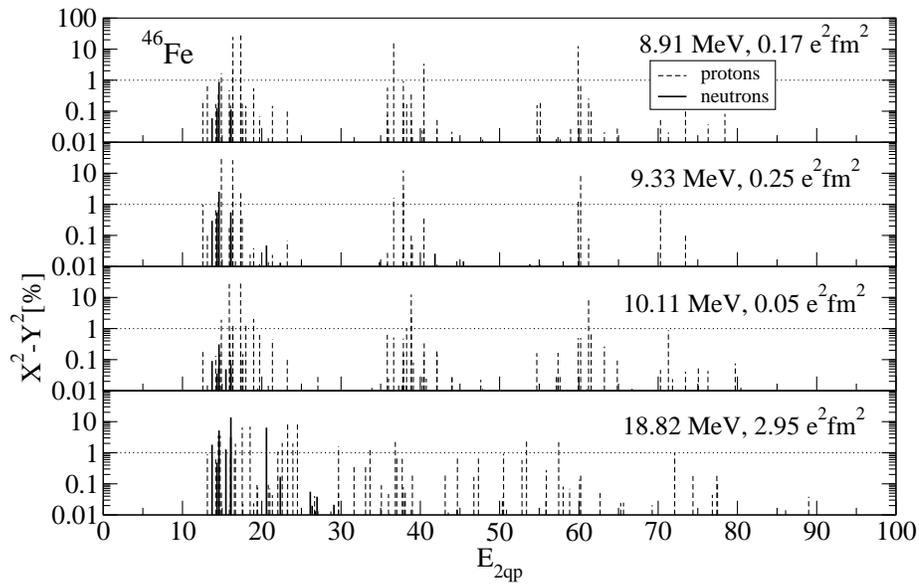}
\vspace*{0.5cm}
\caption{The distributions of the RQRPA amplitudes in $^{46}$Fe, 
plotted as functions of
the unperturbed energy of the respective $2qp$-configurations,
for the main peaks in the low-energy region and for the IV GDR state.}
\label{figFe46}
\end{figure}

In Fig.~\ref{figFe46} we analyse the RQRPA structure of the 
dipole response in $^{46}$Fe. This nucleus is located at the 
proton drip-line, and recently evidence for 
ground-state two-proton radioactivity 
was reported in the decay of $^{45}$Fe~\cite{Gio.02,Pfu.02}.
In the four panels we plot the QRPA
amplitudes of proton and neutron {$2qp$} configurations
\begin{equation}
\xi_{2qp}=\left|X^{\nu}_{2qp}\right|^2-
\left|Y^{\nu}_{2qp}\right|^2
\end{equation}
for the three low-lying states at 8.91, 9.33, and 10.11 MeV, as 
well as for the strongest state in the GDR region at 18.82 MeV. 
For each of the four dipole states, in addition to the excitation energy  
we have also included the corresponding B(E1) value. The 
amplitudes are shown in a logarithmic plot as functions of the 
unperturbed energy of the respective $2qp$-configurations.
Only amplitudes which contribute more than 0.01\% are shown, and 
we also differentiate between proton and neutron configurations. 
We note that, rather than a single
proton {$2qp$} excitation, the low-lying states are 
characterized by a superposition of a number of 
{$2qp$} configurations. Obviously the pygmy states display 
a degree of collectivity that can be directly compared with the QRPA 
structure of the GDR state at 18.82 MeV. In addition, proton 
{$2qp$} configurations account for $\approx$99\% of the 
QRPA amplitude of the pygmy states, whereas the ratio
of the proton to neutron contribution to the GDR 
state is $\approx 2$. For the GDR states the {$2qp$}
configurations predominantly correspond to excitations from the 
$sd$-shell to the $fp$-shell. The structure of the pygmy states,     
on the other hand, is dominated by transitions from the 
$1f_{7/2}$ proton state at -0.21 MeV, and from the $2p_{3/2}$ 
proton state at 3.63 MeV (this state is only bound because 
of the Coulomb barrier). The energy weighted sum of the strength 
below 11 MeV excitation energy corresponds to 2.7\% of the 
TRK sum rule.

\begin{figure}
\vspace*{1cm}
\centering
\includegraphics[scale=0.45]{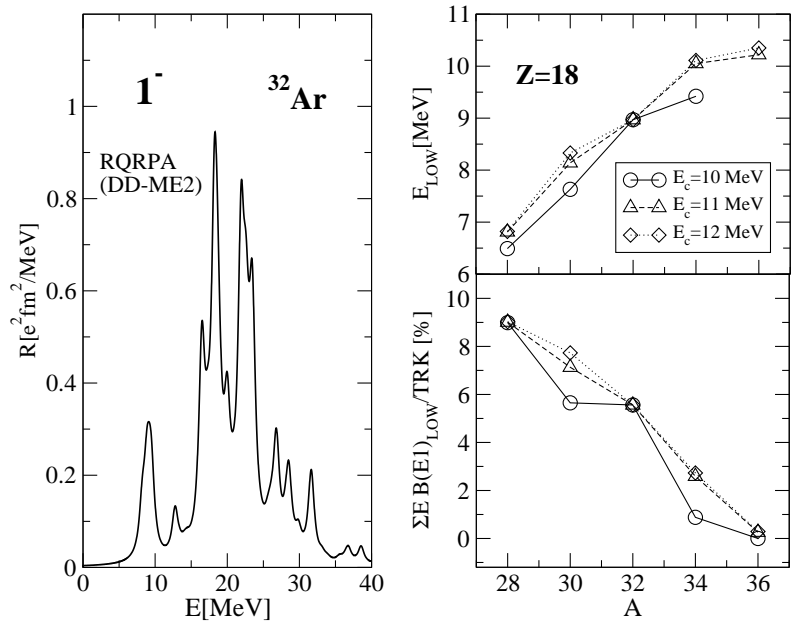}
\vspace*{0.5cm}
\caption{The RHB+RQRPA isovector dipole strength distribution
in $^{32}$Ar (left panel). 
The mass dependence of the PPDR centroid energy for Ar isotopes,
and the corresponding values of the integrated B(E1) strength
below $E_c$=10, 11, and 12 MeV 
are shown in the right panel.}
\label{pp_Ar}
\end{figure}

Another example where a pronounced proton PDR can occur are the 
proton-rich isotopes of Ar. In the left panel of 
Fig.~\ref{pp_Ar} we display the RHB+RQRPA electric dipole strength 
distribution in $^{32}$Ar. In addition to the 
rather fragmented GDR structure at $\approx 20$ MeV, prominent 
proton PDR peaks are calculated at 8.14, 8.79, 9.22, and 9.46 MeV. 
These peaks form the pygmy structure and exhaust 5.7 \% of the TRK
sum rule. The RQRPA amplitudes of the low-lying states present 
superpositions of many  proton {$2qp$} configurations, with the 
neutron contributions at the level of 1\%. The dominant configurations 
correspond to transitions from the proton states $1d_{3/2}$ (-1.94 MeV) 
and $2s_{1/2}$ (-3.98 MeV). In the right panel of Fig.~\ref{pp_Ar}
we display the mass dependence of the centroid energy of the pygmy peaks 
and the corresponding values of the integrated B(E1) strength 
below 10 MeV excitation energy.
In contrast to the case of medium-heavy and heavy neutron-rich isotopes, 
in which both the PDR and GDR are lowered in energy with the increase of the 
neutron number, in proton-rich isotopes the mass dependence of the PDR 
excitation energy and B(E1) strength is opposite to that of the GDR. 
The proton PDR decreases in energy with the development of the proton excess.
This mass dependence is intuitively expected because
the proton PDR is dominated by transitions from weakly-bound
proton orbitals. As the proton drip-line is approached,
either by increasing the number of protons or by decreasing the number of
neutrons, due to the weaker binding of higher proton orbitals one expects
more inert oscillations, i.e. lower excitation energies. The number of
$2qp$ configurations which include weakly-bound proton orbitals increases
towards the drip-line, resulting in an enhancement of the
low-lying B(E1) strength. 

For heavier nuclei the proton drip-line is located in the region of 
neutron-deficient, rather than proton-rich nuclei, and therefore 
one does not expect to find low-lying dipole strength in medium-heavy and 
heavy nuclei close to the proton drip-line.

The effect of the coupling to the continuum on the low-lying dipole strength 
in nuclei close the proton drip-line has also been analyzed in the
non-relativistic continuum RPA (CRPA) framework, 
using Skyrme interactions~\cite{HSZ.98,PPPW.05}.
In the doubly magic nucleus $^{48}$Ni both the CRPA and the RRPA 
predict the occurrence of the proton 
PDR, but the CRPA analysis has shown that, as a result of the coupling 
to the continuum, the PPDR is characterized by a rather large 
escape width~\cite{PPPW.05}.

\subsection{Di-neutron Correlations near the Drip-Line}
\label{dineutron}

The existence of another exotic mode has been suggested in
medium-heavy nuclei close to the neutron drip-line: 
a soft dipole excitation that corresponds to the vibration
of a di-neutron in the nuclear exterior against the 
remaining A-2 subsystem~\cite{ma05,MSM.06}. While in nuclei 
with a pronounced neutron excess the pygmy dipole resonance 
could appear, di-neutron {\it vs} core vibrations may occur in 
very exotic nuclei near the neutron drip-line. 
The latter mode  is strongly influenced
by neutron pairing correlations, and is characterized by a 
large transition density for pair motion of neutrons. 
In the case of light halo-nuclei pairing correlations between 
the loosely-bound neutrons in the halo lead to a strong enhancement 
of the soft dipole excitations. Experimental signatures of di-neutron
correlation in the soft dipole mode have 
been found in $^{11}$Li \cite{Shi.95,Naka.06}.
However, as has recently been shown in Ref.~\cite{ma05}, some features 
of di-neutron correlations may also be present in the ground states of 
exotic medium-mass nuclei, and therefore influence their excitations.

In the HFB description of the nuclear ground state (cf. Sec.~\ref{firstHFB}),
the spatial correlations between a pair of neutrons can be probed by
the neutron two-body correlation density \cite{ma05},
\begin{equation}
\rho_{corr}({\bf r}\sigma,{\bf r'}\sigma')=|\kappa({\bf r}\sigma,{\bf
r'}\bar{\sigma'})|^2-|\rho({\bf r}\sigma,{\bf r'}\sigma')|^2,
\end{equation}
defined by the off-diagonal terms of the pairing tensor (Eq.~(\ref{kappa0})), 
and density matrix (Eq.~(\ref{rho0})). The notation $\bar{\sigma}$ has been 
defined in Eq.~(\ref{eq:ubig}). In the case of medium-heavy
nuclei close to the neutron drip-line, it has been shown that 
the two-body correlation density clearly reflects the presence of 
spatial di-neutron correlations in the pair-correlated ground state, 
and these are especially pronounced on the surface and in the region
of the neutron-skin. The di-neutron correlation originates from a
coherent superposition of quasiparticle orbits with large orbital
angular momenta, which are embedded in the nucleon continuum.

Here we illustrate the role of neutron-pair correlations with the 
example of the dipole response in $^{158}$Sn \cite{MSM.06}. 
The HFB calculation of the ground state has been performed 
with the Skyrme effective interaction SLy4, and the density-dependent 
delta-interaction of Eq.~(\ref{v_pair_dd}) in the pairing channel.
The overall strength parameter of the pairing force has been 
adjusted to reproduce the $^1$S scattering length $a = -18$ fm 
in free space.
By varying the value of the parameter $\eta$, which multiplies the
density-dependent term (see Eq.~(\ref{v_pair_dd})),
one controls the effective pairing interaction. A smaller value
of $\eta$, i.e. a weaker density dependence, results in stronger 
pairing in the interior of the nucleus. For the particular case 
of Sn isotopes, $\eta = 0.71$ has been adjusted so that the
average pairing gap becomes comparable
to the experimental value $\Delta = 1.1 - 1.4$ MeV. The choice $\eta = 0$
corresponds to a density-independent pairing interaction with an extremely
large and unrealistic pairing gap $\Delta \approx 15$ MeV.

The response function has been calculated with the continuum 
QRPA (cf. Sec.~\ref{sec2_cqrpa}), which employs the
quasiparticle Green's function with exact
outgoing boundary condition for neutrons \cite{MSM.06}.
For the $ph$ residual QRPA interaction, the Landau-Migdal
approximation to the Skyrme functional has been used, whereas 
the pairing correlations have been consistently described by the 
same density-dependent delta-interaction~(\ref{v_pair_dd}),
both in the HFB and QRPA. An additional
renormalization factor must be introduced in the $ph$ 
channel in order to remove the spurious center-of-mass
contributions.

\begin{figure}[htb]
\centering
\includegraphics[scale=0.4]{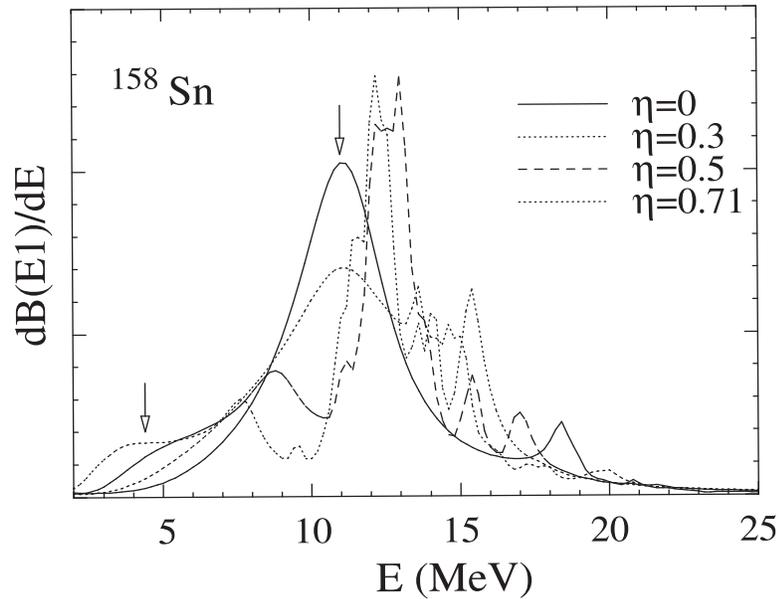}
\caption{The HFB+QRPA isovector dipole strength function in $^{158}$Sn.
The strength distributions obtained for several values of the pairing 
parameter $\eta = 0, 0.3, 0.5$ and 0.71 are shown \cite{MSM.06}.}
\label{fig158Sn}
\end{figure}

In Fig.~\ref{fig158Sn} the HFB+QRPA dipole strength function 
for $^{158}$Sn is shown for different choices of the 
parameter of the pairing interaction: $\eta = 0, 0.3, 0.5$ and 0.71.
For the realistic value of $\eta = 0.71$, in addition to the 
IVGDR in the high-energy region between 10 and 15 MeV, 
the strength distribution exhibits a pronounced low-energy 
structure below 5 MeV. In the case of extremely strong pairing 
($\eta$=0), the transition strength basically contains only a single
broad resonance centered at 11 MeV.

\begin{figure}[htb]
\centering
\includegraphics[scale=0.6,angle=-90]{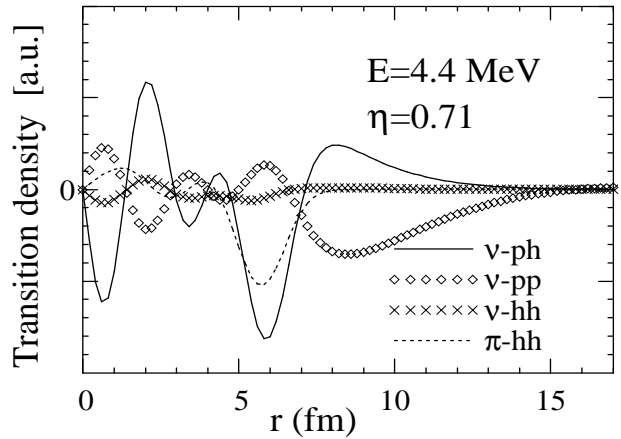}
\caption{The transition densities for the dipole peak at $E=4.4$ MeV
in $^{158}$Sn, calculated with the value $\eta=0.71$ 
for the parameter of the density-dependent pairing interaction \cite{MSM.06}.
The curves correspond to the neutron particle-hole transition density 
(solid), neutron particle-pair transition density (diamonds), 
neutron hole-pair transition density (crosses), and the
proton particle-hole transition density (dashed). 
}
\label{td158Sn}
\end{figure}

The structure of the characteristic low-energy peak at 4.4 MeV,
obtained in the realistic calculation with  $\eta = 0.71$, is 
explored in more details in Fig.~\ref{td158Sn}, where the 
corresponding transition densities are shown \cite{MSM.06}. 
A large contribution of the neutron particle-pair transition 
density is found in the region beyond the nuclear surface. 
It originates from dynamical pairing correlations among
neutrons moving in the external region, i.e., the QRPA
correlation is determined by the pairing interaction.
The dynamics of this soft mode can be interpreted as the vibration
of di-neutrons against the core. Because of the influence of 
neutron pairing correlations, this mode has a dominant
particle-particle character. The two-quasiparticle configurations,
including orbitals in the continuum with orbital angular momenta 
up to $l \approx 10$, contribute coherently to the large particle-pair 
transition density. Let us also note that in the continuum QRPA calculations
of Ref.~\cite{MSM.06}, the quadrupole core {\it vs} dineutron mode has also 
been predicted in the low-energy region, but 
only for a very strong, and therefore unrealistic, pairing interaction.

\section{Isoscalar Modes}
\label{Sec5}

%
\subsection{Isoscalar  
Dipole Compressional and Toroidal Modes}\label{dipole}
%
Experimental excitation energies of compressional (monopole and
dipole) vibrational modes in atomic nuclei can in principle be used 
to deduce the value of the nuclear matter compression
modulus K$_{nm}$~\cite{Bla.80}. This quantity is related
to the curvature of the nuclear matter equation of state at 
the saturation point, and controls basic properties of atomic nuclei, 
the structure of neutron stars, the dynamics
of heavy-ion collisions and of supernovae explosions. The nuclear 
matter compressibility cannot be measured directly, but rather deduced 
from a comparison of experimental excitation energies of isoscalar giant 
monopole resonances (ISGMR), with the corresponding values predicted by 
microscopic nuclear effective interactions characterized by different 
values of K$_{nm}$.

Inelastic $\alpha$-scattering experiments have been used in 
high precision studies of the systematics of ISGMR in nuclei 
with $A \ge 90$. Nuclear structure models provide a consistent
description of the main moments of strength distributions and the
mass dependence of excitation energies, and thus relate the ISGMR  
to the nuclear matter 
compression modulus. There is much less experimental information, 
and only few microscopic theoretical analyses of the 
structure of compressional modes in lighter nuclei with $A < 90$. 
While in heavy nuclei the shape of the ISGMR strength distribution is
typically symmetric, for $A < 90$ the ISGMR display asymmetric shapes 
with a slower slope on the high energy side of the peak, and with 
a further decrease of the mass number the ISGMR strength 
distributions become strongly fragmented.
The determination of K$_{nm}$ is based
on microscopic calculations of ISGMR excitation energies.
Interactions that differ in their prediction
of the nuclear matter compressibility,
but otherwise reproduce experimental data
on ground-state properties reasonably well,
are used to calculate ISGMR in the random phase approximation
or the time-dependent framework. A fully
self-consistent calculation of both ground-state properties and
ISGMR excitation energies restricts the range of possible
values for K$_{nm}$. It has been pointed out,
however, that, since K$_{nm}$ determines
bulk properties of nuclei and, on the other hand,
the ISGMR excitation energies depend also on the surface compressibility,
measurements and microscopic calculations of ISGMR in
heavy nuclei should, in principle, provide a more reliable
estimate of the nuclear matter compressibility~\cite{Bla.80,BBD.95a}.

Recent theoretical studies of nuclear compressional modes  
have employed the fluid dynamics approach~\cite{Kol.00},
the Hartree-Fock plus random phase
approximation (RPA)~\cite{Ham.97,Shl.02,Agr.03,Col.04}, 
the RPA based on separable Hamiltonians~\cite{Kva.03}, linear response
within a stochastic one-body transport theory~\cite{Lac.01}, 
the relativistic transport approach~\cite{Yil.05},
and the self-consistent relativistic RPA~\cite{MGW.01,Pie.01,Pie.02,VNR.03}.
Several analyses have emphasized the importance of a 
fully self-consistent description of ISGMR, and confirmed that
the low value of K$_{nm}= 210-220$ MeV, previously obtained with Skyrme
functionals, is an artefact of the inconsistent implementation of effective
interactions~\cite{Col.04,CG.04}. The excitation energies of the ISGMR 
in heavy nuclei are thus best described with Skyrme and Gogny effective 
interactions with K$_{nm} \approx$ 235 MeV. In Ref.~\cite{Agr.03} it has 
been shown that it is also possible to construct Skyrme forces that fit 
nuclear ground state properties and reproduce ISGMR energies, but with 
K$_{nm}$ $\approx 255$ MeV. In Ref.~\cite{Col.04} a new set of Skyrme forces 
was constructed that spans a wider range of values of K$_{nm}$ and 
the symmetry energy at saturation density $a_4$.  
RPA calculations  with 
these forces have shown that the ISGMR
data can be reproduced either with forces having a softer
density-dependent term (the exponent $\alpha = 1/6$ in 
Eq.~(\ref{Skyrme})) and K$_{nm} \sim 230-240$ MeV, 
or with forces having a stiffer density-dependent 
term ($\alpha = 1/3$) and K$_{nm} \sim  250-260$ MeV. 
Other forces, in particular those characterized 
by larger values of K$_{nm}$, are associated with unrealistic 
values of the effective mass, and do not reproduce
ground-state properties. On the other hand, 
it appears that in the relativistic framework the interval of allowed
values for K$_{nm}$ is more restricted. A recent relativistic RPA 
analysis based on modern effective Lagrangians with explicit 
density dependence of the meson-nucleon vertex functions, has shown that 
only effective interactions with K$_{nm}= 250-270$ MeV reproduce 
the experimental excitation energies of ISGMR in medium-heavy and 
heavy nuclei, and that K$_{nm} \approx 250$ MeV represents the lower 
limit for the nuclear matter compression modulus of relativistic
mean-field interactions~\cite{VNR.03}. 

The isoscalar giant dipole resonance (ISGDR) is a second order effect, 
built on $3 \hbar\omega$, or higher configurations. It corresponds to a 
compression wave traveling back and forth through the nucleus along a 
definite direction. Recent data on the compressional ISGDR 
in $^{90}$Zr, $^{116}$Sn, $^{144}$Sm, and 
$^{208}$Pb~\cite{Tex.01,Tex.04,Uch.03,Uch.04} 
can also be used to constrain the range of allowed values of 
K$_{nm}$~\cite{Garg.04}. The problem, however, is that the isoscalar
E1 strength distributions display a characteristic bimodal structure with 
two broad components: one in the low-energy region close to the isovector
giant dipole resonance (IVGDR) 
($\approx 2 \hbar \omega$), and the other at higher
energy close to the electric octupole resonance ($\approx 3 \hbar \omega$).
Theoretical analyses have shown that only the high-energy component 
represents compressional vibrations~\cite{VWR.00,Colo.00}, whereas
the broad structure in the low-energy region could correspond to vortical 
nuclear flow associated with the toroidal dipole 
moment~\cite{BMS.93,Mis.06,VPRN.02}. However, as has also been pointed
out in the recent study of the interplay between compressional and 
vortical nuclear currents~\cite{Mis.06}, a strong mixing between 
compressional and vorticity vibrations in the isoscalar E1 states 
can be expected up to the highest excitation energies in the region 
$\approx 3 \hbar \omega$. Nevertheless, models which use effective 
interactions with K$_{nm}$ adjusted to ISGMR excitation energies 
in heavy nuclei, also reproduce the overall structure of the high-energy 
portion of ISGDR data~\cite{Shl.02,Uch.03,Uch.04,Itoh.03,Sil.06}.

\begin{figure}[]
\begin{center}
\vspace{1cm}
\includegraphics[scale=0.45]{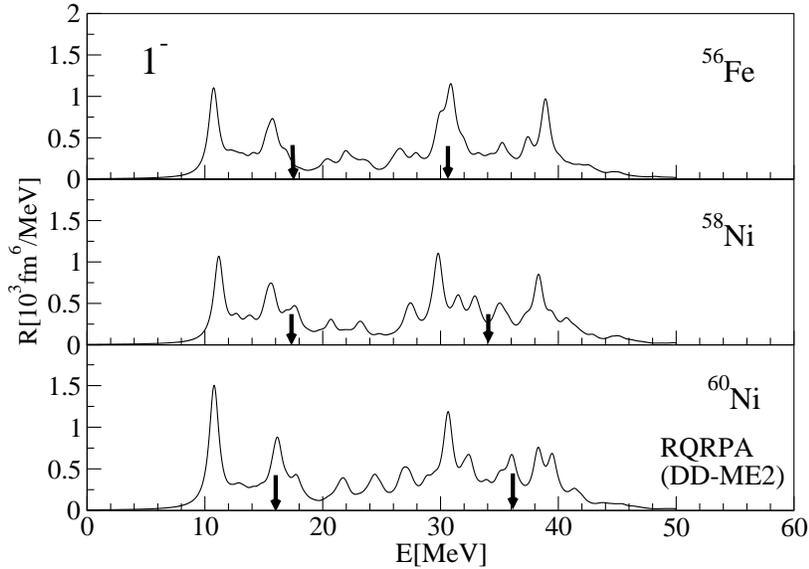}
\vspace{0.75 cm}
\caption{The RHB+RQRPA isoscalar dipole transition strength in
$^{56}$Fe, $^{58}$Ni, and $^{60}$Ni calculated with
DD-ME2 effective interaction. The arrows denote the positions 
of the experimental centroid energies of the low- and high-energy 
components~\protect\cite{Lui.06}.}
\label{29-dip}
\end{center}
\end{figure}

Accurate data on compressional modes are becoming available also 
for lighter nuclei, e.g. $^{56}$Fe, $^{58}$Ni, 
$^{60}$Ni~\cite{Lui.06,Nay.06}. Inelastic $\alpha$-scattering 
data on the isoscalar monopole and dipole strength distributions 
have been analyzed in the relativistic quasiparticle random-phase 
approximation (RQRPA) with the DD-ME2 effective nuclear interaction
in the particle-hole channel and the finite-range 
Gogny force in the particle-particle channel~\cite{PVNR.06}. In 
Fig.~\ref{29-dip} we display the strength functions in 
$^{56}$Fe and $^{58,60}$Ni, for the isoscalar dipole operator:
\begin{equation}
\hat{Q}^{T=0}_{1 \mu}  =   \sum^{A}_{i=1} \, \gamma_0 \, \left(r_i^3
- \frac{5}{3} \, \langle r^2 \rangle_{_0} r_i\right) \ Y_{1
\mu}(\Omega_i), \label{operator}
\end{equation}
where $\langle r^2 \rangle_{_0}$ denotes the ground-state 
expectation value, and the inclusion of the second term in the
operator ensures that
the strength distribution does not contain spurious
components that correspond to the center-of-mass motion.
In all three nuclei the strength 
is strongly fragmented and distrubuted over a wide range of 
excitation energy between 10 MeV and 40 MeV, in agreement 
with the experimental results of Ref.~\cite{Lui.06}. In the experiment 
between 56\% and 72\% of the isoscalar E1 strength has been located in
these nuclei below 40 MeV excitation energy, and some missing strength 
probably lies at higher energies. Similarly to the results obtained for 
heavier nuclei~\cite{VWR.00,Colo.00,VPRN.02}, the E1 strength is 
basically concentrated in two broad structures: one in the region 
10 MeV $\leq$ E$_x$ $\leq$ 20 MeV, and the high-energy component 
above 25 MeV and extending above 40 MeV excitation energy. Only the 
high-energy portion of the calculated E1 strength is sensitive to the 
nuclear matter compression modulus of the effective interaction.
The thick arrows denote the locations of the 
experimental centroid energies ($m_1/m_0$) in the 
low- and high-energy regions of the isoscalar E1 strength in 
$^{56}$Fe, $^{58}$Ni, and $^{60}$Ni~\cite{Lui.06}.
We notice a good qualitative agreement between the calculated 
and experimental centroids in the high-energy region, especially 
taking into account that the isoscalar strength above E$_x = 40$ MeV has 
not been observed in the experiment. In the low-energy region, however,  
the theoretical centroid energies are systematically below 
the experimental values by $\approx$1--4 MeV. This effect is in agreement 
with previous RRPA calculations in heavier nuclei~\cite{VPRN.02}, 
and supports the picture of pronounced mixing between 
compressional and vorticity vibrations in the intermediate 
region of excitation energies. 

The role of toroidal multipole form factors and moments 
in the physics of electromagnetic and weak interactions
has been extensively discussed in Refs.~\cite{DC.75}
and~\cite{DT.83}. They appear in multipole expansions
for systems containing convection and induction 
currents. In particular, the multipole expansion 
of a four-current distribution gives rise to three families 
of multipole moments: charge moments, magnetic moments and 
electric transverse moments. The latter are related to 
the toroidal multipole moments and result from the expansion
of the transverse electric part of the current. The toroidal
dipole moment, in particular, describes a system of 
poloidal currents on a torus. Since the charge density 
is zero for this configuration, and all the turns of the 
torus have magnetic moments lying in the symmetry plane, 
both the charge and magnetic dipole moments of this 
configuration are equal to zero. The simplest model
is an ordinary solenoid bent into a torus.

Vortex waves in nuclei were analyzed in a hydrodynamic
model~\cite{Sem.81}. By relaxing the assumption
of irrotational motion, in this pioneering study solenoidal 
toroidal vibrations were predicted, which correspond
to the toroidal giant dipole resonance at excitation 
energy $E_x \approx (50 - 70)/A^{1/3}$ MeV.
It was suggested that the vortex excitation modes
should appear in electron backscattering.
The isoscalar $1^-$ toroidal dipole states were studied
in the framework of the time-dependent Hartree-Fock theory
by analyzing the dynamics
of the moments of the Wigner transform of the density
matrix~\cite{BM.88}, and excitations with dipole toroidal 
structure were also found in semi-classical studies based
on nuclear fluid dynamics~\cite{BMS.93,Mis.06}.
The first fully microscopic analysis of toroidal dipole 
resonances (TGDR) was performed in the framework of the 
relativistic  RPA~\cite{PPhD.03,VPRN.02}.
Compressional and toroidal dipole modes were
also studied with the Quasiparticle Phonon Model, 
using separable residual interactions
with the Nilsson or Woods-Saxon mean-field potentials~\cite{Kva.03}. 
Continuum RPA calculations with Skyrme interactions in Ni isotopes 
have shown that vortex waves
could also occur for excitations with
higher multi-polarities $2^+$, $3^-$ and
$4^+$~\cite{PWP.04}.

\begin{figure}
\vspace*{1cm}
\centering
\includegraphics[scale=0.45]{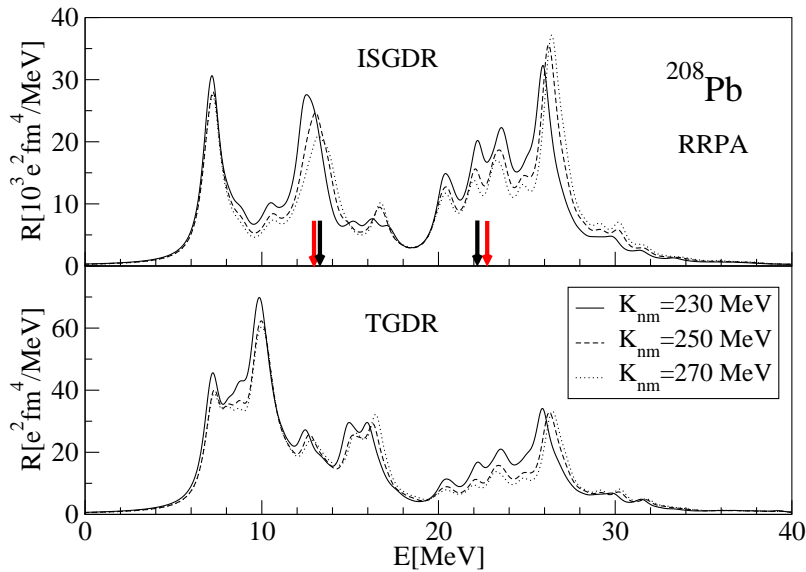}
\vspace*{0.5cm}
\caption{Isoscalar dipole (upper panel)
and toroidal dipole strength distributions (lower panel) for
$^{208}$Pb, calculated with the RRPA based on three 
density-dependent interactions with $K_{nm}$=230, 250, and 270 MeV.
The $(\alpha,\alpha')$ experimental data
for the centroids of the low-energy and high-energy
portions of the isoscalar dipole distribution are
denoted by dark~\protect\cite{You.04} and
light~\protect\cite{Uch.03,Uch.04} arrows.
}
\label{figtor1}
\end{figure}

In Fig.~\ref{figtor1} we display the RRPA dipole strength distributions
in $^{208}$Pb for the isoscalar dipole operator (ISGDR) in the upper 
panel, and for the isoscalar toroidal dipole operator (TGDR)
\begin{equation}
\hspace{-2cm}
\hat{T}^{T=0}_{1 \mu}  =  -\sqrt{\pi} \sum_{i=1}^A \left[~
r^2_i \left ( \overrightarrow{Y}^*_{10\mu} (\Omega_i) + {{\sqrt{2}}\over 5} 
\overrightarrow{Y}^*_{12\mu} (\Omega_i) \right ) \cdot \vec \alpha_i
 - < r^2 >_{_0} \overrightarrow{Y}^*_{10\mu} (\Omega_i) \cdot \vec \alpha_i~ 
\right],
\label{TDR2}
\end{equation}
in the lower panel. $\overrightarrow{Y}_{ll^{\prime} \mu}$ denotes a vector 
spherical harmonic, and $ \vec \alpha$ are the Dirac $\alpha$-matrices.
As in the case of the dipole operator Eq.~(\ref{operator}), the 
second term ensures that the TGDR strength distributions do not 
contain spurious center-of-mass motion components.
To illustrate the correlation between the
nuclear matter compressibility and the isoscalar dipole
response, the strength functions are calculated with 
three different relativistic effective interactions with $K_{nm}$=230, 250, 
and 270 MeV, and the volume asymmetry at saturation $a_4$=32 MeV~\cite{VNR.03}.
The experimental centroids of the low-energy and high-energy 
portions of the dipole strength distributions, 
extracted from small angle $\alpha$-scattering spectra, 
are denoted by arrows~\cite{Uch.04,You.04}.

\begin{figure}
\vspace*{1cm}
\centering
\includegraphics[scale=0.5]{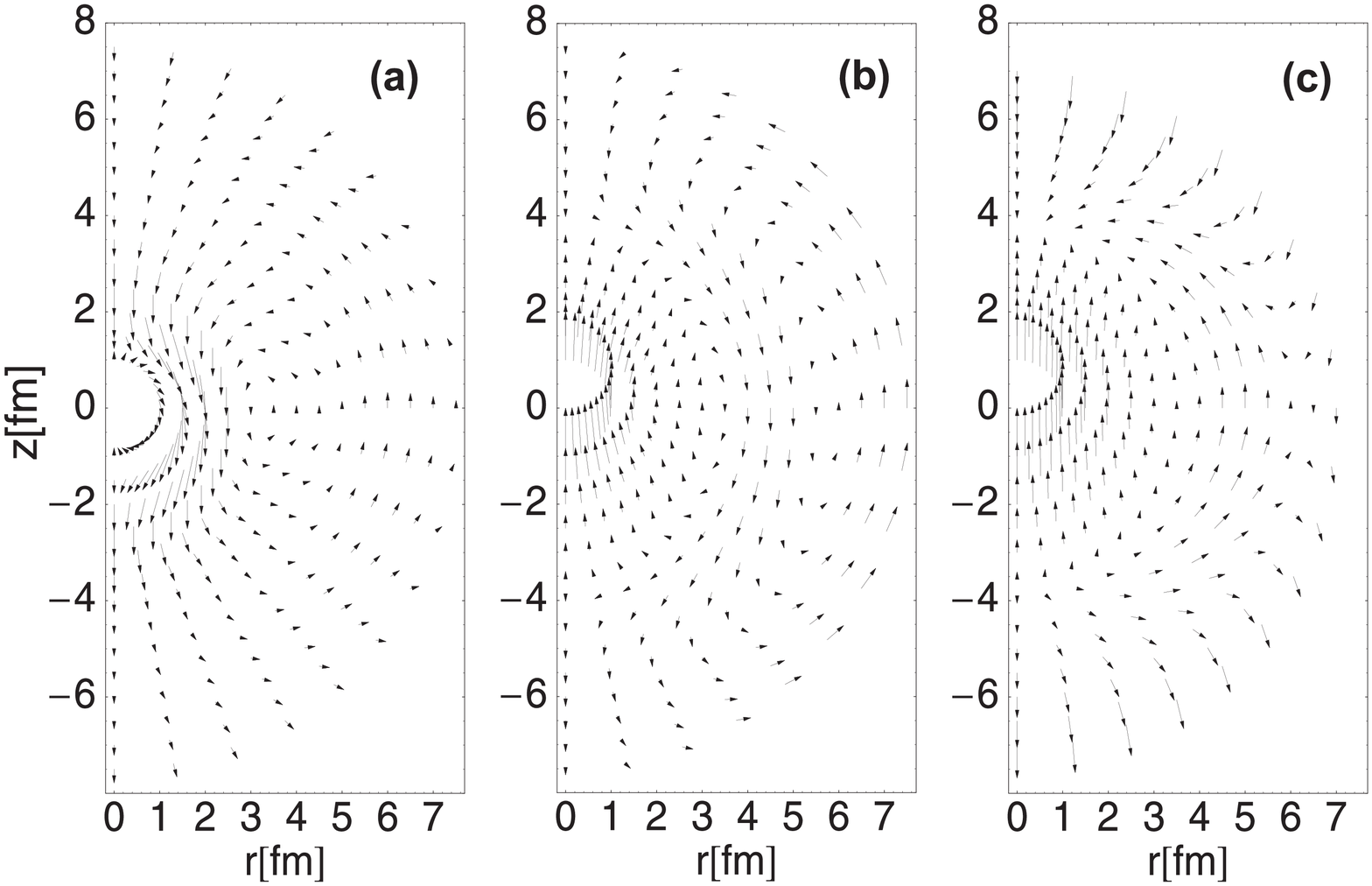}
\vspace*{0.5cm}
\caption{Velocity distributions for the three most pronounced 
peaks in the TGDR response of $^{208}$Pb (calculated with 
the DD-ME2 effective interaction). 
The velocity fields correspond to the peaks at
(a) 7.2 MeV ,(b) 10.0 MeV and (c) 26.1 MeV.}
\label{figvortex}
\end{figure}

\begin{figure}
\centering
\includegraphics[scale=0.6]{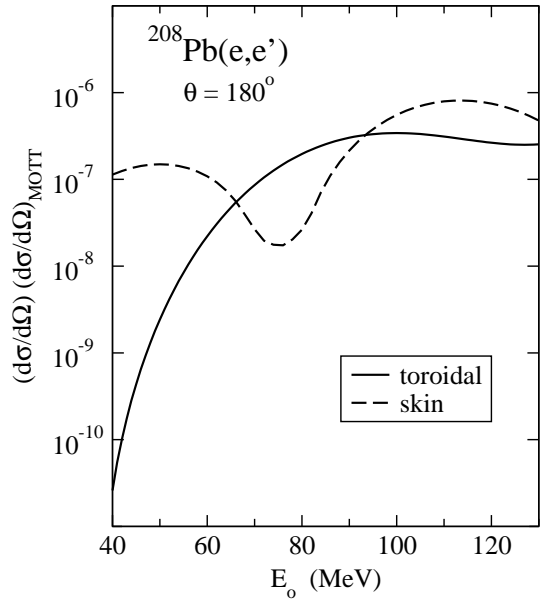}
\caption{Electron scattering form factors of dipole transitions
at 180$^{\circ}$ for $^{208}$Pb, based on calculations with the 
quasiparticle phonon model~\protect\cite{Ric.04}. The two curves 
correspond to predominantly toroidal (solid) and neutron-skin 
(dashed) density oscillations, respectively.}
\label{tor-skin}
\end{figure}

The positions of the calculated peaks in the low-energy region 
(below 20 MeV) depend only weakly on the incompressibility, whereas 
the structures in the high-energy region are much 
more sensitive to the choice of
the compression modulus of the interaction. Both dipole strength
distributions display two broad structures: one at low energies 
between 8 and 20 MeV, and one in the high-energy region $20 - 30$ MeV. 
Obviously, a strong coupling between the two isoscalar $1^-$ modes can 
be expected. This coupling becomes even more
evident if one rewrites the expression in square brackets of
the toroidal operator Eq.~(\ref{TDR2}) as~\cite{Sem.81}
\begin{equation}
\nabla \times ( \vec r \times \nabla ) (r^3 -
\frac{5}{3} \, < r^2 >_{_0}  r) \ Y_{1 \mu},
\end{equation}
and compares it with the isoscalar dipole operator of the compression
mode (\ref{operator}). The relative position of the two resonance 
structures will, therefore, depend on the interaction between the 
toroidal and compression modes. 
In Table~\ref{tabletor} we compare the RQRPA
centroid energies of the low-lying portion of the response to
the toroidal operator, with the corresponding experimental
values for $^{116}$Sn, $^{144}$Sm, and $^{208}$Pb~\cite{Uch.04,You.04}. 
We notice that the theoretical centroids, calculated with the 
DD-ME2 effective interaction, are systematically located $\approx 1-2$ MeV
below the experimental values.
\begin{table}[b]
\centering
\begin{tabular}{lccc} \hline                                                  &  $m_1/m_0$(MeV)  &  $E_x$(MeV)~\cite{Uch.04}  &  $E_x$
(MeV)~\cite{You.04}   \\ \hline\hline                              $^{116}$Sn     &   13.3      &    15.6$\pm$0.5 &  14.38$\pm$0.25
\\ \hline                                                          $^{144}$Sm    &   12.7      &   14.2$\pm$0.2 &  14.00$\pm$0.30 \\
\hline                                                              $^{208}$Pb    &    11.2 &  13.0$\pm$0.1 & 13.26$\pm$0.30
  \\ \hline                                                        \end{tabular}
\caption{The RQRPA centroid energies of TGDR 
calculated with the DD-ME2 effective
interaction in the region below 20 MeV excitation energy,
compared with the corresponding experimental centroid energies
from Refs.~\cite{Uch.04} and~\cite{You.04}.}
\label{tabletor}                                                   
\end{table}
The dynamics of the solenoidal toroidal vibrations is illustrated in
Fig.~\ref{figvortex}, where we plot the velocity fields for the three most
pronounced peaks of TGDR response function in
$^{208}$Pb (calculated with DD-ME2). A vector of unit length is
assigned to the largest velocity. All the other velocity vectors are
normalized accordingly. Since the collective flow is axially
symmetric, we plot the velocity field in cylindrical coordinates.
The $z$-axis corresponds to the symmetry axis of a torus.
The lowest peak at 7.2 MeV is dominated by vortex collective motion. 
The velocity fields in the
$(z,r_{\perp})$ plane correspond to poloidal currents on a torus
with vanishing inner radius. The poloidal currents determine the
dynamical toroidal moment. The high-energy peak at 26.1 MeV
displays the dynamics of dipole compression mode. The ``squeezing"
compression mode is identified by the flow lines which concentrate
in the two ``poles" on the symmetry axis. The velocity field
corresponds to a density distribution which is being compressed in
the upper half-plane, and expands in the lower half-plane. The
centers of compression and expansion are located on the symmetry
axis, at approximately half the distance between the center and the
surface of the nucleus. Finally, the intermediate peak at 10.0 MeV
displays the coupling between the toroidal and compression
dipole modes. A very similar behavior of the velocity distributions
as function of excitation energy is also observed for $^{116}$Sn 
and $^{144}$Sm. 
A direct experimental evidence for the TGDR mode 
remains a challenge for future studies. In principle 
the vortex type of motion could be identified in the measurement 
of transverse electron scattering form factors.
An exploratory study with the quasiparticle
phonon model (QPM) has shown that the cross 
sections in electron back-scattering
could differentiate between the toroidal and 
neutron-skin dipole modes~\cite{Ric.04}. 
The respective QPM electron scattering
form factors at 180$^{\circ}$ are shown in Fig.~\ref{tor-skin},
for transitions dominated by the toroidal and neutron-skin 
oscillations. Information about the nature
of the low-lying dipole excitations could be 
obtained in the range of incident energies 
$40- 90$ MeV, even though the 
predicted values for the
cross sections are low~\cite{Ric.04}.

\subsection{Low-Lying Quadrupole States in Unstable Nuclei}

In most even-even nuclei the first excited state is a $J^{\pi}$=2$^+$.
For magic nuclei the electric transition from the ground state 
to the first 2$^+$ state reflects directly the structure of the shell gap. 
The systematics of the 2$_1^+$ energies is very important in studies of 
shell evolution in nuclei far from stability. In particular, low-lying 
quadrupole states are closely related to the phenomenon of nuclear 
superfluidity: the 2$_1^+$ states are built from nucleonic configurations 
located close to the Fermi surface, which is precisely the energy
region in which the pairing interaction is most effective~\cite{ring80,BB.94}. 

In oxygen isotopes the evolution of shell structure 
can be explored from the stable isotope to the neutron 
drip-line in just a few mass units (cf. also Sec.~\ref{low_oxy_sec}).
A number
of experimental~\cite{ta97,je99,th00,kh00,be01,st04,be06} and 
theoretical~\cite{kh02,ma01,ot01,to02,ri03,br05} studies
of neutron-rich oxygen nuclei
have been reported. Recent theoretical analyses have predicted 
the appearance of new magic numbers in $^{22}$O ($N$=14) \cite{br05}, and
$^{24}$O ($N$=16). Shell-model calculations~\cite{br01} also show 
a strong gap $\approx 4.3$ MeV between the 1$d_{5/2}$ and 2$s_{1/2}$ 
subshells, and thus $^{22}$O appears to be a magic nucleus. 
Accordingly, both QRPA~\cite{kh00} and shell-model~\cite{br05,ut99} 
calculations predict a decrease of the $B(E2;2_1^+ \to 0_1^+)$ 
from $^{20}$O to $^{22}$O. This has been confirmed
by recent experimental results~\cite{th00,st04,be06}.
Table~\ref{tableoxy} summarizes the experimental values of excitation 
energies and $B(E2)$ values for the 2$^+_1$ state in neutron-rich 
isotopes. 

\begin{table}
\centering
\begin{tabular}{lccc} \hline 
&  E$_{exp}$ (MeV)  &  B(E2)$_{exp}$ (e$^2\cdot$fm$^4$)  & $\frac{M_n/M_p}{N/Z}$ 
\\ \hline\hline                          
$^{18}$O     &  2.0      &    45$\pm$2 &  0.9$\pm$0.2
\\ \hline  
$^{20}$O    &   1.7      &   28$\pm$2 &  2.2$\pm$0.5 
\\ \hline    
$^{22}$O    &    3.2     &  21$\pm$8 &  1.4$\pm$0.5
\\ \hline
$^{24}$O    &  $>$ 3.8 &  - & -
\\ \hline
\end{tabular}
\caption{Experimental values of excitation energies, electromagnetic
transition probabilities, and ratios of the transition matrix 
elements, for the first 2$^+$ states in neutron-rich oxygen isotopes.}
\label{tableoxy}
\end{table}

The energy of the first $2^+$ state in $^{22}$O has been measured at
3199(8) keV~\cite{be01}, compared to 1670 keV in $^{20}$O, and its small
$B(E2)$ value of 21(8) e$^2\cdot$fm$^4$~\cite{th00} indicates a strengthening
of the $N$=14 shell gap. Even though the 2$^+_1$ state of $^{24}$O has not
been  observed directly, it has been shown that its energy must lie 
above 3.8 MeV, and this points to $N$=16 as a new shell closure~\cite{st04}. 
$^{28}$O, which is a doubly magic nucleus in the standard shell model, 
was found to be neutron unbound~\cite{ta97}. 

\begin{figure}[]
\centerline{\includegraphics[scale=0.45]{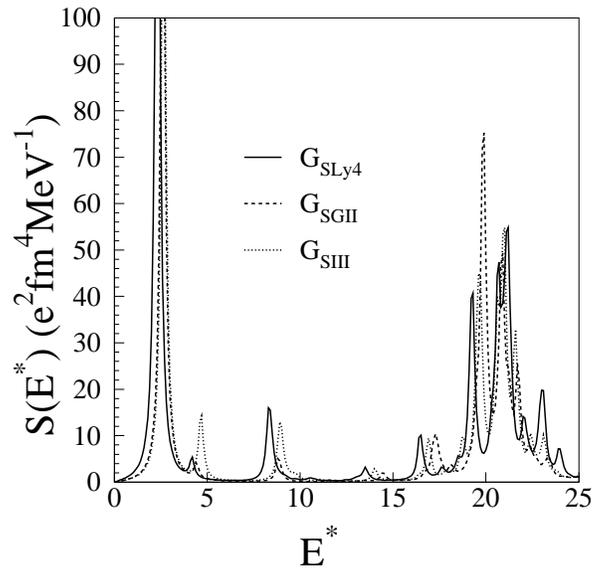}}
\caption{The QRPA isoscalar quadrupole strength function for 
the $^{20}$O nucleus, calculated 
with the SLy4, SGII and SIII Skyrme forces.
\label{fig:skyrme}}
\end{figure}

\begin{figure}[]
\centerline{\includegraphics[scale=0.45]{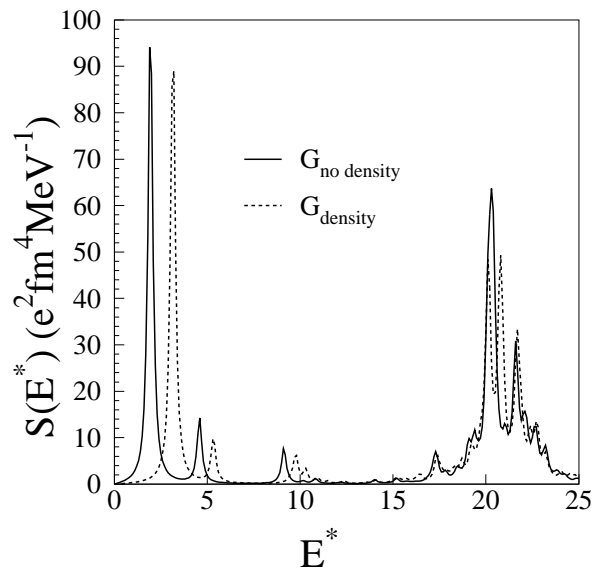}}
\caption{Isoscalar quadrupole strength function of $^{18}$O 
calculated with two types of pairing interaction. See text for description.
\label{fig:densdep}}
\end{figure}

Both the Gogny functionals~\cite{Per.05,gi03}, and Skyrme interactions 
with density dependent pairing
(cf. Eq.~(\ref{v_pair_dd}))~\cite{kh02,ma01,kh00,kh01}, 
have been employed in recent QRPA studies 
of the structure of 2$_1^+$ states in exotic nuclei. In Fig.~\ref{fig:skyrme} 
we display the quadrupole response function in $^{20}$O, calculated 
with the Skyrme interactions SLy4~\cite{chab98}, 
SGII~\cite{gi81} and SIII~\cite{beiner}. In addition to the strong 2$_1^+$ 
state at $\approx 3$ MeV, the pronounced structure above 20 MeV 
corresponds to the isoscalar giant quadrupole resonance (IS GQR). 
Obviously the details of the calculated quadrupole strength function depend 
on the choice of the interaction. The choice of the pairing interaction, 
in particular, plays an important role in the calculation of the 
2$_1^+$ states. This is illustrated in Fig.~\ref{fig:densdep} which 
compares the isoscalar quadrupole strength functions of $^{18}$O, 
calculated with two different pairing interactions Eq.~(\ref{v_pair_dd}).
In addition to the density-dependent (surface-type pairing) interaction of 
Eq.~(\ref{v_pair_dd}), a density-independent interaction (volume-type pairing)
i.e. $\eta$=0 in Eq.~(\ref{v_pair_dd}), has been used in the pairing channel. 
The strength parameter is adjusted to the empirical pairing gap in $^{18}$O.
We note that without any density dependence in the pairing channel, the 
calculated excitation energy of the 2$_1^+$ state is in better agreement with 
the experimental value of $\approx 2$ MeV (see Table~\ref{tableoxy}). 
This shows that an analysis of low-lying quadrupole states in
neutron-rich nuclei can be used to determine the structure and 
medium dependence of effective pairing interactions~\cite{sa05}. 

\begin{figure}[]
\centerline{\includegraphics[scale=0.5]{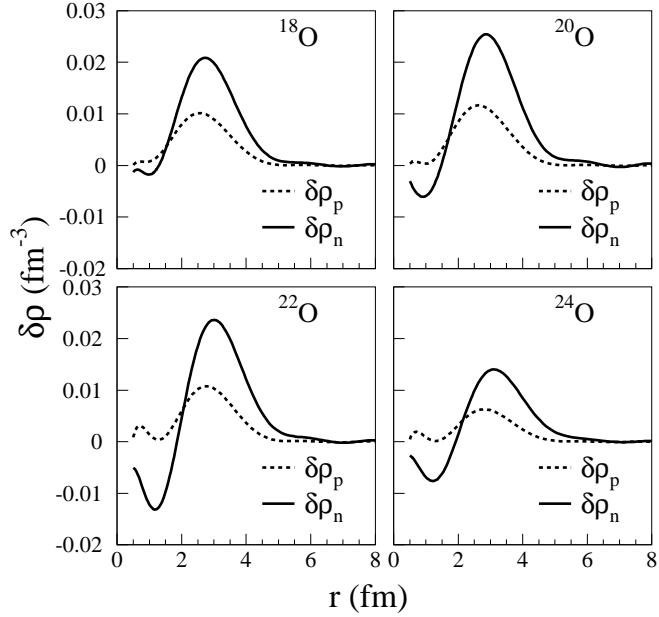}}
\caption{Neutron and proton transition densities for the first $2^+$ state in
$^{18,20,22,24}$O.  
\label{fig:transoxy}}
\end{figure}

The calculated transition densities of the 2$_1^+$ state in 
neutron-rich oxygen isotopes are shown in Fig.~\ref{fig:transoxy}.
While the peak in the proton transition density does not change its 
position with the increase of the number of neutrons, the radial 
dependence of the neutron transition density clearly reflects 
the formation of the neutron skin, especially in $^{22}$O and $^{24}$O.
The decrease in magnitude of the neutron transition density in 
$^{24}$O can be related to the predicted N=16 magic neutron number, 
which appears because of the 2s$_{1/2}$ subshell closure, i.e. 
the 2s$_{1/2}$ state is more bound in $^{24}$O than in $^{22}$O.

The measured $B(E2;2_1^+ \to 0_1^+)$ values are nicely 
reproduced by QRPA calculation, except for the problematic 
$^{18}$O~\cite{kh02}. In this nucleus a large discrepancy 
between the empirical and theoretical $B(E2)$ values has been 
found in several shell-model~\cite{ut99,br88} and QRPA
calculations~\cite{kh00,gi03}. This could be explained by the 
presence of deformed states in the experimental low-lying spectrum of
$^{18}$O. It has been suggested that the low-lying states in 
$^{16,17,18}$O contain sizeable admixtures of highly
deformed states~\cite{br66,fe65}. By taking into account the mixing 
between spherical and deformed states, it should be possible to
simultaneously reproduce the excitation energies and  $B(E2)$ 
values of the low-lying states. For heavier oxygen isotopes  
the deformed states are predicted at higher energies,
and thus the mixing is weaker. This explains why the spherical
QRPA results for the $B(E2)$ values in $^{20,22}$O are in much
better agreement with experiment. The calculation predicts a 
decrease in the $B(E2)$ value in $^{24}$O, and this is due to 
the effect of the 2$s_{1/2}$ subshell closure.

\begin{figure}[htb]
\centerline{\includegraphics[scale=0.45]{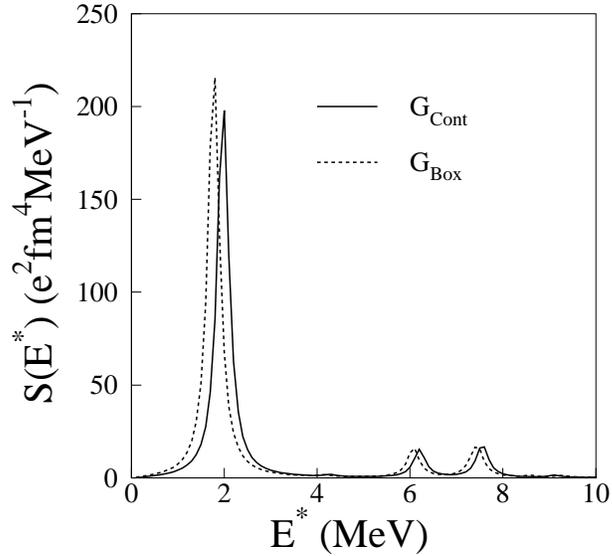}}
\caption{Isoscalar quadrupole strength function of $^{22}$O, 
calculated with the continuum QRPA (solid), and with a box 
discretization procedure (dashed).
\label{fig:contilow}}
\end{figure}

In nuclei at the neutron drip-line the Fermi level of neutron states 
is found close to the continuum and, therefore, one expects that 
continuum effects play an important role also in the calculation 
of low-lying states. This is illustrated in Fig.~\ref{fig:contilow} 
where we plot the low-energy portion of the quadrupole response function 
of $^{22}$O. The two curves correspond to calculations with 
box boundary conditions, and with the exact treatment of continuum 
states (cf. Sec. 2). The latter predicts the 2$_1^+$ state at a 
slightly higher energy, and with a weaker transition
strength. In Ref.~\cite{ma01} a continuum QRPA based on the 
HFB framework in coordinate space has been formulated, and the 
quadrupole response of the drip-line nucleus $^{24}$O has been 
described using density-dependent zero-range forces 
in the particle-hole and particle-particle channels. It has 
been shown that the low-lying isoscalar quadrupole state is 
embedded in the neutron continuum, and its excitation energy 
and strength are very sensitive to the density dependence of 
pairing correlations.

In order to disentangle the proton and neutron contributions to the 
$2^+_1$ excitations, the following reduced matrix elements will be useful:
\begin{eqnarray}
M_p & = & \langle 2^+ || \sum_{i=1}^Z r_i^2 Y_{2}(\hat r_i) || 0 \rangle, \nonumber\\
M_n & = & \langle 2^+ || \sum_{i=1}^N r_i^2 Y_{2}(\hat r_i) || 0 \rangle,
\label{mpmn}
\end{eqnarray}
so that for an electromagnetic probe the $B(E2)$ is $\sim M_p^2$, whereas 
both $M_n$ and $M_p$ contribute in the case of hadron scattering. 
For instance, experimental evidence for the magicity of the N=14 
neutron number in $^{22}$O cannot be conclusive without 
separating the proton and neutron contributions to the $2^+_1$ state.
In Ref.~\cite{th00} the $B(E2)$ value for $2^+_1$ was determined 
by inelastic scattering of $^{22}$O from
$^{197}$Au at an energy of 50 MeV/nucleon. The extracted value, however, 
depends on theoretical predictions for the ratio between $M_n$ and $M_p$,
because both Coulomb and nuclear interactions contribute to the 
reaction. $M_n$ and $M_p$ can be separated  by means of
two experiments which employ different probes. An electromagnetic probe 
is used to measure the $B(E2)$ value directly, whereas the second 
measurement is usually a (p,p$^\prime$) scattering experiment 
at around 50 MeV/nucleon. This combination allows to 
determine both $M_n$ and $M_p$, and therefore to 
probe more directly possible shell closures in exotic nuclei.
Angular distributions for elastic and inelastic proton
scattering to the $2^+_1$ state of $^{22}$O have been measured using a
secondary radioactive beam~\cite{be06}. Proton and neutron contributions 
have been disentangled by a comparison of the (p,p$^\prime$) results
with a heavy ion scattering experiment dominated by electromagnetic
excitation, and evidence for a strong N=14 shell closure has been found. 

In order to compare the
QRPA predictions with proton scattering data, microscopic optical
potentials can be generated from the HFB and QRPA densities using two 
different methods: the folding model~\cite{khoa02}, 
or the optical model potential (OMP) parameterization using the 
Jeukenne, Lejeune and Mahaux (JLM) interaction~\cite{jlm75}. The
folding model analysis uses the CDM3Y6 interaction folded with the HFB
densities to generate the isoscalar and isovector parts of the OMP. The
spin-orbit potential and the transition potentials are determined
by folding the QRPA transition densities with the nucleon-nucleon
interaction. The imaginary part of the OMP is generated with 
the Koning and Delaroche~\cite{ko03} phenomenological parameterization. 
Cross sections are calculated using DWBA with the ECIS97~\cite{ra81} code. 

The elastic angular distribution is nicely reproduced, even at
large angles~\cite{be06,kh01}. Since the $B(E2)$ can be 
described by the proton transition density, the neutron transition density 
is renormalized to reproduce the data. This procedure assumes 
that the QRPA provides a reliable description of the shape of
the transition density for collective states, and provides an
empirical value of the $M_n$/$M_p$ ratio for the 2$^+_1$ state, 
deduced from the combination of the electromagnetic and the 
(p,p$^\prime$) measurements~\cite{kh00}. The resulting 
$M_n$/$M_p$ values are included in Table~\ref{tableoxy}. 
We note that the value of $M_n$/$M_p$ divided by N/Z, is considerably 
smaller for the first 2$^+$ state of $^{22}$O, compared to $^{20}$O. 
In $^{22}$O the contributions of protons and neutrons are comparable, 
because the measured ($M_n$/$M_p$) ratio is close to N/Z. This is
different from $^{20}$O, where the much higher value of 
$M_n$/$M_p$ divided by N/Z shows that neutrons predominantly 
contribute to the quadrupole excitation. Combined with the high energy 
of the 2$_1^+$ state in $^{22}$O, these results confirm the  N=14
shell closure in neutron-rich oxygen nuclei. 
The dependence on the potential used to describe the (p,p') angular
distributions can be checked by using the complex optical and transition
potentials obtained by inserting the calculated ground state and transition
densities into the JLM density-dependent
optical potential~\cite{jlm75}. Renormalizing the neutron transition 
density to reproduce the inelastic data, the same value of the ratio 
$M_n/M_p$ is obtained as with the folding potential. In this way two 
reliable optical potentials are used to test the HFB+QRPA matter and 
transition densities.
%

A similar study of the N=16 sub-shell closure in $^{24}$O could be 
performed in a $^{24}$O(p,p$^\prime$) experiment, but this will have to
wait for the next generation of radioactive beam facilities. The generality
of the method used to microscopically determine the ratio $M_n$/$M_p$ from
two complementary sets of data, will allow to extend our understanding of
neutron-shell closure to regions of heavier nuclei far from stability. 
An ingenious proton-scattering setup with a liquid hydrogen target and 
$\gamma$-detectors opens the possibility to determine $M_n$/$M_p$ 
in nuclei very far from stability, e.g. in $^{28}$Ne~\cite{do06}.

\begin{figure}[htb]
\centerline{\includegraphics[scale=0.55]{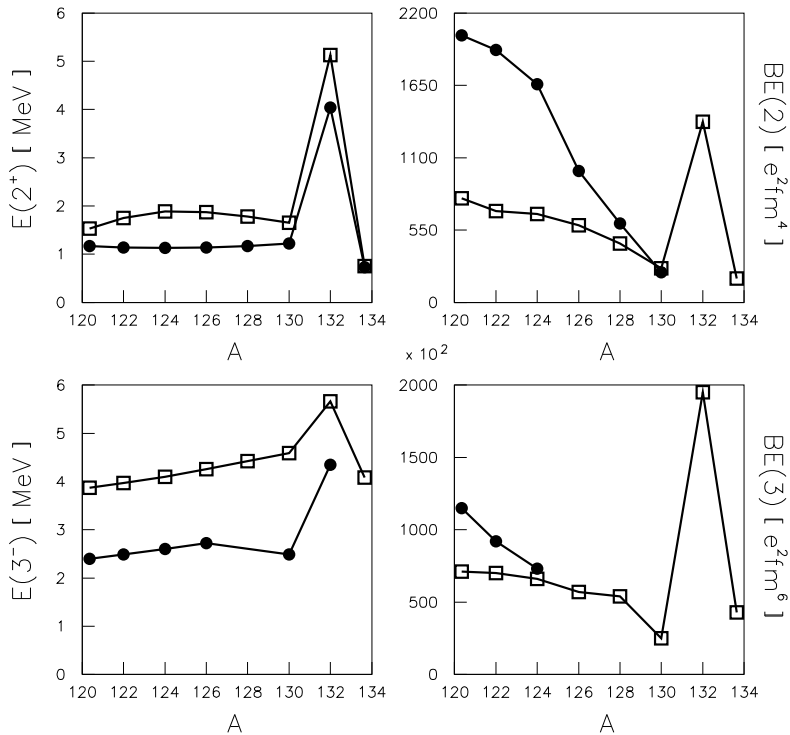}}
\caption{The QRPA excitation energies (left panels) and transition
strength (right panels) for the 2$_1^+$ and 3$_1^+$ 
states in the sequence of even-even Sn isotopes (squares)~\cite{co03},
compared with data (circles).
\label{fig:ll_sn}}
\end{figure}

The Skyrme-HFB plus QRPA framework has been applied to the study of 
low-lying quadrupole states in N = 20 isotones, including the neutron-rich 
nuclei $^{32}$Mg and $^{30}$Ne \cite{YG.04}. The calculation reproduces 
both the excitation energies of the first $2^+$ states and the 
B(E2; $0^+_1 \to 2^+_1$) values, and it has been concluded that 
pairing effects account largely for the anomalous B(E2) value and the 
very low excitation energy in $^{32}$Mg.

QRPA calculations for the low-lying 2$^+$ and 3$^-$ states have also been
performed for heavier neutron-rich nuclei. For example, the QRPA with 
Skyrme forces in the $ph$ channel, and a density-dependent pairing force 
in the $pp$ channels, has been employed in the study of the evolution 
of the 2$_1^+$ and 3$_1^-$ states along the Sn isotopic chain~\cite{co03}.
The results are shown in Fig.~\ref{fig:ll_sn}, in comparison with
experimental data~\cite{expSn}, including recent Coulomb excitation 
measurements of the unstable $^{126,128,130}$Sn isotopes~\cite{ornl_sn}. 
We notice that the calculation nicely reproduces 
the excitation energies of the 2$_1^+$ state, but overestimates the 
experimental energies of the 3$_1^-$ state by more than 50\%. 
On the other hand, the measured $B(E2)$ values in open shell 
nuclei are systematically underestimated by large factors with
the QRPA model, 
whereas this discrepancy is less pronounced for the $B(E3)$ values. 
The reason for such a large difference between the experimental and 
calculated $B(E2)$ values, is that the QRPA does not take into account 
the anharmonicities of the low-lying quadrupole vibrations, i.e.  
the 2$_1^+$ is described as a pure 1-phonon state, whereas in this 
mass region the 2$_1^+$ states contain pronounced admixtures of 
2-phonon states which cannot be described in a simple QRPA framework.

In the recent Skyrme-HFB plus QRPA analysis of 
the $J^{\pi} = 0^+$, $1^-$, and $2^+$ multipoles for the even Ca, Ni and Sn 
isotopes from the proton to the neutron drip-lines \cite{TE.06}, it has been 
shown that the strength functions in the $2^+$ channels are qualitatively 
different from those in the lower-multipole channels. The low-energy strength 
grows with the neutron number, but unlike in the $0^+$ channel, both 
neutrons and protons contribute to the transitions to these states, even 
near the drip line, because the isoscalar (IS) peaks are much larger than 
the isovector (IV) peaks. An interesting result is also that at the neutron 
drip-line the IS and IV strength functions in Ni and Sn have distinct low-energy 
peaks, whereas in Ca the peaks coincided. In all nuclei near the neutron 
drip-line, the states in the low-energy peaks are mostly above the 
neutron-emission threshold, yet the neutron tails cut off at much smaller 
radii than do those in the $0^+$ and $1^-$ channels. Also the transition 
densities to the $2^+$ states are different: they have no real nodes and a 
proton component that is of the same order as the neutron component. 
The strong low-energy states near the neutron drip-line have transition 
densities that resemble those of surface vibrations are often quite 
collective. The detailed analysis of Ref.~ \cite{TE.06} has emphasized 
the complicated relationship among collectivity, strength, and  
transition density in neutron-rich nuclei.

\subsection{Giant Quadrupole Resonance and Higher Multipoles}

The giant quadrupole resonance (GQR) corresponds to a highly collective 
oscillation of the neutron and proton density distributions between 
prolate and oblate ellipsoidal shapes. In the isoscalar mode the 
proton and neutron densities oscillate in phase, with an empirical 
excitation energy: E$_{GQR} \approx 64$ MeV$\cdot$A$^{-1/3}$ and, 
in heavy nuclei, this mode typically exhausts 50 to 100 \%
of the energy weighted sum rule (EWSR)~\cite{harak01}. The 
isovector GQR is found at much higher excitation energies and, 
even in heavy nuclei, it is much more fragmented.

In neutron-rich nuclei far from stability one expects that the 
neutron skin has a pronounced effect on the high-energy 
quadrupole vibrational mode, however, no data on the GQR in 
unstable nuclei are available at present. There have also been 
only few theoretical studies of giant quadrupole resonances in 
exotic nuclei. In Refs.~\cite{kh02,ma01,PRNV.03} the QRPA and 
RQRPA calculations have been performed for the isoscalar GQR 
in oxygen isotopes. Since in light systems this mode can be 
highly fragmented, for a study of the effect of neutron excess 
on the GQR one should consider heavier nuclei, in which the 
GQR displays a single peak at energies between 10 and 15 MeV.
Such studies have been carried out with the continuum-RPA 
using Skyrme forces~\cite{SE.01}, and recently RPA calculations 
with the Gogny interaction have also been performed~\cite{Per.05}.

The isospin dependence of the excitation energy and width of the GQR 
in exotic nuclei is illustrated in Table~\ref{tablegqr} for the 
example of $^{34,40,48,60}$Ca~\cite{SE.01}. In addition to the systematic
lowering of the GQR excitation energies with the increase of the 
number of neutrons ($^{48}$Ca is an exception, because of the neutron 
shell closure at $N=28$), the calculation predicts an enhancement 
of the low-energy quadrupole strength both in the 
proton-rich $^{34}$Ca, and neutron-rich $^{60}$Ca. Continuum-RPA 
calculations predict the escape width of a resonant state, and in 
Table~\ref{tablegqr} we notice a pronounced increase of the 
escape width in the weakly-bound nuclei $^{34}$Ca and $^{60}$Ca. 
The calculated widths, however, do not contain the spreading 
contribution, i.e. the width that results from the coupling of 
the GQR to more complex states like, for instance, two-phonon 
admixtures. In this sense the widths in Table~\ref{tablegqr} 
are not realistic, and only illustrate the effect of the coupling to 
continuum states. No systematic calculation of the GQR spreading 
width in exotic nuclei has been reported so far, the only exception 
is the study of the GQR in $^{28}$O in Ref.~\cite{ghielmetti}, 
which has shown that the spreading width is enhanced with respect 
to the isotopes close to the stability line.
\begin{table}
\centering
\begin{tabular}{lccc} \hline 
&  E$_{GQR}$ (MeV)  & $\Gamma_{FWHM}$ (MeV)
\\ \hline\hline                          
$^{34}$Ca     &  17.5 & 1.6      
\\ \hline  
$^{40}$Ca    &   16.2 & 0.6    
\\ \hline    
$^{48}$Ca    &    16.7 &  0.3
\\ \hline
$^{60}$Ca    &  14.9  & 1.3
\\ \hline
\end{tabular}
\caption{Continuum-RPA results for the GQR centroids and widths 
of Ca isotopes~\cite{SE.01}}
\label{tablegqr}
\end{table}

\begin{figure}[htb]
\centerline{\includegraphics[width=7cm,angle=-90]{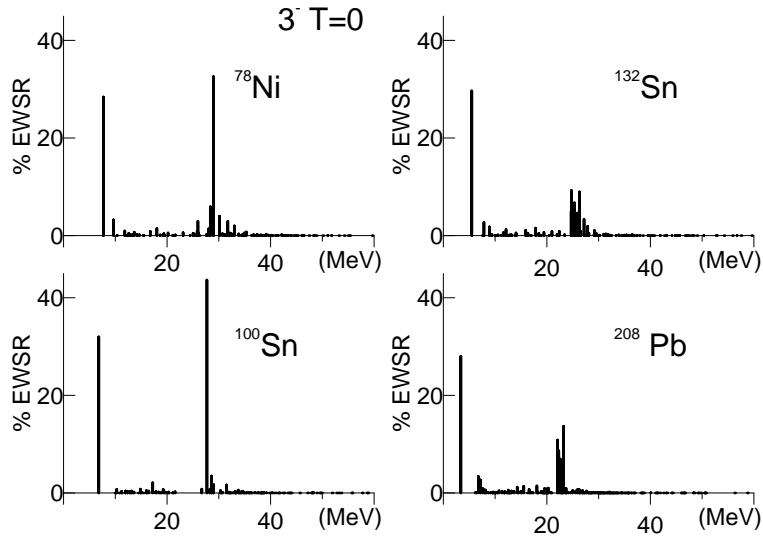}}
\caption{Fraction of the EWSR exhausted by the isoscalar octupole states in
$^{78}$Ni, $^{132}$Sn, $^{100}$Sn and $^{208}$Pb~\protect\cite{Per.05}.} 
\label{fig:octu}
\end{figure}

Giant quadrupole resonances in exotic nuclei have also been calculated 
with the RPA based on the Gogny interaction~\cite{Per.05}. For the 
doubly-magic $^{78}$Ni, $^{132}$Sn and $^{100}$Sn the Gogny-RPA results 
predict GQR excitation energies that are $1- 1.5$ MeV above the 
empirical relation E$_{GQR} \approx 64$ MeV$\cdot$A$^{-1/3}$. 
In Ref.~\cite{Per.05} RPA-Gogny calculations have also been 
performed for the isoscalar and isovector octupole strength 
distributions, and in Fig.~\ref{fig:octu} we display the 
isoscalar octupole states in $^{78}$Ni, $^{132}$Sn, $^{100}$Sn 
and $^{208}$Pb.  The strength is clearly separated into
two regions: the 3$_1^-$ state dominates the low-energy strength, 
whereas the strong peaks above 20 MeV excitation energy correspond 
to the High Energy Octupole Resonance (HEOR). The properties of 
the HEOR in nuclei far from stability ($^{78}$Ni, $^{132}$Sn,
$^{100}$Sn) are not significantly different from those in stable 
nuclei, e.g. $^{208}$Pb, probably because the HEOR corresponds 
to 3$\hbar\omega$ excitations and these high-energy configurations 
may not be very sensitive to changes in the number of neutrons/protons 
with respect to stable nuclei. 

\subsection{Pairing Vibrations in Drip-Line Nuclei}\label{pairvib}

Two-neutron transfer reactions such as $(t,p)$ or $(p,t)$ have been used for
many years in studies of nuclear pairing correlations~\cite{oe01}.
The corresponding pair-transfer modes are usually described in terms of
pairing vibrations or pairing rotations~\cite{be66,br73}. 
Pairing vibrations are $L=0$ modes induced by the addition or 
removal of a pair of neutrons, and can be associated with
the fluctuation of the pairing field $\hat{\Delta}$ 
(see Sec.~\ref{firstHFB}) around its
equilibrium value. Around magic nuclei, such as $^{208}$Pb, the
J$^{\pi}$=0$^+$ spectrum generated by the pair-addition 
($c_{i}^{\dagger}c_{j}^{\dagger}$) and pair-subtraction 
($c_{i}c_{j}$) operators in Eq.~(\ref{eq:extpart}), is harmonic
and corresponds to the so-called pairing
vibration mode, which can be viewed as a vibration in 
an abstract ``gauge'' space, instead of the ordinary 
three-dimensional space in which shape vibrations take place. 

In a microscopic approach the collective two neutrons transition can be
described in the QRPA framework: the excitation operator of 
Eq.~(\ref{eq:extpart}) includes particle-hole, particle-particle and 
hole-hole excitations. In this case 
the non-conservation of the particle number, which is implicit in the 
quasiparticle formalism, can be used as a tool to study particle-violating
transitions. The pairing vibrational state reads
\begin{equation}
\vert A+2,n \rangle= \left( \sum_{k_F<k<l}
X_{kl}^{(n)}c^\dagger_k c^\dagger_l - 
\sum_{k<l<k_F}Y_{kl}^{(n)}c^\dagger_l c^\dagger_k \right) |A> \; .  
\end{equation}

High-energy collective pairing modes -- Giant Pairing Vibrations (GPV) 
have been predicted and studied theoretically~\cite{oe01,br73,fo02,br77,he80}, 
but have never been observed despite a number of $(p,t)$ experiments 
performed in the 70's and 80's~\cite{oe01,br73}. The experimental setup for 
such studies must achieve a proper balance between the low energy of the 
incident proton beam (below 50 MeV) necessary for the excitation of the 
$L=0$ mode, and the energy which is required to populate the energy 
region of the GPV. These experiments are therefore rare~\cite{cr77}, 
and the discovery of deep hole-states of non-collective character 
made the detection of the GPV even harder~\cite{cr81,bo86}. Beams 
of exotic nuclei could provide the solution. Incident beams of a
few MeV/nucleon of weakly-bound projectiles 
(providing high Q-values) could be used to populate 
the GPV~\cite{oe01} using reactions such as ($^6$He,$\alpha$).
However, the intensity of radioactive beams is typically
several orders of magnitude lower than that of stable beams, and
the background resulting from break-up reactions could be large 
with weakly-bound projectiles. There is a renewed interest in
GPV and improved experimental investigations are currently being
planned, both with stable and exotic beams. 

Pairing vibrations in exotic nuclei could provide valuable
structure information~\cite{oe01,fo02,kh04}, in particular on pairing 
correlations in systems far from stability and the effects of the 
coupling to the continuum. Pairing vibrations generally depend on the
strength of the pairing interaction between the two transferred neutrons:  
transfer cross sections are enhanced when the two neutrons form a 
strongly bound pair. The theoretical analysis of two-nucleon transfer 
modes in nuclei far from stability is complicated by the effect of 
continuum coupling. The right tool to study pairing vibrations in exotic
nuclei is the continuum-QRPA, because it provides a consistent microscopic 
treatment of both pairing and continuum effects. The strength function 
which describes the two-particle transfer from the ground state
of a nucleus with A nucleons to the excited states of a nucleus with A+2
nucleons reads:
\begin{equation}\label{eq:stren2}
S(\omega)=-\frac{1}{\pi}Im \int  F^{12*}({\bf r}){\bf{
G}}^{22}({\bf r},{\bf r}';\omega) F^{12}({\bf r}')
d{\bf r}~d{\bf r}'
\end{equation}
where ${\bf{G}}^{22}$ denotes the ($pp$,$pp$) component of the 
Green's function, and $F^{12}$ is the perturbing 
external field (see Sec.~\ref{sec2_cqrpa}).  

\begin{figure}[htb]
\centerline{\includegraphics[scale=0.4]{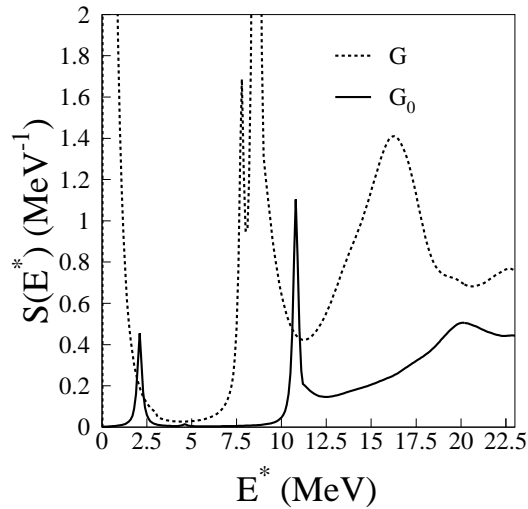}}
\caption{The strength function for the two-neutron transfer on $^{22}$O. 
The solid (dashed) curve corresponds to the unperturbed (QRPA) response.
\label{fig:o22trength}}
\end{figure}

The strength function for the neutron-pair
transfer to $^{22}$O is shown in Fig.~\ref{fig:o22trength}~\cite{kh04}. 
The interaction in the $ph$ channel is Skyrme SLy4, and a zero-range 
density-dependent pairing interaction is used in the $pp$ channel.
The solid curve denotes the unperturbed strength, which
displays peaks characteristic for the addition of two neutrons in specific
configurations. The first peak results from the filling of the 
2$s_{1/2}$ orbital, and the peak at 10.8 MeV corresponds to the two-neutron 
(1$d_{3/2}$)$^2$ quasiparticle configuration. The effect of the 
residual interaction on the pair transfer mode is seen in the 
QRPA strength function (dashed curve), and it demonstrates 
the collective nature of pairing vibrations~\cite{br73}. Namely, 
the residual interaction shifts the position of the two-quasiparticle 
resonant state, located at 10.8 MeV in the unperturbed response, 
to lower energy and increases its strength. The strong peak
at zero-energy corresponds to the pair transfer to the spurious Goldstone mode 
associated with particle-number fluctuations. 
The peak at $\approx 8.6$ MeV represents
a neutron-pair transfer predominantly to the $1d_{3/2}$ states, 
whereas the broad resonant structure around 16 MeV is built mainly on
the neutron resonant state $1f_{7/2}$. This broad two-quasiparticle
resonance is characteristic for a giant pairing vibration~\cite{fo02,br77,he80}.

To compare the results with the measured cross section, one must 
calculate the nuclear structure form factor, which describes the 
wave function of the two transferred neutrons, as well as the reaction part 
which includes the optical potential. For closed-shell nuclei the
two-particle transfer modes are described by the particle-particle 
($pp$) RPA~\cite{bo74,ri69}, whereas the QRPA~\cite{br73,fo02} 
is used for open-shell nuclei. Most of the
cross section calculations employ the Distorted Wave Born Approximation (DWBA).


In nuclear response theory the transition from the ground state to 
the excited state $|\nu\rangle$ in the same nucleus is described by
the transition density
\begin{equation}\label{eq:rhor}
\delta\rho^\nu\left({\bf r}\sigma\right) = 
\left<0|c^{\dagger}\left({\bf r}\sigma\right)
c\left({\bf r}\sigma\right)|\nu\right>
\end{equation}
where $c^{\dagger}\left({\bf r}\sigma\right)$ is
the particle creation operator in coordinate space.
The corresponding quantity used in the description of pair transfer 
processes is the pair transition density defined by
\begin{equation}\label{eq:rhotildr}
\delta\kappa^\nu\left({\bf r}\sigma\right) = 
\left<0|c\left({\bf r}\bar{\sigma}\right)
c\left({\bf r}\sigma\right)|\nu\right>
\end{equation}
where the operator $c^{\dagger}\left({\bf r}\bar{\sigma}\right)$= $-2\sigma
c^{\dagger}\left({\bf r}-\sigma\right)$ creates a particle in the time-reversed
state. The pair transition density determines the transition from the
ground state of a nucleus with A nucleons to the state $|\nu\rangle$ in the 
nucleus with A+2 nucleons. This quantity is calculated in the QRPA.

\begin{figure}[htb]
\centerline{\includegraphics[scale=0.4]{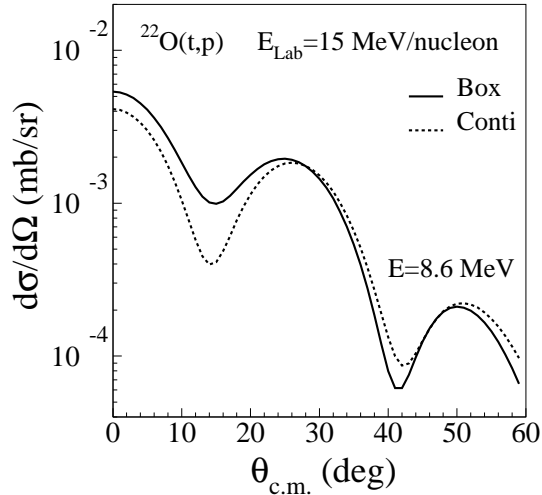}}
\caption{Two-neutron transfer cross section on $^{22}$O, for the state
at $\approx 8.6$ MeV. The solid (dashed) curve corresponds to the QRPA result
obtained with the box (exact) boundary conditions.
\label{fig:microdisconti}}
\end{figure}

The DWBA calculation of the cross section for the two-neutron transfer
requires the form factor which describes the correlation between the two
neutrons and the initial nucleus~\cite{sa83}. Calculations based on the
continuum-QRPA include effects of both pairing correlations and 
continuum coupling. The calculation of the form factor and cross section
will be illustrated for the transfer reaction $^{22}$O(t,p). 
The DWBA cross section is calculated for the $^{22}$O+t Becchetti and 
Greenlees optical potential~\cite{be71} in the entrance channel, and the
$^{22}$O+p Becchetti and Greenlees~\cite{be69} in the exit channel. 
The calculation is performed in the zero-range approximation, 
in which the two-neutrons and the residual fragment are located at 
the same point in space, and therefore the range function is determined
by a simple constant D$_0$~\cite{sa83}. The zero-range approximation 
provides a satisfactory description of the shape of the angular
distribution~\cite{br73,sa83,ig91}, but its magnitude is generally underestimated. 
The form factor for the pair transfer is
obtained by folding the pair transition density $\delta\kappa^\nu$ (Eq.
(\ref{eq:rhotildr})) with the interaction acting between the transferred
pair and the residual fragment~\cite{oe01}. In the zero-range approximation
one uses the $\delta$-force for this interaction, and therefore 
the form factor coincides with the pair transition density 
(\ref{eq:rhotildr})~\cite{sa83}.
In order to illustrate the effect of the continuum on the form factor, 
the (t,p) angular distribution for the mode at $\approx 8.6$ MeV 
(see Fig.~\ref{fig:o22trength}) is calculated using both 
box boundary conditions and with the exact treatment of the 
continuum. The resulting two-neutron transfer cross sections are
shown in Fig.~\ref{fig:microdisconti}, and we notice a pronounced effect
of the continuum in the diffraction minima.  

The continuum-QRPA could be used in the analysis of the forthcoming data 
on the GPV, obtained with beams of exotic nuclei at around 5 MeV/nucleon.
Preliminary calculations~\cite{fo02} have shown that ($^6$He,$\alpha$)
reactions can excite the GPV with a cross section of the order of 
few millibarns, whereas this mode is not excited in ($^{14}$C,$^{12}$C) 
reactions.

\section{\label{secsi}Charge-exchange Resonances}

%
\subsection{Proton-Neutron QRPA}\label{pnqrpa} 
%

Starting from the ground-state of the (N,Z) nuclear system, 
charge-exchange excitations induced by isospin lowering 
$\tau_-$ and raising $\tau_+$ operators correspond to 
transitions to final states in the neighboring 
(N$\mp$1,Z$\pm$1) nuclei, respectively. 
These transitions can occur either spontaneously
in $\beta$-decay, or they can be induced in charge-exchange 
reactions, e.g. $(p,n)$ or ($^3$He,$t$) in the $\tau_-$ case, 
when the final state lies outside the $\beta$-decay 
energy window. Charge-exchange resonances in stable nuclei 
have been the subject of numerous experimental and theoretical
studies (see, for instance, Ref.~\cite{harak01}). However,
a systematic experimental information on these resonances 
is still missing, mainly because of the absence of selective 
probes. For instance, there is very little data on the 
charge-exchange dipole resonance and, despite considerable 
experimental effort, the charge-exchange monopole mode has 
not yet been observed. This resonance is particularly 
interesting, because its excitation energies could  
provide useful information
on the isospin mixing in nuclear ground-states, 
the symmetry term of the nuclear equation of state, and
the isovector terms in effective nucleon-nucleon interactions. 
Charge-exchange transitions are important for 
weak-interaction processes in nuclei, in particular 
charged current processes in nuclear astrophysics and
neutrino physics. These include the $\beta$-decay of nuclei 
that lie on the $r$-process path of stellar nucleosynthesis, 
and neutrino-nucleus scattering. 

The only exception are the Isobaric Analog Resonance
(IAR) and the Gamow-Teller Resonance (GTR), for which 
rather detailed experimental information has become 
available in several mass-regions. They are induced by
the following transition operators
\begin{equation}
\hat{F}_{IAR}=\sum_{i=1}^A {\tau}_-(i),
\end{equation}
and
\begin{equation}
\hat{F}_{GTR}=\sum_{i=1}^A \sigma(i) {\tau}_-(i)\; ,
\end{equation}
respectively. Since these are $L=0$ modes, they can be 
selectively excited by using
zero-degree charge-exchange 
reactions, like $(p,n)$ or ($^3$He,$t$). 
When the incident projectile energy is increased, 
the excitation of the GTR is favored over the IAR.
A comprehensive review of the properties of the IAR 
and GTR in stable nuclei can
be found in Ref.~\cite{Ost.92}.

Charge-exchange resonances in exotic nuclei are virtually unexplored.
The experimental investigation of these modes in nuclei far from 
stability will become possible with the new generation 
of radioactive-beam facilities. So far, experimental studies
have been restricted to the states which are accessible in 
$\beta$-decay. In neutron-rich nuclei, for example, the neutron 
Fermi level is located at much higher energy than the 
proton valence states, and the $\beta^-$-decay becomes 
progressively much faster. In proton-rich nuclei the same happens 
for the $\beta^+$-decay. Closer to the nucleon drip-lines  
more and more single-nucleon states enter the $\beta$-decay 
energy window, and the experiment probes most of the 
charge-exchange strength. The importance of this phenomenon for 
the understanding of the isospin properties of nuclei has been
first pointed out in Ref.~\cite{Ham.93}, where the 
HF-RPA has been employed in the study of charge-exchange transitions 
in weakly-bound systems. 

The extension of the Skyrme-RPA model to the charge-exchange 
channel~\cite{auerbach_proc}, has been used in  
calculations of the response to different operators~\cite{auerbach}. 
The discrete RPA has also been employed in studies of charge-exchange modes. 
In addition, the formalism of the coupling of $p-h$ configurations
to more complex states of $2p-2h$ character, has been 
developed in Refs.~\cite{Col.94,adachi,Col.98}. 
In the case of open-shell nuclei, most of the charge-exchange 
QRPA calculations have used simple separable 
forces as the $ph$ and $pp$ residual interactions. The same functional
form has been used in both channels, with two different
strength parameters $g_{ph}$ and $g_{pp}$, as in the pioneering work of 
Ref.~\cite{Hal.67} where the formalism has been developed for the first 
time. More recently, the PN-QRPA based on 
Skyrme forces in the $ph$ channel, with a simplified 
residual interaction of a separable form, and with 
the BCS treatment of the $pp$ channel,
has also been extended to the description of deformed nuclei.
$\beta$-decay rates in different isotopic chains have been
studied~\cite{Srr.98,Srr.99,Srr.01,Srr.01.b}, in order 
to determine to what extent the intrinsic deformation 
influences $\beta$-decay properties, and whether 
the decay spectra differentiate between spherical, prolate and oblate 
shapes. Medium-heavy nuclei like Kr, Sr  have been studied, 
as well as heavier systems e.g. Hg, Pb and Po~\cite{Srr.06}. 
It has been shown that prolate and
oblate deformations lead to significantly different Gamow-Teller
spectra, but in many cases these signatures of the
intrinsic deformation are not too sensitive to the choice
of the Skyrme force used in the calculation. 

The path of the $r$-process nucleosynthesis is governed by 
delicate balance between neutron-capture reactions and 
$\beta$-decay rates. This has motivated the calculation of the GT 
$\beta$-decay of the so-called ``waiting-point'' nuclei in 
Ref.~\cite{Eng.99}, which was also the first attempt 
to perform a self-consistent PN-QRPA analysis based on the 
HFB treatment of the nuclear ground state. 
A Skyrme interaction and a zero-range pairing interaction 
were used in the mean-field and pairing channels, 
respectively. The associated QRPA equations were solved in the canonical
basis, in which they read as in Eq.~(\ref{rrpaeq}), with the matrices 
$A$ and $B$ given by:
\begin{eqnarray}
A_{pn,p^\prime n^\prime}^{J} &=& H^{11}_{pp^\prime}\delta_{nn^\prime} +
  H^{11}_{nn^\prime}\delta_{pp^\prime}  \nonumber \\ & & +
\lp( u_p v_n u_{p^\prime} v_{n^\prime} +
v_p u_n v_{p^\prime} u_{n^\prime}\rp)
 V_{pn^\prime n p^\prime}^{ph J} \nonumber \\ & & + 
\lp( u_p u_n u_{p^\prime} u_{n^\prime} + v_p v_n v_{p^\prime} v_{n^\prime}\rp) 
 V_{pn p^\prime n^\prime}^{pp J} \nonumber \\
B_{pn,p^\prime n^\prime}^{J} &=& (-1)^{j_{p^\prime}-j_{n^\prime}+J}
\lp( u_p v_n v_{p^\prime} u_{n^\prime} + v_p u_n u_{p^\prime} v_{n^\prime}\rp)
 V_{pp^\prime n n^\prime}^{ph J} \nonumber \\ & &- 
\lp( u_p u_n v_{p^\prime} v_{n^\prime} + v_p v_n u_{p^\prime} u_{n^\prime}\rp) 
 V_{pn p^\prime n^\prime}^{pp J} \; .
\label{abmat}
\end{eqnarray}
Here $p$, $p^\prime$, and $n$, $n^\prime$ denote proton and neutron
canonical states, respectively, $V^{ph}$ is the proton-neutron $ph$ 
residual interaction, and $V^{pp}$ is the corresponding 
$pp$ interaction. Since the canonical basis does not diagonalize the 
mean-field Hamiltonian $\hat{h}$, nor the pairing field $\hat{\Delta}$,
the non-diagonal matrix elements $H^{11}_{nn^\prime}$ and
$H^{11}_{pp^\prime}$ appear in the matrix $A$:
\begin{equation}
H_{\kappa \kappa^\prime}^{11}=(u_{\kappa }u_{\kappa^\prime }
-v_{\kappa }v_{\kappa^\prime
})h_{\kappa \kappa^\prime }-(u_{\kappa }v_{\kappa^\prime }+
v_{\kappa }u_{\kappa^\prime
})\Delta _{\kappa \kappa^\prime }\;.
\label{H11}
\end{equation}

The PN-QRPA calculation of Ref.~\cite{Eng.99} has shown that the 
excited 1$^+$ states and $\beta$-decay half-lives are very  
sensitive to the $T=0$ component of the residual $pp$ 
interaction. Since the $T=0$ pairing is not manifest in the
ground states of nuclei with $N$ different from $Z$,
it has to be introduced in the PN-QRPA independently 
from the $T=1$ channel, and this of course breaks the 
self-consistency of the calculation. In particular, a finite-range
interaction with adjustable parameters has been employed,
and the overall strength was tuned to reproduce some selected
experimental $\beta$-decay half-lives. 
The same approach has been used in Ref.~\cite{Ben.02} 
to analyze properties of GT resonances
predicted by several Skyrme parameterizations, 
in correlation with the corresponding values of 
the Landau parameters in infinite nuclear matter. 

\begin{figure}
\centering
\vspace*{1cm}
\includegraphics[scale=0.4,angle=0]{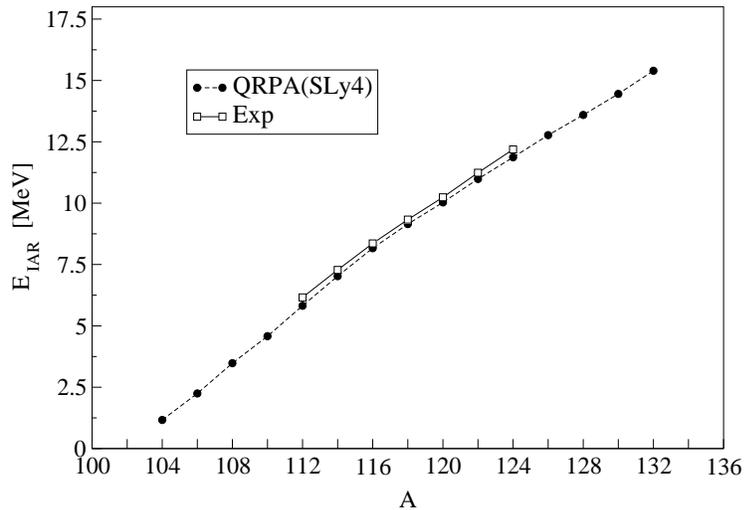}
\vspace*{0.5cm}
\caption{IAR excitation energy in the sequence of Sn isotopes. The
results from the fully self-consistent Skyrme QRPA calculation 
are compared with the experimental excitation 
energies~\protect\cite{Pham.95}.}
\label{iar_sn}
\end{figure}

Very recently a fully self-consistent
non-relativistic charge-exchange QRPA has been developed and
tested in the description of IAR~\cite{Fra.05}. In this case
only the $T=1$ component of the residual pairing interaction 
contributes and, because of isospin invariance, its strength 
must be identical to that used for the calculation of
HF-BCS ground-state. In Fig.~\ref{iar_sn} we show the 
excitation energies of IAR of Sn isotopes, calculated with 
the PN-QRPA based on the Skyrme interaction
SLy4 and compared with the experimental values.
We notice a very good agreement with data, and this shows 
that the model can be systematically employed in 
calculation of charge-exchange processes in stable and exotic
open-shell nuclei.

%
\subsection{Proton-Neutron Relativistic QRPA}\label{pnrqrpa} 
%

In the relativistic mean-field framework 
the spin-isospin dependent interaction terms are generated by the $\rho$-
and $\pi$-meson exchange. Because of parity conservation, 
the one-pion direct contribution vanishes 
in the mean-field calculation of a nuclear ground state. Its inclusion is 
important, however, in calculations of excitations that involve
spin and isospin degrees of freedom.
The particle-hole residual interaction in the PN-RQRPA is derived from the 
Lagrangian density 
\begin{equation}
\mathcal{L}_{\pi + \rho} =
      - g_\rho \bar{\psi}\gamma^{\mu}\vec{\rho}_\mu \vec{\tau} \psi
      - \frac{f_\pi}{m_\pi}\bar{\psi}\gamma_5\gamma^{\mu}\partial_{\mu}
        \vec{\pi}\vec{\tau} \psi \; .
\label{lagrres}
\end{equation}

In Ref.~\cite{PNVR.04} the proton-neutron relativistic quasiparticle 
RPA (PN-RQRPA) has  been formulated in the canonical single-nucleon basis 
of the time-dependent RHB model, and applied in studies of 
charge-exchange excitations in open shell nuclei.
The spin-isospin dependent residual two-body interaction
in the particle-hole channel reads
\begin{eqnarray}
V(\mathbf{r}_1,\mathbf{r}_2) &=& \vec{\tau}_1\vec{\tau}_2 (\beta \gamma^\mu)_1
      (\beta \gamma_\mu)_2 {g_\rho}^2
      D_\rho (\mathbf{r}_1,\mathbf{r}_2) \nonumber \\
      &-&\left(\frac{f_\pi}{m_\pi}\right)^2\vec{\tau}_1\vec{\tau}_2
      (\mathbf{\Sigma}_1\mathbf{\nabla}_1)(\mathbf{\Sigma}_2\mathbf{\nabla}_2)
      D_\pi (\mathbf{r}_1,\mathbf{r}_2)\;,
\end{eqnarray}
where $D_{\rho (\pi)}$ denotes the meson propagator
\begin{equation}
 D_{\rho (\pi)}(\mathbf{r}_1,\mathbf{r}_2) = \frac{1}{4\pi}
        \frac{e^{-m_{\rho (\pi)}|\mathbf{r}_1-\mathbf{r}_2|}}{|\mathbf{r}_1-\mathbf{r}_2|}\;, 
\end{equation}	   
and 
\begin{equation}
{\mathbf{\Sigma}} = \left(
\begin{array}
[c]{cc}%
\mathbf{\sigma} & 0\\
0 & \mathbf{\sigma}%
\end{array}
\right)  \;.
\label{bigsigma}
\end{equation}
Transitions between the $0^+$ ground state of a spherical
even-even parent nucleus, and the state with angular momentum 
and parity $J^\pi$ of the odd-odd daughter nucleus are considered.
With respect to the RHB calculation of the ground state of the 
even-even nucleus, the charge-exchange channel includes the 
additional one-pion exchange contribution. For the pseudovector 
pion-nucleon coupling the standard values: $m_{\pi}=138.0$ MeV, 
and ${f_{\pi NN}^{2}}/{4\pi}=0.08$, are used.
The derivative type of the pion-nucleon coupling necessitates the  
inclusion of the zero-range
Landau-Migdal term, which accounts for the contact part of the 
nucleon-nucleon interaction
\begin{equation}
V_{\delta\pi}(\vec{r}_1,\vec{r}_2) = g^\prime \lp( \frac{f_\pi}{m_\pi} \rp)^2 
\vec{\tau}_1\vec{\tau}_2 \vec{\Sigma}_1 \cdot \vec{\Sigma}_2 
\delta (\vec{r}_1-\vec{r}_2)\; ,
\label{deltapi}
\end{equation}
with the parameter $g^{\prime} \approx 0.6$ usually adjusted
to reproduce data on excitation energies of Gamow-Teller 
resonances~\cite{PNVR.04,Con.00,Ma.03}. In the non-relativistic limit 
the corresponding two-body interaction reduces to 
the familiar form $G_0'\bm{\sigma}_1 \vec{\tau}_1 \cdot \bm{\sigma}_2
\vec{\tau}_2$. We note that a $ph$ 
residual interaction based on $\rho$- and $\pi$-meson 
exchange has also been used in a number of non-relativistic RPA 
studies of charge-exchange excitations~\cite{Kre.81,Bro.81,Kre.88}. 

In the particle-particle channel of the PN-RQRPA equations both the 
$T=1$ and $T=0$ pairing interactions contribute. In Ref.~\cite{PNVR.04}
the finite-range Gogny force with the parameter set D1S~\cite{berg91}
has been used in the $T=1$ channel, both in the RHB calculation of the 
ground state of the even-even system and as the PN-RQRPA $pp$ residual 
interaction. For the $T=0$ proton-neutron pairing force 
a similar interaction has been employed:
a short-range repulsive Gaussian combined with a
weaker long-range attractive Gaussian 
\begin{equation}
\label{eq2}
V_{12}
= - V_0 \sum_{j=1}^2 g_j \; {\rm e}^{-r_{12}^2/\mu_j^2} \;
    \hat\Pi_{S=1,T=0}
\quad ,
\label{pn-pair}
\end{equation}
where $\hat\Pi_{S=1,T=0}$ projects onto states with $S=1$ and $T=0$.
This interaction was also used in the non-relativistic QRPA calculation
of $\beta$-decay rates for spherical neutron-rich 
$r$-process waiting-point nuclei~\cite{Eng.99}.
The ranges of the two Gaussians
$\mu_1$=1.2\,fm and $\mu_2$=0.7\,fm  are taken from the
Gogny interaction Eq.~(\ref{Gogny}), and the choice of the 
relative strengths: $g_1 =1$ and
$g_2 = -2$ makes the force repulsive at small distances.  
The overall strength parameter $V_0$ can be adjusted, for instance, 
to experimental $\beta$-decay half-lives~\cite{Eng.99,Nik.05}.

The two-quasiparticle PN-RQRPA configuration space includes states with
both nucleons in the discrete bound levels, states with one nucleon in the
bound levels and one nucleon in the continuum, and also states with both
nucleons in the continuum. In addition to configurations built from
two-quasiparticle states of positive energy, the RQRPA configuration space
contains pair-configurations formed from the fully or partially occupied
states of positive energy and the empty negative-energy states from the
Dirac sea. The inclusion of configurations built from occupied
positive-energy states and empty negative-energy states is essential 
for the consistency of the relativistic (proton-neutron) QRPA
(current conservation, decoupling of spurious states, sum rules).
The PN-RQRPA model is fully consistent:
the same interactions, both in the $ph$ and $pp$
channels, are used in the RHB equation that determines the
canonical quasiparticle basis, and in the PN-RQRPA equation. 
In both channels the same strength parameters of the interactions 
are used in the RHB and RQRPA calculations.

\subsection{Gamow-Teller Resonances}
Collective spin and isospin excitations in atomic nuclei have 
been the subject of many experimental and theoretical studies
(for an extensive review see Ref.~\cite{Ost.92}).
Nucleons with spin up and spin down can oscillate either in 
phase (spin scalar $S=0$ mode) or out of phase (spin vector $S=1$ 
mode). The spin vector, or spin-flip excitations can be of 
isoscalar ($S=1$, $T=0$) or isovector ($S=1$, $T=1$) nature. These 
collective modes provide direct information on the spin and
spin-isospin dependence of the effective nuclear interaction.

The Gamow-Teller resonance represents a fundamental charge-exchange mode 
and corresponds to a collective spin-isospin state $J^\pi = 1^{+}$ formed
when the excess neutrons coherently change the direction of their
spin and isospin without changing their orbital motion.
This collective mode was predicted already in 1963~\cite{Ike.63}, 
but it was only in 1975 that the first experimental indications of the 
GT resonance were observed in $(p,n)$ charge-exchange reactions at
intermediate energies~\cite{Doe.75}. More recently, advanced 
experiments with ($^3$He,t) reactions have achieved very good 
energy ($\Delta E\approx$~50 keV) and angular 
($\Delta \Theta \approx 0.3^{\circ}$) resolution~\cite{Fuj.02},
allowing not only the detection of the fine structure of 
GT transitions~\cite{Kal.06}, but also
studies of GT strength in exotic proton-rich nuclei~\cite{Fuj.05}.
The detailed knowledge of GT strength distributions in regions away 
from the valley of $\beta$-stability is essential for an
understanding of nuclear processes relevant for
nucleosynthesis. In particular, the low-lying GT strength 
is directly related to $\beta$-decay rates, as well as to 
the electron-capture process leading to the
stellar collapse and supernovae explosion. At present,
however, very little is known about charge-exchange excitations in 
exotic nuclei, and the modeling of nuclear weak-interaction processes in 
stars must rely on theoretical predictions of the GT strength 
distributions. 

Recent theoretical studies of Gamow-Teller excitations and 
$\beta$-decay rates have been based on: (i) the shell-model approach,
(ii) the non-relativistic proton-neutron (Q)RPA, and
(iii) the relativistic proton-neutron (Q)RPA.
Data on charge-exchange excitations in light and medium-mass
nuclei are very successfully reproduced by large-scale shell-model
calculations~\cite{Cau.05}. However, as the number of valence nucleons
increases, the dimension of shell-model configuration space
becomes far too large for any practical calculation. Present shell-model
calculations are thus restricted up to the 
region of $pf$-shell nuclei with $A=45-65$~\cite{Cau.99}.
Both the GT$^-$ and GT$^+$ response, as well as $\beta$-decay rates
in medium-mass nuclei have been successfully described by shell-model 
Monte Carlo calculations (SMMC)~\cite{Cau.95,Rad.97}.
Studies of charge-exchange excitations have only recently been
reported in the relativistic mean-field plus RPA framework. These
include the relativistic RPA analysis of isobaric analog states
and Gamow-Teller resonances (GTR) in the doubly closed-shell nuclei 
$^{48}$Ca, $^{90}$Zr, and $^{208}$Pb~\cite{Con.00,Con.98}, performed
in a restricted configuration space that did not include 
negative-energy states from the Dirac sea.
Although these configurations do not affect 
the excitation energies of the charge-exchange modes, they have 
a pronounced effect on the Gamow-Teller sum rule~\cite{PNVR.04,Kur.03a}.
GT resonances in doubly closed shell nuclei 
have also been studied with the relativistic RPA in the response 
function formalism~\cite{Ma.03}. 

The PN-QRPA can be employed in calculations of charge-exchange  
excitations in mass regions that are presently beyond the 
reach of the most advanced shell-model codes, both in stable
and exotic nuclei~\cite{Bor.03}. Most QRPA studies of the GTR
have been based on Skyrme effective interactions with BCS-type
pairing~\cite{Col.94,Ham.93,Ben.02,Gia.81,Suz.00,Bor.00}.
It has been shown that the choice of the spin-isospin terms
of the Skyrme energy functional affects the 
calculated strength distribution 
and excitation energy of the GT resonance, i.e., the
properties of the GTR are not entirely determined by the
Landau-Migdal residual interaction~\cite{Ben.02}.
In addition, the inclusion of particle-particle
correlations in the QRPA residual interaction is important in 
calculations of the GT transition strength~\cite{Cha.83,Eng.88,Che.95},
$\beta$-decay rates~\cite{Eng.99,Bor.95,Bor.96}
and double $\beta$-decay amplitudes~\cite{Che.93,Hir.97}.
The inclusion of proton-neutron pairing changes
significantly the rates of the neutrinoless double $\beta$-decay,
allowing for larger values of the expectation value of
light neutrino masses~\cite{Pan.96}. Both the
two-neutrino and neutrinoless double $\beta$-decay matrix
elements are suppressed by the particle-particle 
interaction~\cite{Eng.88}. The importance of including 
proton-neutron pairing in calculations of GT excitations has also 
been illustrated in relativistic QRPA studies~\cite{PNVR.04}. 
Pairing correlations have recently also been included in the 
proton-neutron continuum QRPA based on a phenomenological 
mean-field potential and the isovector part of the 
Landau-Migdal $ph$ interaction~\cite{Rod.03}.
In addition to pairing, deformation plays an important
role in the description of GT$^{\pm}$ transitions in many
nuclei~\cite{Srr.01,Ste.04}. The sensitivity of GT strength
distributions and double $\beta$-decay matrix elements to
the deformed mean-field has recently been analyzed with 
the QRPA~\cite{Alv.04}, and it has been shown that nuclear 
deformation could result in a suppression of two-neutrino
double $\beta$-decay rates.

\begin{figure}
\centering
\vspace*{1cm}
\includegraphics[scale=0.45,angle=0]{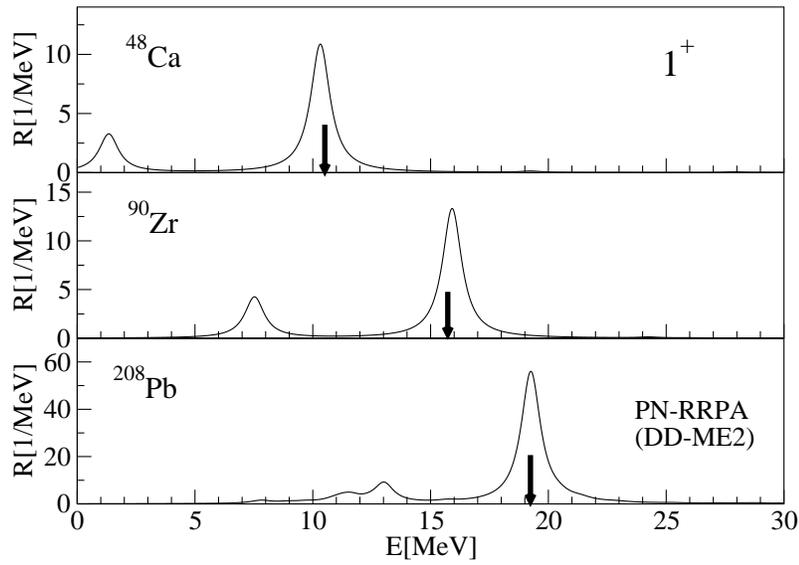}
\vspace*{0.5cm}
\caption{Gamow-Teller
strength distributions for
$^{48}$Ca, $^{90}$Zr, and $^{208}$Pb. PN-RRPA results are
shown in comparison with experimental data (arrows) for the
GTR excitation energies in $^{48}$Ca~\protect\cite{And.85},
$^{90}$Zr~\protect\cite{Bai.80,Wak.97},
and $^{208}$Pb~\protect\cite{Kra.01,Aki.95,Hor.80}.}
\label{gtmagic}
\end{figure}

The GT resonance represents a coherent superposition of 
high-lying $J^{\pi}=1^{+}$ proton-particle -- neutron-hole
configurations. In the relativistic formalism
the GT operator reads 
\begin{equation}
Q_{\beta^{\pm}}^{GT}=\sum_{i=1}^{A}\mathbf{\Sigma}\tau_{\pm} \; ,
\label{gtopera}
\end{equation}
where $\mathbf{\Sigma}$ is defined in Eq.~(\ref{bigsigma}).
In Fig.~\ref{gtmagic} we display the GT$^-$ strength
distributions in the magic nuclei $^{48}$Ca, $^{90}$Zr, and $^{208}$Pb, 
calculated in the PN-RRPA with the DD-ME2 effective interaction and,
for comparison, data on the excitation energies of the GTR are 
also included in the figure.
In addition to the high-energy GT 
resonance -- a collective superposition of direct spin-flip 
($j = l + \frac{1}{2}$ $\rightarrow$ $j = l - \frac{1}{2}$) 
transitions, the response functions display a concentration of
strength in the low-energy tail. The low-lying GT excitations 
correspond to core-polarization 
($j = l \pm \frac{1}{2}$ $\rightarrow$ $j = l \pm \frac{1}{2}$), 
and back spin-flip 
($j = l - \frac{1}{2}$ $\rightarrow$ $j = l + \frac{1}{2}$)
neutron-hole -- proton-particle transitions.
The strength parameter of the zero-range Landau-Migdal 
force Eq.~(\ref{deltapi}) has been adjusted to reproduce
the GTR excitation energy in $^{208}$Pb ($g^{\prime}= 0.52$),
but we notice a very good agreement with data also for 
$^{48}$Ca and $^{90}$Zr. The adjusted value of $g^{\prime}$
in general depends on the choice of the effective interaction. 
By employing a set of RMF effective interactions with density-dependent 
meson-nucleon couplings~\cite{NVR.02},
in Ref.~\cite{PNVR.04} it has been shown that there is 
a linear correlation between the value of the nuclear asymmetry 
energy at saturation $a_4$, and the value of $g^{\prime}$ adjusted to 
reproduce the GTR excitation energies: effective interactions with higher 
values of $a_4$ require higher values of $g^{\prime}$.  
\begin{figure}
\centering
\vspace*{1cm}
\includegraphics[scale=0.45]{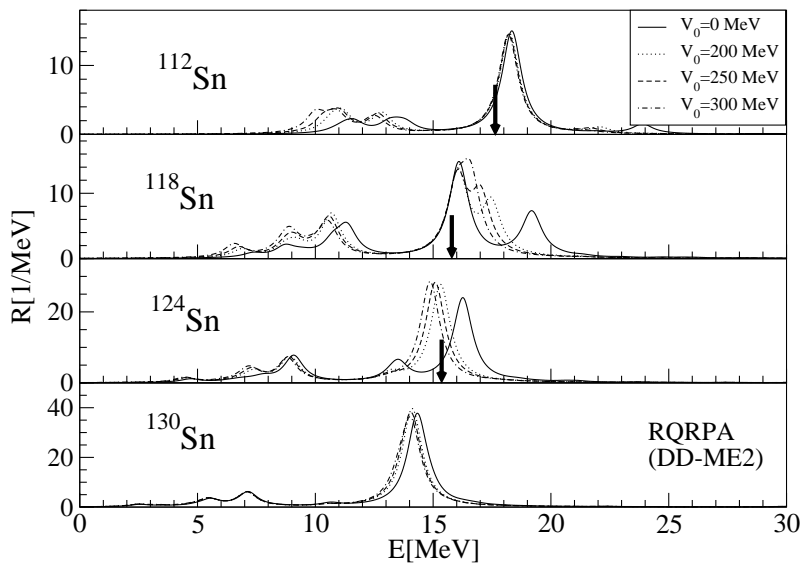}
\vspace*{0.5cm}
\caption{The GT$^-$ 
strength distribution in $^{112,118,124,130}$Sn, calculated for different
values of the strength parameter $V_0$ of the $T=0$ pairing
interaction Eq.~(\protect\ref{pn-pair}).
The experimental excitation energies of the GTR 
are denoted by arrows~\cite{Pham.95}.}
\label{gtr_Sn}
\end{figure}

As an example of Gamow-Teller resonances in open shell nuclei,
in Fig.~\ref{gtr_Sn} we plot the PN-RQRPA strength distributions
for the $^{112,118,124,130}$Sn target nuclei, in comparison with the
data for the centroids of the high-energy direct spin-flip strength, 
obtained in the Sn($^3$He,t)Sb charge-exchange reactions~\cite{Pham.95}.
Direct spin-flip transitions 
dominate the high-energy region above 10 MeV. The low-energy tail
of the strength distribution corresponds to core-polarization, 
and back spin-flip transitions. 
The solid curves have been calculated without including 
the $T=0$ proton-neutron pairing in the RQRPA residual interaction.
The resulting high-energy GT strength in $^{118}$Sn and $^{124}$Sn
is divided into two main components, because of
the splitting between different $ph$ configurations.
GTR configuration splitting (an appearance of two or more collective
bumps with comparable intensities in the GTR strength function) was 
investigated in Ref.~\cite{GNU.89} in the framework of the shell 
optical model. For Sn nuclei, in particular, this effect was predicted
to occur as the valence neutron start to occupy the level with the 
highest $j$ in the shell: $h_{11/2}$. The configuration splitting 
was attributed to the fact that the unperturbed energies
of the $(1g^\pi_{7/2})(1g^\nu_{9/2})^{-1}$ and 
$(1h^\pi_{9/2})(1h^\nu_{11/2})^{-1}$ configurations are almost degenerate.
The residual interaction removes this degeneracy and, as a result,
the main GT component separates in two distinct peaks. 
The ground-state pairing correlations have a strong influence
on the occupation of the $1h^\nu_{11/2}$ level, and therefore
the energy spacing between the two peaks will depend on 
$T=1$ pairing. For $^{118}$Sn the calculated energy splitting 
of the GTR was 2.6 MeV~\cite{GNU.89}. 
Subsequently, the
fragmentation and splitting of the GTR in Sn nuclei was experimentally
investigated in Ref.~\cite{Pham.95}. The theoretically predicted 
configuration splitting of the main GT component could not 
be observed, however, because the total widths of the resonances 
$\approx 5-6$ MeV exceed the predicted splitting. 
The splitting of the main GT component 
shown in Fig.~\ref{gtr_Sn} (solid line) of $\approx 3$ MeV is in 
agreement with the result of Ref.~\cite{GNU.89}.

The other curves shown in Fig.~\ref{gtr_Sn} 
(dotted, dashed and dot-dashed) have been calculated 
for different values of the 
strength-parameter of the $T=0$ proton-neutron pairing interaction 
in Eq.~(\ref{pn-pair}):  $V_0$ = 200, 250, and 300 MeV. 
In Ref.~\cite{Eng.99} the overall strength parameter $V_0$ of the 
interaction was adjusted to the measured half-lives of neutron-rich 
nuclei in regions where the $r$-process nucleosynthesis 
path comes closest to the valley of stability: 
$V_0 = 230$ MeV near $N=50$, and $V_0$ = 170 MeV in the $N=82$ region. 
In our illustrative calculation of Sn nuclei
the inclusion of the $T=0$ pairing has a strong influence 
on the low-energy tail of the GT distribution in $^{112,118,124}$Sn, 
and the configuration splitting between the two high-energy
peaks in $^{118}$Sn and $^{124}$Sn disappears. 
This happens because the $T=0$ pairing interaction does 
not affect configurations based on the $(1g^\nu_{9/2})$ orbital 
(fully occupied), whereas it lowers configurations based on 
$(1h^\nu_{11/2})$ and $(2d^\nu_{5/2})$ (partially occupied). This calculation
therefore demonstrates that the $T=0$ proton-neutron pairing strongly 
reduces the predicted configuration splitting of the main high-energy 
GT component. In addition to the main GTR which decreases in energy 
with increasing mass number,
part of the strength associated with direct spin-flip transitions 
is concentrated at $\approx 10$ MeV in $^{112}$Sn and $^{118}$Sn. 
In Ref.~\cite{PNVR.04} it has been shown that the centroid of
the GT strength composed of direct spin-flip transitions 
practically does not depend on the strength of the 
$T=0$ proton-neutron $pp$ residual interaction.

The structure and evolution of the low-energy tail of the GT distribution determines the
$\beta$-decay rates of very neutron-rich nuclei, and thus sets the time scale 
of the $r$-process nucleosynthesis. Since the vast majority of nuclides which lie on 
the path of the $r$-process are out of experimental reach,  
nuclear structure models must be developed that can provide 
predictions of weak-interaction rates of thousands of nuclei with large 
neutron to proton asymmetry. Two microscopic approaches can be employed 
in large-scale calculations of $\beta$-decay rates: the interacting shell model and the QRPA. The advantage of using the shell model is the ability to take into account the 
detailed structure of the $\beta$-strength function \cite{LM.03}, 
whereas the QRPA approach is based on global effective interactions and provides a systematic description of $\beta$-decay properties of arbitrarily heavy nuclei along 
the $r$-process path. In a recent review of modern QRPA calculations of $\beta$-decay rates for astrophysical applications \cite{Bor.06}, the importance of performing 
calculations based on self-consistent mean-field models has been emphasized, rather than on empirical mean-field potentials, e.g. the Woods-Saxon potential. In a 
self-consistent framework both the nuclear ground states, i.e. the masses which determine the possible $r$-process path, and the corresponding 
$\beta$-decay properties are calculated from the same energy 
density functional or effective nuclear interaction. 
This approach ensures the consistency of the nuclear structure input for 
astrophysical modeling, and allows reliable extrapolations of the nuclear 
spin-isospin response to regions of very neutron-rich nuclei.

%
\subsection{\label{gtrsumr}Effect of the Dirac Sea on the Gamow-Teller Sum Rule}
%
In nuclei all over the periodic table the
GT strength distribution, when measured in the excitation energy region 
where the most pronounced GT peaks occur, is quenched by more than
20\% with respect to the model independent Ikeda sum rule~\cite{Ike.63}
\begin{equation}
\left(  S_{\beta^{-}}^{GT}-S_{\beta^{+}}^{GT}\right)  =3(N-Z),\label{gtsrul
e}%
\end{equation}
where $S_{\beta^{\pm}}^{GT}$ denotes the total sum of
Gamow-Teller strength for the $\beta^{\pm}$ transition.
In the early $(p,n)$ studies, for instance, only $\approx$~60\% 
of the sum rule was observed~\cite{Gaa.83}. Two physically 
different mechanisms had been suggested as a possible 
explanation of the quenching of the total GTR strength:
(i) nuclear configuration mixing -- the high-lying 
$2p-2h$ states mix with the $1p-1h$ GT states and 
shift the GT strength to high-energy region 
far beyond the resonance~\cite{Shi.74,Ber.82,DOS.86};
(ii) excitation of a nucleon into the high-energy
$\Delta$-isobar~\cite{Rho.74,Oht.74}, with the 
$\Delta$-isobar -- nucleon-hole 
configurations ($\Delta-h$) coupling to the GT mode and 
removing part of the strength from the low-lying 
excitation spectrum~\cite{Eric.73,Knu.80}. In more
recent $(p,n)$ scattering experiments data on GT strength 
in the high energy region up to 50 MeV became available~\cite{Wak.97,Wak.99}.
It has been shown that the measured GT strength exhausts 
88\% and 84\% of Ikeda sum rule in $^{90}$Nb and $^{27}$Si, respectively.
Most of the GT strength missing in early experimental
studies is, therefore, recovered in the energy region where
the multiconfiguration spreading mechanism is effective, and 
only a small fraction of the GT quenching may have
its origin in $\Delta-h$ transitions lying high above the 
ordinary $ph$ excitations. 

\begin{figure}
\centering
\vspace*{1cm}
\includegraphics[scale=0.4]{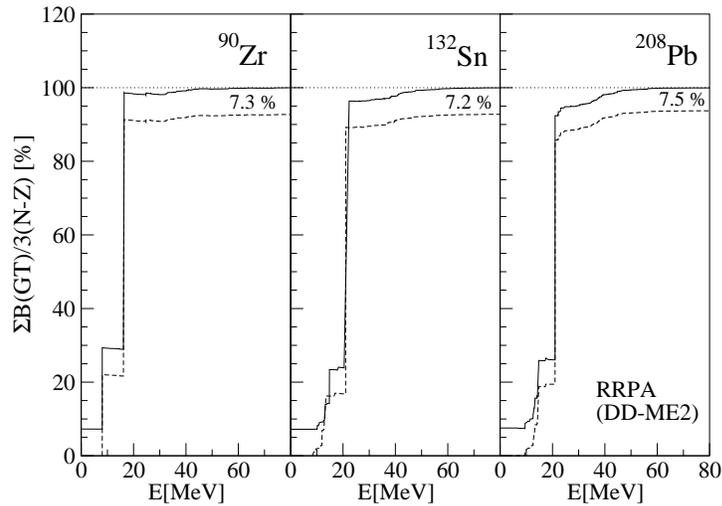}
\vspace*{0.5cm}
\caption{The running sum of the GTR strength for $^{90}$Zr,
$^{132}$Sn, and $^{208}$Pb. 
The dashed lines corresponds to PN-RRPA calculations with 
only positive-energy $ph$ configurations. For the calculation
denoted by the solid lines the RRPA space includes 
configurations formed from occupied 
states in the Fermi sea and empty negative-energy 
states in the Dirac sea. The
total sum of the GT strength is compared to the model 
independent Ikeda sum rule
(dotted lines).}
\label{sr1}
\end{figure}

When the Gamow-Teller strength is calculated in the relativistic 
RPA framework, the total GT strength in the nucleon sector is reduced by 
$\approx 12$\% in nuclear matter, and by $\approx 6$\% in finite 
nuclei, as compared to the Ikeda sum 
rule~\cite{PNVR.04,Ma.03,Kur.03a,Kur.03b,Kur.04,Kur.06}.
This reduction has been attributed to the effect of Dirac sea negative-energy 
states, i.e. the missing part of the sum rule is taken by particle-hole 
excitations formed from ground-state configurations of occupied 
states in the Fermi sea and empty negative-energy states in the Dirac sea.
The effect is illustrated in Fig.~\ref{sr1}, where we display the 
running sum of GTR strength relative to the Ikeda sum rule for $^{90}$Zr,
$^{132}$Sn, and $^{208}$Pb, evaluated in the PN-RRPA with the
DD-ME2 effective interaction. The horizontal 
dotted lines denote the value $3(N-Z)$ of the Ikeda sum rule.
The solid and dashed lines correspond to the values of the GTR sum 
calculated from $- \infty$ to the excitation energy denoted on the 
abscissa. The big jump in the calculated GTR sum occurs, of course,
when the main GTR peak is included. The PN-RRPA calculation represented 
by the dashed lines includes only positive energy $ph$ configurations. 
Even by extending the sum up to 80 MeV, the total sum is reduced
$\approx$~7--8\% with respect to the Ikeda sum rule. The Ikeda sum
rule is completely exhausted by the calculated GT strength only when the 
relativistic RPA/QRPA space includes $ph$ excitations formed from ground-state 
configurations of the fully or partially occupied states of
positive energy, and the empty negative-energy states from the Dirac sea
(solid lines in Fig.~\ref{sr1}).

\begin{figure}
\centering
\vspace*{1cm}
\includegraphics[scale=0.4]{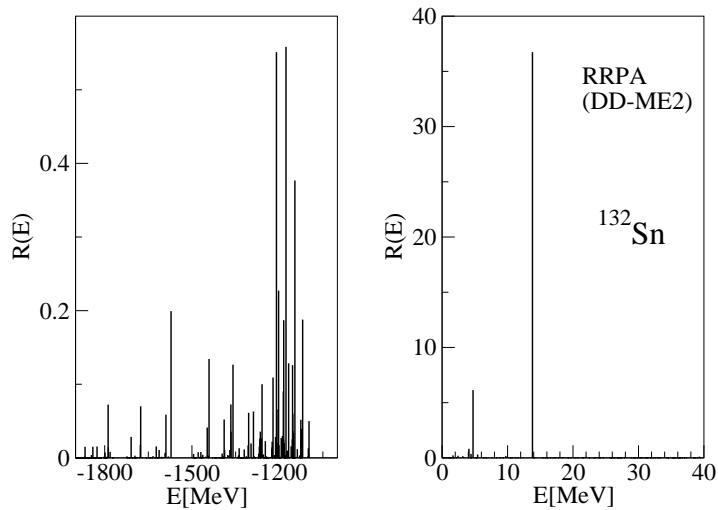}
\vspace*{0.5cm}
\caption{
The PN-RQRPA strength distribution of discrete
GT$^{-}$ states for $^{132}$Sn.  
The panel on the right-hand side contains the positive energy
$\pi p-\nu h$ strength.
The panel on the left displays the negative energy spectrum 
built from transitions to the empty states in the Dirac sea.}
\label{sr2}
\end{figure}

In Fig.~\ref{sr2} we plot the discrete GT spectrum 
for $^{132}$Sn. Two regions of excitation energies are shown. 
The panel on the right contains the positive energy
$\pi p-\nu h$ strength, 
with a pronounced Gamow-Teller resonance peak.
The panel on the left displays the negative energy spectrum 
built from $\pi\alpha-\nu h$ transitions ($\alpha$ denotes a 
negative energy state). Even though these transitions 
are much weaker than the GTR (notice that the vertical scales 
are different for the two panels), there are many of them and 
their overall sum represents the strength missing  
in the positive energy sector.

%
\subsection{Spin-Isospin Resonances and the Neutron Skin of Nuclei}
%

In medium-heavy and heavy neutron-rich nuclei the ground-states are
often characterized by an extended neutron density distribution, 
and in some cases evidence has been reported for the formation of 
a neutron skin on the surface of the nucleus. The determination of
neutron density distributions provides not only basic
nuclear structure information, but it also places important
additional constraints on effective interactions used in nuclear
models. Extremely accurate data on charge densities, and therefore on proton 
distributions in nuclei, have been obtained from elastic scattering of 
electrons~\cite{Nad.94}. Data of comparable precision on neutron density 
distributions are, however, not yet available. It is much more difficult 
to measure the distribution of neutrons, though experimental information 
on the differences between radii of the neutron and proton density distributions 
has been reported~\cite{Kra.91,Kra.99,Suz.95}. Various experimental methods 
have been used, or suggested, for the determination of the neutron density 
in nuclei~\cite{Bat.89}, but no existing measurement of neutron densities 
or radii has an accuracy better than a few percent.

One of the modern approaches that provides information about the neutron
skin in nuclei is based on studies of giant resonances. In particular, 
in Ref.~\cite{Kra.91} excitations of the giant dipole
resonance (GDR) were analyzed, and the spin-dipole resonance 
(SDR) was studied in \cite{Kra.99}. The GDR cross section strongly depends on 
the difference between neutron and proton density 
distributions~\cite{Shl.89,Nak.87,Pet.87}. In Ref.~\cite{Kra.99} it has been 
demonstrated that there is a predictable correlation between the SDR 
cross section and the difference
between the rms radii of the neutron and proton density distributions.
By normalizing the results to $^{120}$Sn, data on neutron-skin
thickness along the stable Sn isotopic chain were obtained, in good
agreement with theoretical predictions.
 
\begin{figure}
\centering
\vspace*{1cm}
\includegraphics[scale=0.4]{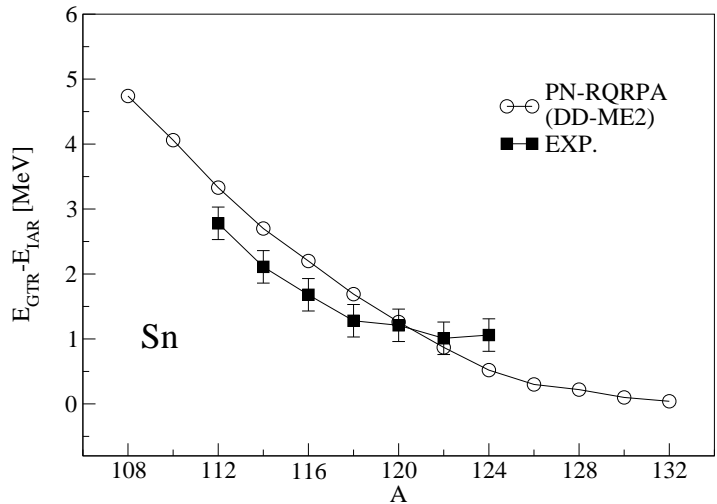}
\vspace*{0.5cm}
\caption{ RHB plus proton-neutron RQRPA results for the energy spacings
between the Gamow-Teller
resonances and the respective isobaric analog resonances for the
sequence of even-even $^{112 - 124}$Sn target nuclei.
The experimental data are from Ref.~\protect\cite{Pham.95}.}
\label{gt-iar}
\end{figure}
Recently a new method has been suggested for determining the difference 
between the radii of the neutron and proton density distributions along 
an isotopic chain, based on measurement of the excitation energies of the 
Gamow-Teller resonances relative to the isobaric analog states~\cite{VPNR.03}.
The Gamow-Teller resonance (GTR) $J^\pi = 1^{+}$ represents a
collective spin-isospin oscillation with the excess neutrons 
coherently changing the direction of their spins and isospins 
without changing their orbital motion. The simplest 
charge-exchange excitation mode, however, does not require 
the spin-flip and corresponds to the 
well known isobaric analog state (IAS) $J^\pi = 0^{+}$. 
The spin-isospin characteristics of the GTR and the IAS are
related through the Wigner supermultiplet scheme. The Wigner 
SU(4) symmetry implies the degeneracy of the GTR and IAS, and 
furthermore the resonances would completely exhaust the corresponding
sum rules. The Wigner SU(4) symmetry is, however, 
broken by the spin-orbit term of the effective nuclear potential. 
The energy difference between the GTR and the IAS decreases with 
increasing asymmetry $(N-Z)/A$.
It is implicit, therefore, that the energy difference between the 
GTR and the IAS reflects the magnitude of the effective spin-orbit
potential. A number of relativistic mean-field calculations have shown
that the magnitude of the spin-orbit potential is considerably reduced 
in neutron-rich nuclei~\cite{LVR.97}, and this is 
reflected in the larger spatial extension of the neutron density,
which becomes very diffuse on the surface. The neutron-skin increases 
correspondingly. The energy spacings between neutron spin-orbit partner
states decrease, and this reduction is quantitatively 
in accordance with the gradual weakening of the 
spin-orbit term of the effective potential.
\begin{figure}
\centering
\vspace*{1cm}
\includegraphics[scale=0.5]{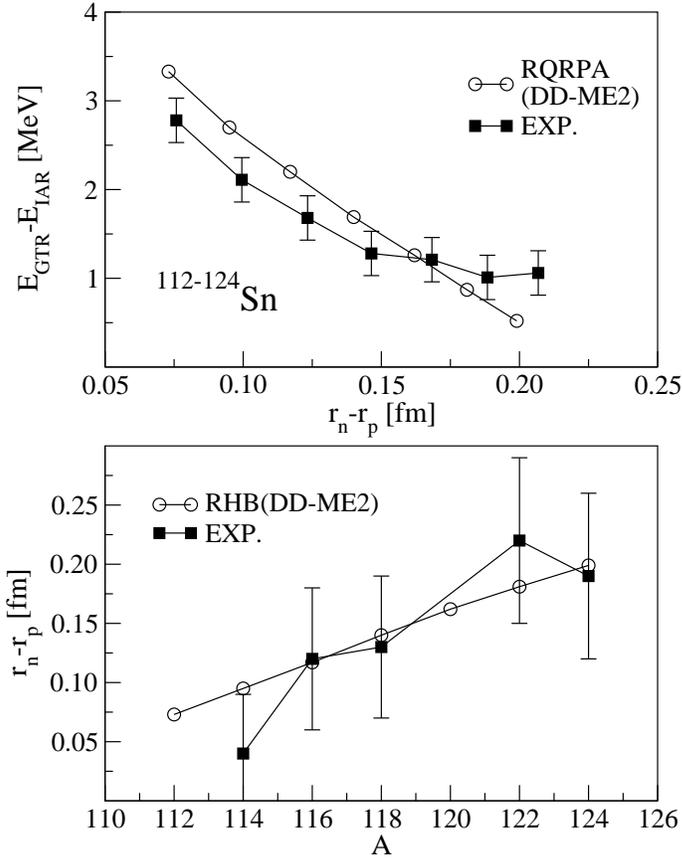}
\vspace*{0.5cm}
\caption{The PN-RQRPA and experimental~\protect\cite{Pham.95}
differences between the excitation energies
of the GTR and IAR, as a function of the calculated
differences between the rms radii of the neutron and proton
density distributions of even-even Sn isotopes (upper panel).
The calculated differences $r_n - r_p$
are compared with experimental data~\protect\cite{Kra.99} (lower
panel).}
\label{gt-iarskin}
\end{figure}

There is a direct connection between the increase of the neutron-skin
thickness in neutron-rich nuclei, and the decrease of
the energy difference between the GTR and the IAS.
In Fig.~\ref{gt-iar} we display the calculated differences between 
the centroids of the direct spin-flip GT strength and
the respective isobaric analog states for the
sequence of even-even Sn target nuclei. For $A=112 - 124$
the results of RHB plus PN-RQRPA
calculation  (DD-ME2 density-dependent effective interaction,
Gogny $T=1$ pairing, $T=0$ pairing interaction Eq. (\ref{pn-pair}) 
with $V_0 =$ 250 MeV, the Landau-Migdal parameter $g^{\prime}= 0.52$
adjusted to reproduce the excitation energy of the GT
resonance in $^{208}$Pb), are compared with the experimental values 
obtained in a systematic study of the ($^3$He,t) charge-exchange 
reaction over the entire range of stable Sn isotopes~\cite{Pham.95}. 
The calculated energy spacings are in very good agreement with 
the data, although for the lighter Sn isotopes it 
appears that the calculated values differ
somewhat from the experimental trend. However, the theoretical 
energy spacings might depend on the details of the effective
interaction and, in fact the data shown in Fig.~\ref{gt-iar}
provide valuable information that can be used to constrain
the spin-isospin channel of the effective interaction. 

In Fig.~\ref{gt-iarskin} the calculated and experimental 
energy spacings between the 
GTR and IAS are plotted as a function of the calculated 
differences between the rms radii of the neutron and proton 
density distributions of even-even Sn isotopes (upper panel). 
We note the remarkable uniform dependence of the energy 
difference between the GTR and IAS on the size of the neutron-skin.
This means that, in principle, the value of $r_n - r_p$ can be 
directly determined from the theoretical curve for a given value of 
$E_{\rm GTR} - E_{\rm IAS}$.                                                                               
In the lower panel the calculated differences between neutron and  
proton rms radii are compared with available experimental
data~\cite{Kra.99}. The agreement between the theoretical and                                                                                 
experimental values suggests that the
neutron-skin thickness can be determined from the measurement 
of the excitation energies of the GTR relative to IAS.                                                                                                                                                                                                                                                                                                                                                                                                       
This method is, of course, not completely model
independent, but it does not require additional assumptions. 
Since the neutron-skin thickness is determined in an indirect 
way from the measurement of the GTR and IAS excitation 
energies in a sequence of isotopes, in practical applications
at least one point on the theoretical curve should be checked 
against independent data on $r_n - r_p$.

\section{Exotic Nuclear Modes in Astrophysical Processes}
\label{Sec7}

Collective modes play an important role in astrophysical processes
that involve both stable and exotic nuclei far from stability. 
In particular, spin-isospin excitations such as the Gamow-Teller 
resonance are essential for weak-interaction processes,
e.g. $\beta$-decay of exotic nuclei on the path of r-process 
nucleosynthesis, electron capture on neutron-rich nuclei at 
temperatures and densities characteristic for stellar core collapse, 
neutrino-induced reactions on heavy neutron-rich nuclei in 
the post-collapse supernova environment, including the 
process of neutrino nucleosynthesis. A recent review of 
nuclear weak-interaction processes in stars can be found in 
Ref.~\cite{LM.03}.

In this section we will review the role of non-charge-exchange 
nuclear collective modes in astrophysical processes.  
Particularly interesting is the effect of the low-lying 
dipole transition strength on the r-process nucleosynthesis, 
and in the propagation of ultra high-energy cosmic
rays. We will also discuss exotic collective modes
in the crust of a neutron star.

\subsection{Low-Energy Dipole Strength and the r-process}
\label{ssect_rprocess}

Approximately half of the nuclides with $A>60$ 
found in nature are formed in the
rapid neutron-capture process (r-process) nucleosynthesis.
The nuclear input for r-process 
calculations necessitates the knowledge of the properties of 
thousands of nuclei far from stability, including the 
characteristics of strong, electromagnetic and weak interaction 
processes~\cite{go04}. Most of these nuclei are not accessible 
in experiments and, therefore, many 
nuclear astrophysics calculations crucially depend on accurate 
theoretical predictions for the nuclear masses, bulk properties, 
nuclear excitations, ($n,\gamma$) and ($\gamma,n$) rates, 
$\alpha$- and $\beta$-decay half-lives, fission probabilities, 
electron and neutrino capture rates, etc. 
Early calculations of the nuclear processes relevant for astrophysical 
applications were based only on phenomenological models 
\cite{go98,mc81,ka83,ko87}. Only more recently large-scale microscopic 
calculations became standard in the prediction of nuclear masses \cite{ms03b}, 
and dipole strength distributions~\cite{Gor.02,Gor.04}. The availability  
of large scale microscopic calculations opens the possibility  
for global predictions of the nuclear ingredients for the r-process, 
based on an universal nuclear energy density functional. However, fully
microscopic calculations of nuclear observables over the whole 
isotope chart are not yet feasible, especially for excited states.
Only the excitation spectra of all even-even spherical nuclei can be obtained
from fully consistent microscopic Skyrme-QRPA calculations~\cite{kh02},
whereas for deformed, odd-even, and odd-odd nuclei
a series of approximations must be designed specifically for 
astrophysical applications, and implemented in the microscopic QRPA.

Two principal candidates have been suggested for the astrophysical site
of the r-process. In the first scenario the r-process takes place
in explosive stellar events, such as the core collapse
supernova~\cite{go04}, in an environment characterized by 
high neutron densities ($N_n \simeq 10^{20}~{\rm cm^{-3}}$), 
so that successive neutron-captures proceed into regions of 
neutron-rich nuclei far away from the valley of $\beta$-stability.  
If the temperature or the neutron density that
characterize the r-process are low enough to break the
(n,$\gamma$) -- ($\gamma$,n) equilibrium, the
waiting-point approximation is not valid any more, and the
r-abundance distribution directly depends on the neutron-capture 
rates on exotic neutron-rich nuclei~\cite{go98}. 

Recently an alternative scenario for the r-process has attracted 
renewed interest. It is related to the decompression of 
cold neutron star matter, in particular its crust 
(see~\cite{go04} for a more detailed description). In this
scheme the production of heavy nuclei follows a completely different path 
from the core-collapse supernovae scenario. The inner crust of a 
neutron star is composed of nuclear clusters immersed in a neutron gas.
When a decompression of this crust occurs, nuclear clusters and the 
neutron gas are both ejected, and this leads to a decrease of the
matter density. The $\beta$-equilibrium is broken and nuclei with 
Z in the range between 40 and 70 are produced. At very low density 
drip-line nuclei are formed, immersed in a neutron flux 
of $N_n \simeq 10^{35}~{\rm cm^{-3}}$. The
competition between the neutron-capture and the $\beta$-decay, starting
from these drip-line nuclei, leads to the production of heavy
nuclear systems. Because this type of nucleosynthesis 
starts from drip-line nuclei, where the waiting-point approximation 
cannot be applied, theoretical predictions of various 
structure phenomena are essential ingredients for r-process modeling. 

\begin{figure}[htb]
\centerline{\includegraphics[width=8cm,height=4cm]{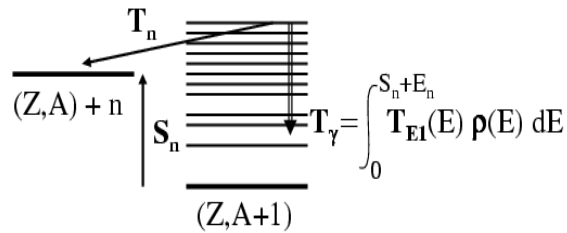}}
\caption{Illustration of the statistical radiative (n,$\gamma$)
neutron-capture. T$_n$ is the neutron-capture coefficient, 
and T$_{\gamma}$ is the photo-transmission coefficient. 
T$_{\gamma}$ is determined by the E1 strength
T$_{E1}$(E), and the level density $\rho$(E).} 
\label{fig:ncapt}
\end{figure}

Several types of nuclear observables are therefore required for the 
description of r-process abundances. In addition to the $\beta$-decay rates, 
it is also necessary to describe the (n,$\gamma$)
rates. This process can be divided into two steps: the neutron capture and the
photo-deexcitation~\cite{go00}. Nuclear masses and neutron-nucleus
optical potentials enter into the calculation of the neutron-capture rates. 
The description of the photo-deexcitation process necessitates 
predictions of the E1 strength functions, as well the level densities 
in daughter nuclei. Neutron capture rates are evaluated in the framework 
of the Hauser-Feshbach statistical model, which is based on the fundamental
assumption that the capture process occurs with the intermediary
formation of a compound nucleus in thermodynamic equilibrium. In this
approach the Maxwellian-averaged (n,$\gamma$) rate, at temperatures 
characteristic for the r-process environment, strongly depends on 
the electromagnetic interaction, i.e. on the photo-deexcitation probability. 
The modeling of the r-process abundances requires a reliable
extrapolation of the E1-strength functions towards the neutron-drip line.
Figure \ref{fig:ncapt} shows a schematic picture of the (n,$\gamma$) reaction
in the statistical model framework.

Low-energy dipole modes are expected to play a more
important role in the r-process than the higher lying strength, 
including the giant dipole resonance (GDR). Namely, the
low-lying modes located close to neutron separation energy 
S$_n$ are directly sustained by neutron-capture reactions, 
as shown in Fig. \ref{fig:ncapt}. The importance
of the soft dipole modes has been emphasized in Ref.~\cite{go98}, 
where it has been shown that the presence of a low-lying 
resonant component of the E1 strength leads to an increase of 
the radiative neutron-capture rate by factors 10 -- 100, 
for nuclei with S$_n$ between 2 MeV and 4 MeV. The r-abundance
distribution is affected because the presence of low-energy dipole 
resonances accelerates neutron capture and allows the production 
of heavy nuclei around $A=130$.

Large-scale calculations of E1-strength functions 
for astrophysical applications are usually performed 
using phenomenological Lorentzian models~\cite{go98}. 
Several refinements can be introduced, such as including 
the energy dependence and/or the temperature dependence
of the width of the Lorentzian~\cite{go98,mc81,ka83,ko87}. 
The Lorentzian GDR approach presents, however, 
several problems. On one hand, in this framework it is 
not possible to predict the enhancement of the E1 strength at 
energies close the neutron separation energy. 
On the other hand, even if a Lorentzian-type 
function provides a suitable description of the E1 strength in stable
nuclei, the location of its maximum and its width for each nucleus
remain to be predicted from some systematics or underlying model. For
astrophysical applications these properties have often been obtained from a
droplet-type model~\cite{my77}. This approach is clearly not reliable when
dealing with exotic nuclei, and this was already demonstrated in
Refs.~\cite{Gor.02,Cat.97}. In order to achieve a better description of
the r-process, one has to improve the nuclear structure modeling.
Generally speaking, the more microscopic the underlying theory, 
the more reliable will be the extrapolations towards the neutron drip-line.
Microscopic calculations of the E1-strength functions for
the whole nuclear chart have recently been reported~\cite{Gor.02,Gor.04}. 
In a first step the dipole response was calculated with the QRPA 
based on the HF-Skyrme plus BCS description of nuclear ground 
states~\cite{Gor.02}. In neutron-rich nuclei pronounced E1 strength
was  predicted in the low-energy region below the giant dipole resonance.

The dipole strength can also be calculated with the QRPA based on HFB
ground states~\cite{kh02,gr01}. In this microscopic approach 
photoabsorption cross sections were determined~\cite{Gor.04}, and one was
able to judge the ability of the different forces to reproduce 
experimental data. In Refs.~\cite{ms03b,ms02,sg02,ms03} nuclear deformation 
was also taken into account. Based on the
Skyrme-HFB approach, a number of new effective forces have recently been
introduced~\cite{ms03b,sg02,ms03}. Their parameters were {\it exclusively} 
adjusted to the 2135 experimental masses~\cite{aw01}, 
with some additional constraints related to the stability
of neutron matter and the incompressibility of nuclear matter. 
The new effective interactions BSk2-BSk7 are summarized in
Ref.~\cite{go04}. 

\begin{figure}[htb]
\centerline{\includegraphics[scale=0.45]{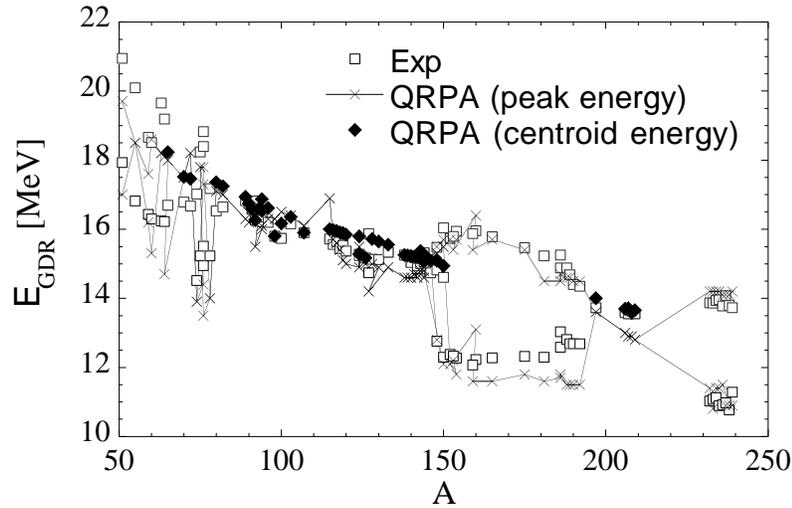}}
\caption{Comparison between the experimental GDR energies and the HF-BCS+QRPA
calculations with the SLy4 force~\cite{Gor.02}.} 
\label{fig:gdrexp}
\end{figure}

Photo-induced reaction cross sections were compiled in 
Refs.~\cite{di89,iaea00}, and represent the most reliable source of
data with which HFB+QRPA predictions can be compared. 
These include the GDR parameters (the peak energy, peak cross
section, and the full width at half maximum) observed in photonuclear
reactions measured by bremsstrahlung, quasimonoenergetic and tagged photons
for 84 nuclei. Among these, 48 nuclei are spherical and can be used to
test the (HF-BCS or HFB)+QRPA predictions. Figure \ref{fig:gdrexp} shows the
calculated GDR centroid energies of these 48 spherical nuclei using 
the HF-BCS+QRPA with the SLy4 parameterization, in comparison with the data.

For the BSk6 and BSk7 interactions, in particular, 
the comparison shows that these functionals not only reproduce the 
experimental masses with great accuracy (the rms deviation is only 
$0.676$ MeV for the set of 2135 known masses), 
but also are suited for the description of E1 collective 
excitations~\cite{Gor.04}. The agreement with data is of the same
level of accuracy as in the case of phenomenological models, 
i.e. the rms deviation for the GDR excitation energies is around 500 keV. 
However, since the microscopic HFB+QRPA approach
potentially encompasses a wide range of phenomena, including the soft dipole
mode, it is clearly preferable in large-scale calculations.

Turning now to astrophysical applications, microscopic predictions need to be
extended on the whole nuclear chart, including also nuclei 
with an odd number of protons and/or neutrons, and deformed systems.
The approximations that are currently used in these calculations
include: the filling approximation for odd-A and odd-odd nuclei,
a phenomenological damping of the E1 strength, and a phenomenological
treatment of deformation. These estimates may be improved in the near
future, for instance by developing a deformed QRPA model that could be
used in large-scale calculations of thousands of nuclei. 

The QRPA provides a reliable description of the GDR centroid
energy, but it is necessary to go beyond this approximation 
to describe the damping of collective motion.  The GDR are empirically 
known to have rather large widths and therefore finite lifetimes, which
can be described by several models~\cite{Dro.90,CB.01,sc01}. In
Ref.~\cite{Gor.02}, for instance, the QRPA strength function was folded by an 
arbitrary Lorentzian adjusted to the empirical GDR width. 
The damping of the E1 strength can also be described by the 
approximate procedure developed in Ref.~\cite{Dro.90}. In the large-scale 
calculation of Ref.~\cite{Gor.04} the QRPA strength was folded by a 
Lorentzian function representing the self-energy operator~\cite{Dro.90,sm88}. 
It is also necessary to introduce a temperature-dependent correction 
factor in the expression for the GDR width~\cite{ka83,ko90,bo01}. 
In deformed axially symmetric nuclei the GDR splits into two major
components as a result of the different resonance conditions characterizing
the oscillations of protons against neutrons along the axis of rotational
symmetry, and an arbitrary axis perpendicular to it. In the phenomenological
approach, the Lorentzian-type formula is generalized to a sum of two
Lorentzian functions~\cite{th83}. 

\begin{figure}[htb]
\centerline{\includegraphics[scale=0.4]{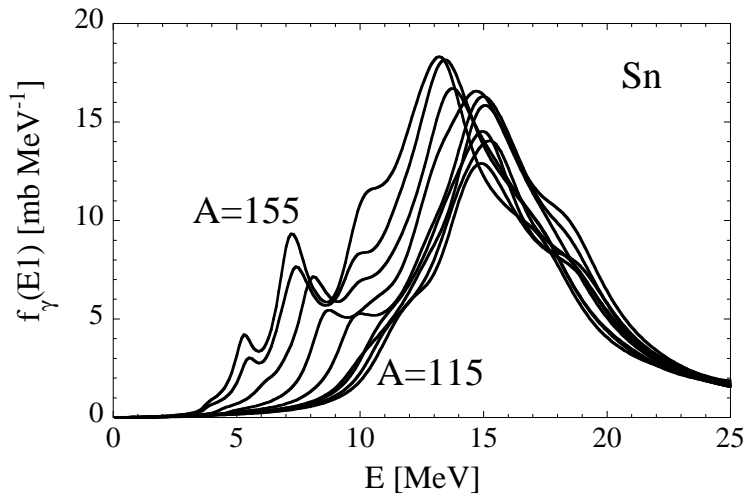}}
\caption{E1 strength functions for the Sn
isotopic chain, calculated in the HFB+QRPA model 
with the BSk7 force. Only isotopes between A=115 and A=150, 
with a step of $\Delta A = 5$ are shown~\cite{Gor.04}.} 
\label{fig_sn}
\end{figure}

Large-scale QRPA calculations based on the BSk7 Skyrme interaction have 
recently been performed for some 8300 nuclei with $8 \le Z\le 110$, 
extending between the proton and neutron drip-lines~\cite{Gor.04}.  
In the region of neutron-deficient nuclei, as well as
along the valley of $\beta$-stability, the calculated E1-strength functions
are very similar to the empirical Lorentzian approximation. However,
in neutron-rich nuclei the QRPA predictions start to deviate from
the simple Lorentzian shape. In particular, low-lying transitions 
are found at excitation energies well
below the GDR, and their strength increases with neutron excess. This
effect has been discussed in detail in Sec.~\ref{ssec_isopygmy}. 
In Fig.~\ref{fig_sn} we plot the E1 strength function
in Sn isotopes, calculated in the HFB+QRPA model with the BSk7 effective 
interaction. For $A\ge 140$ a significant portion of the strength is 
concentrated at low energies: $ E \simeq 5-7$ MeV. It should be noted that 
phenomenological models cannot predict these low-energy components. 
For $^{150}$Sn, for instance, all phenomenological systematics that are 
used in calculations of neutron-capture cross sections, 
predict a $\gamma$-ray strength peaked around 15 MeV with a full width at
half maximum of about 4.5~MeV~\cite{ripl2}. This is obviously very different
from the microscopic prediction shown in Fig.~\ref{fig_sn}. More generally, 
HFB+QRPA calculations confirm that the neutron excess affects the spreading
of the isovector dipole strength, as well as the centroid of the strength
function. The energy shift is larger than predicted by the usual $A^{-1/6}$
or $A^{-1/3}$ dependence given by the phenomenological liquid drop
approximations~\cite{my77}. The basic features of the QRPA
E1 strength function for nuclei with a large neutron excess are 
qualitatively independent of the choice of the effective interaction.

The radiative neutron capture cross sections are calculated
with the Hauser-Feshbach statistical model~\cite{go01a}, 
starting from nuclear ground state properties determined 
consistently in the microscopic HFB model with the 
same BSk7 Skyrme force~\cite{ms03b}. The calculation
includes the improved nuclear level density prescription based on
the microscopic statistical model, also used to estimate the nuclear
temperature~\cite{dem01}. The direct-capture contribution, as well as the
possible overestimate of the statistical predictions for resonance-deficient
nuclei, could have a significant effect on the radiative
neutron capture by exotic nuclei~\cite{go98}. However, we will focus on
the role of the dipole strength, which is almost exclusively probed by
the statistical model. 

\begin{figure}[htb]
\centerline{\includegraphics[scale=0.4]{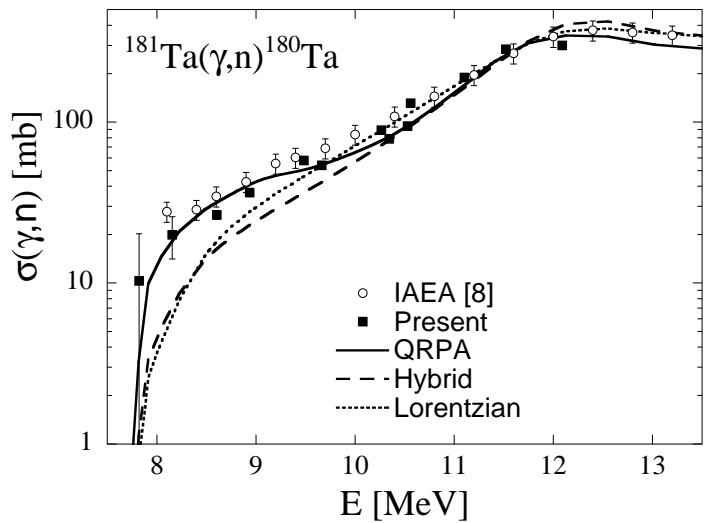}}
\caption{Comparison of the experimental photoneutron cross section for
$^{181}$Ta, with the Hauser-Feshbach predictions obtained with the QRPA,
hybrid, and Lorentzian models~\cite{ut03}.} 
\label{fig:utsu}
\end{figure}

The results of the Hauser-Feshbach calculation can be compared to the
experimental (n,$\gamma$) cross-sections. However, it is more convenient
to measure the inverse ($\gamma$,n) reaction. Such an experiment has
recently been performed on $^{181}$Ta: 
$^{181}$Ta($\gamma$,n)$^{180}$Ta~\cite{ut03}. In Fig.\ref{fig:utsu} we
show the comparison between the data and the Hauser-Feshbach calculations,
obtained using either the QRPA E1 strength, or the phenomenological 
strength distributions. We notice that at low energies ($ E < 10$ MeV) 
the microscopic calculation of the dipole strength produces results which 
are in excellent agreement with data, but significantly 
different from those obtained with the phenomenological approach. 
The tail of the cross section between 7.5 and 10 MeV is attributed to 
the presence of the low-energy dipole (pygmy) mode in the QRPA calculation, 
which does not appear in the phenomenological models. 

\begin{figure}[htb]
\centerline{\includegraphics[scale=0.4]{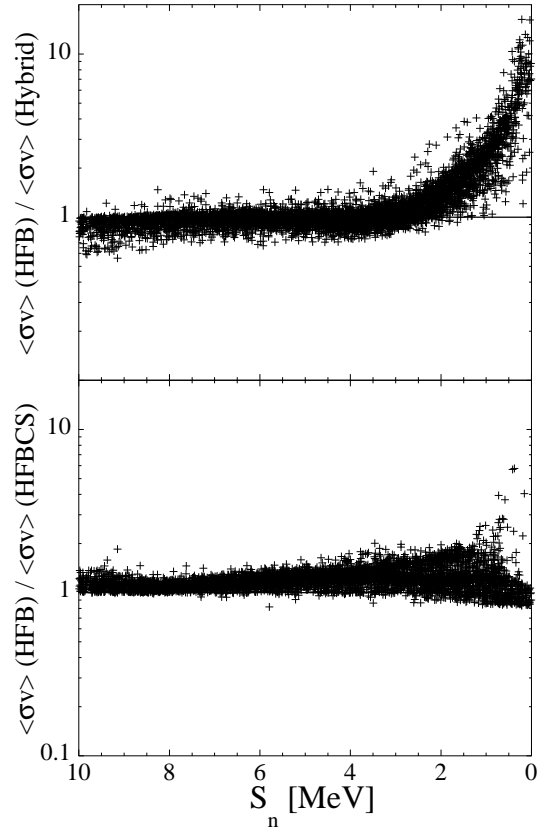}}
\caption{Upper panel: ratio of the Maxwellian-averaged (n,$\gamma$) 
rates (at a temperature of $1.5~10^9$~K) obtained
with the HFB+QRPA E1 strength to those calculated with the Hybrid
formula~\cite{go98}, as a function of the neutron separation
energy $S_n$ for all nuclei with $8\le Z \le 110$. 
Lower panel: the ratios of the 
HFB+QRPA neutron-capture rates and those calculated with the HF-BCS+QRPA 
model of Ref.~\cite{Gor.02}.} 
\label{fig_rate}
\end{figure}

The Maxwellian-averaged neutron-capture rates calculated 
with the HFB+QRPA E1 strength functions
are compared in Fig.~\ref{fig_rate} with those based on the Hybrid
phenomenological formula, for all nuclei with $8 \le Z \le 110$. 
The Hybrid E1 strength differs from the QRPA estimate in the
location of the centroid energy, as well as in the low-energy tail. 
Obviously the E1 strength obtained with QRPA enhances the 
capture rates by a factor up to 10 close to the neutron drip-line. 
For r-process nuclei characterized by neutron separation 
energies $S_n \lesssim 3$~MeV, neutron capture proceeds 
much faster than predicted by the phenomenological
Hybrid formula. This is due to the shift of the GDR to lower energies
as compared to the usually adopted liquid-drop $A^{-1/3}$ rule, as well
as to the appearance of dipole modes at low energies.
Both effects tend to enhance the E1 strength at energies below the
GDR, i.e in the energy window relevant for the neutron capture
process. For less exotic nuclei this effect is much smaller, and
the differences are mainly due to the predicted location of the GDR
and the strength of the low-energy tail. When compared to the HF-BCS+QRPA
results~\cite{Gor.02}, the HFB+QRPA model~\cite{Gor.04} predicts very
similar neutron-capture rates, even close to the neutron drip-line
(see the lower panel in Fig.~\ref{fig_rate}). This demonstrates the
consistency of the results obtained with various microscopic
approaches.

Many further improvements may be useful, but this will require intensive 
theoretical and computational advances. For instance, large-scale 
microscopic QRPA predictions of the E1 strength in deformed nuclei, 
the inclusion of the particle-vibration coupling effects on the
low-energy strength, an improved treatment of odd  and odd-odd nuclei, etc. 
The aim is to achieve a fully microscopic description
of the r-process based on an universal nuclear energy density functional.

\subsection{Nuclei as Ultra-High Energy Cosmic Rays}

\begin{figure}[htb]
\centering \includegraphics[scale=0.6]{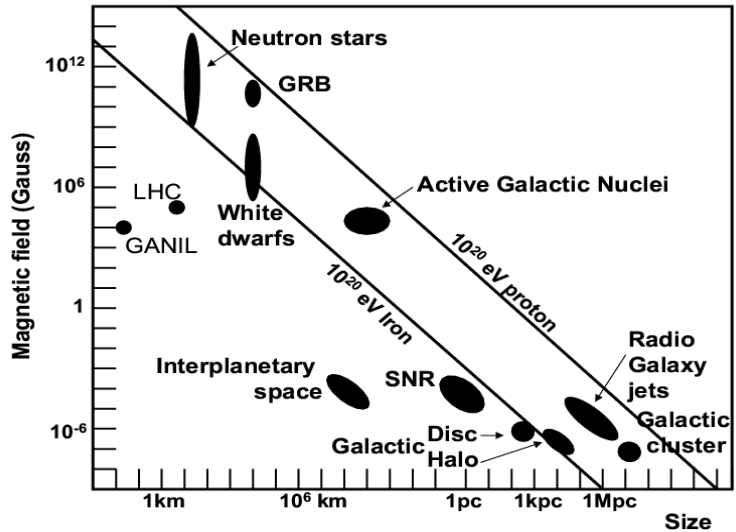}
\caption{Various accelerators in the Universe, located on the plot
of their magnetic field versus their size.}
\label{fig:acc}
\end{figure}

Cosmic rays are energetic particles that originate
in the Universe, with observed energies up to $\sim
3\,10^{20}$~eV~\cite{ay94,bi95}. Astrophysical sites able to accelerate
particles to such ultra high energies are currently under discussion. Among
them, violent processes related to neutron stars are possible candidates,
because they generate strong magnetic fields able to confine protons of
energy 10$^{20}$~eV. Figure \ref{fig:acc} illustrates various 
natural accelerators in the Universe, and displays their typical 
magnetic field B with respect to their curvature radius $\rho$.
The product B$\rho$ value gives the energy
which can be reached by the accelerated particles, from the well known
relation B$\rho$ = p/Q where p is the momentum of the particle, and Q its
charge. Sites which could possibly accelerate protons to 10$^{20}$~eV are
located in the shaded band.

The composition of Ultra High Energy Cosmic Rays (UHECR) is not known. 
According to two principal hypotheses, they are composed either only of
protons, or a mixture of protons and nuclei, ranging from hydrogen 
to iron. It has been
known for almost four decades that UHECR interact with the 2.7 K cosmic
microwave background radiation (CMB), leading for instance to a spectacular
decrease in their flux above energies around $10^{20}$~eV - the so-called
Greisen, Zatsepin and Kuzmin (GZK) suppression~\cite{gr66,za66}. The
following UHECR issues need to be addressed: the initial composition, the
acceleration mechanism, and the intergalactic propagation. Going through these
steps, the UHECR may reach an Earth based detection setup, such as the Auger
detector~\cite{ma06}. The first two issues include many unsolved questions 
about the composition and acceleration processes of UHECR. It is therefore
necessary to describe very accurately their propagation in order to provide
a test of the composition and acceleration scenarios by comparing
the predictions with data measured on Earth. 

\begin{figure}[htb]
\centering \includegraphics[scale=0.7]{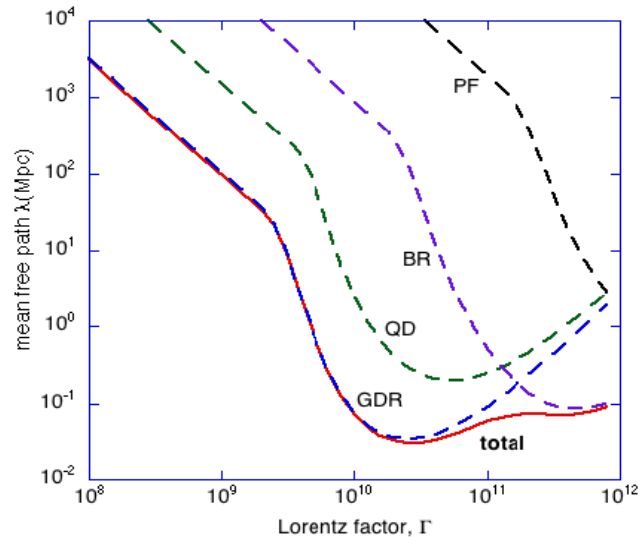}
\caption{Mean free path of  $^{56}$Fe nuclei against the various processes in
the CMB: Giant Dipole Resonance (GDR), Quasi-Deuteron process (QD),
Baryonic Resonance (BR) and Pion Photoproduction (PF)~\cite{al05}.}
\label{fig:nucproc}
\end{figure}

In the rest frame of a nucleus, at typical 
UHECR energies of 10$^{19}$-10$^{21}$
eV, the CMB photons are boosted to the energy range between a few
hundred keV, up to a few hundred MeV. The interaction process between
the UHECR and the CMB is dominated by the giant dipole resonance (GDR) at
photon energies below 30-50~MeV, and to a lesser extent by the quasideuteron
emission at intermediate energies (between 50 MeV and 150 MeV), and the pion
photoproduction at energies above 150 MeV \cite{pu76,ra96}. Figure
\ref{fig:nucproc} displays the mean free path of $^{56}$Fe nuclei for
various photonuclear processes, and shows the predominant role of the GDR.
Nuclei photodesintegrate by emitting nucleons through ($\gamma$,n),
($\gamma$,p), ($\gamma$,2n), ... reactions. It is therefore necessary to
accurately describe the dipole strength for the nuclei on the
photodisintegration path from Fe to protons. It should be noted that many 
nuclei along this path are unstable, for instance $^{44}$Ti.

In order to describe the changes in the abundance of heavy
nuclei as a result of the
interaction of the UHECR with the CMB, a nuclear reaction network that
includes all interactions of interest must be used.  The chosen set of nuclear
species are coupled by a system of differential equations corresponding to
all the reactions affecting each nucleus, i.e. mainly photodisintegrations
and $\beta$-decays~\cite{kh05}. All nuclei lighter than the seed nuclei and
located between the valley of stability and the proton drip-line must be
included in the network. Under the most natural astrophysical assumptions,
UHECR are accelerated out of the ambient gas, possibly enriched in Fe close
to neutron stars or depleted in metals (i.e. nuclei heavier than H) if
significant photodisintegration occurs during the acceleration stage itself.
Therefore, if nuclei are indeed present among the UHECR, it is expected
that they typically include the most abundant elements found in the
interstellar medium, i.e. essentially those lighter than Fe. The interaction of
UHECR with the CMB is thus expected to include all possible nuclei
resulting from the photodisintegration of the heaviest species and therefore
involve all stable and neutron-deficient unstable isotopes with A $\le$
56.

The UHECR photodisintegration was originally investigated by Puget, Stecker
and Bredekamp (PSB)~\cite{pu76,st99}. However, in the PSB model two major
approximations were employed to estimate the intergalactic UHECR
propagation. The first is related to the total photoabsorption cross section
which is parameterized as a simple Gaussian function~\cite{pu76,st99},
abruptly cut below the theoretical reaction threshold. The second is
based on the use of a reduced reaction network, including only one nuclide 
for each value of A, to estimate the time evolution of the UHECR
composition. More precisely, assuming that the $\beta$-decay of the unstable
nuclei produced by photodisintegration is always faster than the
corresponding photoemission rate, a unique nuclear path is followed from the
initial $^{56}$Fe source to the final protons~\cite{pu76}, neglecting the
contribution of unstable nuclei. 
It is necessary to achieve an accurate description of the
photodisintegration rates using photoreactions that are extensively 
studied in nucleosynthesis, where phenomenological parameterizations of the
photoabsorption cross sections have been optimized during the last decades,
and where large-scale microscopic predictions have also become 
available, as described in section \ref{ssect_rprocess}~\cite{Gor.02,Gor.04}. 
New compilations of experimental
photoabsorption data also help to determine the degree of accuracy with
which the present reaction models predict the corresponding cross sections.
Important progress has also been made in the field of
nucleosynthesis by solving large reaction networks on the nuclear chart
and thus following the time evolution of the composition of the
material at given astrophysical sites. Similar tools can therefore be used
in the field of UHECR in order to take into account the contribution of
unstable nuclei during the photodisintegration path.

The total photon transmission coefficient characterizing the probability to
populate by photoabsorption a compound nucleus excited state is obviously one
of the key ingredients for the evaluation of the photoreaction rates. In the
specific astrophysical conditions considered for UHECR energies of
$10^{19-21}$~eV, this function is dominated by the $E1$ transition. To
estimate the accuracy of the different methods available for the
evaluation of an $E1$-strength function, four models have been considered:
the Lorentzian~\cite{ax62}, the generalized Lorentzian~\cite{ko90},
the HF-BCS+QRPA~\cite{Gor.02}, and the HFB+QRPA~\cite{Gor.04}. 
The former two are
phenomenological, and the latter two are microscopic.

The photoreaction cross sections are estimated with the Talys nuclear
reaction code~\cite{ko03,ko04}, which takes into account all types of direct,
pre-equilibrium and compound mechanism to estimate the total reaction
probability, as well as the competition between the various open channels.
The quasideuteron process is neglected because of the limited photon energy
range~\cite{ra96}. The predictions are compared with available experimental
data~\cite{ia00} for nuclei with A$\le$56. It should be noted that, even
for stable nuclei, the data on such nuclei are scarce. For instance, total
photoabsorption cross sections around the GDR peak energy are available for
only 10 nuclei~\cite{ia00}. An extensive study has been performed to
compare the predictions with the data~\cite{kh05}. Both the microscopic and
the Lorentzian approaches correctly describe the data, because
the nuclei involved lie close to the valley of stability. 

\begin{figure}[htb]
\centering \includegraphics[width=15cm]{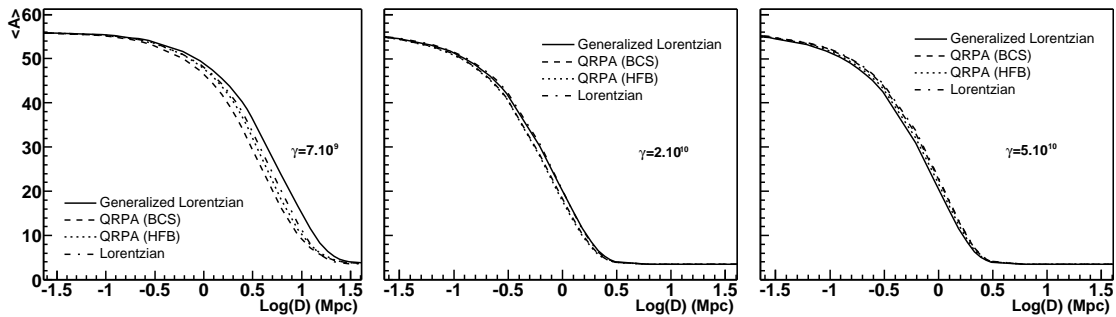}
\caption{Evolution of the average mass number $<A>$ with
respect to the distance from the $^{56}$Fe source for three different energies.
Left: $E=3.6\times 10^{20}$ eV; center: $E=10^{21}$eV; right:
$E=2.61\times 10^{21}$eV~\cite{kh05}.}
\label{fig:prop}
\end{figure}

The intergalactic UHECR propagation is calculated considering the
interaction with the CMB. For illustrative purposes, we will restrict
ourselves in a first step to study the propagation of a UHECR source made of
$^{56}$Fe only. All stable and neutron-deficient unstable nuclei with
A$\le$56 are included in the reaction network.  Fig.
\ref{fig:prop} shows the evolution of the average mass number $<A>$ as a
function of the distance from the $^{56}$Fe source, calculated with the four
GDR prescriptions. For a given source distance, $<A>$ is the average value
of the calculated nuclei abundances. The full reaction network is solved at
each time-step, taking into account all open photoemission channels, i.e
($\gamma$,n), ($\gamma$,p), ($\gamma$,$\alpha$), ($\gamma$,2n),
($\gamma$,2p), ($\gamma$,2$\alpha$), ($\gamma$,np), ($\gamma$,n$\alpha$),
($\gamma$,p$\alpha$). In other words, the abundance of each nuclide
is derived by taking into account the contribution of all production
channels, from the source nucleus downwards the table of nuclides, with the
appropriate weight determined according to the corresponding cross sections.
The values obtained are thus equilibrium values, representing the
composition which would result from the propagation of an infinite number of
nuclei up to the time considered. 

To illustrate the role of the soft dipole mode, let us consider a $^{56}$Fe
nucleus with an energy of $3.6\times 10^{20}$ eV. In this regime
only the lowest energy part of the $E1$-strength overlaps with the photon
density n($\epsilon$). The distance of propagation is mainly sensitive to
the low-energy component of the $E1$-strength function, and using
different prescriptions leads to significant differences in the propagation
distance. This emphasizes the necessity accurately describe 
the low energy tail of the E1 strength, even for nuclei close to the valley
of stability. In contrast, results for $^{56}$Fe at higher energy mainly 
depend on the location of the GDR peak or integrated photoabsorption, 
and for this reason the propagation distance is less sensitive to the 
photoreaction details. 

\begin{figure}[htb]
\centering \includegraphics[width=14cm]{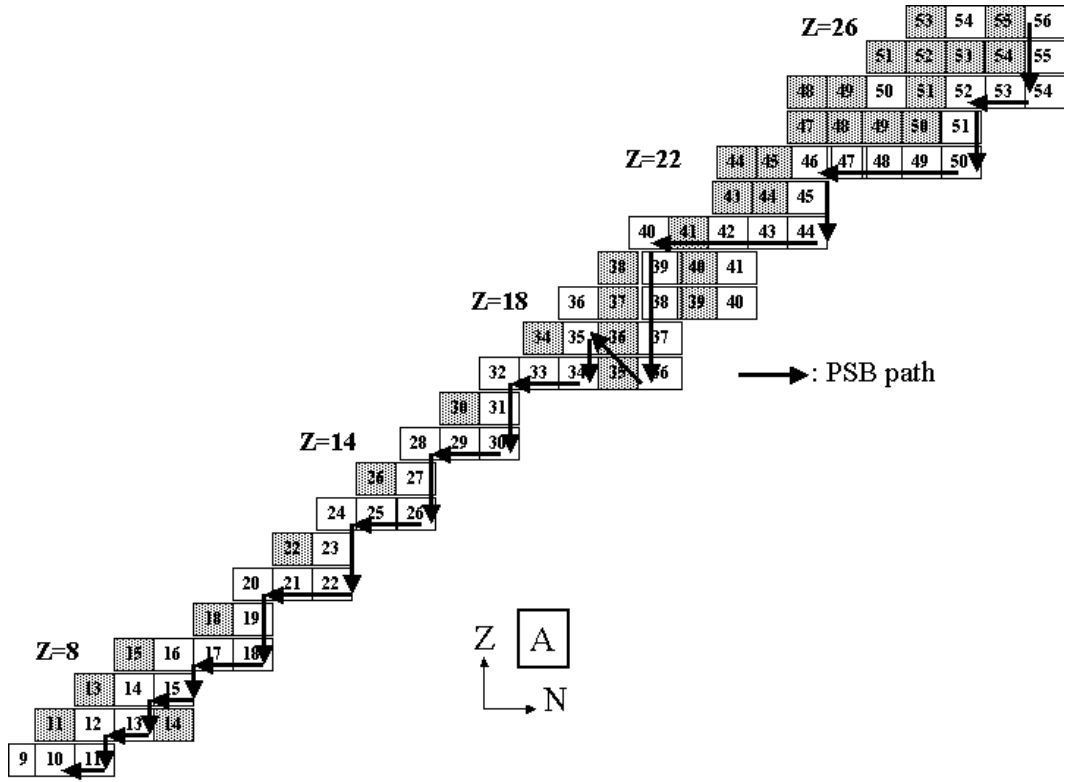}
\caption{Nuclei involved in the
photodisintegration process of $^{56}$Fe at $E = 10^{21}$eV.
Unstable nuclei are denoted by shaded squares, and the PSB path 
is indicated by arrows. The mass number A of each nucleus is 
written in the corresponding square~\cite{kh05}.}
\label{fig:chart}
\end{figure}

The UHECR propagation distance has been estimated using the complete
reaction network. The initial PSB calculations were based on the reduced PSB
path illustrated in Fig.~\ref{fig:chart}. In this approximation, as
explained above, only one stable isotope is considered for an isobaric chain
and the corresponding isobars are not affected
by competitive channels. However, as shown in Fig.~\ref{fig:chart}, about 85
nuclei are involved in the $^{56}$Fe photodisintegration at $E = 10^{21}$ eV,
and a number open channels, including $\beta$-decay, can compete (the Lorentz
dilation of time allows $\beta$-unstable nuclei with half-lives of the order
of an hour to survive over a Mpc scale, and thus have a chance to interact
with a CMB photon). Most of the stable nuclei involved in the
photodisintegration process have more neutrons than protons. Neutron
photo-emissions are therefore favored and the corresponding unstable nuclei will
$\beta^+$-decay towards the valley of stability. 
\bigskip

Significant discrepancies are therefore expected between the recent
calculation in Ref.~\cite{kh05}, and the original PSB results based on the
reduced path and the Gaussian parameterization of GDR strengths. In
particular, as seen in Fig.~\ref{fig:chart} for A$\ge$45,
about 70\% of the nuclei are shortcut by the simplified PSB path. For
$E=3.6\times 10^{20}$eV, a significant difference is found between the
PSB predictions and the one using an accurate E1 strength description, such
as the QRPA or the generalized Lorentzian. This effect is due to the prediction
of the low-energy part of the dipole strength, which in the
PSB case does not agree with the data. 

\begin{figure}[htb]
\centering \includegraphics[scale=0.6]{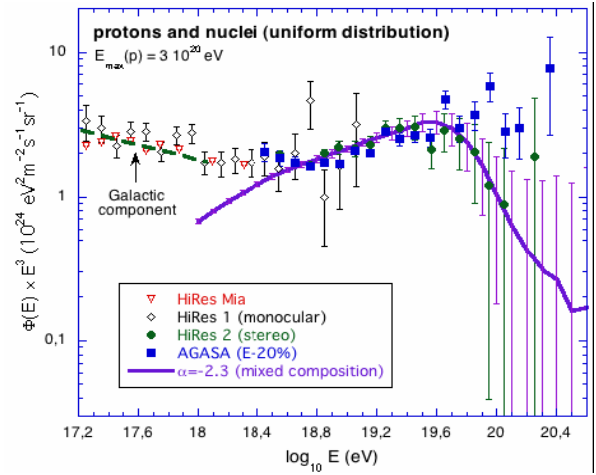}
\caption{Propagated spectra from a source composed of protons and nuclei
(solid line), compared with available
data \cite{al05}.}
\label{fig:spec}
\end{figure}

A full propagation calculation based on a Monte-Carlo simulation, has 
been performed in Ref.~\cite{al05}. A mixed source of protons and 
nuclei with A$<$56 has been considered, as well as
quasideuteron, pion photoproduction, and pair-production processes.
As shown in Fig.~\ref{fig:spec}, the propagated spectra are in good 
agreement with the cosmic ray data.
The effect of the prediction of the dipole strength compared to the PSB
parameterization is non-negligible, and the sensitivity to the 
description of the E1 strength will become even more important, because the
data provided by the Auger detector will drastically reduce the statistical
error bars at high energy.

In Ref.~\cite{al05} it has also been shown that a mixed composition of protons
and nuclei provides a reasonable interpretation of the high energy part of 
the CR spectrum around the so-called ankle area. Better agreement with
observed data is obtained with composite source models, than with 
just a uniform proton source. Moreover, the work of Ref.~\cite{al06} has
shown that by including nuclei in UHECR a much better agreement with
complementary experimental observables is obtained.
All these studies support the hypothesis that UHECR may 
be composed of nuclei.

\subsection{Supergiant Resonances in the Inner Crust of Neutron Stars}
\label{supergiant}

Some of the most exotic nuclear excitations could arise in the
crust of neutron stars. In microscopic calculations the inner 
crust matter is usually described in the Wigner-Seitz (WS) 
approximation~\cite{pe95,ne73}, i.e. the inner crust is modeled 
by non-interacting cells that contain a neutron-rich nucleus 
immersed in a dilute gas of neutrons and relativistic
electrons. For baryonic densities in the range from 1.4 10$^{-3} \rho_0$ 
about 0.5$\rho_0$, where $\rho_0$=0.16 fm$^{-3}$ is the nuclear matter
saturation density, nuclear clusters can be considered
spherical~\cite{ne73,do00}. At higher densities, the inner crust 
matter can develop various non-spherical phases 
(e.g. rods, slabs, tubes, bubbles)~\cite{pe95}.
The structure of the WS cells has been analyzed as a function
of their density~\cite{ne73}, which is related to the distance of
the WS cell from the center of the neutron star. The
equilibrium condition for the WS cells has been derived, 
with the values of Z and N that minimize the energy of the system. 
The resulting values of Z are typically
located between 10 and 50, depending on the density of the WS cell. The 
corresponding number of neutrons N is about several hundreds. 
Therefore the cell with Z=50 and N=1750 is quoted as $^{1800}$Sn. 
In this cell the density of the neutron
gas far from the nuclear cluster is $\approx 0.018$ fm$^{-3}$, and the radius
of the cell is 27.6 fm. The WS cell can therefore be
represented as a drip-line nucleus immersed in a neutron gas, and 
considered as a system between a finite nucleus and the uniform neutron
matter. Nuclear collective modes are expected to develop in such systems, 
and several interesting questions arise: What is the structure of collective
states in these extremely neutron-rich systems? What is the relationship 
between the nuclear cluster and the neutron gas? Can the study of 
very neutron-rich nuclei be helpful in understanding these excitations?

The cooling of low-mass neutron stars is strongly influenced by the
superfluid properties of inner crust matter~\cite{yako04}. These
properties and their effect on the specific heat have been analyzed 
in various theoretical frameworks, e.g. semiclassical pairing 
models~\cite{br94,el96}, Bogoliubov-type calculations
based on a Woods-Saxon mean-field~\cite{ba98,pi02}, and the 
self-consistent Hartree-Fock-Bogoliubov (HFB) approach~\cite{sa04,sa04b}. 
These calculations have shown that pairing correlations can 
reduce by orders of magnitude the specific heat of baryonic 
matter in the inner crust. It must be emphasized, however, that 
the specific heat was evaluated for a system of 
non-interacting quasiparticles. This quantity
can also be strongly affected by collective modes determined by the
residual interaction between quasiparticles, especially if these modes
appear at low-excitation energy. This effect has been studied
in Ref.~\cite{gori04,kh05b}. It should be noted that the specific heat 
of the inner crust is also determined by the motion of electrons and,
to a lesser extent, by lattice vibrations~\cite{pe95,yako04,pi02,ma04}. 
These degrees of freedom of the inner crust matter will not be 
taken into account in the present discussion. 
 
\begin{figure}[htb]
\centering \includegraphics[scale=0.48]{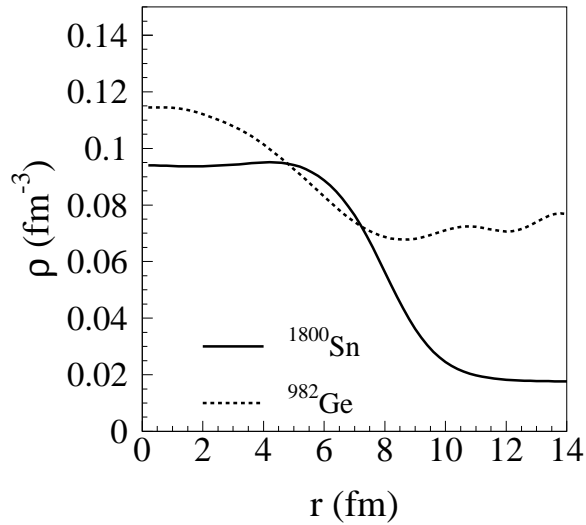}
\caption{Particle densities for the WS cells $^{1800}$Sn and  $^{950}$Ge, 
calculated with the HFB model \cite{kh05b}.}
\label{fig:wsdens}
\end{figure}

In Ref.~\cite{gori04} RPA calculations were performed for $^{580}$Sn,
and a pronounced low-lying quadrupole peak was obtained. 
In order to take into account pairing effects,
it is convenient to calculate the collective response with the HFB+QRPA 
model formulated in the coordinate representation~\cite{kh02}. 
This representation is particularly suited to describe systems with a
large number of quasiparticle states, such as WS cells.
In the first step of the calculation the HFB
equations for the ground-state of the given WS cell are solved,
considered as an isolated system. The HFB calculations are performed in
the coordinate representation and the Dirichlet-Neumann boundary conditions
are imposed at the contour of the cell~\cite{ne73}.
These are the only discretization conditions which can produce a 
constant density around the outer boundary of the WS cell. 
Figure \ref{fig:wsdens} displays the HFB results
for the particle densities of the $^{1800}$Sn and $^{982}$Ge WS
cells~\cite{kh05b}. The density profiles are very diffuse, because of
the presence of the neutron skin on the surface of the nuclear cluster
($\simeq$ 7 fm). At larger radii the densities remain constant and 
correspond to the neutron gas component. 

The response of several cells has been calculated within the HFB+QRPA model,
e.g. $^{1500}$Zr, $^{1800}$Sn and $^{950}$Ge \cite{kh05b}.
In these cells a very collective low-lying state has been predicted,
located in the energy region between 2 and 4 MeV. This 
SuperGiant Resonance (SGR) typically exhausts around 70 \% of the
EWSR \cite{kh05b}. The analysis of the SGR structure
has shown that the contributions of the nuclear cluster and
the neutron gas to the neutron transition density are comparable.
However, the magnitude of the strength (but
not the energy position) is mainly determined by the neutron gas,
because it is given by $\int dr~r^{L+2} \delta \rho(r)$ for a given 
multipole L (cf. Eq~(\ref{eq:td1})), 
and this strongly favors effects located at large radii. 

\begin{figure}[htb]
\centering \includegraphics[width=5.1cm]{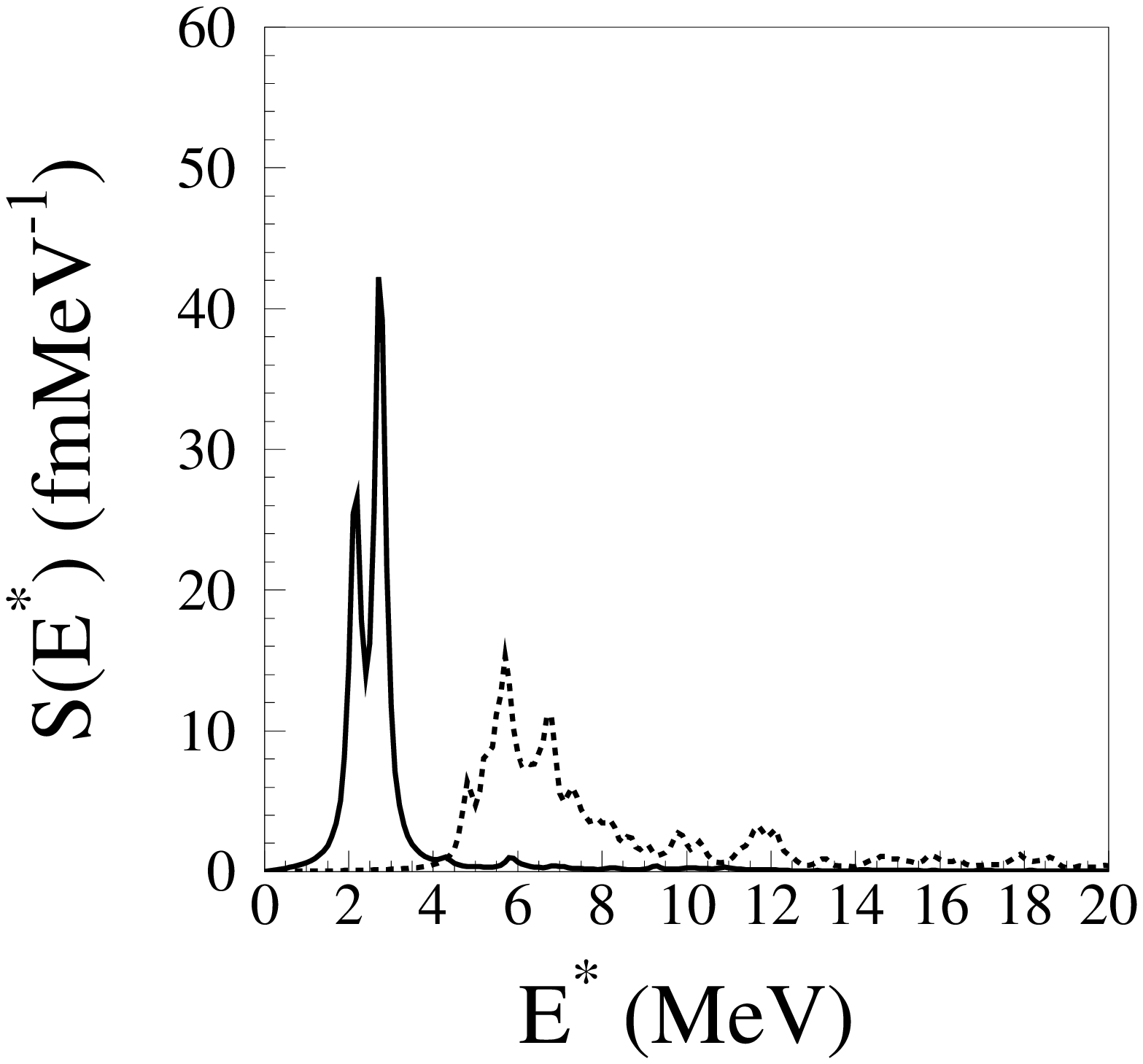}
\includegraphics[width=5.1cm]{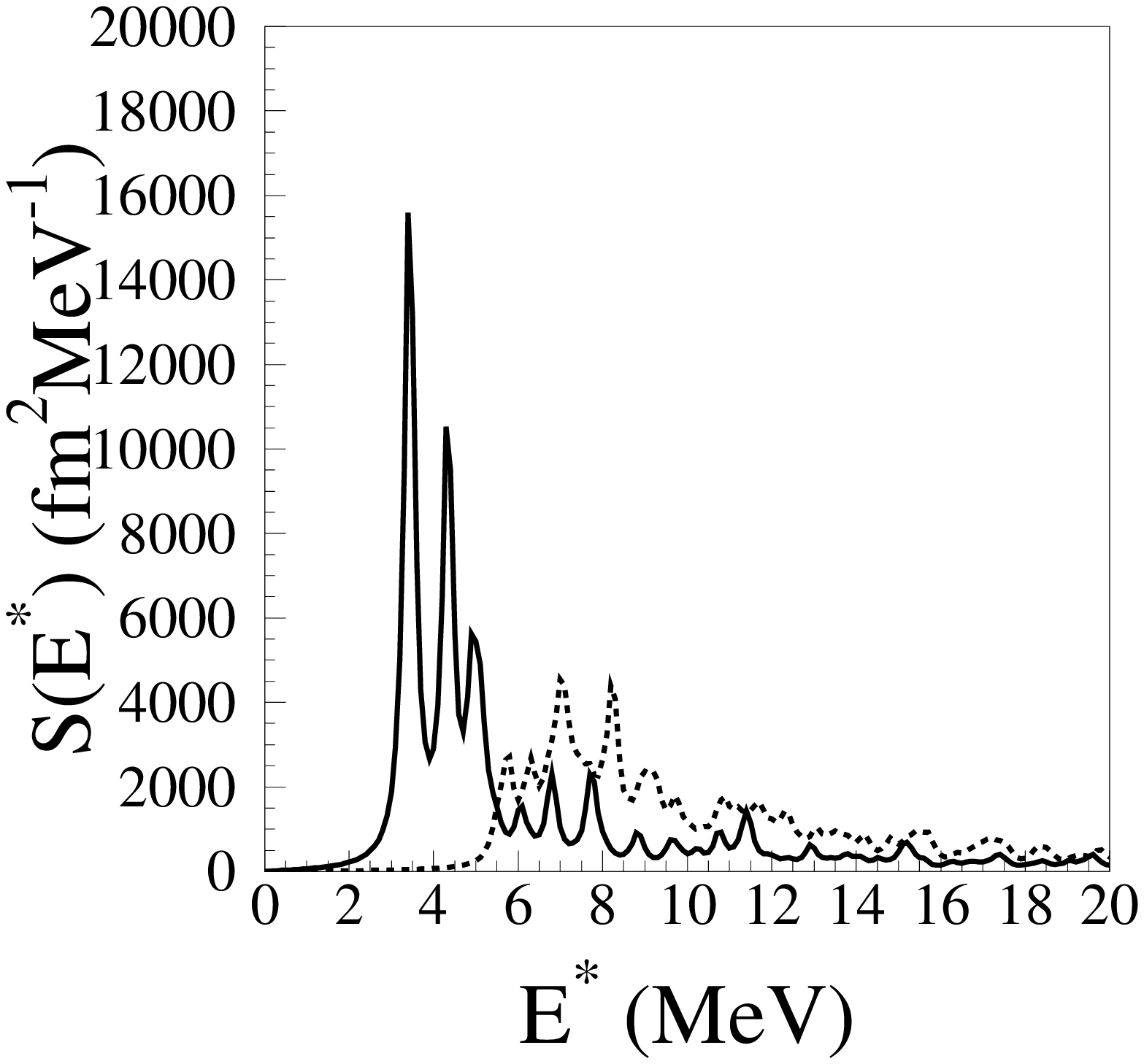}
\includegraphics[width=5.1cm]{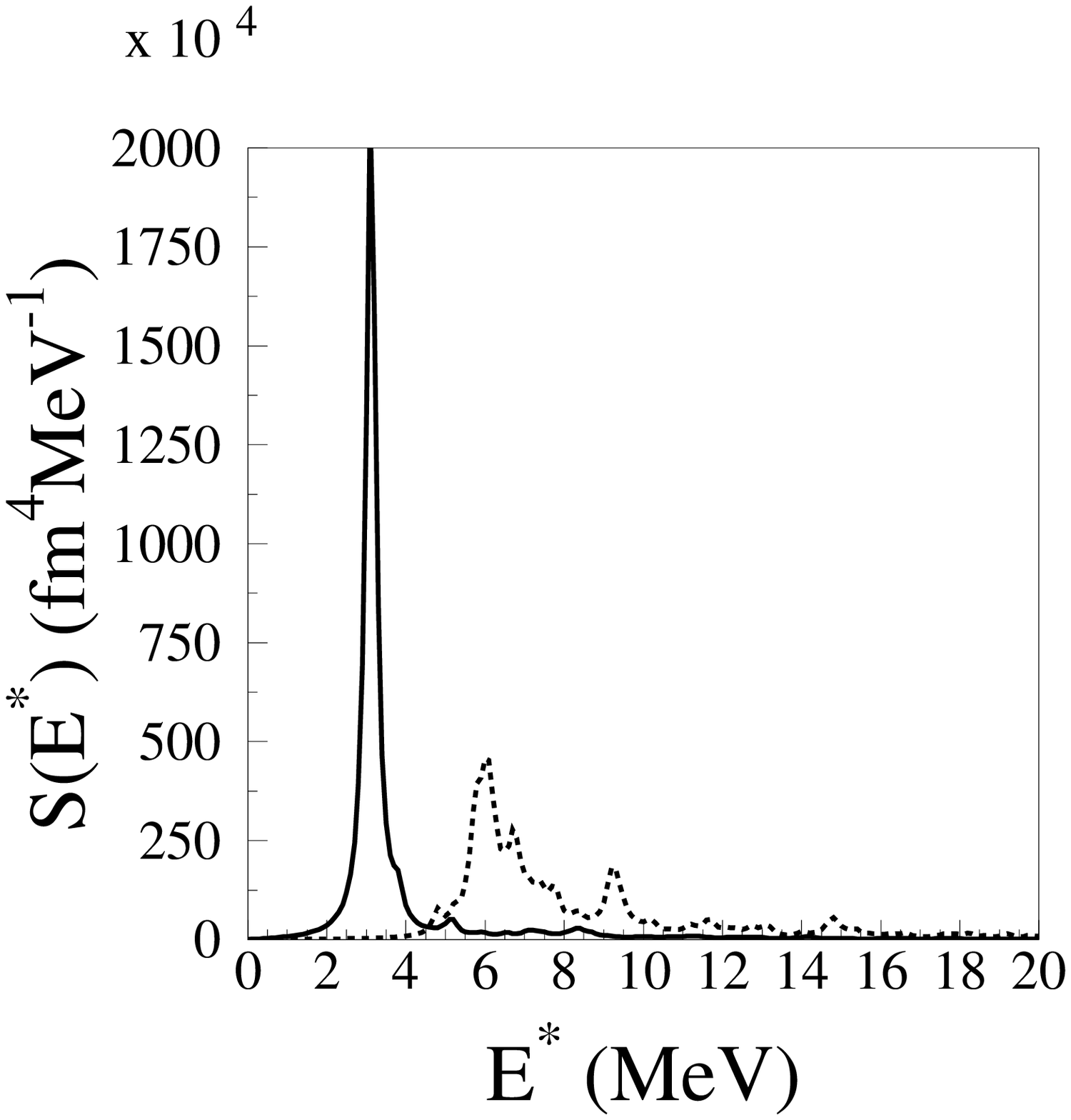}
\caption{Monopole (left), dipole (center) and quadrupole (right) strength
distributions for the cell $^{1800}$Sn. The solid and dashed curves
represent the QRPA strength and the HFB unperturbed strength, respectively.}
\label{fig:sgr}
\end{figure}

The monopole, dipole, and quadrupole neutron response of the 
cell $^{1800}$Sn are shown in Fig.~\ref{fig:sgr}. We display both 
the unperturbed HFB response built from non-interacting quasiparticle 
states, and the QRPA response function. When the residual interaction 
between quasiparticles is turned on, the unperturbed spectrum, 
which is distributed over a large energy region, is collected in a 
strong peak located at about 3 MeV. All multipolarities exhibit the SGR, 
and in the case of the quadrupole response the SGR peak collects more than
99$\%$ of the total quadrupole strength. This mode is extremely collective,
there are more than one hundred two-quasiparticle configurations
contributing to the QRPA amplitude, and its reduced transition probability 
B(E2) $\approx 25 \times 10^3$ Weisskopf units. This value is two
orders of magnitude larger than the B(E2)'s found in ordinary nuclei. 
It should be noted that the extrapolation of the energy position based 
on the Giant Quadrupole Resonance systematics in finite nuclei, 
i.e. 65A$^{-1/3}$ MeV~\cite{harak01}, predicts the low-energy peak at about 5 MeV.
The additional lowering of the peak, which is due to the 
comparable contributions of the nuclear cluster states and
the neutron gas, shows that the WS cells 
cannot be simply considered as giant nuclei.

In order to study collective excitations at higher
baryonic densities of the inner crust matter, the response of the cell
$^{982}$Ge has also been calculated in Ref.~\cite{kh05b}. 
In this cell the cluster and the neutron gas are less separated than 
in the case of  $^{1800}$Sn (see Fig. \ref{fig:wsdens}). 
Because of the large degeneracy of states located
close to the Fermi level, many particle-hole configurations add coherently to
form the SGR. It is also interesting to investigate whether the contribution
of the free neutron gas to the collective mode could be eventually
described in a semiclassical picture by using the hydrodynamic model. 
This requires a coherence length that is much smaller than the size of the
system~\cite{ge89}, and is only fulfilled for the most external WS cells, 
such as $^{1800}$Sn. For the WS cells of higher density, 
such as $^{1500}$Zr and $^{982}$Ge, the hydrodynamic model 
cannot reproduce the energy position of the SGR~\cite{kh05b}.

The predicted energies of the SGR are of the same order of magnitude as the
average pairing gap of the neutron superfluid and, consequently these 
modes could have a significant effect on the entropy and the 
specific heat of baryonic matter in the inner crust. A quantitative 
estimate of these effects will require finite-temperature HFB+QRPA 
calculations \cite{kh04b}. There are, however, open questions about 
the validity of the non-interacting WS approximation in the 
description of collective modes. Their amplitude is strongly sensitive 
to the surface, where the neutron gas connects two contiguous
cells. Tests have been performed by modifying the Dirichlet-Neumann boundary
conditions, but no pronounced deviation of the strength has been found.
The role of the coupling between two contiguous WS cells through the neutron
gas should be investigated. The cells which have an impact on the
specific heat remain also to be determined and, finally, 
neutron-rich nuclei which will be produced by
the next-generation radiocative-beam facilities,
could provide data that will constrain models and
functionals which describe WS cells. 

\section{Concluding Remarks and Outlook}\label{Sec8}

Experimental studies with radioactive isotope beams have disclosed 
a wealth of structure phenomena in nuclei far from the line of 
$\beta$-stability. Among the most interesting results 
are those on the evolution of collective modes of excitation in 
regions of unstable nuclei, and the possible existence of entirely 
new types of excitation in weakly-bound systems. The best studied 
example is the evolvement of the dipole response in neutron-rich 
nuclei, and the possible occurrence of the pygmy dipole resonance: 
the resonant oscillation of the weakly-bound neutron skin against 
the isospin saturated proton-neutron core. However, even 
in this case the available data on the low-energy dipole 
response do not discriminate between various theoretical 
interpretations. In nuclei near the neutron drip-line new 
types of excitation could be induced by the strong surface pairing
properties as, for instance, the soft dipole pairing mode.
On the proton-rich side, even if the protons do not develop 
a pronounced skin structure, in relatively light nuclei close
to the proton drip-line the weakly-bound proton orbitals, 
including states which are bound only because of the presence 
of the Coulomb barrier, could contribute to the evolution of 
the proton pygmy dipole resonance. The existence of exotic 
isoscalar modes has also been predicted: a low-lying dipole mode 
which could correspond to the toroidal dipole resonance, giant 
pairing vibrations, quadrupole excitations and higher multipoles.
The increasing interest in exotic collective modes  
has led to important advances both in the experimental methods 
and theoretical tools that are used in the study and interpretation
of these phenomena.

A modern microscopic time-dependent mean-field theory of collective 
excitations in nuclei far from stability necessitates a fully 
self-consistent implementation of effective 
nuclear interactions: both the equations which determine 
the ground state of a nucleus, and the residual interaction 
which governs the small-amplitude vibrations around the 
equilibrium, must be derived from the same effective interactions 
in the particle-hole and particle-particle channels.
Only fully self-consistent calculations ensure the
separation of spurious states from the physical excitations, 
and provide reliable microscopic predictions of collective 
modes directly based on ground-state properties. 
The presently available models which meet these conditions are:
the HFB plus QRPA based on Skyrme or Gogny effective interactions, 
and in the relativistic framework the RHB plus RQRPA based on 
effective Lagrangians with density-dependent meson-nucleon couplings.
Even though some of these state-of-the-art models include the 
coupling to the particle continuum, i.e. they take into account 
the escape width of vibrational states that lie above the particle 
emission threshold, one really must go beyond the 
mean-field approximation in order to quantitatively describe 
decay properties of collective excitations. At present only few 
theoretical approaches take into account in a 
fully consistent way the coupling of the simple $1p - 1h$ 
(or two-quasiparticle) states to more complex $2p - 2h$ 
configurations, which basically determines the spreading width 
of resonances. The way the damping mechanism is modified in 
nuclei far from stability, e.g. neutron-rich nuclei, is 
therefore largely unexplored.

Studies of low-energy collective excitations in nuclei with
relatively moderate neutron excess provide crucial
information on the manifestation of exotic modes
in neutron-rich nuclei, and their subsequent 
evolution towards systems with large isospin asymmetry near 
the nucleon drip-lines. Evidence for the existence of 
low-lying dipole strength in neutron-rich nuclei, which
might indeed correspond to the pygmy dipole resonance (PDR), 
has become available from several experiments
based on $(\gamma,\gamma')$ resonant scattering, and 
data on low-lying E1 strength in exotic nuclei has 
recently been obtained in studies with radioactive ion beams.
Experimentally, however, very little is known about the nature
of the observed low-lying dipole states, the degree of 
collectivity and the isospin character, and 
extensive investigations have to be carried out. For instance,
studies of low-energy transition strength below the particle 
emission threshold in $(\alpha,\alpha'\gamma)$ 
coincidence experiments, which enable a clear separation
of E1 excitations from  states of other multipolarities,
will allow a more detailed analysis of the structure of 
these states \cite{Sav.06}.

A number of theoretical studies have recently been devoted to
properties of the PDR, and some models predict an enhancement 
of collectivity for these states, and pronounced mixing 
of isoscalar and isovector components. An important issue 
in neutron-rich nuclei is also the relationship between the 
PDR excitation energy and the neutron separation threshold. 
It appears that in nuclei not so far from stability 
the PDR is located below or very close to the neutron 
separation energy, whereas in weakly-bound systems with  
large neutron excess the PDR energy is high above the neutron threshold. 
Only part of the low-lying E1 strength is observed in current experiments.
While the $(\gamma,\gamma')$ spectra are composed of peaks below the 
threshold energy, the Coulomb excitation of fission fragments gives 
the transition strength above the neutron separation energy. 
Future experiments therefore need to provide complete low-lying 
dipole spectra, both below and above the neutron threshold energy. 
A comparison of the complete spectra of measured low-lying
strength in nuclei far from stability with the predictions of 
self-consistent microscopic theories, will present a very
sensitive test for the isovector channel of nuclear effective
interactions. Namely, it is difficult to adjust the isovector 
terms of effective interactions only to data in stable nuclei.  
Stringent constraints on the microscopic approach to nuclear dynamics 
and effective nuclear interactions will emerge from studies of 
the structure and stability of exotic nuclei with extreme isospin 
values.

Except for the PDR, few data have been reported so far on other 
possible exotic modes of excitation, but dedicated experiments 
are being planned and designed at radioactive-beam laboratories.
For instance, in the near future more information could become 
available on modes which arise in nuclei near the drip-lines: 
the proton PDR, and the di-neutron {\it vs} core oscillations.
An important question which has not been addressed so far 
is the evolution of low-energy modes with temperature in hot nuclei, 
a topic that could be important for astrophysical applications. 
The dipole toroidal mode could be probed by the measurement of 
transverse electron scattering form factors at 180$^{\circ}$. 
An entirely unexplored field is the evolution of modes that involve 
the spin and/or isospin degrees of freedom in nuclei far from stability. 
It must also be emphasized that while most theoretical studies of 
exotic modes of excitation have assumed spherical symmetry, the 
evolution of deformation in unstable neutron-rich nuclei could give 
rise to interesting collective phenomena such as, for instance, 
the low-energy scissors mode of oscillation of the neutron 
skin \cite{War.97}.

Of particular importance is the role that new exotic modes or,
more generally collective excitations in nuclei far from stability, 
play in astrophysical processes: the description of Gamow-Teller
resonances is essential in calculations of $\beta$-decay, electron
capture and neutrino-nucleus interaction rates; the low-energy
dipole transition strength in neutron-rich nuclei has a pronounced 
effect on the calculated r-process abundances
and on the propagation of ultra-high energy 
cosmic rays; on the proton-rich 
side the proton pygmy dipole resonance could contribute to the 
nucleosynthesis in rapid proton capture processes, as well as 
in the two-proton capture in astrophysical conditions 
characteristic for explosive hydrogen burning in novae and x-ray 
bursts \cite{Gri.06}. Theoretical studies of electron capture rates 
on neutron-rich nuclei, at temperatures and densities characteristic 
for core collapse, have recently shown that these rates can be so large 
that electron capture on nuclei dominates over capture on free protons. 
Further studies of capture rates, particularly in a self-consistent 
approach extended to finite temperatures, are clearly desirable.
The effect of giant resonances in extremely neutron-rich systems 
on the cooling time of neutron stars should also be investigated.
Neutrino-nucleus reactions in the low-energy range 1-100 
MeV play an important role in many astrophysical processes, 
including stellar nucleosynthesis, and the study of 
low-energy neutrino reactions on medium-heavy and heavy nuclei is of 
great current interest. Since the description of a neutrino-nucleus 
reaction becomes increasingly complicated as the target mass number 
increases, accurate self-consistent mean-field approaches must 
be developed and applied in calculations of cross sections for 
all relevant neutrino-induced reactions. More generally, 
microscopic nuclear structure theory must be integrated 
into various astrophysical models of nucleosynthesis processes, 
supernova dynamics, and neutrino-induced reactions, by 
providing accurate global predictions for bulk nuclear properties 
and nuclear excitations. 

\bigskip
\leftline{\bf ACKNOWLEDGMENTS}
Many students and colleagues have contributed to the work reviewed in
this article. In particular, we would like to thank 
D. Allard, T. Aumann, 
F. Barranco, K. Bennaceur, K. Boretzky, P. F. Bortignon, R. A. Broglia,
H. Emling, P. Finelli, S. Fracasso, F. Ghielmetti, Nguyen Van Giai, G. Giambrone,  
S. Goriely, M. Grasso, I. Hamamoto, 
D. T. Khoa, A. Klimkiewicz, G. A. Lalazissis,
Zhong-yu Ma, J. Margueron, M. Matsuo,  J. Meyer,
P. von Neumann-Cosel, T. Nik\v si\' c, B. {\" O}zel, 
P. Papakonstantinou, V. Yu. Ponomarev, M. R. Quaglia, 
F. Ramponi, P. Ring, R. Roth,  H. Sagawa, N. Sandulescu, D. Sarchi, 
H. Utsunomiya, E. Vigezzi, A. Vitturi, and S. Volz.
This work has been supported in part by the Croatian
Ministry of Science and Education under project 1191005-1010, 
by the Deutsche
Forschungsgemeinschaft (DFG) under contract SFB 634, 
and by the INFN.

\end{document}